\documentclass[a4paper,11pt]{scrbook}

\usepackage[utf8]{inputenc}
\usepackage[T1]{fontenc}
\usepackage[portuguese]{babel}
\usepackage{geometry}
\usepackage{microtype}
\usepackage{hyperref}
\usepackage{parskip}
\usepackage{graphicx}
\usepackage{amsmath, amssymb}
\usepackage{longtable, booktabs}
\usepackage{upgreek}   
\usepackage{algorithm, algpseudocode}
\usepackage[acronym]{glossaries}
\makeglossaries

\usepackage{tikz}
\usetikzlibrary{positioning,fit,arrows.meta,backgrounds,calc}
\usepackage{physics}

\title{Fundamentos, Avanços Recentes e Desafios sobre Algoritmos Criptográficos para a Era da Computação Quântica}
\author{Darlan Noetzold and Valderi Reis Quietinho Leithardt}
\date{}  

\addto\captionsportuguese{}

\newglossaryentry{pqc}{
    name={PQC},
    description={Post-Quantum Cryptography — criptografia pós-quântica; família de algoritmos resistentes a ataques por computadores quânticos, executáveis em hardware clássico}
}

\newglossaryentry{qkd}{
    name={QKD},
    description={Quantum Key Distribution — distribuição quântica de chaves; protocolos que geram uma chave secreta compartilhada com segurança fundamentada nas leis da mecânica quântica}
}

\newglossaryentry{nist}{
    name={NIST},
    description={National Institute of Standards and Technology — instituto dos EUA que conduz a padronização de algoritmos de criptografia, incluindo o processo de PQC}
}

\newglossaryentry{rsa}{
    name={RSA},
    description={Rivest–Shamir–Adleman — sistema de chave pública baseado na fatoração de inteiros grandes; usado para assinatura e encapsulamento com esquemas seguros de padding}
}

\newglossaryentry{ecc}{
    name={ECC},
    description={Elliptic Curve Cryptography — criptografia de curva elíptica; baseia-se no problema do logaritmo discreto em curvas elípticas (ECDLP)}
}

\newglossaryentry{aes}{
    name={AES},
    description={Advanced Encryption Standard — cifra de bloco simétrica padrão (Rijndael), com chaves de 128/192/256 bits}
}

\newglossaryentry{chacha20}{
    name={ChaCha20},
    description={Cifra de fluxo ARX (adição, rotação, XOR), eficiente em software; frequentemente usada com o MAC Poly1305 em AEAD}
}

\newglossaryentry{poly1305}{
    name={Poly1305},
    description={MAC do tipo Carter–Wegman; quando combinado com ChaCha20 compõe o AEAD ChaCha20-Poly1305}
}

\newglossaryentry{gcm}{
    name={GCM},
    description={Galois/Counter Mode — modo AEAD que combina CTR com autenticação GHASH em $ \mathbb{F}_{2^{128}} $; requer unicidade de nonce por chave}
}

\newglossaryentry{aead}{
    name={AEAD},
    description={Authenticated Encryption with Associated Data — cifra autenticada com dados associados que garante confidencialidade e integridade}
}

\newglossaryentry{ghash}{
    name={GHASH},
    description={Função de autenticação polinomial usada no AES-GCM sobre o campo $ \mathbb{F}_{2^{128}} $}
}

\newglossaryentry{hkdf}{
    name={HKDF},
    description={HMAC-based Key Derivation Function — KDF em duas fases (extract/expand) que suporta separação de domínios via rótulos de contexto}
}

\newglossaryentry{hmac}{
    name={HMAC},
    description={Hash-based Message Authentication Code — MAC construído sobre funções hash, muito usado em autenticação e KDFs}
}

\newglossaryentry{sha2}{
    name={SHA-2},
    description={Família de funções hash (SHA-256/512) padronizadas no FIPS 180-4 (Merkle–Damgård)}
}

\newglossaryentry{sha3}{
    name={SHA-3},
    description={Keccak — funções hash por construção esponja; inclui variantes SHAKE (XOFs)}
}

\newglossaryentry{argonn}{
    name={Argon2},
    description={Função de derivação de senha com custo em memória; resistente a paralelização em GPU/ASIC}
}

\newglossaryentry{kem}{
    name={KEM},
    description={Key Encapsulation Mechanism — mecanismo de encapsulamento de chaves para estabelecer segredos compartilhados}
}

\newglossaryentry{dem}{
    name={DEM},
    description={Data Encapsulation Mechanism — componente simétrico (geralmente AEAD) usado junto ao KEM (paradigma KEM-DEM)}
}

\newglossaryentry{hpke}{
    name={HPKE},
    description={Hybrid Public Key Encryption — padrão (RFC 9180) que organiza KEM, KDF e AEAD de forma modular}
}

\newglossaryentry{tls}{
    name={TLS},
    description={Transport Layer Security — protocolo de segurança de transporte; a versão 1.3 incorpora PFS e AEADs modernos}
}

\newglossaryentry{quic}{
    name={QUIC},
    description={Protocolo de transporte sobre UDP que integra TLS 1.3, rekeying e baixa latência}
}

\newglossaryentry{ipsec}{
    name={IPsec},
    description={Conjunto de protocolos para VPN em nível de rede; oferece confidencialidade e autenticação de pacotes IP}
}

\newglossaryentry{ssh}{
    name={SSH},
    description={Secure Shell — protocolo para acesso remoto seguro, com suporte a chaves assimétricas e AEADs}
}

\newglossaryentry{hn_dlf}{
    name={Harvest Now, Decrypt Later},
    description={Ameaça em que dados são coletados hoje para decriptação futura quando recursos (ex.: computadores quânticos) estiverem disponíveis}
}

\newglossaryentry{bb84}{
    name={BB84},
    description={Protocolo de QKD de prepare-and-measure (Bennett-Brassard, 1984) usando bases mutuamente incompatíveis}
}

\newglossaryentry{e91}{
    name={E91},
    description={Protocolo de QKD baseado em emaranhamento (Ekert, 1991), com segurança ligada à violação de desigualdades de Bell}
}

\newglossaryentry{sarg04}{
    name={SARG04},
    description={Variação do BB84 mais robusta a certas imperfeições de fonte e ataques práticos}
}

\newglossaryentry{decoystate}{
    name={Decoy-State},
    description={Técnica em QKD que alterna intensidades para estimar ganhos/erros de fótons únicos e mitigar ataques PNS}
}

\newglossaryentry{dps}{
    name={DPS},
    description={Differential Phase-Shift QKD — codificação de bits em diferenças de fase entre pulsos consecutivos}
}

\newglossaryentry{cow}{
    name={COW},
    description={Coherent One-Way QKD — protocolo de variável discreta com estrutura simples e robusta para longas distâncias}
}

\newglossaryentry{rrdps}{
    name={RRDPS},
    description={Round-Robin Differential Phase-Shift QKD — reduz dependência da estimação de taxa de erro para segurança}
}

\newglossaryentry{mdi}{
    name={MDI-QKD},
    description={Measurement-Device-Independent QKD — remove confiança nos detectores deslocando a medição para um nó intermediário}
}

\newglossaryentry{tfqkd}{
    name={TF-QKD},
    description={Twin-Field QKD — ultrapassa limites taxa–distância sem repetidores, adequado para enlaces de longa distância}
}

\newglossaryentry{cvqkd}{
    name={CV-QKD},
    description={Continuous-Variable QKD — codifica informação em quadraturas (X,P) e usa detecção homo/heteródina}
}

\newglossaryentry{diqkd}{
    name={DI-QKD},
    description={Device-Independent QKD — segurança baseada apenas na violação de Bell, independente de modelos de dispositivo}
}

\newglossaryentry{qds}{
    name={QDS},
    description={Quantum Digital Signatures — assinaturas digitais com segurança baseada em princípios quânticos}
}

\newglossaryentry{qber}{
    name={QBER},
    description={Quantum Bit Error Rate — taxa de erro observada nas chaves brutas de QKD, usada para decidir extração de chave}
}

\newglossaryentry{plob}{
    name={Limite PLOB},
    description={Pirandola–Laurenza–Ottaviani–Banchi — limite fundamental de capacidade secreta de canais quânticos sem repetidores}
}

\newglossaryentry{snspd}{
    name={SNSPD},
    description={Superconducting Nanowire Single-Photon Detector — detectores de fóton único com baixa contagem escura e alta eficiência}
}

\newglossaryentry{spdc}{
    name={SPDC},
    description={Spontaneous Parametric Down-Conversion — processo óptico não linear para geração de pares emaranhados de fótons}
}

\newglossaryentry{chsh}{
    name={CHSH},
    description={Desigualdade de Clauser–Horne–Shimony–Holt — critério para testar não-localidade e violação de Bell}
}

\newglossaryentry{nocloning}{
    name={Teorema da Não-Clonagem},
    description={Não existe operador unitário que clone estados quânticos arbitrários sem perturbação; base de segurança em QKD}
}

\newglossaryentry{superposicao}{
    name={Superposição},
    description={Princípio quântico segundo o qual estados podem coexistir como combinações lineares antes da medição}
}

\newglossaryentry{emaranhamento}{
    name={Emaranhamento},
    description={Correlação quântica não local entre sistemas; base para E91, DI-QKD e teleportação}
}

\newglossaryentry{born}{
    name={Regra de Born},
    description={Probabilidades de medição dadas por projeções do estado; fundamenta a geração aleatória em QKD}
}

\newglossaryentry{rho}{
    name={Matriz densidade},
    description={Descrição de estados mistos e correlações (por exemplo, $\rho_{ABE}$) em segurança de QKD}
}

\newglossaryentry{vnentropy}{
    name={Entropia de von Neumann},
    description={Medida de incerteza quântica $S(\rho)=-\mathrm{Tr}(\rho\log\rho)$; usada em amplificação de privacidade}
}

\newglossaryentry{fidelidade}{
    name={Fidelidade de canal},
    description={Métrica de qualidade entre estado transmitido e original; correlaciona-se com QBER e taxa de chave}
}

\newglossaryentry{teleportacao}{
    name={Teleportação quântica},
    description={Transferência de estados quânticos via pares emaranhados e comunicação clássica autenticada}
}

\newglossaryentry{lwe}{
    name={LWE},
    description={Learning With Errors — problema central em reticulados; base de segurança de muitos esquemas PQC}
}

\newglossaryentry{rlwe}{
    name={RLWE},
    description={Ring-LWE — versão em anéis de LWE que habilita eficiência via NTT em $R_q=\mathbb{Z}_q[x]/(x^n+1)$}
}

\newglossaryentry{svp}{
    name={SVP},
    description={Shortest Vector Problem — problema do vetor mais curto em reticulados; relacionado a reduções de segurança}
}

\newglossaryentry{gapsvp}{
    name={GapSVP},
    description={Versão decisória do SVP; aparece em reduções de segurança para LWE/RLWE}
}

\newglossaryentry{sivp}{
    name={SIVP},
    description={Shortest Independent Vectors Problem — problema de encontrar vetores linearmente independentes curtos}
}

\newglossaryentry{ntt}{
    name={NTT},
    description={Number Theoretic Transform — transformada discreta modular que acelera multiplicações polinomiais}
}

\newglossaryentry{rns}{
    name={RNS},
    description={Residue Number System — representação por restos que acelera aritmética modular grande}
}

\newglossaryentry{crt}{
    name={CRT},
    description={Chinese Remainder Theorem — teorema que permite decomposição e recomposição modular eficiente}
}

\newglossaryentry{relinearizacao}{
    name={Relinearização},
    description={Técnica em HE para reduzir o grau do ciphertext após multiplicações, usando chaves de apoio}
}

\newglossaryentry{keyswitch}{
    name={Key switching},
    description={Troca da “base” de encriptação de um ciphertext para evitar crescimento descontrolado de termos}
}

\newglossaryentry{packing}{
    name={Packing (batching)},
    description={Codificação de múltiplos slots de dados em um único ciphertext para SIMD homomórfico}
}

\newglossaryentry{galoiskeys}{
    name={Galois keys},
    description={Chaves auxiliares que permitem aplicar automorfismos (rotações/permutas) em packing sem a chave secreta}
}

\newglossaryentry{modswitch}{
    name={Modulus switching},
    description={Técnica em HE para reduzir o módulo e controlar o crescimento do ruído}
}

\newglossaryentry{rescaling}{
    name={Rescaling},
    description={Em CKKS, divisão da escala e do módulo para manter precisão e controlar erro aproximado}
}

\newglossaryentry{bootstrapping}{
    name={Bootstrapping (HE)},
    description={Avaliação homomórfica da própria função de decifrar para “resetar” o ruído e permitir profundidade ilimitada}
}

\newglossaryentry{bgv}{
    name={BGV},
    description={Esquema homomórfico (Brakerski–Gentry–Vaikuntanathan) sobre RLWE com leveled FHE}
}

\newglossaryentry{bfv}{
    name={BFV},
    description={Esquema HE (Brakerski–Fan–Vercauteren) para aritmética exata modular sobre inteiros}
}

\newglossaryentry{ckks}{
    name={CKKS},
    description={Esquema HE aproximado para reais/complexos com controle de escala e erro (rescaling)}
}

\newglossaryentry{tfhe}{
    name={TFHE},
    description={Esquema HE otimizado para portas booleanas com bootstrapping por porta muito eficiente}
}

\newglossaryentry{fhew}{
    name={FHEW},
    description={Esquema HE booleano pioneiro em bootstrapping rápido de portas}
}

\newglossaryentry{fhe}{
    name={FHE},
    description={Fully Homomorphic Encryption — criptografia homomórfica totalmente; permite computar funções arbitrárias sobre cifrados}
}

\newglossaryentry{she}{
    name={SHE},
    description={Somewhat Homomorphic Encryption — suporta um número limitado de multiplicações e somas}
}

\newglossaryentry{leveled}{
    name={Leveled FHE},
    description={HE dimensionado para profundidade fixa sem bootstrapping durante a execução}
}

\newglossaryentry{kyber}{
    name={Kyber},
    description={KEM baseado em reticulados (ML-KEM) selecionado pelo NIST para padronização de troca de chaves pós-quânticas}
}

\newglossaryentry{dilithium}{
    name={Dilithium},
    description={Esquema de assinatura baseado em reticulados (ML-DSA) selecionado pelo NIST}
}

\newglossaryentry{sphincs}{
    name={SPHINCS+},
    description={Assinatura baseada em hash, stateless, resistente a quânticos; alternativa padronizada pelo NIST}
}

\newglossaryentry{mceliece}{
    name={Classic McEliece},
    description={Esquema de criptografia/encapsulamento baseado em códigos, com chaves públicas grandes e segurança consolidada}
}

\newglossaryentry{bike}{
    name={BIKE},
    description={Bit Flipping Key Encapsulation — KEM baseado em códigos quase-cíclicos binários}
}

\newglossaryentry{hqc}{
    name={HQC},
    description={Hamming Quasi-Cyclic — KEM baseado em códigos; candidato PQC analisado para níveis de segurança práticos}
}

\newglossaryentry{ntru}{
    name={NTRU},
    description={Família de esquemas baseados em reticulados (cifração/KEM) com eficiência e parâmetros consolidados}
}

\newglossaryentry{isogenias}{
    name={Isogenias},
    description={Mapeamentos entre curvas elípticas usados em propostas de PQC; ramo impactado por ataques recentes}
}

\newglossaryentry{falcon}{
    name={Falcon},
    description={Assinatura baseada em reticulados com amostragem gaussiana; alto desempenho para verificação}
}

\newglossaryentry{hashbased}{
    name={Assinaturas baseadas em hash},
    description={Família (XMSS, LMS, SPHINCS+) baseada em árvores de Merkle e funções hash}
}

\newglossaryentry{xmss}{
    name={XMSS},
    description={eXtended Merkle Signature Scheme — assinatura baseada em hash com estado (stateful)}
}

\newglossaryentry{lms}{
    name={LMS},
    description={Leighton–Micali Signatures — família de assinaturas baseadas em árvore de Merkle (stateful)}
}

\newglossaryentry{ed25519}{
    name={Ed25519},
    description={Assinatura EdDSA sobre Curve25519; verificação rápida e codificação canônica}
}

\newglossaryentry{x25519}{
    name={X25519},
    description={ECDH no formato Montgomery sobre Curve25519; amplamente usado em TLS, Signal e WireGuard}
}

\newglossaryentry{curve25519}{
    name={Curve25519},
    description={Curva elíptica segura e eficiente usada para ECDH (X25519) e assinaturas (Ed25519)}
}

\newglossaryentry{ecdsa}{
    name={ECDSA},
    description={Elliptic Curve Digital Signature Algorithm — assinatura sobre ECC; requer nonces seguros/determinísticos}
}

\newglossaryentry{pss}{
    name={RSA-PSS},
    description={Probabilistic Signature Scheme para RSA; recomendada para assinaturas com segurança moderna}
}

\newglossaryentry{oaep}{
    name={RSA-OAEP},
    description={Optimal Asymmetric Encryption Padding para RSA; recomendado para encapsulamento/criptografia}
}

\newglossaryentry{pfs}{
    name={PFS},
    description={Perfect Forward Secrecy — propriedade de sigilo futuro perfeito em handshakes (ex.: (EC)DHE no TLS 1.3)}
}

\newglossaryentry{indcpa}{
    name={IND-CPA},
    description={Indistinguishability under Chosen Plaintext Attack — noção de segurança para cifração sem acesso a decifragem}
}

\newglossaryentry{indcca}{
    name={IND-CCA2},
    description={Indistinguishability under Adaptive Chosen Ciphertext Attack — noção forte de segurança contra decifragem adaptativa}
}

\newglossaryentry{ufcma}{
    name={UF-CMA},
    description={Unforgeability under Chosen Message Attack — segurança de assinaturas contra forja com oráculo de assinatura}
}

\newglossaryentry{domainsep}{
    name={Separação de domínios},
    description={Técnica de rotular contextos/propósitos em KDF/PRF/AEAD para evitar colisões semânticas e confusões de chave}
}

\newglossaryentry{nonce}{
    name={Nonce/IV},
    description={Valor único por chave usado em AEADs (ex.: GCM/ChaCha20-Poly1305); reutilização compromete segurança}
}

\newglossaryentry{siv}{
    name={AES-SIV},
    description={AEAD com misuse-resistance contra repetição de nonce; usa S2V para sintetizar IV a partir de dados}
}

\newglossaryentry{xts}{
    name={AES-XTS},
    description={Modo de disco com tweak por setor; provê confidencialidade, mas não integridade (exige MAC/AEAD na camada superior)}
}

\newglossaryentry{sidechannel}{
    name={Canais laterais},
    description={Vazamentos por tempo, cache, potência, EM, falhas; mitigados por código em tempo constante, masking/blinding}
}

\newglossaryentry{blinding}{
    name={Blinding},
    description={Técnica para randomizar operações (ex.: RSA/ECC) e mitigar ataques por tempo/falhas}
}

\newglossaryentry{masking}{
    name={Masking},
    description={Particionamento/aleatorização de segredos internos para reduzir correlação com leak físico}
}

\newglossaryentry{mps}{
    name={MPC},
    description={Secure Multi-Party Computation — computação multipartidária segura que distribui confiança entre participantes}
}

\newglossaryentry{tee}{
    name={TEE},
    description={Trusted Execution Environment — ambiente de execução confiável por hardware (ex.: Intel SGX) para proteger código/dados}
}

\newglossaryentry{hsm}{
    name={HSM},
    description={Hardware Security Module — módulo de hardware para geração, armazenamento e uso de chaves com proteção física}
}

\newglossaryentry{pki}{
    name={PKI},
    description={Public Key Infrastructure — infraestrutura de chaves públicas (CAs, certificados, políticas de revogação)}
}

\newglossaryentry{ocsp}{
    name={OCSP/CRL},
    description={Mecanismos de verificação de status e revogação de certificados em PKIs}
}

\newglossaryentry{blockchain}{
    name={Blockchain},
    description={Livro-razão distribuído e imutável; demanda migração para primitivas resistentes a quânticos}
}

\newglossaryentry{iot}{
    name={IoT},
    description={Internet of Things — rede de dispositivos com recursos limitados; exige criptografia leve e eficiente}
}

\newglossaryentry{governanca}{
    name={Governança criptográfica},
    description={Políticas e processos para ciclo de vida de chaves, parametrização, auditoria e conformidade}
}

\newglossaryentry{isdpns}{
    name={PNS},
    description={Photon-Number Splitting — ataque em QKD com fontes coerentes; mitigado por Decoy-State}
}

\newglossaryentry{qrin}{
    name={QRNG},
    description={Quantum Random Number Generator — gerador de números verdadeiramente aleatórios por fenômenos quânticos}
}

\newglossaryentry{hpkehybrid}{
    name={KEM-DEM (híbrido)},
    description={Composição que combina KEM clássico e KEM pós-quântico para transições seguras}
}

\newglossaryentry{diffhellman}{
    name={Diffie–Hellman},
    description={Protocolo de acordo de chaves baseado no problema do logaritmo discreto (ou ECDH em ECC)}
}

\newglossaryentry{ecdhe}{
    name={(EC)DHE},
    description={Diffie–Hellman efêmero (em GF(p) ou ECC) com sigilo futuro perfeito}
}

\newglossaryentry{tls13}{
    name={TLS 1.3},
    description={Versão do TLS que simplifica ciphersuites, exige PFS e usa AEADs modernas}
}

\newglossaryentry{hpkectx}{
    name={Contexto (HPKE)},
    description={Rótulos, info e AAD usados em KDF/AEAD para vincular chaves a finalidades (domain separation)}
}

\newglossaryentry{lattice}{
    name={Reticulado},
    description={Estrutura discreta em $ \mathbb{R}^n $ gerada por combinações inteiras de vetores base; fundamento de LWE/RLWE}
}

\newglossaryentry{codebased}{
    name={Code-based},
    description={Abordagens de PQC baseadas em códigos corretores de erros (ex.: McEliece, BIKE, HQC)}
}

\newglossaryentry{multivariado}{
    name={Multivariado},
    description={Sistemas de assinatura baseados em polinômios multivariados sobre corpos finitos}
}

\newglossaryentry{hashfamily}{
    name={Hash-based (família)},
    description={Esquemas de assinatura baseados unicamente em funções hash (XMSS, LMS, SPHINCS+)}
}

\newglossaryentry{isogenyfamily}{
    name={Isogeny-based},
    description={Abordagens PQC baseadas em isogenias entre curvas elípticas}
}

\newglossaryentry{ctr}{
    name={CTR (Counter Mode)},
    description={Modo de operação de cifra de bloco que transforma uma cifra de bloco em uma cifra de fluxo, utilizando um contador incrementado para cada bloco. Permite paralelização e acesso aleatório, sendo amplamente utilizado em combinação com modos de autenticação como o GCM}
}

\newglossaryentry{rfc}{
    name={RFC (Request for Comments)},
    description={Série de documentos técnicos que descrevem padrões, protocolos e práticas da Internet, publicados pela IETF (Internet Engineering Task Force). Exemplos relevantes incluem RFC 5297 (AES-SIV) e RFC 9180 (HPKE)}
}

\newglossaryentry{vpn}{
    name={VPN (Virtual Private Network)},
    description={Rede privada virtual que estabelece túneis criptografados sobre redes públicas, garantindo confidencialidade, integridade e autenticidade das comunicações. Implementações comuns utilizam IPsec, OpenVPN ou WireGuard}
}

\newglossaryentry{simd}{
    name={SIMD (Single Instruction Multiple Data)},
    description={Paradigma de paralelismo em que uma única instrução opera sobre múltiplos dados simultaneamente. Em criptografia homomórfica, refere-se ao empacotamento (packing) de múltiplos valores em um único ciphertext}
}

\newglossaryentry{hestandard}{
    name={HE-Standard},
    description={Conjunto de diretrizes e práticas recomendadas para parametrização segura de esquemas de criptografia homomórfica, incluindo terminologia, níveis de segurança (128/192/256 bits) e procedimentos de relato}
}

\newglossaryentry{hebench}{
    name={HEBench},
    description={Framework de benchmark para criptografia homomórfica, permitindo comparação reprodutível de latências (mul/add/rot/ks), throughput, bootstrapping e consumo de memória entre diferentes implementações}
}

\newglossaryentry{chet}{
    name={CHET},
    description={Compilador para criptografia homomórfica que mapeia operações de alto nível (tensores, convoluções) para circuitos HE otimizados, gerenciando automaticamente escalas, layout de slots e rotações}
}

\newglossaryentry{eva}{
    name={EVA/HEAX},
    description={Ferramentas de co-design hardware/software para criptografia homomórfica, incluindo representação intermediária (IR) e kernels acelerados para GPU/FPGA}
}

\newglossaryentry{helib}{
    name={HElib},
    description={Biblioteca pioneira de criptografia homomórfica desenvolvida pela IBM, com suporte a BGV/BFV, packing, automorfismos e operações avançadas sobre slots}
}

\newglossaryentry{seal}{
    name={Microsoft SEAL},
    description={Biblioteca robusta de criptografia homomórfica desenvolvida pela Microsoft, com suporte a BFV e CKKS, implementação RNS/NTT otimizada e API estável. Amplamente adotada em pesquisa e produção}
}

\newglossaryentry{openfhe}{
    name={OpenFHE},
    description={Biblioteca de código aberto para criptografia homomórfica, sucessora do PALISADE, com suporte a BGV, BFV, CKKS e TFHE. Inclui bootstrapping moderno e ferramentas de pesquisa}
}

\newglossaryentry{tfhelib}{
    name={TFHE lib},
    description={Implementação de referência do esquema TFHE, focada em operações booleanas e bootstrapping rápido por porta lógica}
}

\newglossaryentry{privacyamplification}{
    name={PA-Privacy Amplification},
    description={Processo de pós-processamento em QKD e protocolos criptográficos que reduz a informação potencialmente acessível a um adversário, aplicando funções hash universais para comprimir a chave bruta em uma chave final indistinguível de aleatória}
}

\newglossaryentry{informationreconciliation}{
    name={IR-Information Reconciliation},
    description={Etapa de correção de erros em QKD, na qual Alice e Bob utilizam códigos corretores (LDPC, Polar) para alinhar suas chaves brutas, minimizando o vazamento de informação no canal público}
}

\newglossaryentry{belltest}{
    name={Bell Test},
    description={Experimento que verifica violações de desigualdades de Bell, demonstrando correlações não-locais incompatíveis com teorias de variáveis ocultas locais. Essencial para certificação de emaranhamento em QKD}
}

\newglossaryentry{bbm92}{
    name={BBM92},
    description={Variante do BB84 baseada em emaranhamento, proposta por Bennett, Brassard e Mermin (1992). Demonstra a equivalência entre protocolos prepare-and-measure e baseados em emaranhamento}
}

\newglossaryentry{bkz}{
    name={BKZ (Block Korkine-Zolotarev)},
    description={Algoritmo de redução de base de reticulados, utilizado em ataques contra esquemas baseados em LWE/RLWE. Variantes com pruning e quantum sieving determinam a segurança prática de parâmetros pós-quânticos}
}

\newglossaryentry{pruning}{
    name={Pruning},
    description={Técnica de otimização em algoritmos de redução de reticulados (como BKZ) que reduz o espaço de busca, acelerando ataques mas com trade-off em taxa de sucesso}
}

\newglossaryentry{quantumsieving}{
    name={Quantum Sieving},
    description={Algoritmo quântico para busca em reticulados que oferece aceleração quadrática sobre métodos clássicos. Considerado na estimativa de segurança de esquemas pós-quânticos}
}

\newglossaryentry{iso4922}{
    name={ISO/IEC 4922},
    description={Padrão internacional emergente para interoperabilidade e segurança em criptografia homomórfica e tecnologias de preservação de privacidade}
}

\newglossaryentry{nistpqc}{
    name={NIST PQC Frameworks},
    description={Conjunto de diretrizes, processos de padronização e frameworks de avaliação conduzidos pelo NIST para seleção e adoção de algoritmos criptográficos pós-quânticos}
}

\begin{document}

\maketitle
\thispagestyle{empty} 
\clearpage             

\section*{Prefácio}
\addcontentsline{toc}{section}{Prefácio}
Este livro surge da necessidade de proporcionar uma visão clara e atualizada sobre os impactos da computação quântica na criptografia. 
Ao longo dos capítulos, apresentamos fundamentos, discutimos algoritmos clássicos e pós-quânticos, avaliamos padrões emergentes e apontamos desafios de implementação no mundo real.
O objetivo inicial é servir como guia para estudantes, pesquisadores e profissionais que precisam compreender não apenas a matemática envolvida, mas também suas implicações práticas em sistemas e políticas de segurança. Para os profissionais mais avançados o objetivo principal e apresentar conteúdo e ideias para que possam avaliar as mudanças e perspectivas na era dos algoritmos criptográficos quânticos.

Para tanto, a estrutura do texto foi pensada para ser progressiva: iniciamos com os conceitos essenciais, avançamos para algoritmos quânticos e suas consequências (com ênfase no algoritmo de Shor), apresentamos questões com foco nas "famílias" de esquemas pós-quânticos (baseados em reticulados, códigos, funções hash, multivariados, isogenias), analisamos o estado da arte em padronização (com destaque para o processo do NIST) e, por fim, discutimos migração, interoperabilidade, desempenho e \gls{governanca}.

Esperamos que esta obra auxilie na formação de um pensamento crítico e na tomada de decisões técnicas informadas, fomentando estratégias de transição seguras para a era pós-quântica.

\vspace{1em}
\noindent Darlan Noetzold, Valderi Leithardt

\clearpage 

\tableofcontents
\clearpage 

\section*{Resumo}
\addcontentsline{toc}{section}{Resumo}
Na iminência da era da computação quântica, este livro oferece um guia essencial sobre os algoritmos criptográficos que moldarão a segurança digital do futuro. Explorando desde os fundamentos da criptografia clássica até os avanços mais recentes em criptografia quântica e pós-quântica. A obra aborda as vulnerabilidades impostas por computadores quânticos e apresenta soluções inovadoras, incluindo esquemas baseados em reticulados, códigos, funções hash e isogenias. Apresenta também uma investigação sobre o estado da arte da literatura atual sobre os temas que envolvem criptográfica, comparando e tendo por base trabalhos publicados nas conferências e revistas mais prestigiadas da área. Portanto, apresenta uma análise aprofundada dos processos de padronização e dos desafios de implementação no mundo real, este livro é indispensável para estudantes, pesquisadores e profissionais que buscam compreender e navegar pela complexa transição para um cenário de segurança cibernética resistente a ameaças quânticas, garantindo a proteção de dados em um futuro cada vez mais interconectado.

\clearpage 

\chapter{Introdução}
A segurança da informação é um componente central da sociedade contemporânea, sustentando desde comunicações diplomáticas e operações militares até transações financeiras e interações cotidianas realizadas por bilhões de usuários \cite{Pirandola2023}. A disciplina que garante esse pilar essencial é a criptografia, cuja função é assegurar confidencialidade, integridade, autenticidade e irrefutabilidade na transmissão e no armazenamento de dados. Portanto, se faz necessário e extremamente relevante o gosto e por que não dizer paixão pela matemática, que é a base para trabalhar com criptografia. Entretanto, a criptografia não é estática: trata-se de um campo em constante evolução, moldado por um processo de competição entre criptógrafos, que projetam sistemas de proteção, e cripto analistas, que desenvolvem métodos de quebra de algoritmos. Cada nova técnica criptográfica surgiu como uma resposta a vulnerabilidades que ameaçavam a estabilidade das comunicações em seu tempo.

Durante milênios, antes do advento da computação, a criptografia se fundamentava em técnicas essencialmente manuais ou mecânicas. O exemplo clássico é a Cifra de César, que se baseia em um deslocamento simples no alfabeto:

\[
C \equiv (P + k) \pmod{26}
\]

onde $P$ é uma letra da mensagem em forma numérica, $k$ é o deslocamento e $C$ a letra cifrada. Apesar de inovadora em sua época, esse método é trivialmente vulnerável à análise de frequência. No século XX, avanços mecânicos culminaram na produção da máquina \textit{Enigma}, cuja complexidade de rotores permitia bilhões de combinações de chaves. No entanto, a utilização sistemática da análise estatística e o desenvolvimento de máquinas eletromecânicas (como a \textit{Bombe}, em Bletchley Park) levaram à quebra do sistema, mudando o rumo da Segunda Guerra Mundial. Esse episódio ilustrou brutalmente a vulnerabilidade de sistemas criptográficos quando confrontados com adversários de poder tecnológico equivalente ou superior.

A revolução seguinte ocorreu na década de 1970, quando a criptografia começou a se consolidar como ciência matemática aplicada. Em 1976, Diffie e Hellman introduziram o conceito inovador de {criptografia de chave pública}, solucionando o problema crítico da distribuição de chaves secretas em larga escala. Seu protocolo utiliza operações modulares baseadas em números primos. As partes $A$ e $B$ combinam publicamente um número primo grande $p$ e uma base $g$. Cada uma escolhe um segredo ($a$ e $b$) e troca valores $g^a \pmod{p}$ e $g^b \pmod{p}$. A chave secreta compartilhada é dada por:

\[
K \equiv g^{ab} \pmod{p}
\]

que era protegida pela dificuldade computacional do \textit{Problema do Logaritmo Discreto}. Quase simultaneamente, em 1977, Rivest, Shamir e Adleman introduziram o \gls{rsa}, que utiliza a aritmética de números primos e a dificuldade da fatoração de grandes inteiros. Assim, uma mensagem $M$ pode ser cifrada como:

\[
C \equiv M^e \pmod{n}, \quad
M \equiv C^d \pmod{n}
\]

onde $n = pq$ (produto de dois primos grandes), $e$ é o expoente público e $d$ o expoente privado. Até hoje, esse método constitui o pilar das infraestruturas de chave pública (\gls{pki}) em internet banking, certificados digitais e assinaturas eletrônicas. A década de 1980 trouxe ainda a {criptografia de curvas elípticas} (\gls{ecc}), que oferece segurança equivalente ao RSA com chaves menores, explorando propriedades do grupo de pontos de uma curva elíptica definida sobre corpos finitos:

\[
E(\mathbb{F}_p): y^2 \equiv x^3 + ax + b \pmod{p}
\]

Na prática, ECC tornou-se particularmente atrativa em dispositivos de baixo consumo, como cartões inteligentes, smartphones e \gls{iot} (\textit{Internet das Coisas}), reduzindo custos energéticos e de transmissão. Com a massificação da internet nos anos 1990, os sistemas criptográficos clássicos passaram a ser incorporados em protocolos globais como SSL/\gls{tls}, \gls{ipsec} e \gls{vpn}, sustentando o crescimento do comércio eletrônico, da banca digital e da comunicação via e-mail. Entretanto, a escalada da capacidade computacional expôs fragilidades: o \textbf{DES}, dotado de chaves de 56 bits, foi quebrado por força bruta já em 1998, em menos de três dias. Isso levou a comunidade internacional a recorrer ao concurso organizado pelo \gls{nist}, que culminou na seleção do \gls{aes} em 2001, com suporte a chaves de 128, 192 e 256 bits, tornando-se imediatamente o novo padrão global.

Nesse contexto de confiança e robustez matemática, surgiu a maior ruptura conceitual da história da criptografia: a computação quântica. Peter Shor, em 1994, apresentou um algoritmo que mostra que a fatoração de grandes números e o cálculo do logaritmo discreto, até então intratáveis, podendo ser resolvidos em tempo polinomial em computadores quânticos. Se implementado em grande escala, o algoritmo de Shor permitiria quebrar RSA e ECC, pilares da segurança digital, em questão de horas ou minutos. Pouco depois, em 1996, Lov Grover demonstrou um algoritmo capaz de reduzir exponencialmente o esforço de ataques de busca exaustiva, diminuindo a complexidade efetiva de $2^n$ para $2^{n/2}$. Isso implica, por exemplo, que uma chave simétrica de 128 bits ofereceria uma segurança efetiva equivalente a apenas 64 bits diante de um adversário quântico.

Para responder a essa ameaça, pesquisadores recorreram à própria mecânica quântica como recurso de defesa. Em 1984, Bennett e Brassard introduziram o \gls{bb84}, o primeiro protocolo de distribuição de chaves quânticas (\gls{pqc}), cuja segurança é garantida pelo fato de que a medição de um estado quântico perturba o próprio estado, permitindo a detecção imediata de eavesdroppers. O protocolo \gls{e91}, proposto por Ekert em 1991, introduziu o uso de \gls{emaranhamento} quântico e da violação das desigualdades de Bell como fundamentos de segurança. Desde então, dezenas de variações foram propostas, incluindo o \gls{sarg04} (2004), o \gls{mdi} (2009), que elimina vulnerabilidades de dispositivos de detecção, e o \gls{cvqkd} (baseado em variáveis contínuas). Em 2017, o satélite Micius, lançado pela China, comprovou a viabilidade de transmissões intercontinentais baseadas em QKD, estabelecendo um novo marco tecnológico.

Apesar disso, a implementação prática da criptografia quântica enfrenta dificuldades de larga escala, como perdas ópticas em fibras acima de centenas de quilômetros, necessidade de alinhamento preciso e altos custos de infraestrutura. Tais limitações impulsionaram uma segunda linha de defesa: a {criptografia pós-quântica (\gls{pqc}). Essa abordagem objetiva desenvolver algoritmos criptográficos que sejam resistentes tanto a computadores clássicos quanto quânticos, mas que funcionem em hardware padrão. Entre as famílias mais promissoras estão os algoritmos baseados em \gls{lattice} (\gls{kyber}, \gls{dilithium}), códigos corretores (McEliece), árvores de Merkle e funções hash (\gls{sphincs}). Desde 2016, o NIST conduz um processo público de padronização internacional desses algoritmos, cujo resultado final servirá como referência para governos e indústrias globais.

Diante desse panorama, a comunidade criptográfica caminha atualmente para soluções {híbridas}, que combinam criptografia clássica com pós-quântica, ou mesmo PQC com QKD, em uma estratégia de defesa multicamadas. O objetivo é proteger informações críticas contra adversários presentes e futuros, mitigando o risco do ataque \gls{hn_dlf}, em que dados interceptados hoje poderiam ser decriptados no futuro após o advento de um computador quântico universal.

Assim, a evolução histórica da criptografia deve ser entendida como um processo contínuo e adaptativo em resposta às ameaças emergentes, indo das cifras manuais às redes quânticas, dos logaritmos discretos aos reticulados pós-quânticos. A Tabela \ref{tab:evolucao_criptografia} sintetiza essa trajetória em uma linha temporal que mostra como cada ameaça impulsionou a criação ou adoção de novas soluções criptográficas.

\begin{table}[h!]
\centering
\scriptsize
\caption{Evolução histórica das técnicas criptográficas}
\label{tab:evolucao_criptografia}
\begin{tabular}{p{2cm}p{2cm}p{10cm}}
\hline
\textbf{Período/Ano} & \textbf{Tipo de Criptografia} & \textbf{Descrição} \\
\hline
Pré-1970 & Clássica (manual) & Sistemas artesanais como cifra de César e a máquina Enigma, baseados em substituição simples e mecanismos eletromecânicos, vulneráveis à análise estatística e ao poder computacional crescente. \\

1976 & Assimétrica & Protocolo Diffie-Hellman cria o paradigma de chave pública, resolvendo o problema da distribuição segura de segredos em larga escala. \\

1977 & Simétrica & DES, primeiro padrão oficial, marco da criptografia moderna, mas quebrado décadas depois por ataques de força bruta. \\

1977 & Assimétrica & RSA introduz a primeira implementação prática de criptografia de chave pública, baseando-se na fatoração de inteiros. \\

1984 & Quântica & BB84 inaugura a criptografia quântica e a distribuição de chaves com segurança incondicional fundamentada na mecânica quântica. \\

1985 & Assimétrica & ECC surge como alternativa mais eficiente ao RSA, explorando propriedades matemáticas de curvas elípticas. \\

1991 & Quântica & E91 (Ekert) propõe uso de emaranhamento e desigualdades de Bell como mecanismo de segurança. \\

1994-1996 & Ameaça Quântica & Algoritmos de Shor e Grover revelam que computadores quânticos poderão comprometer RSA, ECC e reduzir a segurança de cifras simétricas. \\

2001 & Simétrica & AES substitui o DES, oferecendo segurança robusta contra ataques clássicos e resistência relativa a Grover. \\

2004-2009 & Quântica & Protocolos como SARG04 e MDI-QKD aprimoram QKD lidando com vulnerabilidades de dispositivos e ataques práticos. \\

2017 & Quântica & Satélite Micius comprova a primeira rede QKD intercontinental, conectando continentes com segurança quântica. \\

2016-2022 & Pós-Quântica & NIST lidera padronização de algoritmos resistentes a ataques quânticos: Kyber, Dilithium e SPHINCS+. \\

2020-2025 & Híbrida & Aproximação entre PQC, QKD e algoritmos clássicos em esquemas multicamadas de segurança para transição à era quântica. \\
\hline
\end{tabular}
\end{table}

Em síntese, a trajetória da criptografia pode ser compreendida como uma linha de resposta histórica a ameaças crescentes: da análise manual de cifras clássicas aos computadores digitais, da escalada da internet ao avanço da computação quântica. Cada marco representou tanto uma quebra de paradigma quanto uma antecipação a riscos futuros. A Tabela \ref{tab:evolucao_criptografia} evidencia essa sequência, funcionando como guia visual que conecta problemas, soluções e a contínua necessidade de inovação.

Para complementar a compreensão da evolução e das interconexões entre as diferentes áreas da criptografia, apresentamos a Figura \ref{fig:linha_tempo_criptografia}, que oferece uma representação visual da linha do tempo dos principais marcos criptográficos. Esta figura ilustra de forma concisa os momentos-chave, desde as técnicas clássicas até as soluções híbridas, destacando a progressão e a resposta às ameaças emergentes.
\begin{figure}[h!]
    \centering
    \begin{tikzpicture}[
        event/.style={circle, fill=blue!20, draw=blue!60, inner sep=1pt, font=\tiny}, 
        labelnode/.style={font=\tiny, align=center, text width=2.2cm}, 
        arrow/.style={->, thick, blue!50}
    ]
    \draw[thick, gray!50] (0,0) -- (15,0);

    \node[event] (c1970) at (1,0) {};
    \node[labelnode, above=0.1cm of c1970] {Pré-1970: Clássica (manual/mecânica)}; 

    \node[event] (dh) at (3,0) {};
    \node[labelnode, above=0.1cm of dh] {1976: Diffie-Hellman (Chave Pública)};

    \node[event] (rsa) at (4,0) {};
    \node[labelnode, below=0.1cm of rsa] {1977: RSA (Fatoração)};

    \node[event] (bb84) at (5,0) {};
    \node[labelnode, above=0.1cm of bb84] {1984: BB84 (QKD)};

    \node[event] (ecc) at (6,0) {};
    \node[labelnode, below=0.1cm of ecc] {1985: ECC (Curvas Elípticas)};

    \node[event] (shor) at (7.5,0) {};
    \node[labelnode, above=0.1cm of shor] {1994-1996: Shor/Grover (Ameaça Quântica)};

    \node[event] (aes) at (9,0) {};
    \node[labelnode, below=0.1cm of aes] {2001: AES (Padrão Simétrico)};

    \node[event] (micius) at (11,0) {}; 
    \node[labelnode, above=0.1cm of micius] {2017: Micius (QKD Intercontinental)}; 

    \node[event] (nistpqc) at (12.5,0) {};
    \node[labelnode, below=0.1cm of nistpqc] {2016-2022: NIST PQC (Padronização)};

    \node[event] (hybrid) at (14,0) {};
    \node[labelnode, above=0.1cm of hybrid] {2020s: Soluções Híbridas};

    \draw[arrow] (c1970) -- (dh);
    \draw[arrow] (dh) -- (rsa);
    \draw[arrow] (rsa) -- (bb84);
    \draw[arrow] (bb84) -- (ecc);
    \draw[arrow] (ecc) -- (shor);
    \draw[arrow] (shor) -- (aes);
    \draw[arrow] (aes) -- (micius);
    \draw[arrow] (micius) -- (nistpqc);
    \draw[arrow] (nistpqc) -- (hybrid);


    \end{tikzpicture}
    \caption{Linha do tempo da evolução das técnicas criptográficas.}
    \label{fig:linha_tempo_criptografia}
\end{figure}
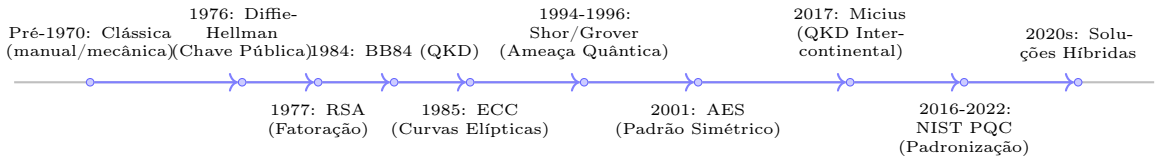

Adicionalmente, a Figura \ref{fig:interdependencia_criptografia} complementa essa visão histórica ao ilustrar a interdependência e a evolução das diferentes áreas da criptografia frente à ameaça quântica. Ela demonstra como a criptografia clássica, ao ser confrontada pela ameaça quântica, impulsionou o desenvolvimento tanto da criptografia pós-quântica (PQC) quanto da criptografia quântica (QKD), que por sua vez convergem para a formação de soluções híbridas, representando uma estratégia de defesa multicamadas para o futuro.

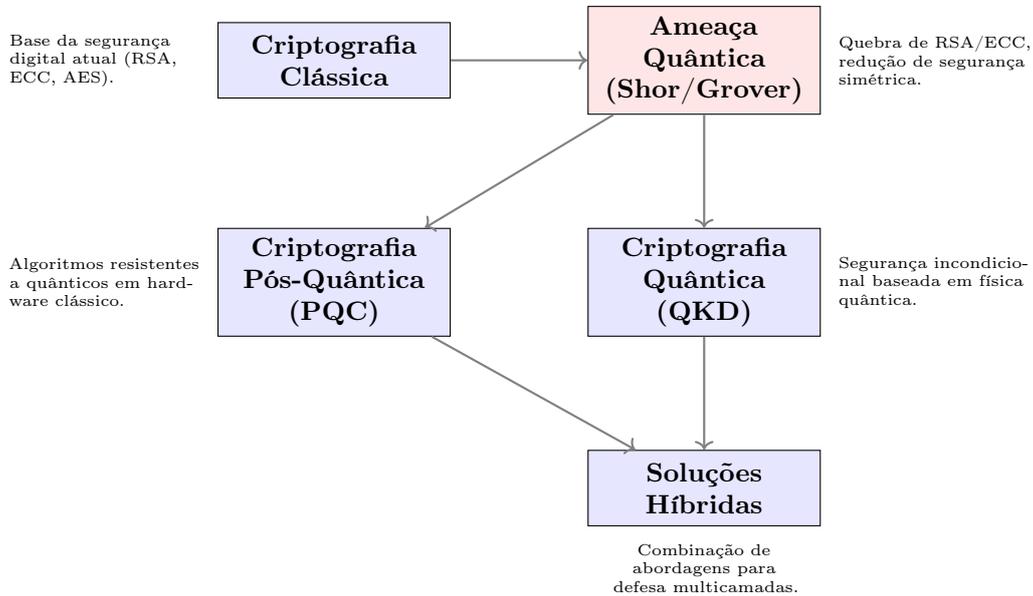
\begin{figure}[h!]
    \centering
    \begin{tikzpicture}[
        box/.style={rectangle, draw, fill=blue!10, text width=2.8cm, minimum height=1cm, text centered, font=\bfseries\small},
        arrow/.style={->, thick, black!50},
        threat/.style={rectangle, draw, fill=red!10, text width=2.8cm, minimum height=1cm, text centered, font=\bfseries\small},
        expl_text/.style={font=\tiny, text width=2.5cm, align=left}
    ]

    \node[box] (classic) {Criptografia Clássica};
    \node[threat, right=1.8cm of classic] (threat_quantum) {Ameaça Quântica (Shor/Grover)};
    \node[box, below=1.5cm of threat_quantum] (qkd) {Criptografia Quântica (QKD)}; 
    \node[box, below=1.5cm of qkd] (hybrid) {Soluções Híbridas}; 

    \node[box, left=1.8cm of qkd] (pqc) {Criptografia Pós-Quântica (PQC)}; 

    \draw[arrow] (classic) -- (threat_quantum);
    \draw[arrow] (threat_quantum) -- (qkd);
    \draw[arrow] (threat_quantum) -- (pqc); 
    \draw[arrow] (pqc) -- (hybrid);
    \draw[arrow] (qkd) -- (hybrid);

    \node[expl_text, left=0.1cm of classic.west, anchor=east] {Base da segurança digital atual (RSA, ECC, AES).};
    \node[expl_text, right=0.1cm of threat_quantum.east, anchor=west] {Quebra de RSA/ECC, redução de segurança simétrica.};
    \node[expl_text, left=0.1cm of pqc.west, anchor=east] {Algoritmos resistentes a quânticos em hardware clássico.};
    \node[expl_text, right=0.1cm of qkd.east, anchor=west] {Segurança incondicional baseada em física quântica.};
    \node[expl_text, below=0.1cm of hybrid.south, align=center] {Combinação de abordagens para defesa multicamadas.};

    \end{tikzpicture}
    \caption{Interdependência e evolução das áreas criptográficas frente à ameaça quântica.}
    \label{fig:interdependencia_criptografia}
\end{figure}

Por fim, este livro está organizado da seguinte forma: após esta introdução (Seção 1), apresenta-se na Seção 2 a criptografia clássica, seus fundamentos e vulnerabilidades perante algoritmos quânticos. A Seção 3 discute a criptografia quântica, incluindo seus principais protocolos, métodos experimentais e desafios. A Seção 4 aborda a criptografia pós-quântica, com foco em suas classes algorítmicas e no processo de padronização internacional. A Seção 5 introduz as propostas híbridas, que combinam soluções clássicas, quânticas e pós-quânticas. Por fim, na Seção 6 realiza-se uma comparação entre as abordagens e discutem-se as tendências futuras, sendo a Seção 7 reservada à conclusão.

\section{Estado da Arte}

A literatura de criptografia reúne contribuições que vão de exposições introdutórias a monografias especializadas e relatórios de padronização. Em um extremo, livros-texto apresentam modelos de segurança, construções canônicas e provas de redução, formando a base para raciocínio formal sobre confidencialidade, integridade e autenticidade. Em outro, trabalhos dirigidos a subáreas tratam de hipóteses específicas, novos esquemas, refinamentos de eficiência e análises de implementação. Entre esses polos, surveys e documentos técnicos sintetizam resultados, comparam abordagens, reúnem métricas de desempenho e destacam questões em aberto. A seguir posiciona-se o leitor nesse panorama, descrevendo o escopo típico de cada classe de referência e explicitando o enfoque adotado nesta análise.

Textos introdutórios tendem a organizar o campo por primitivas e objetivos de segurança. É comum encontrar capítulos dedicados a cifração simétrica (redes SPN, ARX, modos de operação e segurança em IND-CPA), cifração pública (RSA, esquemas baseados em DLP e ECC), funções hash (modelos Merkle–Damgård e esponja) e assinaturas digitais (UF-CMA). Geralmente apresentam modelos de ameaça e noções de redução que conectam a resistência a ataques com a dificuldade de problemas matemáticos. Handbooks com foco aplicado incluem seções sobre geração de números aleatórios, gerenciamento de chaves, padronização e práticas de implementação, além de notas sobre ataques por \gls{sidechannel} e parâmetros recomendados.

Publicações sobre criptografia homomórfica usualmente descrevem as classes de esquemas (parcial, nívelado e totalmente homomórfico), o papel do ruído, técnicas de \gls{relinearizacao} e de empacotamento, e a relação com \gls{lwe}/\gls{rlwe}. São frequentes análises de custo para somas e multiplicações sobre cifrados, avaliações de bootstrapping e comparações entre aritmética exata (BFV/\gls{bgv}) e aproximada (\gls{ckks}). Em trabalhos voltados para implementação, encontram-se descrições da utilização de \gls{ntt}, vetorização, uso de GPU e práticas de escolha de parâmetros sob metas de segurança específicas.

Na área de criptografia pós-quântica (PQC), os surveys apresentam as famílias por suposições de segurança: reticulados (LWE/RLWE), códigos corretores, sistemas multivariados, \gls{isogenias} e abordagens baseadas em funções hash. Esses materiais discutem impactos de algoritmos quânticos sobre RSA, DLP e chaves simétricas, e detalham critérios de avaliação como tamanho de chaves, latência de encapsulamento/decapsulamento, custo de assinatura/verificação e memória. Relatórios de padronização, como os do NIST, organizam rodadas de avaliação, propõem níveis de segurança e registram decisões de seleção e justificativas técnicas.

Documentos sobre criptografia quântica (QKD) usualmente apresentam os princípios físicos que suportam a segurança, as diferenças entre protocolos discretos e de variáveis contínuas, as implicações de perdas e ruído em canais ópticos e as considerações de implementação envolvendo detectores e fontes. Esses textos ressaltam o papel de QKD como parte de arquiteturas de comunicação que permanecem dependentes de camadas clássicas de autenticação e de transporte.

A Tabela~\ref{tab:trabalhos_relacionados} lista referências representativas, segmentadas por categoria. Os comentários de escopo indicam o que tais obras tendem a abordar e como podem auxiliar na construção de uma base para estudos e avaliações técnicas.

\scriptsize
\begin{table}[h!]
\centering
\scriptsize
\caption{Referências representativas por categoria e tópicos usualmente abordados}
\label{tab:trabalhos_relacionados}
\begin{tabular}{p{1.5cm} p{4cm} p{1.8cm} p{6.9cm}}
\toprule
\textbf{Categoria} & \textbf{Referência} & \textbf{Âmbito principal} & \textbf{Conteúdo frequentemente tratado} \\
\midrule
Livro-texto &
Katz e Lindell, \emph{Introduction to Modern Cryptography} \cite{KatzLindell} &
Fundamentos clássicos &
Modelos de segurança (IND-CPA/CCA, UF-CMA), provas de redução, esquemas de cifra e assinatura, protocolos e composição. \\

Handbook &
Menezes, van Oorschot e Vanstone, \emph{Handbook of Applied Cryptography} \cite{HAC} &
Aplicada clássica &
AES, RSA, DLP/ECC, funções hash, geração de aleatoriedade, PKI, padrões e recomendações de implementação. \\

Survey/ Coletânea &
Bernstein e Lange, \emph{Post-Quantum Cryptography} \cite{BernsteinPQ} &
Pós-quântica &
Famílias PQC (reticulados, códigos, multivariados, isogenias, hash-based), impactos de Shor/Grover, parâmetros e desafios. \\

Relatórios &
NIST PQC (chamadas e relatórios) \cite{NISTPQC} &
Padronização &
Critérios e níveis de segurança, métricas de desempenho, decisões sobre KEMs e assinaturas (Kyber/ML-KEM, Dilithium/ML-DSA, SPHINCS+). \\

Artigos/ Survey &
Gentry (\gls{fhe}) e esquemas BGV/BFV/CKKS \cite{Gentry2009FHE,BGV,BFV,CKKS} &
Homomórfica &
Bootstrapping, gestão de ruído, relinearização, empacotamento, aritmética exata e aproximada, complexidade e casos de uso. \\

Bibliotecas &
Microsoft SEAL; OpenFHE/HElib \cite{SEAL,OpenFHE,HElib} &
Implementação HE &
Interfaces, escolhas de parâmetros, NTT, vetorização, exemplos, notas de segurança e medições de custo por operação. \\

Survey &
Pirandola et al. \cite{Pirandola2023} &
Criptografia quântica &
Protocolos BB84/E91/MDI, segurança de dispositivo, limitações físicas, cenários de implantação e integração com camadas clássicas. \\

Artigos clássicos &
\gls{diffhellman}; RSA \cite{DiffieHellman,RSA} &
Chave pública &
Troca de chaves em canais abertos, cifração e assinatura baseadas em fatoração e logaritmo discreto. \\

Padrões &
AES; SHA-2; SHA-3 \cite{AES,SHA2,SHA3} &
Simétrica e hash &
Redes SPN, ARX, modos de operação (\gls{gcm}), modelos de compressão e esponja, propriedades de segurança e requisitos de uso. \\
\bottomrule
\end{tabular}
\end{table}

\normalsize

Os livros de fundamentos são adequados para estabelecer terminologia, hipóteses de segurança e mecanismos de prova. A exposição detalha como propriedades como indistinguibilidade e inquebrável sob texto cifrado escolhido são formalizadas e como se relacionam com problemas de difícil solução computacional. Handbooks e guias aplicados reúnem recomendações de parâmetros, alertas sobre armadilhas de implementação, descrições de APIs e alinhamento com padrões. Na área homomórfica, a literatura apresenta estratégias para equilibrar profundidade de circuito e orçamento de ruído, além de métodos de empacotamento que viabilizam paralelismo. Em PQC, surveys mapeiam a transição de suposições baseadas em fatoração e logaritmo discreto para reticulados e outras estruturas, e relatórios de padronização documentam critérios comparativos, decisões e implicações para migração. Em QKD, a ênfase recai sobre aspectos físicos, sobre a modelagem de dispositivos e sobre a integração com pilhas de comunicação.

A análise adota como eixo comparativo técnicas estabelecidas e o estado atual de cada família criptográfica discutida. O objetivo é apresentar, para cada tema, uma camada histórica, uma camada de fundamentos formais e uma camada de técnicas contemporâneas, de modo a facilitar a avaliação de compromissos entre segurança, desempenho e viabilidade de implantação. Em cifração simétrica, descrevem-se redes baseadas em substituição-permutação e construções ARX, e discutem-se ajustes de parâmetros quando se considera a existência de adversários com recursos quânticos, como implicações do algoritmo de Grover sobre tamanhos de chaves. Em cifração e assinatura de chave pública, a análise parte de RSA e ECC e segue para alternativas que adotam LWE/RLWE, códigos e funções hash, observando efeitos sobre tamanhos de chave, latência, memória e integração com infraestruturas já existentes.

Na parte dedicada à criptografia homomórfica, a exposição considera o custo das operações básicas sobre cifrados, a diferença entre aritmética exata e aproximada e os requisitos para aplicações como agregação estatística e inferência de modelos com dimensão moderada. Apresentam-se elementos que afetam a prática, como uso de transformadas número-teóricas, vetorização, chaves de rotação e procedimentos de relinearização.

Em PQC, descrevem-se esquemas indicados em processos de seleção pública e as razões técnicas que justificam tais escolhas, com notas sobre níveis de segurança, propriedades de robustez e pontos de atenção para migração. Os impactos operacionais de substituição de algoritmos em protocolos como TLS, \gls{ssh} e IPsec são discutidos com foco em compatibilidade e caminhos de transição.

No caso de QKD, apresentam-se os princípios físicos dos protocolos, a necessidade de autenticação clássica para mitigar ataques de intermediário e os custos de implantação que influenciam o uso em redes metropolitanas e enlaces de longa distância. Indicam-se formas de empregar QKD em conjunto com mecanismos clássicos e pós-quânticos, destacando quais funções de rede são atendidas em cada camada.

Ao articular essas áreas, propõe-se um quadro comparativo que auxilia na tomada de decisão: para uma dada aplicação, quais suposições de segurança são aceitáveis, quais custos de execução são compatíveis com o ambiente alvo e quais ajustes são requeridos para atender a requisitos regulatórios. A comparação é apresentada com métricas e exemplos de configuração, de modo a permitir a reprodução e a avaliação de alternativas de forma informada.

\chapter{Criptografia Clássica}

A criptografia clássica corresponde ao conjunto de técnicas desenvolvidas antes da era da computação quântica, cuja segurança reside na complexidade de problemas matemáticos considerados intratáveis para máquinas convencionais. Compreende, de forma geral, duas grandes categorias: algoritmos {simétricos}, que utilizam a mesma chave para cifrar e decifrar uma mensagem, e algoritmos {assimétricos}, que se baseiam em pares de chaves distintas (pública e privada). Ao longo das últimas cinco décadas, essas técnicas estruturaram a segurança digital mundial, garantindo a confidencialidade de comunicações pessoais, diplomáticas e militares, além de sustentar sistemas críticos como a infraestrutura bancária, comércio eletrônico, redes privadas virtuais (VPNs), assinaturas digitais e certificados eletrônicos.

A importância da criptografia clássica se dá também pelo seu papel histórico na transição entre as cifras manuais e mecânicas, utilizadas até meados do século XX, e os algoritmos matemáticos modernos. A quebra da máquina Enigma durante a Segunda Guerra Mundial demonstrou que técnicas baseadas apenas em obscuridade não poderiam sobreviver diante do avanço tecnológico. Foi justamente na década de 1970, com a emergência da computação de propósito geral e posteriormente da internet, que a criptografia clássica ganhou forma científica consolidada, tornando-se disciplina matemática aplicada e fundamental para a segurança global.

\section{Conceitos e Motivação}

A consolidação da criptografia clássica decorreu da necessidade crescente de proteger volumes cada vez maiores de dados transmitidos por redes digitais emergentes. Inicialmente, governos e forças armadas buscavam preservar a segurança de comunicações estratégicas, porém, rapidamente, bancos, empresas e depois usuários individuais passaram a demandar mecanismos robustos de privacidade e autenticação. Essa necessidade se intensificou especialmente a partir dos anos 1980 e 1990, com a expansão da internet e da digitalização de serviços.

No paradigma {simétrico}, uma mesma chave secreta é compartilhada entre emissor e receptor, de modo que as operações de codificação e decodificação são matematicamente inversas. Um exemplo clássico é o \textit{Data Encryption Standard} (DES), publicado em 1977, e posteriormente substituído pelo \textit{Advanced Encryption Standard} (AES) \cite{AES}, que permanece como padrão mundial. Esse tipo de criptografia é extremamente eficiente em termos de velocidade e consumo computacional, sendo amplamente utilizado em proteção de dados em trânsito (e.g., redes sem fio, VPNs) e em repouso (e.g., discos criptografados). Sua limitação central, no entanto, está na distribuição segura da chave secreta entre as partes.

Esse desafio foi superado pelo paradigma {assimétrico}, inaugurado com os trabalhos de Diffie e Hellman (1976) e posteriormente com o algoritmo RSA (1977). Nesse modelo, cada usuário possui um par matemático de chaves: uma pública, divulgada livremente, e outra privada, mantida em sigilo. A chave pública é usada para cifrar mensagens ou verificar assinaturas, enquanto a chave privada é necessária para decifrar ou assinar documentos digitalmente. Esse esquema solucionou o problema até então insolúvel da troca segura de chaves em canais abertos, permitindo a construção de protocolos seguros como SSL/TLS, que sustentam as comunicações na web até hoje.

A motivação para o desenvolvimento da criptografia clássica pode, portanto, ser contextualizada em três grandes fatores: 
(i) a necessidade governamental e militar por métodos seguros de comunicação frente à evolução da criptanálise; 
(ii) a demanda crescente do setor financeiro por meios confiáveis de realizar transações digitais sem risco de fraudes; 
(iii) a transformação social trazida pela internet, que expandiu a segurança da informação para a vida cotidiana, tornando a criptografia invisível porém onipresente no uso de smartphones, aplicações de mensagens, comércio eletrônico e autenticação em múltiplos serviços.

Além disso, a criptografia clássica consolidou não apenas algoritmos, mas também modelos formais de segurança, como as noções de indistinguibilidade, resistência a ataques de texto escolhido (\textit{chosen-plaintext attacks, CPA}) e de texto cifrado escolhido (\textit{chosen-ciphertext attacks, CCA}). Esses critérios passaram a guiar não só a construção, mas também a avaliação científica dos esquemas, estabelecendo as bases para futuras transições tecnológicas.

\section{Metodologias}

Os algoritmos clássicos fundamentam sua segurança na complexidade computacional de determinados problemas matemáticos, assumindo que certas operações são fáceis em uma direção, mas extremamente difíceis na direção inversa. Essa assimetria computacional é a essência do paradigma clássico da criptografia. A robustez desses sistemas decorre do fato de que, para problemas específicos de matemática discreta, nenhuma solução em tempo polinomial é conhecida até hoje em computadores convencionais.

No caso de algoritmos assimétricos como o \textbf{RSA}, a dificuldade fundamental deriva da fatoração de inteiros. Dada a construção de uma chave pública a partir do produto de dois primos grandes $p$ e $q$:

\[
N = p \cdot q, \quad \text{com $p$ e $q$ primos de centenas ou milhares de bits},
\]

Com isso, é trivial calcular $N$ a partir de $p$ e $q$, mas o processo inverso, encontrar $p$ e $q$ conhecendo apenas $N$, é computacionalmente inviável em máquinas clássicas quando $N$ atinge tamanhos maiores (2048 ou 4096 bits, por exemplo). A segurança do RSA depende também da dificuldade de calcular o inverso modular da exponenciação, o chamado \textit{Problema RSA} \cite{RSA_Longevity}. Uma mensagem $M$ é cifrada segundo a congruência:

\[
C \equiv M^e \pmod{N},
\]

e decifrada via:

\[
M \equiv C^d \pmod{N},
\]

onde $e$ é o expoente público e $d$ o expoente privado, obtido pela relação $ed \equiv 1 \pmod{\varphi(N)}$, sendo $\varphi(N) = (p-1)(q-1)$ a função totiente de Euler. A geração de chaves envolve a seleção de dois primos grandes $p$ e $q$, o cálculo de $N$ e $\varphi(N)$, a escolha de $e$ tal que $1 < e < \varphi(N)$ e $\text{mdc}(e, \varphi(N)) = 1$, e o cálculo de $d$ como o inverso modular de $e$ modulo $\varphi(N)$.

De forma complementar, o protocolo \textbf{Diffie-Hellman} fundamenta-se no \textbf{Problema do Logaritmo Discreto} (DLP). Dados um número primo grande $p$ e uma base $g$ (gerador de um subgrupo cíclico de $\mathbb{Z}_p^*$), o cálculo da exponenciação modular $g^x \pmod{p}$ é computacionalmente eficiente. No entanto, o cálculo inverso, encontrar $x$ dado $y = g^x \pmod{p}$, permanece intratável para valores grandes de $p$. Assim, duas partes, Alice e Bob, podem compartilhar publicamente $g^a \pmod{p}$ e $g^b \pmod{p}$ (onde $a$ e $b$ são seus segredos privados), mas apenas elas conseguem calcular a chave comum $K \equiv (g^b)^a \equiv (g^a)^b \equiv g^{ab} \pmod{p}$, o que garante sigilo contra adversários externos.

Outra metodologia fundamental é a {criptografia de curvas elípticas} (ECC), que baseia sua segurança no {Problema do Logaritmo Discreto em Curvas Elípticas} (ECDLP). Os pontos de uma curva elíptica definida sobre corpos finitos formam grupos abelianos com operações bem estruturadas. A operação central é a "multiplicação escalar", onde $Q = kP$ representa a adição do ponto $P$ a si mesmo $k$ vezes. O desafio de determinar o escalar $k$ em $Q = kP$, dado um ponto base $P$ e outro ponto $Q$ na curva, oferece a mesma segurança de RSA, porém com chaves significativamente menores (e.g., uma chave ECC de 256 bits oferece segurança comparável a uma chave RSA de 3072 bits), aumentando a eficiência em dispositivos com recursos computacionais limitados, como dispositivos móveis e IoT \cite{Curve25519}. As curvas mais utilizadas são as de Weierstrass da forma $y^2 = x^3 + Ax + B \pmod{p}$ ou curvas de Montgomery e Edwards para maior eficiência.

Na perspectiva da criptografia simétrica, algoritmos como o \textbf{AES} (Advanced Encryption Standard) operam em blocos de 128 bits de dados, utilizando chaves de 128, 192 ou 256 bits. A transformação de um bloco de texto claro em texto cifrado ocorre através de um número fixo de rodadas (10, 12 ou 14, respectivamente), cada uma composta por uma sequência de quatro operações distintas:
\begin{enumerate}
    \item \textit{SubBytes}: Uma substituição não linear byte a byte, onde cada byte do estado é substituído por outro byte de uma S-box (tabela de substituição) predefinida. Esta operação introduz a propriedade de \textit{confusão}, tornando a relação entre a chave e o texto cifrado o mais complexa possível.
    \item \textit{ShiftRows}: Uma permutação cíclica de bytes dentro das linhas da matriz de estado. A primeira linha permanece inalterada, a segunda linha é rotacionada em um byte, a terceira em dois bytes, e a quarta em três bytes. Esta operação promove a \textit{difusão}, espalhando a influência de um único bit de entrada por múltiplos bits de saída.
    \item \textit{MixColumns}: Uma transformação linear que opera em cada coluna da matriz de estado. Cada coluna é tratada como um polinômio sobre o corpo finito $\mathbb{F}_{2^8}$ e multiplicada por um polinômio fixo modulo $x^4+1$. Esta operação garante que cada byte de saída de uma coluna dependa de todos os quatro bytes de entrada daquela coluna, aumentando significativamente a difusão.
    \item \textit{AddRoundKey}: Uma operação XOR bit a bit do estado com uma subchave derivada da chave principal. Uma subchave única é gerada para cada rodada através de um algoritmo de expansão de chave. Esta etapa é crucial para a introdução da chave secreta em cada estágio da cifragem.
\end{enumerate}
A segurança do AES advém da composição iterativa dessas etapas, que proporcionam as propriedades de {confusão} e {difusão} conforme descritas por Claude Shannon em 1949. A confusão refere-se à complexidade de relacionar o texto claro com o texto cifrado, enquanto a difusão diz respeito à dispersão da influência de um bit de entrada em múltiplos bits de saída.

Para ilustrar os princípios fundamentais de confusão e difusão, essenciais para a segurança de cifras simétricas como o AES, podemos recorrer ao modelo conceitual de Shannon. A Figura \ref{fig:shannon_model} apresenta o diagrama de um sistema de criptografia secreta, onde a chave secreta ($K$) e o texto claro ($M$) são combinados por uma função de cifragem ($E_K$) para produzir o texto cifrado ($C$). A confusão busca obscurecer a relação entre a chave e o texto cifrado, enquanto a difusão visa espalhar a redundância do texto claro por todo o texto cifrado.

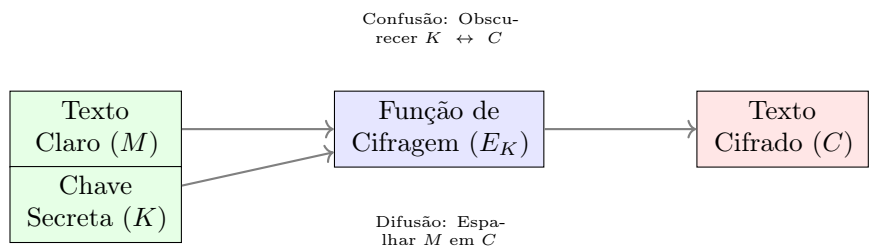
\begin{figure}[h!]
    \centering
    \begin{tikzpicture}[
        node distance=2cm,
        block/.style={rectangle, draw, fill=blue!10, text width=2.5cm, minimum height=1cm, text centered, font=\small},
        input/.style={rectangle, draw, fill=green!10, text width=2cm, minimum height=0.8cm, text centered, font=\small},
        output/.style={rectangle, draw, fill=red!10, text width=2cm, minimum height=0.8cm, text centered, font=\small},
        arrow/.style={->, thick, black!50}
    ]
    \node[input] (M) {Texto Claro ($M$)};
    \node[input, below=1cm of M.west, anchor=west] (K) {Chave Secreta ($K$)};
    \node[block, right=2cm of M.south east, anchor=south west] (Enc) {Função de Cifragem ($E_K$)};
    \node[output, right=2cm of Enc] (C) {Texto Cifrado ($C$)};

    \draw[arrow] (M) -- (Enc);
    \draw[arrow] (K) -- (Enc);
    \draw[arrow] (Enc) -- (C);

    \node[font=\tiny, text width=3cm, align=center, above=0.5cm of Enc.north] {Confusão: Obscurecer $K \leftrightarrow C$};
    \node[font=\tiny, text width=3cm, align=center, below=0.5cm of Enc.south] {Difusão: Espalhar $M$ em $C$};

    \end{tikzpicture}
    \caption{Diagrama de Shannon para um sistema de criptografia secreta, princípios de confusão e difusão.}
    \label{fig:shannon_model}
\end{figure}

O \gls{chacha20}, introduzido por Daniel J. Bernstein \cite{ChaCha}, ilustra a linha das cifras de fluxo modernas que utilizam apenas operações de adição modular, rotação e XOR (modelo ARX). Este design evita a dependência em tabelas de substituição (S-boxes) e operações de multiplicação, o que aumenta tanto a robustez contra ataques de tempo e cache quanto a velocidade em implementações de software puro. O ChaCha20 opera em um bloco de estado de 512 bits (16 palavras de 32 bits), que é inicializado com uma chave de 256 bits, um \gls{nonce} de 96 bits e um contador de 32 bits. As rodadas de transformação aplicam uma série de operações ARX (quarto de rodada) para gerar um fluxo de chaves pseudoaleatório. Cada quarto de rodada envolve quatro operações ARX em quatro palavras do estado, garantindo uma rápida difusão dos bits. O fluxo de chaves gerado é então XORado com o texto claro para produzir o texto cifrado. Sua eficiência e simplicidade o tornaram o padrão de fato em conexões TLS para dispositivos móveis, especialmente em ambientes onde não há aceleração AES em hardware.

Além das cifras de bloco e de fluxo, outro eixo central das metodologias clássicas é constituído pelas {funções hash criptográficas}. Essas funções mapeiam dados de tamanho arbitrário para um valor de tamanho fixo (o "resumo" ou "hash"), com propriedades de segurança específicas:
\begin{enumerate}
    \item \textit{Resistência à pré-imagem}: É computacionalmente inviável encontrar uma mensagem $M$ que produza um hash $H(M)$ dado apenas o hash.
    \item \textit{Resistência à segunda pré-imagem}: É computacionalmente inviável encontrar uma mensagem $M' \neq M$ que produza o mesmo hash que uma mensagem $M$ dada.
    \item \textit{Resistência à colisão}: É computacionalmente inviável encontrar duas mensagens distintas $M_1 \neq M_2$ que produzam o mesmo hash, ou seja, $H(M_1) = H(M_2)$.
\end{enumerate}
O \gls{sha2} (Secure Hash Algorithm 2), estabelecido como FIPS 180-4 \cite{SHA2}, inclui funções como SHA-256 e SHA-512. Ele segue o modelo Merkle–Damgård, processando blocos de dados de forma iterativa. Cada bloco de mensagem é comprimido com o estado interno anterior para produzir um novo estado, que é então usado na próxima iteração. A função de compressão interna do SHA-2 é uma função uniderecional baseada em operações bit a bit (AND, OR, XOR, NOT) e rotações, aplicada a oito variáveis de trabalho. A segurança do SHA-2 é baseada na complexidade de encontrar colisões ou pré-imagens para suas funções de compressão.

O \gls{sha3} (Secure Hash Algorithm 3), também conhecido como Keccak \cite{SHA3}, adota uma construção de esponja, que difere fundamentalmente do modelo Merkle–Damgård. A construção de esponja opera em um estado interno de $b$ bits, dividido em uma parte de taxa $r$ e uma parte de capacidade $c$ ($b = r+c$). A função de permutação $f$ (Keccak-f) é aplicada iterativamente. Durante a fase de "absorção", blocos da mensagem são XORados com a parte de taxa do estado e $f$ é aplicada. Na fase de "compressão", a parte de taxa do estado é extraída como saída, e $f$ é aplicada novamente para gerar o próximo bloco de saída. A capacidade $c$ é uma medida de segurança, garantindo que ataques de colisão exijam $2^{c/2}$ operações. Funções hash fornecem integridade de mensagens, derivação de chaves e suporte a assinaturas digitais.

Por fim, destacam-se as metodologias de {assinatura digital}, que oferecem autenticidade, integridade e não-repúdio em comunicações. Sistemas clássicos incluem \gls{pss} (Probabilistic Signature Scheme) e \gls{ecdsa} (Elliptic Curve Digital Signature Algorithm):
\begin{itemize}
    \item O \textbf{RSA-PSS} combina o algoritmo RSA com um esquema de preenchimento probabilístico para aumentar a segurança e evitar ataques de forja. A assinatura de uma mensagem $M$ envolve o cálculo de um hash $H(M)$, a aplicação de uma função de geração de máscara (MGF) para criar um preenchimento aleatório, a combinação do hash e do preenchimento com um valor aleatório (salt) para formar uma mensagem formatada, e a aplicação da função de assinatura RSA com a chave privada do signatário sobre essa mensagem formatada. A aleatoriedade introduzida pelo salt e pela MGF torna cada assinatura única, mesmo para a mesma mensagem.
    \item O \textbf{ECDSA} utiliza as propriedades das curvas elípticas para gerar assinaturas digitais. Dada uma mensagem $M$, um hash $H(M)$ é calculado. Um número aleatório $k$ é escolhido, e um ponto aleatório $R = kG$ (onde $G$ é o ponto gerador da curva) é calculado. As coordenadas $x_R$ e $y_R$ de $R$ são usadas para derivar $r = x_R \pmod{n}$, onde $n$ é a ordem do ponto $G$. Em seguida, $s$ é calculado usando a chave privada do signatário $d_A$, o hash $H(M)$, e $r$ através da fórmula $s = k^{-1}(H(M) + d_A r) \pmod{n}$. A assinatura é o par $(r, s)$. A verificação envolve o uso da chave pública do signatário $Q_A$ para reconstruir um ponto na curva e verificar se suas coordenadas correspondem a $r$, utilizando $u_1 = H(M)s^{-1} \pmod{n}$ e $u_2 = rs^{-1} \pmod{n}$ para calcular $R' = u_1 G + u_2 Q_A$. A assinatura é válida se $x_{R'} \equiv r \pmod{n}$.
\end{itemize}
Evoluções modernas exploram construções baseadas em funções hash \cite{Hash_Signatures}, como XMSS (eXtended Merkle Signature Scheme) e SPHINCS+ (Stateless Practical Hash-based Signatures), que são resistentes inclusive a adversários quânticos e já começam a ser integradas a PKIs convencionais, representando uma ponte para a era pós-quântica.

Essas metodologias permeiam protocolos fundamentais da segurança moderna: \textit{key exchange} por RSA/DH/ECC, confidencialidade via AES/ChaCha20, integridade por SHA-2/SHA-3 e autenticação por assinaturas digitais. Estas abordagens formaram a base do comércio eletrônico, das comunicações seguras e da infraestrutura da Internet, constituindo o núcleo do que chamamos hoje de criptografia clássica.

\section{Estado da Arte}

A criptografia clássica permanece onipresente, sendo constantemente atualizada com algoritmos mais seguros e eficientes em resposta a vulnerabilidades descobertas e ao aumento do poder computacional disponível. Atualmente, a literatura especializada reúne uma série de contribuições que configuram o estado da arte desse campo, abrangendo algoritmos simétricos, assimétricos e funções hash.

O sistema RSA, proposto em 1977 por Rivest, Shamir e Adleman, foi a primeira implementação prática de criptografia de chave pública. Sua segurança baseia-se na dificuldade da fatoração de números inteiros grandes: dada uma chave pública $(n,e)$, um emissor cifra uma mensagem pela operação $C \equiv M^e \pmod{n}$, sendo a recuperação possível apenas ao detentor da chave privada $d$, com $M \equiv C^d \pmod{n}$. Apesar da sua robustez histórica, diversos trabalhos recentes avaliam sua longevidade frente à crescente capacidade computacional e a ameaça quântica \cite{RSA_Longevity}. Na prática contemporânea, recomenda-se RSA-PSS para assinatura e \gls{oaep} para encapsulamento, com implementações em tempo constante e \gls{blinding} no decifrar/assinar para mitigar ataques de \emph{timing} e \emph{faults}; verificações estritas de preenchimento evitam \emph{oracle attacks} no estilo Bleichenbacher.

No campo da criptografia simétrica, o \textbf{Advanced Encryption Standard (AES)} consolidou-se como o padrão global após ser escolhido pelo NIST em 2001. Desenvolvido por Daemen e Rijmen \cite{AES}, o AES estrutura-se em blocos de 128 bits e utiliza uma rede de substituição-permutação em $\mathbb{F}_{2^8}$, garantindo propriedades de confusão e difusão. É atualmente empregado largamente em protocolos de rede (TLS, IPsec, WPA2/3), sistemas operacionais e dispositivos embarcados. 

Em produção, o AES é usado predominantemente em modos \gls{aead}: o \textbf{AES-GCM} combina contagem, \gls{ctr}, com autenticação \gls{ghash} sobre $\mathbb{F}_{2^{128}}$, com provas formais que vinculam a vantagem do adversário ao número de consultas e ao volume total autenticado \cite{AESGCMProof}. Requisito operacional crítico é a unicidade de \emph{nonce} por chave; reutilização compromete confidencialidade e integridade. Quando tolerância a erro de \emph{nonce} é necessária, \gls{siv} (RFC~5297/9146) fornece \emph{misuse resistance} ao custo de desempenho. Para criptografia de disco, \gls{xts} introduz \emph{tweak} por setor, mas não autentica; em cenários que exigem integridade, usa-se MAC ou AEAD em camada superior. Resultados de criptanálise como ataques de \emph{biclique} reduzem marginalmente a complexidade teórica contra AES-128, porém sem impacto prático com parâmetros recomendados e chaves frescas.

Complementando esse cenário, Bernstein apresentou em 2008 o \textbf{ChaCha20} \cite{ChaCha}, uma cifra de fluxo derivada da Salsa20. Diferente do AES, o ChaCha20 utiliza apenas operações de adição modular, rotação e XOR (modelo ARX). Sua eficiência em software, segurança analisada e ausência de dependência em aceleração AES de hardware fizeram-no ser adotado em larga escala em dispositivos móveis (Android, iOS), navegadores web e no protocolo \gls{tls13}. Em combinação com \gls{poly1305}, forma o AEAD \textbf{ChaCha20-Poly1305}, cujo MAC é do tipo Carter–Wegman com chave efêmera; a análise cobre não só a cifra e o MAC isoladamente, mas a composição e a gestão de \emph{nonces} e chaves em cenários de alto tráfego \cite{ChaChaTLS}. A seleção prática entre AES-GCM e ChaCha20-Poly1305 é guiada por latência, disponibilidade de AES-NI/ARMv8 Crypto Extensions, consumo energético e estabilidade de desempenho em microarquiteturas variadas.

As funções hash também evoluíram. O \textbf{SHA-2}, refinado em 2010 \cite{SHA2}, consolidou-se como função de compressão iterativa do tipo Merkle–Damgård, permitindo gerar resumos criptográficos de 256 ou 512 bits. Posteriormente, Bertoni et al. introduziram o \textbf{SHA-3 (Keccak)} \cite{SHA3}, padrão oficializado em 2015 após competição internacional no NIST. Diferentemente do SHA-2, utiliza uma construção baseada em \emph{função esponja}, onde $Z = \text{Trunc}_\ell(f(P))$, com estado $b=r+c$ (taxa $r$ e capacidade $c$); a resistência a colisões escala como $2^{c/2}$. Na prática de protocolos, a derivação de chaves emprega amplamente \gls{hkdf} (\gls{hmac}-based KDF), separando extração e expansão e habilitando \emph{domain separation} por rótulos de contexto; para proteção de senhas, \gls{argonn} (custeio em memória) reduz a viabilidade de paralelização em GPUs/ASICs.

No campo das curvas elípticas, Bos et al. \cite{Curve25519} destacaram a \gls{curve25519} (2017) como candidata segura de próxima geração, utilizada em protocolos como Signal, TLS 1.3 e WireGuard. Esta curva oferece operações rápidas em dispositivos móveis, baseando-se no logaritmo discreto em curvas elípticas, com a equação $E: y^2 = x^3 + 486662x^2 + x$ sobre $\mathbb{F}_p$. Na prática atual, ECDH sobre \gls{x25519}/X448 e assinaturas \textbf{EdDSA} (\gls{ed25519}/Ed448) consolidaram-se pela codificação canônica, fórmulas eficientes e verificações simplificadas em tempo constante. Em ECDSA, a segurança depende criticamente de \emph{nonces} efêmeros imprevisíveis (ou determinísticos RFC~6979); falhas nesse componente levam à extração direta de chaves. Ataques de \emph{invalid-curve} e \emph{twist} motivam validação de pontos ou o uso de curvas com propriedades que minimizam superfícies de ataque; em X25519, o \emph{clamping} do escalar e as verificações de codificação contribuem para robustez.

Paralelamente, Albrecht et al. \cite{SymmAttacks} conduziram, em 2018, uma análise abrangente de ataques diferenciais e lineares contra cifras simétricas modernas, avaliando o efeito de avanços em técnicas criptanalíticas sobre cifras reduzidas. Esse estudo reforça a importância de esquemas com número de rodadas elevadas para resistir a ataques práticos. Entre contribuições mais recentes, Halderman e Heninger \cite{RSA_Longevity} estudaram, em 2020, a resiliência de chaves RSA de diferentes comprimentos diante do aumento do poder computacional. Em 2021, Gueron e Lindell apresentaram uma prova formal da segurança do AES no modo GCM \cite{AESGCMProof}, demonstrando que a autenticação via GHASH ($X \cdot H \pmod{2^{128}}$) tem limites de segurança explicitamente quantificados em função de \emph{nonces}, consultas e comprimento de mensagens. Em 2022, Mateus et al. \cite{AES_GCM} propuseram otimizações específicas de \textbf{AES-GCM} para dispositivos embarcados de baixo consumo, explorando paralelismo de GHASH, reuso seguro de subchaves de rodada e agendamento eficiente em pipelines restritos. Já em 2023, Hülsing et al. \cite{Hash_Signatures} avançaram na pesquisa de assinaturas digitais baseadas em hash (XMSS, SPHINCS+), empregando árvores de Merkle como $\text{root} = H(H(L_1) || H(L_2) || \dots || H(L_n))$ e integrando essas construções a PKIs clássicas como alternativa resistente a adversários quânticos. Ainda em 2023, Grothe et al. \cite{ChaChaTLS} detalharam o uso prático de \textbf{ChaCha20-Poly1305} em TLS 1.3 e \gls{quic}, medindo desempenho, latências e resiliência operacional em redes heterogêneas e confirmando sua adequação em ambientes móveis onde AES pode ser menos eficiente.

No plano de protocolos e interoperabilidade, o \textbf{TLS 1.3} impõe \gls{pfs} via (EC)DHE, reduz a superfície de negociação removendo ciphersuites legadas e fixa um conjunto enxuto de AEADs; \textbf{QUIC} incorpora TLS 1.3 no transporte, introduzindo rekeying frequente e baixa latência. O \gls{hpke} (RFC~9180) padroniza encapsulamento híbrido no paradigma \gls{hpkehybrid}: um \gls{kem} (p.ex., X25519) estabelece um segredo $Z$; HKDF extrai e expande materiais ($\mathsf{key}$, $\mathsf{nonce}$, $\mathsf{ad}$); e um AEAD (AES-GCM ou ChaCha20-Poly1305) protege dados. Essa modularidade facilita migração futura para KEMs pós-quânticos preservando o \gls{dem} e práticas operacionais existentes.

Do ponto de vista de implementação segura, o estado da arte enfatiza código \emph{constant-time} (sem ramos/acessos de memória dependentes da chave), \gls{masking}/\gls{blinding} em RSA/ECC, geradores de entropia conformes (NIST SP 800-90), validações rigorosas de entrada (pontos ECC, formatos, comprimentos), \gls{domainsep} e rotação de chaves baseada em volume/tempo. Para AEADs, políticas de orçamento definem limites de mensagens e bytes por chave a fim de manter a probabilidade de forja abaixo de alvos explícitos; contadores/IVs monotônicos evitam colisões de \emph{nonce}. Para ECC, auditorias incluem checagem de \emph{cofactor}, rejeição de pontos fora do subgrupo e codificações canônicas.

Em síntese, a prática contemporânea converge para AEADs com provas e limites operacionais explícitos, ECC moderna (X25519/Ed25519) com implementações em tempo constante e validações robustas, protocolos com PFS e desenho modular \gls{hpkectx} e governança operacional de chaves e \emph{nonces} para preservar garantias teóricas em ambientes reais. Na Tabela \ref{tab:estado_arte_classica} esses artigos e contribuições são sintetizados, oferecendo uma visão cronológica comparativa do estado da arte da criptografia clássica.

\begin{table}[h!]
\centering
\scriptsize
\caption{Principais artigos representativos do estado da arte em criptografia clássica}
\label{tab:estado_arte_classica}
\begin{tabular}{p{1cm}p{4cm}p{9cm}}
\hline
\textbf{Ano} & \textbf{Algoritmo/Estudo} & \textbf{Descrição} \\
\hline
1977 & Rivest, Shamir \& Adleman (RSA) \cite{RSA_Longevity} & Primeiro sistema prático de chave pública; cifra $C \equiv M^e \pmod{n}$ e decifra $M \equiv C^d \pmod{n}$ com segurança baseada na fatoração. \\
2001 & Daemen \& Rijmen (AES) \cite{AES} & Padrão global de cifra em blocos de 128 bits; rede de permutação-substituição com operações em $\mathbb{F}_{2^8}$. \\
2008 & Bernstein (ChaCha20) \cite{ChaCha} & Cifra de fluxo ARX. Em AEAD com Poly1305, fornece desempenho consistente sem AES-NI. \\
2010 & NIST (SHA-2 refinement) \cite{SHA2} & SHA-256/512 por compressão iterativa Merkle–Damgård; base para HMAC/HKDF. \\
2015 & Bertoni et al. (SHA-3/Keccak) \cite{SHA3} & Esponja com capacidade $c$; XOFs (SHAKE) e propriedades úteis para construção de primitivos. \\
2017 & Bos et al. (Curve25519) \cite{Curve25519} & ECDH eficiente em Montgomery; base para X25519/Ed25519 em protocolos modernos. \\
2018 & Albrecht et al. (Attacks on Symmetric Ciphers) \cite{SymmAttacks} & Análises diferenciais/lineares em ciphers reduzidos; informam margens de segurança e número de rodadas. \\
2020 & Halderman et al. (RSA longevity) \cite{RSA_Longevity} & Atualiza viabilidade de tamanhos de chave RSA frente a poder computacional crescente. \\
2021 & Gueron \& Lindell (AES-GCM proofs) \cite{AESGCMProof} & Vínculos formais entre vantagem do adversário, consultas e comprimento autenticado. \\
2022 & Mateus et al. (AES-GCM optimizations) \cite{AES_GCM} & Paralelismo de GHASH e melhorias para hardware embarcado. \\
2023 & Hülsing et al. (\gls{hashfamily}) \cite{Hash_Signatures} & XMSS/SPHINCS+ integráveis a PKIs clássicas, caminho resistente a quânticos. \\
2023 & Grothe et al. (ChaCha20-Poly1305 in TLS/QUIC) \cite{ChaChaTLS} & Estudos comparativos de desempenho/segurança operacional em redes e dispositivos móveis. \\
2024 & Lee et al. (AES-CTR/ECB on GPUs) \cite{Lee2024AES} & Implementação bit-sliced de AES-CTR e AES-ECB em GPUs com aplicações para FrodoKEM e busca exaustiva de chaves em alta velocidade. \\
2024 & Sahu \& Mazumdar (Quantum threats analysis) \cite{Sahu2024Quantum} & Análise do estado da arte sobre ameaças quânticas à criptografia clássica (RSA, ECC, AES); revisão de QKD e algoritmos pós-quânticos. \\
2025 & Selvi \& Sakthivel (SymECCipher) \cite{Selvi2025Hybrid} & Framework híbrido ECC-AES para proteção de dados em nuvem; redução de 25-40\% no overhead computacional comparado a sistemas tradicionais. \\
2025 & Jiang et al. (AES quantum circuits) \cite{Jiang2025AES} & Circuitos quânticos eficientes para AES com técnica CCQC; redução de 2.3\% em TDW e 45.2\% em FDW para implementação do S-box. \\
\hline
\end{tabular}
\end{table}

\section{Questões para Reflexão e Pesquisa Futura}

Ao concluir este capítulo sobre criptografia clássica, é importante não apenas assimilar os conceitos e as técnicas aqui apresentadas, mas também questionar criticamente suas premissas, limitações e implicações práticas. Considere, por exemplo, as seguintes provocações: quais são as hipóteses de segurança (assunções sobre o adversário, entropia das chaves, modelos de erro) que sustentam o uso de RSA, ECC e AES em seu ambiente, e em que cenários essas hipóteses deixam de ser adequadas? Como mudarão essas decisões caso o horizonte temporal de confidencialidade dos dados se estenda para décadas, considerando a ameaça de “harvest now, decrypt later”?

Reflita sobre a relação entre teoria e implementação: até que ponto as propriedades formais de um algoritmo permanecem válidas quando ele é traduzido para código otimizado e executado em hardware real? Que tipos de vulnerabilidades por canais laterais (timing, cache, consumo energético, emissões) são mais prováveis nas implementações correntes e que práticas de engenharia, tanto de software quanto de hardware, são necessárias para mitigá-las sem sacrificar desempenho? Em cenários onde se utiliza \gls{hsm} ou TPM, quais compromissos operacionais e de confiança delegada estamos implicitamente aceitando, e como auditar de forma eficaz esses componentes?

Questione também as escolhas arquiteturais: como a estrutura de autoridade de certificação (PKI), políticas de revogação e ciclos de vida de chaves influenciam a resiliência de um sistema distribuído diante de comprometimentos parciais? Quais são os custos reais (técnicos, operacionais e legais) de rotacionar chaves e atualizar algoritmos em larga escala, e que mecanismos de automação e orquestração precisam existir para que essas operações sejam viáveis sem causar interrupções? Em que situações a simplicidade de uma solução clássica justifica postergar a migração, e em que situações esse adiamento aumenta o risco de maneira inaceitável?

No nível de protocolos, indague sobre as interações entre primitivas: como falhas em domain separation, encodings ou no tratamento de erros em um handshake podem abrir vetores de ataque imprevistos? De que forma decisões aparentemente pequenas, escolha de KDF, formato de padding, gestão de tickets de sessão, afetam a segurança composicional de uma pilha criptográfica completa? Quais métricas e benchmarks deveriam ser considerados prioritários ao avaliar mudanças: latência de estabelecimento, custo por operação, impacto em MTU/fragmentação, robustez a perda de pacotes e custos energéticos em dispositivos móveis?

Pense também nas dimensões de governança e conformidade: como os requisitos regulatórios e contratuais (por exemplo, exigências de proteção de dados pessoais ou de setores regulados) influenciam decisões técnicas sobre quais algoritmos adotar e quando? Que processos internos e evidenciais (logs imutáveis, relatorios de conformidade, testes de regressão) são necessários para demonstrar que uma migração ou um patch criptográfico não criou retrocessos de segurança? Em contextos multinacionais, como balancear diferenças regulatórias e garantir interoperabilidade entre jurisdições?

Finalmente, abra questões de caráter investigativo e ético: quais são as implicações sociais e de privacidade se capacidades de descriptografia retroativa forem efetivamente acessíveis no futuro? Que responsabilidades têm estados, organizações e fornecedores de tecnologia em mitigar riscos decorrentes da obsolescência de esquemas hoje considerados seguros? E, por fim, quais linhas de pesquisa aplicada, por exemplo, verificações formais de implementações, suites padronizadas de testes de canais laterais, ferramentas automáticas de migração e validação contínua, teriam maior impacto prático ao reduzir a distância entre o estado da arte teórico e a operação segura em produção?

Estas perguntas não pretendem esgotar o tema, mas servir de ponto de partida para investigações, estudos de caso e experimentos práticos que aprofundem a compreensão do leitor e fomentem um pensamento crítico orientado à ação.

\chapter{Criptografia Homomórfica}

A criptografia homomórfica (CH) descreve uma família de esquemas de cifra que preservam a possibilidade de computação útil diretamente sobre dados cifrados. Em termos práticos, um provedor de serviços consegue executar operações aritméticas ou booleanas sobre um conjunto de cifras e produzir como resultado outro cifrado, sem jamais observar os textos claros subjacentes. Apenas o detentor da chave privada consegue, ao final, decifrar o resultado e obter o valor correto em claro. Essa propriedade resolve um dilema tradicional em segurança da informação: como extrair valor de dados sensíveis em ambientes potencialmente não confiáveis, como a nuvem pública, sem expô-los a operadores, administradores de sistema ou terceiros?

A ideia central pode ser expressa por uma igualdade conceitual. Sejam $\mathsf{Enc}$ e $\mathsf{Dec}$ as funções de cifragem e decifragem, e seja $\circ$ uma operação desejada no espaço dos textos claros, como soma ou produto. Um esquema é homomórfico para $\circ$ se existir uma operação correspondente $\star$ no espaço dos cifrados tal que:
\[
\mathsf{Dec}\!\left(\mathsf{Enc}(m_1)\ \star\ \mathsf{Enc}(m_2)\right) \;=\; m_1 \circ m_2.
\]
Historicamente, os primeiros exemplos foram \emph{parcialmente} homomórficos, como Paillier (aditivo) e RSA (multiplicativo), suficientes para tarefas de agregação simples ou voto eletrônico. O salto qualitativo ocorreu com a construção de Gentry (2009), que introduziu o \gls{bootstrapping}: o processo de avaliar, \emph{sobre o próprio cifrado}, um circuito que emula a decifragem, ``refrescando'' o ruído e permitindo computações de profundidade essencialmente ilimitada. A partir daí, a computação sob sigilo tornou-se um objetivo tangível em larga escala.

\section{Fundamentos Matemáticos}

A segurança moderna em CH fundamenta-se, em grande medida, em problemas difíceis sobre reticulados, notadamente LWE (\textit{Learning With Errors}) e sua versão em anéis, RLWE. Em linhas gerais, mensagens são embutidas em estruturas algébricas, como o anel quociente:
\[
R_q \;=\; \mathbb{Z}_q[x]/(x^N + 1),
\]
e cifradas em formas que incorporam um ruído discreto pequeno, cuidadosamente controlado. Uma forma simplificada de cifra pode ser escrita como $c(x) = (a(x), b(x)) \in R_q^2$, com:
\[
b(x) \;=\; a(x)\cdot s(x) + \Delta \cdot m(x) + e(x) \pmod{q},
\]
onde $s(x)$ é a chave secreta, $\Delta$ é um fator de escala e $e(x)$ é o ruído. Operações homomórficas sobre $c(x)$ promovem operações correspondentes sobre $m(x)$, mas também fazem crescer o ruído. Enquanto esse ruído permanecer abaixo de um certo limiar, a decifragem recupera exatamente a mensagem. Quando o orçamento de ruído se aproxima do limite, emprega-se o \textit{bootstrapping} para restaurar a margem de segurança.

Dentro desse panorama, destacam-se três famílias de esquemas. BGV e BFV suportam aritmética exata sobre inteiros modulares e são preferidos em cenários de contagem, somas, multiplicações e estatística discreta. CKKS, por sua vez, introduz aritmética \emph{aproximada} em números reais/complexos, com um mecanismo de \gls{rescaling} que controla as escalas numéricas e o crescimento do erro, sendo particularmente adequado a aprendizado de máquina e inferência numérica. Em outro eixo, \gls{fhew} e \gls{tfhe} priorizam a avaliação rápida de portas booleanas, com \textit{bootstrapping} muito eficiente aplicado porta a porta, o que facilita a execução de circuitos lógicos finos com latências menores por operação.

Do ponto de vista de suposições de segurança, LWE e RLWE reduzem-se, sob parâmetros apropriados, à dureza de problemas geométricos em reticulados como \gls{gapsvp} e \gls{sivp} em média, através de reduções clássicas e em anel. A hipótese RLWE explora a estrutura de $R=\mathbb{Z}[x]/(f)$ (com $f$ tipicamente ciclotômico) e a aritmética em $R_q$, possibilitando convoluções via produto polinomial modular com custos quase-lineares usando NTT (\textit{Number Theoretic Transform}). A NTT substitui a FFT complexa por uma transformada discreta sobre $\mathbb{Z}_q$ utilizando raízes primitivas da unidade, permitindo multiplicações de polinômios de grau $N-1$ em $O(N \log N)$ e amortizando o custo de operações bateladas, \gls{simd}, por slots.

A \emph{amostragem de erro} é componente crítico: distribuições discretas subgaussianas (tipicamente discretizações de gaussianas com desvio $\sigma$) controlam a entropia e o crescimento do ruído. As implementações empregam amostragem por tabelas cumulativas (\textit{Cumulative Distribution Table}), rejeição (\textit{rejection sampling}) ou métodos de Knuth–Yao, calibrados para garantir tanto a correção (taxas de erro desprezíveis) quanto a segurança de entropia. Em RLWE, o erro é um polinômio com coeficientes independentes, e sua norma afeta diretamente os limites de decifragem e a profundidade multiplicativa suportável antes do \textit{bootstrapping}.

Codificações de mensagem adotam duas estratégias principais: codificação por coeficientes (inteiros reduzidos modulo $t$ embutidos nos coeficientes de $m(x)$) e codificação por avaliação (\gls{crt} packing/\emph{slot packing}), em que o espaço $R_t$ é decomposto em um produto de corpos/campos residuais e a mensagem é inserida por slots paralelos. Essa última habilita operações vetoriais homomórficas (rotações e permutações via \gls{galoiskeys}) com grande ganho de \emph{throughput} em aplicações de ML e estatística.

A profundidade multiplicativa de um circuito homomórfico determina o orçamento de ruído necessário. Em BGV/BFV, composições de multiplicações exigem \emph{relinearization} para reduzir o grau do \emph{ciphertext} (de grau 2 para 1), via chaves de relinearização pré-computadas. Em CKKS, multiplicações elevam a escala numérica e o ruído; aplica-se \emph{rescaling} (divisão por um fator $p$ embutido no módulo) para manter escalas dos operandos alinhadas e o erro sob controle. A cadeia de módulos (\emph{modulus switching chain}) é definida previamente (por exemplo, $q=q_L q_{L-1}\dots q_0$), e cada operação consome níveis, determinando a profundidade máxima antes de necessário \textit{bootstrapping} ou recriptografia.

O \textit{bootstrapping} realiza uma “auto-decriptação” homomórfica: avalia-se a função de decifragem sobre o próprio \emph{ciphertext} usando chaves especiais (por exemplo, \emph{Galois keys}, \emph{rotation keys}, \emph{key-switching keys}). Em esquemas de portas (FHEW/TFHE), o \textit{bootstrapping} é formulado como uma transformação rápida que mapeia um \emph{ciphertext} ruidoso de uma porta para um \emph{ciphertext} com ruído “reinicializado”, usando \emph{FFT/NTT}-like e \emph{accumulators} no domínio de Fourier toroidal. Em esquemas aritméticos (BGV/BFV/CKKS), \textit{bootstrapping} envolve aproximações polinomiais de funções não lineares, trocas de base e \gls{keyswitch} encadeado, com custo significativamente maior, mas que viabiliza avaliação de circuitos de alta profundidade. 
A escolha de parâmetros segue linhas mestras: o grau $N$ do polinômio (geralmente potência de 2 com polinômios ciclotômicos $x^N+1$) define a dimensão do reticulado subjacente; o módulo $q$ é fatorado em primos adequados à NTT (raiz primitiva de ordem $2N$) e calibrado para suportar a profundidade alvo; o tamanho do ruído inicial, via $\sigma$, equilibra taxa de erro e segurança; e o módulo de mensagem $t$ controla precisão (BGV/BFV) ou escala (CKKS). A segurança clássica é estimada com ferramentas como LWE Estimator, sintonizando $(N, q, \sigma)$ para níveis de 128, 192 ou 256 bits. Em CKKS, além da segurança, é necessário projetar o “orçamento de precisão”: o erro relativo acumulado deve ficar abaixo do erro numérico tolerado pelo algoritmo alvo (por exemplo, tolerâncias em inferência).

Operações avançadas, como rotações e somas entre slots, são implementadas por automorfismos do anel $R_q$ (geralmente mapeando $x \mapsto x^g \bmod (x^N+1)$ com $g$ coprimo a $2N$) acompanhados de \emph{key switching}. As \emph{Galois keys} encapsulam a habilidade de aplicar esses automorfismos sem acesso à chave secreta, ao custo de expansão de chaves públicas auxiliares e aumento de ruído controlado. De modo similar, a relinearização é um caso particular de \emph{key switching} que reduz a representação do \emph{ciphertext} após multiplicações.

Quanto ao custo computacional, multiplicações polinomiais com NTT dominam o tempo de execução em BGV/BFV/CKKS; otimizações exploram segmentação de cache, paralelismo \emph{intra-NTT}, versões \emph{in-place} e seleção de primos “amigáveis” (por exemplo, 60 bits em arquiteturas 64-bit). Em FHEW/TFHE, o gargalo desloca-se para a etapa de \textit{bootstrapping} por porta, onde implementações de FFT/NTT de precisão mista, pré-computação de \emph{accumulators} e vetorização por SIMD (AVX2/AVX-512/Neon) são determinantes para latência.

No eixo de correção e segurança, a decriptação correta requer que a norma do ruído permaneça dentro de uma “janela de decodificação”. Em BFV/BGV, isso se expressa como um limite sobre $|e(x)|_\infty$ relativo ao \emph{modulus} de mensagem e ao fator de \emph{scaling} utilizado nos estágios de \gls{modswitch}. Em CKKS, o erro é relativo à escala e acumula-se multiplicativamente/adiativamente; projetam-se pipelines com \emph{rescaling} e normalização de escalas para minimizar perda de bits de precisão significativos. A segurança, por sua vez, herda a robustez de LWE/RLWE contra ataques de \emph{lattice reduction} (\gls{bkz}, variantes com \gls{pruning} e \gls{quantumsieving}); parâmetros modernos visam margens alinhadas às recomendações de segurança de 128 bits ou superiores.

Finalmente, a escolha entre as famílias depende do tipo de computação alvo: BGV/BFV para aritmética modular exata, com suporte eficiente a comparações e agregações discretas; CKKS quando o problema aceita erro aproximado e escalas numéricas (aprendizado de máquina, estatística, métodos numéricos); FHEW/TFHE quando o circuito é booleano granular e a latência por operação de porta é crítica. Em todos os casos, o projeto requer orquestrar NTTs, amostragem de erros, \emph{key switching}/relinearização e, quando necessário, \textit{bootstrapping}, sob um orçamento de ruído e segurança predefinidos que garantam decifragem correta e resistência contra adversários clássicos e quânticos.

\section{Modelos de Capacidade e Desempenho}

A literatura classifica os esquemas segundo a profundidade de circuito suportada e o tipo de operação. Esquemas parcialmente homomórficos oferecem uma única operação (soma \emph{ou} produto) em profundidade arbitrária; são úteis, por exemplo, quando apenas agregações são necessárias. Esquemas \emph{somewhat homomorphic} \gls{she} suportam somas e multiplicações, porém com profundidade limitada pelo orçamento de ruído. Esquemas \gls{leveled} ampliam essa capacidade para uma profundidade previamente escolhida, sem recorrer a \textit{bootstrapping} durante a execução; os parâmetros criptográficos são dimensionados para aquele nível de complexidade. Por fim, esquemas totalmente homomórficos (FHE) permitem computações de profundidade não limitada, usando \textit{bootstrapping} sempre que necessário para “zerar” o ruído acumulado.

O desempenho depende de três fatores principais. O primeiro é a aritmética de anéis, em que convoluções polinomiais são aceleradas por transformadas número-teóricas (NTT), explorando vetorização (AVX2/AVX-512/Neon) e, em alguns casos, GPU ou FPGA. O segundo é o empacotamento (\gls{packing}), que codifica vetores de dados em um único cifrado e habilita paralelismo do tipo SIMD: uma única operação homomórfica atua em diversos \emph{slots} simultaneamente; rotações e permutações de slots, realizadas via automorfismos do anel acompanhados de \textit{key switching}, permitem implementar somas prefixo, convoluções 1D/2D e camadas lineares de modelos de ML com custos amortizados. O terceiro é a gestão do ruído, que envolve relinearização após multiplicações (reduzindo o grau do \emph{ciphertext} e controlando o crescimento de erro), troca de módulo (\textit{modulus switching}) e, quando necessário, \textit{bootstrapping}. Essas técnicas, combinadas, determinam a latência por operação, o \emph{throughput} e o consumo de memória.

Em BGV/BFV, a profundidade efetiva é governada pela cadeia de módulos ($q = q_L q_{L-1}\dots q_0$): cada multiplicação consome níveis e requer relinearização; \textit{rescaling} não é necessário para exatidão modular, mas a troca de módulo controla a magnitude do ruído para manter a janela de decodificação. Em CKKS, por operar com números aproximados, cada multiplicação eleva a escala e o ruído relativo; \textit{rescaling} alinha as escalas dos operandos e reduz o módulo de trabalho, consumindo níveis da cadeia e impondo um orçamento de precisão que deve ser planejado de acordo com as camadas do circuito numérico. Nos esquemas de portas (FHEW/TFHE), a granularidade é por operação booleana, com \textit{bootstrapping} eficiente porta a porta; a latência por porta é o gargalo dominante, mitigado por FFT/NTT otimizadas e paralelismo SIMD.

A análise de desempenho costuma separar custos em quatro blocos: (i) multiplicações polinomiais via NTT (maior contribuinte de tempo em BGV/BFV/CKKS), (ii) key-switching/relinearização (custos proporcionais ao tamanho de chaves auxiliares e ao número de termos no \emph{ciphertext}), (iii) rotações/permutas (automorfismos + key-switching, críticas para SIMD eficiente), e (iv) \textit{bootstrapping} (quando necessário), que reintroduz uma etapa intensiva, porém que libera a profundidade. Estratégias de engenharia incluem escolher primos “amigáveis” à NTT (com raízes primitivas de ordem $2N$), organizar a memória para maximizar localidade de cache, aplicar NTT \emph{in-place} e explorar paralelismo por níveis e por coeficiente. Em hardware, GPUs beneficiam kernels de NTT/butterflies massivamente paralelos, enquanto FPGAs podem reduzir latência com \emph{pipelines} profundos para \textit{key switching} e \textit{bootstrapping}.

A dimensão do anel ($N$), o orçamento de módulo ($\log q$) e a variância do erro ($\sigma^2$) determinam a segurança estimada (ex.: 128/192/256 bits contra reduções em reticulados) e os recursos de computação/memória. Para uma meta de profundidade $D$ sem \textit{bootstrapping}, dimensiona-se a cadeia de módulos para suportar $D$ multiplicações com margens para relinearização e rotações. Em CKKS, adiciona-se o orçamento de precisão alvo (por exemplo, 30–40 bits de precisão efetiva para inferência), que orienta a seleção das escalas iniciais, dos fatores de \textit{rescaling} e da largura dos primos. Para cargas \emph{data-parallel}, o \emph{packing} maximiza o \emph{throughput} amortizando custos fixos de NTT e \textit{key switching}; o desenho de circuitos deve minimizar rotações e alinhar padrões de acesso para reduzir chaves de Galois necessárias.

A Figura~\ref{fig:modelo-capacidade-desempenho} ilustra um pipeline simplificado que relaciona capacidade (profundidade, com ou sem \textit{bootstrapping}) e custos dominantes por etapa (NTT, relinearização, rotações e \textit{bootstrapping}). Ela serve como guia prático para mapear um circuito alvo em escolhas de parâmetros e otimizações.

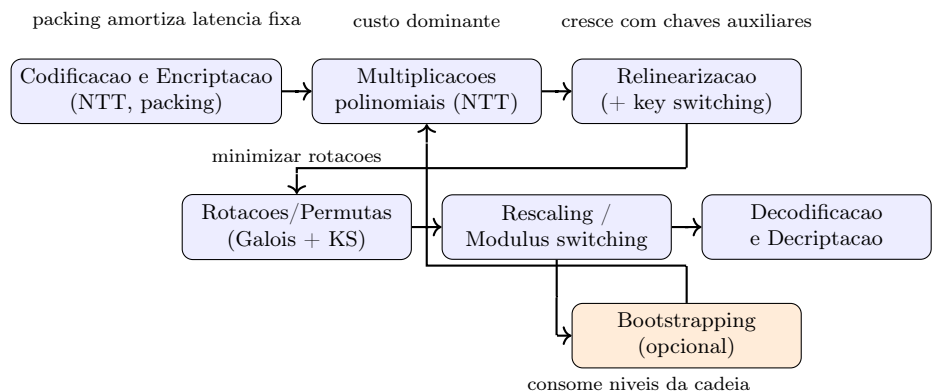
\begin{figure}[h!]
  \centering
  \begin{tikzpicture}[
    scale=0.90, transform shape,
    box/.style={rectangle, draw, rounded corners,
                fill=blue!7, minimum width=3.35cm, minimum height=0.95cm,
                align=center, font=\footnotesize},
    box2/.style={rectangle, draw, rounded corners,
                 fill=orange!15, minimum width=3.35cm, minimum height=0.95cm,
                 align=center, font=\footnotesize},
    port/.style={circle, inner sep=0pt, minimum size=0pt},
    note/.style={font=\scriptsize, align=center}
  ]
    \node[box] (enc) at (-0.3,0)    {Codificacao e Encriptacao\\(NTT, packing)};
    \node[box] (mul) at (3.8,0)  {Multiplicacoes\\polinomiais (NTT)};
    \node[box] (rel) at (7.6,0)  {Relinearizacao\\(+ key switching)};

    \node[box]  (rot) at (1.9,-2.0) {Rotacoes/Permutas\\(Galois + KS)};
    \node[box]  (res) at (5.7,-2.0) {Rescaling /\\Modulus switching};
    \node[box]  (dec) at (9.5,-2.0) {Decodificacao\\e Decriptacao};
    \node[box2] (boot) at (7.6,-3.6) {Bootstrapping\\(opcional)};

    \node[port] (EncOut)  at ($(enc.east)+(0.25,0)$) {};
    \node[port] (MulIn)   at ($(mul.west)+(-0.25,0)$) {};
    \node[port] (MulOut)  at ($(mul.east)+(0.25,0)$)  {};
    \node[port] (RelIn)   at ($(rel.west)+(-0.25,0)$) {};
    \node[port] (RelDown) at ($(rel.south)+(0,-0.65)$) {};

    \node[port] (RotIn)   at ($(rot.north)+(0,0.30)$) {};
    \node[port] (RotOut)  at ($(rot.east)+(0.25,0)$) {};
    \node[port] (ResIn)   at ($(res.west)+(-0.25,0)$) {};
    \node[port] (ResOut)  at ($(res.east)+(0.25,0)$)  {};
    \node[port] (DecIn)   at ($(dec.west)+(-0.25,0)$) {};
    \node[port] (ResDown) at ($(res.south)+(0,-0.55)$) {};
    \node[port] (BootW)   at ($(boot.west)+(-0.22,0)$) {};
    \node[port] (BootN)   at ($(boot.north)+(0,0.55)$) {};
    \node[port] (MulS)    at ($(mul.south)+(0,-0.28)$) {};

    \draw[->, thick] (enc.east) -- (EncOut) -- (MulIn) -- (mul.west);
    \draw[->, thick] (mul.east) -- (MulOut) -- (RelIn) -- (rel.west);

    \draw[->, thick] (rel.south) -- (RelDown) -| (RotIn) -- (rot.north);

    \draw[->, thick] (rot.east) -- (RotOut) -- (ResIn) -- (res.west);
    \draw[->, thick] (res.east) -- (ResOut) -- (DecIn) -- (dec.west);

    \draw[->, thick] (res.south) -- (ResDown) -| (BootW) -- (boot.west);

    \draw[->, thick] (boot.north) -- (BootN) -| (MulS) -- (mul.south);

    \node[note] at (0,1.05)    {packing amortiza latencia fixa};
    \node[note] at (3.8,1.05)  {custo dominante};
    \node[note] at (7.6,1.05)  {cresce com chaves auxiliares};
    \node[note] at (1.9,-0.95) {minimizar rotacoes};
    \node[note] at (6.9,-4.3) {consome niveis da cadeia};
  \end{tikzpicture}
  \caption{Pipeline compacto em dois andares para capacidade e desempenho em HE, com roteamento por portas e rótulos fora das rotas.}
  \label{fig:modelo-capacidade-desempenho}
\end{figure}

\section{Aplicações e Arquiteturas}

A motivação prática para CH emerge de requisitos regulatórios e de negócios que demandam análise de dados sensíveis sem exposição operacional, um problema central para organizações que precisam conciliar privacidade, conformidade jurídica e valor informacional. Em setores regulados, as barreiras ao compartilhamento de dados são substanciais: legislações como o GDPR, HIPAA e LGPD restringem o acesso direto e exigem mecanismos de pseudonimização ou cifragem. Nesse contexto, a criptografia homomórfica passa a viabilizar análises avançadas sobre dados cifrados, reduzindo risco de vazamento e simplificando auditorias de conformidade.

Em saúde, um hospital pode terceirizar o cálculo de indicadores clínicos sobre prontuários cifrados, recebendo de volta apenas resultados também cifrados para posterior decifragem local. Além de proteger identidades de pacientes, esse modelo permite formar \textit{data pools} colaborativos entre instituições, viabilizando estudos multicêntricos sem violar sigilo médico. CH também é empregada em análises genômicas, nas quais operações sobre vetores de DNA e RNA são executadas homomorficamente, permitindo comparações e inferências sem revelar o material biológico original. No setor financeiro, carteiras de clientes podem ser pontuadas por modelos de risco aplicados a atributos cifrados, reduzindo a superfície de ataque associada a ambientes de análise de crédito, \textit{know-your-customer} (KYC) e detecção de fraude. Bancos e seguradoras já demonstram viabilidade de cálculos como regressões logísticas, médias ponderadas e verificações de conformidade sobre dados cifrados, o que possibilita compartilhar dados entre filiais e parceiros sob garantias criptográficas.

Em telecomunicações, CH pode ser aplicada à análise de registros de uso (\textit{Call Detail Records - CDRs}) e de métricas de qualidade de rede, permitindo realizar agregações sob sigilo sem expor o comportamento de usuários individuais. Já em aprendizado de máquina, modelos lineares, árvores rasas e pequenas redes neurais podem executar inferência homomórfica, frequentemente com aproximações polinomiais para funções não lineares, como ReLU, \textit{sigmoid} e \textit{softmax}. O esquema CKKS destaca-se nesses cenários, pois opera com números reais aproximados e suporta cálculo vetorial, sendo utilizado em inferência de modelos CNN compactos e regressões lineares em dados sensíveis. Nos domínios públicos e governamentais, há interesse crescente em aplicações de CH para censos, sistemas eleitorais, validação de políticas sociais e dashboards epidemiológicos, nos quais os dados brutos permanecem cifrados e apenas resultados agregados são revelados. Esses casos evidenciam a capacidade de CH em alinhar exigências éticas e legais com eficiência computacional aceitável, especialmente quando combinada com técnicas de compressão de modelos, quantização de parâmetros e paralelismo de operações vetoriais.

Em arquiteturas de produção, CH raramente opera isolada: tende a ser combinada com outras tecnologias de proteção de dados, formando ecossistemas híbridos de segurança. Computação multipartidária segura (\gls{mps}) fornece garantias baseadas em distribuição de confiança e é útil quando partes distintas colaboram sem revelar insumos individuais. Enclaves de execução confiável (\gls{tee}) reduzem o custo de execução em trechos críticos  geralmente para operações sequenciais de baixa profundidade ou pós-processamento, enquanto técnicas de \textit{differential privacy} mitigam o risco de reidentificação ao publicar estatísticas derivadas. 

O desenho de uma solução industrial frequentemente intercala essas camadas: partes da lógica são avaliadas homomorficamente, enquanto etapas de \emph{pre-} e \emph{post-processing} residem em enclaves ou pipelines MPC, equilibrando custo, latência e garantias criptográficas. Essa composição permite que diferentes componentes adotem graus distintos de segurança,  por exemplo, CH assegura confidencialidade em meio a ambientes não confiáveis, MPC distribui confiança entre nós distintos, e TEE garante integridade de execução. A interoperabilidade entre essas abordagens é um vetor de pesquisa ativo, envolvendo padronização de formatos de chave, representação de dados (como tensores cifrados) e APIs seguras para troca de contextos.

\section{Desafios de Implementação}

Apesar dos avanços, a implantação prática de CH apresenta diversos pontos de atenção. O primeiro é a seleção de parâmetros de segurança e a calibração do orçamento de ruído. Graus polinomiais $N$, módulos $q$ e distribuições de erro devem atender a alvos de segurança comparáveis a 128, 192 ou 256 bits, garantindo resistência a ataques de redução de reticulado (BKZ, Sieving, pruning). Ao mesmo tempo, esses parâmetros devem suportar a profundidade de circuito exigida pela aplicação, equilibrando desempenho e precisão. Pequenos ajustes nesses valores podem multiplicar por 10 o custo computacional, exigindo engenharia fina na escolha de parâmetros e formatos de mensagem (coeficiente ou slot).

O segundo desafio é o custo computacional. Mesmo com NTT vetorizada, \textit{batching} e paralelismo, as latências ainda são ordens de magnitude superiores às do processamento em claro. A execução típica envolve dezenas de multiplicações polinomiais e operações de relinearização, cada uma exigindo transformadas rápidas, key-switching e rotinas de rotação. Assim, o desenho do circuito deve ser otimizado para minimizar multiplicações, reutilizar estados intermediários e explorar ao máximo a linearidade de operações. Em inferência homomórfica, a escolha de topologia de rede e de funções polinomiais aproximadas (\textit{Chebyshev}, truncadas ou spline) influencia diretamente a profundidade e o erro final.

O terceiro ponto é a engenharia de precisão em esquemas aproximados como CKKS, que requer uma gestão criteriosa das escalas e pontos de \textit{rescaling} para manter o erro numérico sob controle. Cada multiplicação reduz o número efetivo de bits de precisão disponíveis,  tipicamente entre 30 e 45 bits após algumas camadas,  e implica decisões de arredondamento que precisam ser consistentes entre partes. Erros acumulados podem inviabilizar a decifragem, especialmente quando operações não lineares e normalizações numéricas são avaliadas homomorficamente.

Há ainda desafios significativos de integração e operação. Bibliotecas como \gls{seal}, \gls{openfhe}, \gls{helib}, Concrete e TFHE fornecem primitivas criptográficas em baixo nível, mas a tradução de consultas SQL, operadores de fluxo de dados ou grafos de inferência para planos de execução homomórficos eficientes requer compiladores especializados. \textit{Frameworks} de alto nível como \gls{chet}, EVA e Concrete ML vêm amadurecendo, mas seu ecossistema é fragmentado: diferentes esquemas suportam subconjuntos distintos de tipos de dados e operações. A ausência de um padrão unificado de compatibilidade e o tamanho volumoso das chaves e parâmetros tornam a portabilidade e implantação em larga escala tarefas desafiadoras.

Outro fator crítico é a governança criptográfica. Ambientes produtivos demandam políticas de ciclo de vida de chaves (emissão, rotação, revogação), segregação de funções e sistemas de auditoria integrados a PKIs e módulos HSM (\textit{Hardware Security Modules}). Operações de relinearização e rotação requerem chaves de apoio volumosas, às vezes centenas de megabytes, e a sua gestão torna-se um componente administrativo e de risco. Além disso, mecanismos de controle de acesso aos contextos cifrados precisam ser integrados a sistemas de identidade corporativos para assegurar rastreabilidade e não repúdio. Essas camadas de governança são essenciais para conformidade em ambientes regulados e para operacionalizar CH no contexto de ecossistemas corporativos.

O estado atual de maturidade tecnológica ainda impõe restrições: operações complexas como transformadas de Fourier, regressões polinomiais de grau alto ou inferência em modelos neurais profundos permanecem impraticáveis por custo. Entretanto, há esforços notáveis de aceleração, seja via paralelização em GPU, seja por síntese em FPGA, que demonstram ganhos de 30x a 100x sobre implementações puramente em CPU.

\section{Síntese e Perspectivas}

A criptografia homomórfica redefine o equilíbrio entre utilidade e privacidade ao permitir o processamento de dados cifrados de forma segura e verificável. Diferentemente de abordagens tradicionais de proteção, que se baseiam em isolamento ou pseudonimização, CH preserva a confidencialidade mesmo em cenários de processamento terceirizado, tornando-se um componente-chave em infraestruturas de dados sensíveis na era pós-quântica.

Com esquemas baseados em LWE/RLWE (BGV, BFV, CKKS) e abordagens otimizadas para portas booleanas (FHEW, TFHE), já é possível viabilizar cenários concretos de agregação, análise estatística e inferência de modelos compactos. O gargalo histórico do \textit{bootstrapping} vem sendo mitigado por avanços algorítmicos, como amostragem de erro estruturada e decomposição em frequências toroidais, além de implementação em hardware com memória hierárquica otimizada. Técnicas de \textit{batching}, relinearização e compressão de chaves públicas reduzem cada vez mais a sobrecarga operacional.

A médio prazo, espera-se que CH coexista de forma complementar com MPC, TEEs e outras \textit{privacy-enhancing technologies}, compondo arquiteturas resilientes e distribuídas, capazes de proteger tanto o dado quanto o processo analítico. A convergência entre essas tecnologias deverá originar plataformas de análise cifrada interoperáveis, compatíveis com padrões emergentes de interoperabilidade, como \gls{gapsvp} e \gls{nistpqc}. 

Perspectivas de pesquisa futura incluem mecanismos de \textit{verifiable homomorphic computation}, nos quais a parte que realiza o processamento pode provar formalmente que o cálculo cifrado foi executado corretamente, sem necessitar decifrar. Além disso, a integração com \gls{blockchain} e atestação remota tende a consolidar CH como módulo de confiança para contratos inteligentes e sistemas descentralizados de processamento de dados.

Em síntese, a criptografia homomórfica deixa de ser uma curiosidade teórica e passa a se consolidar como ferramenta prática de engenharia de privacidade e segurança de dados, sinalizando uma mudança de paradigma: dados podem ser simultaneamente úteis e confidenciais, rompendo a antiga dicotomia entre segurança e usabilidade.

\section{Estado da Arte}

Além desses marcos, há um corpo crescente de resultados que refinam componentes críticos do pipeline homomórfico. Na multiplicação polinomial, variações de NTT \emph{in-place}, \emph{lazy reduction} e esquemas de decomposição mista (\gls{rns} + CRT de base estendida) reduziram significativamente o número de acessos à memória e o custo de normalização modular, deslocando o gargalo para o \emph{key switching}. Para este, decomposições híbridas (por exemplo, base-$w$ combinada com RNS) e tabelas pré-computadas amortizam a expansão de chaves e diminuem o número de multiplicações por termo do \emph{ciphertext}, impactando diretamente relinearização e rotações. No contexto de CKKS, políticas de escala (\emph{scale scheduling}) co-projetadas com a cadeia de módulos e estratégias de avaliação polinomial (Paterson–Stockmeyer, \emph{baby-step/giant-step}) permitem implementar aproximações de funções não lineares com menor profundidade efetiva, preservando 30–40 bits de precisão útil para inferência prática.

O \emph{bootstrapping} aritmético, historicamente a etapa mais custosa em BGV/BFV/CKKS, tem sido acelerado por trocas de base otimizadas, variantes de Bluestein para transformadas fora do conjunto clássico de raízes NTT e aproximações polinomiais mais rasas da função de decodificação. Abordagens recentes exploram particionamento do espaço de coeficientes e \emph{hoisting} para compartilhar NTTs entre múltiplos \emph{ciphertexts}, reduzindo latência em cenários de alto \emph{throughput}. Em FHEW/TFHE, o avanço contínuo de FFT/NTT toroidais, \emph{accumulators} compactos e pipelines SIMD trouxe latências por porta para a casa de dezenas de microssegundos em CPUs modernas, com demonstrações consistentes de paralelismo \emph{multi-core}.
Do lado de codificação e \emph{packing}, mapeamentos cuidadosos de tensores (por canais, por características ou por blocos) minimizam o número de rotações e a cardinalidade de chaves de Galois necessárias. Esse co-design entre dados e esquema é determinante para camadas lineares densas (GEMV/GEMM) e para convoluções 1D/2D em ML homomórfico; técnicas como \emph{diagonal/hoisted rotations} e \emph{replication-by-rotation} transformam padrões de acesso em sequências curtas de automorfismos.

Em termos de \emph{tooling}, observa-se uma convergência para camadas intermediárias (IRs) que abstraem operações de alto nível, soma, produto, convolução, polinômios aproximados, para depois materializá-las em um alvo específico (BFV/BGV/CKKS/TFHE) segundo um \emph{backend} e um conjunto de políticas (precisão, profundidade, memória). Compiladores como CHET e IRs relacionados a \gls{eva}, aliados a bibliotecas generalistas (SEAL, OpenFHE, HElib), compõem um \emph{toolchain} no qual o usuário declara objetivos (erro e latência alvo), enquanto o sistema escolhe parâmetros $(N, q, \sigma)$, escalas, cadeias de módulos, layout de slots e agenda rotações/relinearizações. Esse movimento, já consolidado em \emph{ML compilers} tradicionais, tende a reduzir a assimetria entre especialistas criptográficos e engenheiros de dados.

No plano de segurança e padronização, o \gls{hestandard} fixou terminologia, procedimentos de relato e perfis de segurança (por exemplo, 128/192/256 bits) ancorados em estimadores de complexidade para LWE/RLWE, considerando BKZ com \emph{pruning}, \emph{sieving} clássico e quântico. A normalização de métricas, aliada a \emph{benchmarks} como o \gls{hebench}, tornou comparáveis artefatos antes heterogêneos, permitindo estudos reprodutíveis de custo por operação (mul/add/rot/ks) e de \emph{bootstrapping}, além de perfis de memória e tamanhos de chaves auxiliares. Em paralelo, cresce o interesse por extensões como \emph{verifiable FHE}, que incorporam provas de correção à computação cifrada, e por integrações com MPC/TEE em arquiteturas híbridas que conciliam custos, latência e garantias de sigilo.

\begin{table}[h!]
\centering
\scriptsize
\caption{Síntese de trabalhos e bibliotecas representativos na evolução de CH, com foco em tipo de esquema, domínio algébrico e contribuições técnicas-chave.}
\label{tab:sota-he}
\begin{tabular}{p{3.2cm} p{2cm} p{2.5cm} p{6.5cm}}
\hline
\textbf{Trabalho/Biblioteca} & \textbf{Tipo} & \textbf{Domínio} & \textbf{Contribuição principal} \\
\hline
Gentry \cite{Gentry2009FHE} & FHE (teoria) & Ideais/reticulados & Primeiro FHE e \emph{bootstrapping} para profundidade ilimitada. \\
Regev  \cite{regev2005} & Fundamentos & LWE & Reduções para LWE; base de segurança para esquemas modernos. \\
LPR \cite{lpr2010} & Fundamentos & RLWE & LWE em anéis; eficiência via NTT e estrutura ciclotômica. \\
BGV  \cite{bgv2012} & Leveled FHE & RLWE & \emph{Modulus switching}, relinearização e leveled FHE prático. \\
CKKS  \cite{ckks2017} & Aproximado & RLWE & Reais/complexos com \emph{rescaling}; ML/inferência com erro controlado. \\
FHEW  \cite{fhew2015} & Booleano & LWE (torus) & \emph{Bootstrapping} por porta sub-segundo; FFT/NTT em domínio toroidal. \\
TFHE  \cite{tfhe2016} & Booleano & LWE (torus) & \emph{Bootstrapping} <0.1 s/porta; acelerações e chaves de conversão. \\
Microsoft SEAL \cite{seal} & Biblioteca & BFV/CKKS & Engenharia robusta, RNS/NTT, API estável, ampla adoção. \\
OpenFHE \cite{openfhe} & Biblioteca & BGV/BFV/CKKS /TFHE & Suporte amplo, pesquisa/produção, \emph{bootstrapping} moderno. \\
HElib \cite{helib} & Biblioteca & BGV/BFV & Implementações pioneiras, \emph{packing}/automorfismos. \\
\gls{tfhelib} \cite{tfhe-lib} & Biblioteca & TFHE & Foco em portas booleanas e \emph{bootstrapping} rápido. \\
PALISADE \cite{palisade} & Biblioteca (legado) & Diversos & Base para OpenFHE; suporte a múltiplas famílias e RNS. \\
CHET \cite{chet2018} & Ferramental & Compilador/RT & Mapeamento de tensores para CH; escalas/layout/rotações automáticos. \\
HEAX/EVA \cite{eva2020} & Ferramental & Arquitetura/IR & Co-design HW/SW para CH; IR e kernels acelerados. \\
HE-Standard \cite{he-standard} & Padrão & Diretrizes & Parametrização segura, terminologia e práticas recomendadas. \\
HEBench \cite{hebench} & Benchmark & Metodologia & Comparabilidade de latências (mul/add/rot/ks) e \emph{bootstrapping}. \\
\hline
\end{tabular}
\end{table}

Do ponto de vista de aplicações, a literatura demonstra rotas viáveis para agregações estatísticas privadas, \emph{analytics} sobre bases sensíveis, junções seguras entre organizações e inferência de modelos compactos (regressões, árvores rasas, CNNs pequenas) com latências de segundos a dezenas de segundos por consulta, variando conforme profundidade, grau de \emph{packing} e orçamento de precisão. Em cenários industriais, a decisão entre BGV/BFV (exatidão modular), CKKS (aproximação real/complexa) e TFHE/FHEW (booleano de alta granularidade) é guiada por requisitos de erro, perfil de circuito (linear vs. não linear), padrão de acesso (rotações/permutas) e disponibilidade de aceleração por GPU/FPGA. À medida que \emph{bootstrapping} aritmético se torna mais barato e que pipelines de \emph{key switching} e NTT são offloadados para hardware especializado, projeta-se uma expansão do escopo de CH em serviços interativos, orquestrados por padrões e \emph{toolchains} maduros e sustentados por práticas de governança criptográfica (rotação de chaves, segregação de funções, auditoria e integração com PKI/HSM).

Em síntese, o estado da arte em CH combina solidez teórica (reduções LWE/RLWE), famílias de esquemas com perfis bem compreendidos (BGV/BFV/CKKS/TFHE/FHEW), otimizações algorítmicas e de implementação (NTT/RNS, \emph{key switching}, \emph{packing}, \emph{bootstrapping} moderno) e um ecossistema de bibliotecas e compiladores que aproxima a tecnologia de requisitos produtivos. A Tabela~\ref{tab:sota-he} oferece um mapa compacto dessas contribuições e serve como referência cruzada para o desenho de circuitos, a escolha de parâmetros e a avaliação comparativa de desempenho e precisão em aplicações reais.

\section{Questões para Reflexão e Pesquisa Futura}

Ao concluir o capítulo sobre criptografia homomórfica, o leitor deve ser instigado a não aceitar a tecnologia apenas pelo rótulo de "privacidade preservada", mas a questionar suas limitações práticas, hipóteses de segurança e adequação a cenários reais. Comece por perguntar: quais são exactamente as garantias que cada variante (BFV, BGV, CKKS, TFHE, e suas derivações) fornece no contexto do seu problema? Em particular, quais compromissos entre exatidão e eficiência você está disposto a aceitar, por exemplo, CKKS oferece computação aproximada valiosa para machine learning, mas introduz erro numérico acumulado; BFV/BGV preservam aritmética inteira exata até limites de capacidade, mas podem exigir parâmetros e overhead maiores. Levante também a questão do horizonte temporal de proteção: até que ponto os parâmetros escolhidos hoje resistirão a avanços em criptanálise clássica e quântica, e como isso afeta a decisão de usar FHE para dados que exigem confidencialidade por décadas?

Interrogue as suposições de implementação: como o crescimento de ruído (noise growth) se manifesta para o circuito que você pretende executar? Qual é a profundidade multiplicativa máxima prática antes de precisar de bootstrapping, e qual a penalidade real em latência e recursos quando o bootstrapping é acionado? Em cenários com muitos multiplicadores ou composições repetidas, será mais eficiente reescrever a aplicação (por exemplo, linearizar polinomiais, reorganizar ordem de operações, usar técnicas de baby-step/giant-step, ou aproveitar batching/packing) do que aceitar grandes custos de bootstrapping? Pergunte-se também sobre o custo de memória e I/O: quantos bytes ocupa um ciphertext para os parâmetros necessários, e qual o impacto disso em ambientes distribuídos ou em dispositivos com limitados recursos de RAM?

Considere as implicações de precisão e utilidade para casos aproximados: ao empregar CKKS para inferência de modelos de machine learning, qual será o erro numérico final (RMSE, bias, degradação de acurácia) comparado ao modelo não cifrado? Como o processo de quantização/rescaling influencia a estabilidade numérica em pipelines com normalização, funções não lineares (aproximadas por polinômios) e agregações? Que estratégias de pré-processamento, ajuste de escala e representação de dados (fix-point vs float simplificado) reduzem melhor o erro sem inflar demais os parâmetros?

Questione a integração com arquiteturas existentes e com outros primitivos de privacidade: quando FHE é mais vantajoso que MPC, enclaves de hardware (TEE) ou abordagens híbridas? Em que cenários a combinação FHE+MPC (por exemplo, usar FHE para offload pesado de computação e MPC para interatividades sensíveis) oferece melhor custo-benefício? Como conciliar requisitos de auditoria e verificabilidade: é suficiente confiar no resultado homomórfico, ou é necessário prover provas de correção (verifiable computing, SNARKs) que adicionam ainda mais overhead e complexidade?

Investigue riscos práticos de segurança e operação: que vetores de leak ainda permanecem (padrões de acesso a dados, tempos de resposta, tamanhos das mensagens) que podem ser explorados para inferir informação sensível mesmo quando os dados estão cifrados? Como incorporar mitigação adicional (por exemplo, padding, latência artificial, agregação diferencial) sem tornar a solução inviável? Além disso, examine a cadeia de suprimentos e a segurança de implementação: bibliotecas e otimizações em assembly/AVX/NEON aceleram FHE, mas metodologias de mitigação de canais laterais precisam ser integradas; qual o custo de certificação e auditoria para um produto FHE em produção?

Do ponto de vista experimental, proponha-se reproduzir benchmarks cuidadosos e contextualizados: meça latência de operação (p50/p95/p99), throughput (operações por segundo), consumo de memória, tamanho dos artefatos em trânsito e tempo de bootstrapping para diferentes parâmetros (grau polinomial N, tamanho do módulo q, níveis, escala em CKKS). Compare resultados em múltiplas bibliotecas (por exemplo, SEAL, HElib, PALISADE, TFHE, Lattigo, Concrete) e em diferentes plataformas (CPU, GPU, FPGA), documentando scripts e datasets para permitir reprodução. Registre também métricas de utilidade: erro numérico final, acurácia de modelos cifrados, e custo econômico estimado (tempo CPU × custo hora) para cargas reais. Pergunte: os benchmarks refletem cenários reais ou apenas micro-benchmarks; se não refletem, que adaptações são necessárias para torná-los representativos?

Interpele aspectos organizacionais e regulatórios: FHE muda o modelo de ameaça, mas não elimina a necessidade de governança de chaves, políticas de acesso e conformidade. Quem gerencia as chaves de descriptografia (se houver), como se impõe separação de funções, e como se municia auditoria sem comprometer privacidade? Em setores regulados (saúde, financeiro), que evidências e certificações serão exigidas para homologar soluções FHE? Como lidar com requisitos legais de retenção, acesso por autoridade competente e obrigações contratuais quando os dados permanecem cifrados durante o processamento?

Finalmente, abra questões de pesquisa técnica e aplicada: quais avanços em bootstrapping e reduções de custo poderiam transformar casos de uso hoje impraticáveis em adotáveis? Como melhorar compiladores e otimizadores que traduzam linguagens de alto nível em circuitos homomórficos que minimizem profundidade e consumo de recursos? Que papel têm aceleradores (GPUs, FPGAs, ASICs) e suporte nativo em cloud para viabilizar serviços FHE? E como podemos construir frameworks de avaliação, benchmarks, testbeds, datasets públicos e protocolos experimentais, que permitam comparações justas entre propostas e facilitem a transição de protótipos a produções confiáveis?

\chapter{Criptografia Quântica}

A criptografia quântica emerge como resposta direta às vulnerabilidades projetadas sobre a criptografia clássica, especialmente diante da ameaça representada pelo advento da computação quântica. Enquanto os algoritmos clássicos fundamentam sua segurança na complexidade de problemas matemáticos, como a fatoração de inteiros e o logaritmo discreto,, a criptografia quântica adota uma abordagem radicalmente diferente: baseia-se em princípios fundamentais da mecânica quântica, explorando as próprias leis físicas do universo como recurso intransponível de proteção. 

Essa mudança de paradigma busca oferecer, ao menos em teoria, {segurança incondicional}. Isso significa que a segurança não se apoia em suposições sobre o poder computacional de adversários, mas sim em princípios físicos universalmente aceitos, como o \gls{nocloning}, a \gls{superposicao} e o emaranhamento quântico. Tais propriedades garantem que qualquer tentativa de interceptação ou clonagem de informação quântica inevitavelmente altera o estado original, produzindo erros detectáveis. Assim, ao contrário da criptografia clássica, cuja confiabilidade pode ser posta em xeque pelo avanço tecnológico, a criptografia quântica promete proteção baseada em fundamentos universais da natureza.

O objetivo central dessa abordagem não é substituir integralmente todos os sistemas criptográficos tradicionais. Ao contrário, a criptografia quântica atua de maneira complementar, sendo particularmente adequada a tarefas cujo requisito principal é a distribuição segura de chaves secretas em canais de comunicação potencialmente inseguros. A {Distribuição Quântica de Chaves (QKD)} é o exemplo mais proeminente dessa aplicação: por meio da troca de fótons polarizados em diferentes bases, duas partes podem estabelecer uma chave partilhada com a certeza de que qualquer tentativa de espionagem pode ser detectada com alta confiabilidade.

Ao longo das últimas quatro décadas, a criptografia quântica evoluiu de propostas teóricas pioneiras, como o protocolo BB84 (1984) e o E91 (1991), para experimentos práticos de larga escala, incluindo transmissões metropolitanas em fibras ópticas, enlaces quânticos via satélite (\textit{Micius}, 2017) e, mais recentemente, as primeiras arquiteturas de redes quânticas integradas em ambientes urbanos. Essa trajetória demonstra não apenas a solidez conceitual da proposta, mas também a viabilidade tecnológica de aplicações práticas da mecânica quântica na segurança da informação. Contudo, é importante ressaltar que a criptografia quântica também enfrenta significativos desafios de implementação. A necessidade de alinhamento fino em fibras ópticas, a atenuação de sinais em longas distâncias, a dependência de detectores sensíveis e os altos custos de infraestrutura ainda limitam sua adoção ampla. Além disso, a criptografia quântica não se propõe a resolver todas as demandas do ecossistema de cibersegurança: algoritmos de autenticação, assinaturas digitais e armazenamento seguro continuam a depender de construções clássicas ou híbridas. 

Portanto, a relevância da criptografia quântica reside em sua capacidade de {complementar} os esquemas tradicionais nas áreas mais críticas, especialmente onde se busca sigilo absoluto de longo prazo, oferecendo uma camada de segurança fundamentada não em hipóteses matemáticas, mas nas leis mais fundamentais da física. Nesse sentido, ela representa não apenas um avanço técnico, mas também uma mudança epistemológica no modo como compreendemos e praticamos a segurança digital em escala global.

\section{Fundamentos Teóricos}

A criptografia quântica apoia-se em princípios centrais da mecânica quântica, que oferecem propriedades únicas para a segurança da informação. Além da superposição, do emaranhamento e do teorema da não-clonagem, existem outros conceitos indispensáveis para a compreensão do funcionamento e das limitações dos protocolos quânticos. A seguir, revisamos esses fundamentos com maior detalhamento conceitual e implicações práticas para protocolos como BB84, E91, MDI-QKD e suas variantes modernas, realçando como cada princípio influencia a geração de chaves, a detecção de eavesdroppers, a modelagem de canais e o tratamento de ruído em implementações reais.

\begin{itemize}
    \item \textbf{Superposição:} 
    Um qubit pode estar em combinação linear dos estados $|0\rangle$ e $|1\rangle$, formalmente:
    \[
    |\psi\rangle = \alpha |0\rangle + \beta |1\rangle, \quad |\alpha|^2 + |\beta|^2 = 1.
    \]
    Esse estado não tem análogo clássico e garante aleatoriedade fundamental nas medições. Na prática, a superposição é explorada para codificar símbolos em diferentes bases; por exemplo, no BB84, a escolha aleatória de bases e de estados superpostos garante que qualquer tentativa de medição adversária perturbe estatisticamente os resultados. Em sistemas fotônicos, superposição manifesta-se em graus de liberdade como polarização, fase e tempo de chegada, viabilizando modulações compatíveis com enlaces ópticos de longa distância.
    
    \item \textbf{Complementaridade e Bases Mutuamente Incompatíveis:}
    As medidas em duas bases distintas (e.g., $\{|0\rangle,|1\rangle\}$ e $\{ |+\rangle,|-\rangle \}$) não podem ser realizadas simultaneamente com precisão. 
    Formalmente:
    \[
    |+\rangle = \frac{1}{\sqrt{2}}(|0\rangle + |1\rangle), \quad |-\rangle = \frac{1}{\sqrt{2}}(|0\rangle - |1\rangle).
    \]
    A escolha aleatória de bases é o núcleo da detecção de espionagem no protocolo BB84. Quando um adversário mede na base “errada”, introduz erros que elevam a taxa de erro quântico (\gls{qber}); acima de um limiar teórico, a extração de chave segura torna-se impossível. Em implementações práticas, a incompatibilidade de bases também serve como teste operacional de qualidade do canal e dos alinhamentos ópticos.
    
    \item \textbf{Emaranhamento:} 
    Dois qubits podem ser preparados em estados correlacionados não separáveis, como os estados de Bell:
    \[
    |\Phi^\pm\rangle = \frac{1}{\sqrt{2}} \big(|00\rangle \pm |11\rangle\big), 
    \quad
    |\Psi^\pm\rangle = \frac{1}{\sqrt{2}} \big(|01\rangle \pm |10\rangle\big).
    \]
    Propriedades de não-localidade são confirmadas pela violação da desigualdade de \gls{chsh}:
    \[
    S = E(A,B) + E(A,B') + E(A',B) - E(A',B') \leq 2 \quad \text{clássico}, \quad S \leq 2\sqrt{2} \quad \text{quântico}.
    \]
    O emaranhamento é base para protocolos tipo E91 e para esquemas device-independent (\gls{diqkd}), nos quais as garantias de segurança podem ser estabelecidas mesmo com dispositivos potencialmente não confiáveis, desde que haja violação estatisticamente significativa das desigualdades de Bell. Na prática, a geração e distribuição de pares emaranhados impõe requisitos rigorosos de fonte, acoplamento e estabilidade de fase, sendo viabilizada por processos como conversão paramétrica descendente (\gls{spdc}) e por multiplexação espectral/espacial para aumentar taxa de chaves.
    
    \item \textbf{Teorema da Não-Clonagem:}
    Não existe operador unitário $U$ que satisfaça:
    \[
    U(|\psi\rangle \otimes |0\rangle) = |\psi\rangle \otimes |\psi\rangle, \quad \forall \, |\psi\rangle.
    \]
    Assim, um adversário não pode copiar estados quânticos arbitrários sem ser detectado. Esse resultado limita ataques baseados em interceptação e cópia perfeita, diferentemente do cenário clássico no qual dados podem ser replicados sem custo informacional. Em contextos práticos, embora clones perfeitos sejam proibidos, ataques aproximados (clonagem universal aproximada) ainda introduzem ruído detectável que se reflete na QBER, permitindo que os protocolos negociem cancelamento de chaves ou reforcem pós-processamento.
    
    \item \textbf{Regra de Born (Medição):}
    As probabilidades de medição são dadas pela projeção:
    \[
    P(i) = |\langle i|\psi\rangle|^2.
    \]
    Após a medição, o estado colapsa para $|i\rangle$. Esse processo inviabiliza medições não invasivas. Do ponto de vista operacional, a \gls{born} fundamenta a modelagem estatística da geração de chaves, define as taxas de acerto/erro esperadas e orienta a estimação de parâmetros (parameter estimation) durante a sifting. Além disso, vincula intrinsecamente a segurança à aleatoriedade quântica, que deve ser preservada por geradores de números quânticos (\gls{qrin}) confiáveis.
    
    \item \textbf{Decoerência e Ruído:} 
    A interação com o ambiente degrada estados quânticos superpostos. O modelo de canal ruidoso pode ser descrito por operadores de Kraus, por exemplo, para o canal de despolarização:
    \[
    \mathcal{E}(\rho) = (1-p)\rho + \frac{p}{3}(X\rho X + Y\rho Y + Z\rho Z),
    \]
    onde $X, Y, Z$ são as matrizes de Pauli. A decoerência impõe limitações práticas à distância das transmissões em protocolos QKD. Em fibras ópticas, perdas exponenciais com a distância e dispersão limitam taxas úteis; em canais livres ou satelitais, turbulência atmosférica e apontamento afetam fidelidade. Tais efeitos exigem correção de erros clássica eficiente (\gls{informationreconciliation}) e amplificação de privacidade (\gls{privacyamplification}) para transformar cadeias brutas em chaves finais seguras.
    
    \item \textbf{Estado Misto e Matriz Densidade:} 
    Sistemas não puramente definidos são descritos pela \gls{rho}:
    \[
    \rho = \sum_i p_i |\psi_i\rangle\langle\psi_i|, \quad \text{com} \quad \mathrm{Tr}(\rho) = 1.
    \]
    A descrição por $\rho$ é essencial para lidar com ruído e perdas em canais quânticos. Em segurança de QKD, argumenta-se contra adversários arbitrariamente poderosos analisando estados conjuntos $\rho_{ABE}$, contemplando correlações entre Alice (A), Bob (B) e o eavesdropper (E). O formalismo de densidade permite derivar limites de informação de Eve e calcular taxas assimptóticas (e finitas) de chaves seguras sob hipóteses de canal e de dispositivo.
    
    \item \textbf{Entropia Quântica (\gls{vnentropy}):}
    A incerteza sobre um estado é quantificada pelo operador densidade:
    \[
    S(\rho) = - \mathrm{Tr}(\rho \log \rho).
    \]
    Essa entropia quantifica a informação segura e auxilia em procedimentos de amplificação de privacidade. Em protocolos práticos, usa-se a entropia condicional e min-entropia suave para quantificar, em cenários de amostra finita, o montante de compressão hash necessária para garantir que a chave final seja indistinguível de aleatória para o adversário, incorporando margens estatísticas provenientes da estimação de parâmetros.
    
    \item \textbf{Canal Quântico e Fidelidade:}
    A transmissão de estados quânticos é descrita por canais quânticos $\mathcal{E}$. A fidelidade entre o estado original $|\psi\rangle$ e o transmitido $\rho$ é:
    \[
    F(|\psi\rangle, \rho) = \langle \psi | \rho | \psi \rangle.
    \]
    Fidelidades altas são requisito fundamental para protocolos práticos de QKD. Valores de fidelidade correlacionam-se diretamente com QBER, perdas e taxa de chaves secretas, guiando escolhas de topologia (fibras vs. free-space), taxas de repetição, potência/atenuação e técnicas de estabilização de fase. Em arquiteturas futuras com repetidores quânticos, a fidelidade também determina a viabilidade de operações de purificação e troca de emaranhamento.
    
    \item \textbf{Teleportação Quântica:}
    Recurso fundamental que permite transmitir estados quânticos (\gls{teleportacao}) desconhecidos através de canais clássicos e pares emaranhados: 
    \[
    |\psi\rangle_A | \Phi^+\rangle_{BC} \xrightarrow{\text{medição de Bell}} |\Phi^k\rangle_{AB} \otimes U_k|\psi\rangle_C,
    \]
    onde $U_k$ é uma porta unitária aplicada por Bob após receber informações clássicas de Alice. Essa técnica é essencial para futuros repetidores quânticos em redes de longa distância. No contexto de segurança, a teleportação ilustra como informação quântica pode ser transferida sem transporte físico do portador original, preservando sigilo desde que os recursos emaranhados e os canais clássicos autenticados sejam confiáveis. Em protocolos MDI-QKD, medições de Bell intermediárias mitigam vulnerabilidades em detectores, com impacto positivo em provas de segurança realistas.
\end{itemize}
Em um plano conceitual, os recursos quânticos elevam a segurança enquanto os efeitos físicos do canal degradam a qualidade do enlace. Nas regiões de maior qualidade, mecanismos como superposição, complementaridade e emaranhamento viabilizam detecção de espionagem e argumentos baseados em desigualdades de Bell, ampliando a margem de segurança. À medida que a distância e as perdas aumentam, a fidelidade do canal tende a decair e o QBER a crescer, o que impõe maior esforço de pós-processamento, reconciliação de informações e amplificação de privacidade, para manter a taxa de chave útil dentro de limites operacionais. Técnicas como MDI-QKD e teleportação, situadas em faixas intermediárias de qualidade, atuam como mitigadores práticos: ao deslocarem a superfície de ataque para o nó de medição (ou ao habilitarem arquiteturas de repetidores quânticos), preservam garantias de segurança mesmo na presença de ruído e imperfeições de detecção. Esses compromissos operacionais se alinham às implicações sumarizadas na Tabela~\ref{tab:principios_qkd}, que relaciona cada princípio a métricas observáveis (QBER, taxa e alcance) e orienta escolhas de topologia (fibra vs. espaço livre), taxa de repetição e orçamento de perdas.

Do ponto de vista de engenharia, é útil separar explicitamente os fatores que reforçam a segurança (recursos quânticos e decisões de protocolo) daqueles que a pressionam para baixo (perdas, ruído e limitações do canal). Com o aumento da distância, a tendência natural é de elevação do QBER e redução da fidelidade de transmissão; por isso, os limiares de abortamento e as janelas de estimação de parâmetros devem ser calibrados ao regime de operação de cada enlace. Por outro lado, abordagens como MDI-QKD e técnicas de teleporte/quase-repetição deslocam os gargalos para pontos controlados (nós de medição confiáveis ou verificáveis), mitigando ataques a detectores e viabilizando escalabilidade. O balanço entre ganhos e perdas, portanto, determina a taxa final de chave e o alcance efetivo, como detalhado na Tabela~\ref{tab:principios_qkd}.

\begin{table}[h!]
\centering
\scriptsize
\caption{Princípios e efeitos relevantes em QKD, com implicações operacionais e impacto em métricas observáveis.}
\label{tab:principios_qkd}
\begin{tabular}{p{3.8cm} p{6.1cm} p{4.5cm}}
\hline
\textbf{Princípio / Efeito} & \textbf{Implicação operacional} & \textbf{Impacto em métricas} \\
\hline
Superposição e complementaridade &
Detecção de espionagem pela perturbação estatística em bases incompatíveis; fundamenta o sifting. &
QBER como indicador de ataque; taxa de sifting e taxa líquida de chave. \\

Emaranhamento e não-localidade (E91/DI) &
Segurança baseada em violações de Bell; maior robustez a imperfeições e modelagem adversária. &
Limiar de violação CHSH; tolerância a falhas de dispositivo; taxa sob hipótese DI. \\

Teorema da não-clonagem &
Cópia perfeita é impossível; tentativas de clonagem deixam vestígios detectáveis. &
Elevação do QBER; limiares de abortamento; análise de finitos. \\

Perdas e ruído (decoerência, despolarização, ruído de fundo) &
Limitam alcance e taxa; exigem filtros, alinhamento e compensação ativa. &
Trade-off distância–taxa; eficiência de reconciliação (IR) e amplificação de privacidade (PA). \\

Medição (regra de Born) e estimação &
Geração de aleatoriedade (QRNG); janelas de amostragem e estimação de parâmetros. &
Garantias de aleatoriedade; intervalos de confiança; orçamento de segurança. \\

\gls{fidelidade} e estabilidade &
Qualidade de transmissão e alinhamento (polarização/fase); guia tuning do enlace. &
Otimização de taxa final; estabilidade temporal; jitter e drift. \\

MDI-QKD / Teleportação / caminho a repetidores &
Mitigam ataques a detectores e deslocam superfície de ataque; permitem redes escaláveis. &
Segurança contra side-channels; alcance ampliado; integração com nós intermediários. \\
\hline
\end{tabular}
\end{table}

\section{Protocolos}

Ao longo de quatro décadas, a literatura de QKD evoluiu de esquemas conceituais com fontes ideais para arquiteturas robustas e escaláveis em cenários reais. Uma forma útil de organizar esse ecossistema é por “gerações” tecnológicas: a primeira geração abrange protocolos de \textit{prepare-and-measure}, nos quais Alice prepara estados em bases mutuamente incompatíveis e Bob os mede (ex.: BB84, B92); a segunda geração enfatiza recursos de emaranhamento para certificar correlações não locais e segurança baseada em testes de Bell (ex.: E91, BBM92); gerações mais recentes introduzem mecanismos explícitos de mitigação de vulnerabilidades físicas e de engenharia, como estados \textit{decoy}, tolerância a fontes não ideais e independência dos dispositivos de detecção (ex.: SARG04, \gls{decoystate}, MDI-QKD, Twin-Field, CV-QKD, \gls{rrdps}, \gls{cow}, \gls{dps}), além de propostas de assinaturas digitais quânticas (\gls{qds}) e abordagens \textit{Device-Independent} (DI-QKD).

Além da taxonomia histórica, é conveniente classificar os protocolos pelo “modelo de informação” que utilizam. Em variáveis discretas (DV-QKD), os bits são codificados em bases ortogonais (polarização, tempo, caminho), enquanto em variáveis contínuas (CV-QKD) a informação está nas quadraturas do campo ($X,P$) e é extraída por detecção homó/heteródina. Outra dimensão ortogonal é o “ponto de confiança”: esquemas tradicionais assumem detectores confiáveis; o MDI-QKD desloca a superfície de ataque para um nó de medição potencialmente não confiável; o DI-QKD busca eliminar suposições sobre todo o dispositivo, ancorando a segurança na violação observada de desigualdades de Bell.

Do ponto de vista de desempenho, três métricas sintetizam as decisões de projeto: (i) taxa secreta de chave por segundo, que depende de perdas, eficiência de reconciliação e ruído; (ii) alcance, limitado pela atenuação de fibra/espaço livre e pela sensibilidade dos detectores; e (iii) robustez operacional, refletida na tolerância a QBER, estabilidade de fase/polarização e resistência a \textit{side-channels}. Protocolos como Decoy-State elevam taxas e alcance em DV-QKD com fontes coerentes; MDI-QKD endurece o sistema contra ataques a detectores; Twin-Field melhora a escalabilidade de longa distância ao ultrapassar a barreira PLOB; e CV-QKD facilita integração com infraestruturas ópticas legadas.

Por fim, a maturidade experimental varia entre famílias. BB84/Decoy-State e MDI-QKD já foram consolidados em redes metropolitanas com fibras padrão e componentes comerciais; \gls{tfqkd} é o estado da arte para longas distâncias em fibra; CV-QKD avança em compatibilidade com telecom e altas velocidades; DI-QKD e QDS, embora mais exigentes, caminham de demonstrações laboratoriais para pilotos de campo. Nas subseções seguintes, discutimos os protocolos representativos, seus modelos de segurança e implicações práticas de implementação.

\subsection{BB84 (Prepare-and-Measure).}

Proposto por Bennett e Brassard em 1984 \cite{BB84}, o protocolo BB84 é considerado o marco fundador da criptografia quântica moderna e permanece como a proposta mais estudada, implementada e fundamentada teoricamente. Seu esquema explora diretamente o princípio da complementaridade quântica, segundo o qual não é possível medir simultaneamente um estado em duas bases não compatíveis sem introduzir perturbações observáveis. No protocolo, Alice codifica cada bit da chave em um fóton, escolhendo de forma aleatória entre duas bases mutuamente não complementares:
\[
Z=\{\ket{0}, \ket{1}\}, \quad X=\{\ket{+}, \ket{-}\}, \quad \ket{\pm}=\frac{1}{\sqrt{2}}(\ket{0}\pm\ket{1}).
\]
A base $Z$ pode ser representada pela polarização vertical/horizontal dos fótons, enquanto a base $X$ corresponde à polarização diagonal. Bob mede cada fóton recebido em uma base também escolhida aleatoriamente entre $Z$ e $X$ e, após muitas transmissões, ambos anunciam publicamente apenas as bases utilizadas, sem revelar os resultados. Os eventos com bases incompatíveis são descartados, originando a chave bruta (\textit{raw key}); em média, metade dos bits é eliminada nessa etapa de \textit{sifting}.

A segurança emerge do fato de que qualquer tentativa de interceptação altera estatisticamente as correlações entre Alice e Bob. Ao medir numa base errada, a adversária Eve colapsa o estado e introduz erros que se manifestam no \textbf{Quantum Bit Error Rate} (QBER). Uma amostra pública de bits permite estimar o QBER e decidir se o protocolo prossegue: na formulação original, valores acima de aproximadamente $11\%$ inviabilizam a extração de chave. Em regimes ideais de assimetria simples, a taxa secreta pode ser aproximada por:
\[
R \ge q \left[1 - 2h_2(Q)\right],
\]
onde $q$ é a fração de bits preservados após o sifting e $h_2(Q)$ é a entropia binária. Este resultado conecta-se à prova de segurança de Shor–Preskill, que estabelece a extração de chave desde que o erro observado permaneça abaixo do limiar adequado.

Na prática, fontes coerentes atenuadas substituem fontes ideais de fóton único e introduzem emissões multiphotônicas raras, viabilizando ataques \textit{photon-number splitting} (\gls{isdpns}). Para neutralizá-los, o \textbf{Decoy-State BB84} alterna intensidades (sinal e \textit{decoy}) e permite estimar rigorosamente os ganhos e erros associados a fótons únicos, limitando a informação de Eve. Esse endurecimento físico se soma a contramedidas de implementação contra \textit{side-channels} (temporização, potência, blinding de detectores) e à escolha de codificações mais robustas em fibra (fase/tempo). A engenharia de sistema conecta-se diretamente ao desempenho: a atenuação da fibra em 1550 nm ($\sim 0{,}2$ dB/km), a eficiência de detecção, as contagens escuras e a taxa de repetição determinam a probabilidade de detecção por pulso e, portanto, a chave líquida após reconciliação ($f(Q)\!\approx\!1$ com LDPC/Polar) e amplificação de privacidade. Em regime de \textit{finite-key}, flutuações estatísticas exigem margens adicionais e reduzem a taxa efetiva, o que demanda planejamento de janelas de estimação e amostragem adequadas.

Esse fluxo operacional completo está representado na Figura \ref{fig:bb84_schematic}, onde se ilustram as fases de preparação, transmissão quântica, medição, sifting e pós-processamento. O BB84 foi inicialmente proposto em um cenário idealizado com fontes de fótons individuais, mas implementações práticas geralmente utilizam lasers atenuados que produzem, ocasionalmente, múltiplos fótons, possibilitando ataques do tipo \textit{photon-number splitting} (PNS). Para lidar com isso, foram criadas variações como o \textbf{Decoy-State BB84}, em que Alice altera aleatoriamente a intensidade dos pulsos enviados, permitindo estimar a fração de eventos genuínos de fóton único e frustrar estratégias de ataque de Eve.

\begin{figure}[h!]
\centering
\includegraphics[width=0.6\textwidth]{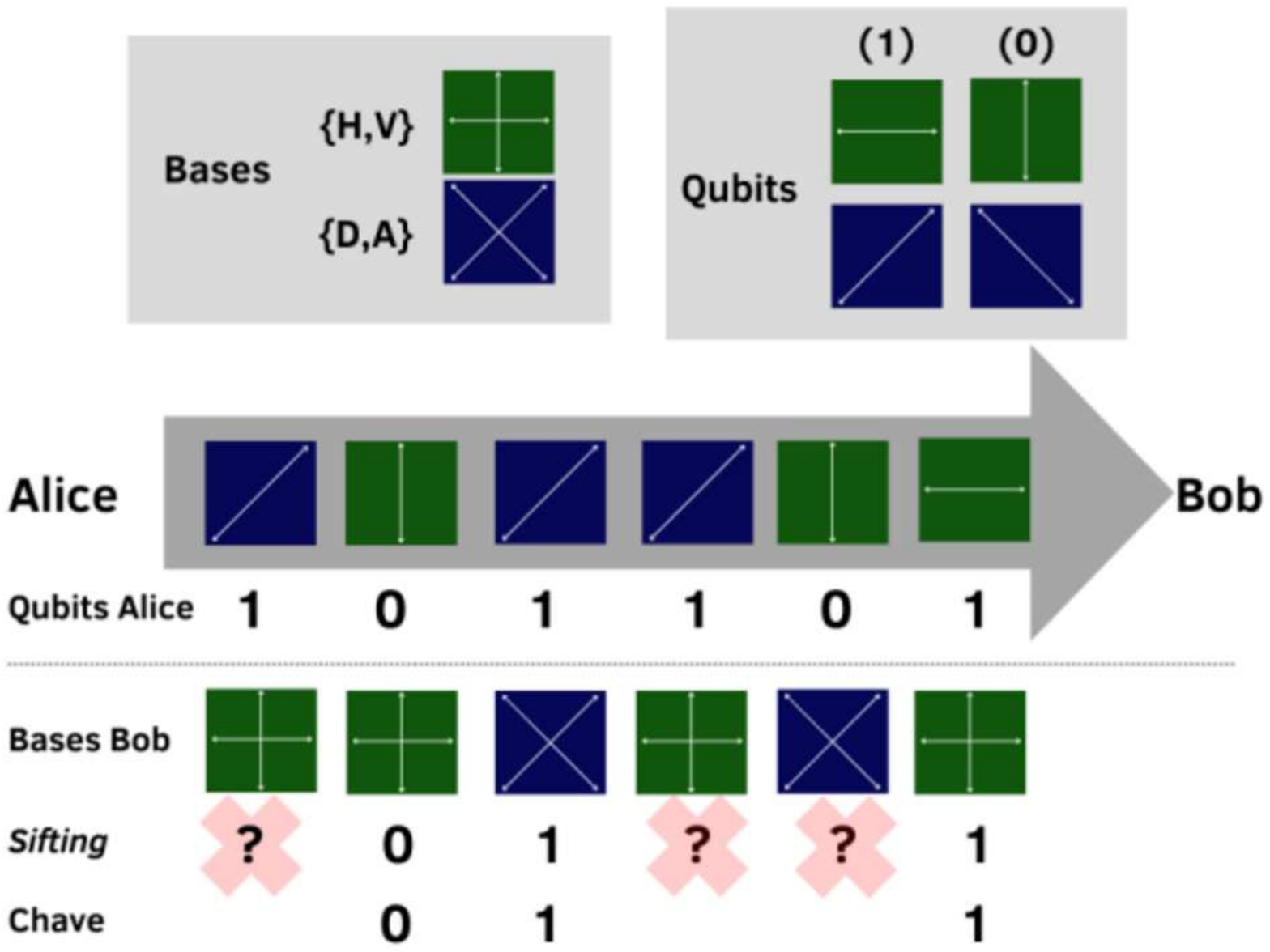}
\caption{Esquema do BB84.}
\label{fig:bb84_schematic}
\end{figure}

Outra variação relevante é o \gls{bbm92}, equivalente ao BB84 mas no cenário emaranhado, no qual uma fonte central distribui pares de fótons emaranhados a Alice e Bob, que medem em bases mutuamente incompatíveis; essa formulação demonstra a equivalência entre prepare-and-measure e protocolos baseados em emaranhamento e aproxima o BB84 do E91. Em enlaces de longa distância, detectores \gls{snspd} com baixíssimas contagens escuras e alta eficiência ampliam significativamente o alcance, enquanto estabilização ativa de fase/polarização reduz derivas e mantém o QBER em níveis compatíveis com taxas líquidas positivas. Em implementações reais, o desempenho do BB84 resulta de um balanço entre perdas do canal, eficiência de detecção e qualidade do pós-processamento. A Tabela~\ref{tab:bb84_param} sintetiza os principais parâmetros operacionais em fibras ópticas e o respectivo impacto em métricas observáveis. Em particular, observa-se que a atenuação a 1550~nm (tipicamente $\sim\!0{,}2$~dB/km) determina a transmissividade $\eta$ e, portanto, o alcance máximo; a eficiência dos detectores e as contagens escuras influenciam diretamente o QBER em longas distâncias; e a eficiência da correção de erros $f(Q)$, quando próxima do limite de Shannon (códigos LDPC/Polar), aumenta a fração de chave líquida recuperável. Também se destaca o papel de \textit{decoy states} para estimar os eventos de fóton único ($Q_1,e_1$) e a importância de estabilização ativa de fase/polarização para manter a fidelidade do estado em enlaces extensos.

\begin{table}[h!]
\centering
\scriptsize
\caption{BB84 em fibra: parâmetros típicos de operação, implicações e impacto em métricas.}
\label{tab:bb84_param}
\begin{tabular}{p{5cm} p{10cm}}
\hline
\textbf{Parâmetro} & \textbf{Implicação/Impacto} \\
\hline
Atenuação da fibra (1550 nm, $\sim$0,2 dB/km) & Define a transmissividade $\eta$ e o alcance útil; perdas adicionais em conectores e emendas entram no orçamento. \\
Eficiência do detector ($\eta_d$) e \emph{dark counts} & Aumentam/diminuem o $p_{\text{click}}$; contagens escuras elevam o QBER em longas distâncias. \\
Taxa de repetição (100 MHz–1 GHz) & Escala a \emph{raw key}; requer controle de jitter e \emph{dead time}. \\
Decoy states (2–3 intensidades) & Estimam $Q_1$ e $e_1$ (fóton único), mitigando PNS; melhoram taxa segura e alcance. \\
Correção de erros (LDPC/Polar), $f(Q)\!\approx\!1$ & Eficiência próxima do limite de Shannon reduz vazamento de informação. \\
Estabilização de polarização/fase & Mantém fidelidade do estado e QBER baixo; essencial em longas fibras. \\
Codificação (polarização vs. fase/tempo) & Fase/tempo é mais robusta a birrefringência em fibras; polarização é conveniente em espaço livre. \\
SNSPD vs. InGaAs SPAD & SNSPD: alta $\eta_d$ e baixo ruído para longas distâncias; InGaAs: custo e operação mais simples. \\
\hline
\end{tabular}
\end{table}

Embora o BB84 ideal ofereça uma narrativa limpa para provas de segurança, a realidade experimental com fontes coerentes demanda técnicas adicionais. A Tabela~\ref{tab:bb84_ideal_decoy} compara lado a lado o BB84 com fontes de fóton único (modelo ideal) e o BB84 com \textit{Decoy-State}, hoje padrão de fato em fibras. A tabela evidencia três pontos práticos: (i) o papel das intensidades múltiplas em transformar a segurança em estimativas verificáveis ($Q_1,e_1$); (ii) a mitigação efetiva de ataques PNS sem degradação significativa de taxa; e (iii) a troca entre simplicidade conceitual e complexidade de implementação, compensada pelo ganho substancial de alcance e robustez.

\begin{table}[h!]
\centering
\scriptsize
\caption{BB84 ideal vs. BB84 com Decoy-State: suposições, ataques mitigados e efeito prático.}
\label{tab:bb84_ideal_decoy}
\begin{tabular}{p{3.5cm}|p{4.6cm} p{5cm}}
\hline
 & \textbf{BB84 ideal (fóton único)} & \textbf{BB84 com Decoy-State (fontes coerentes)} \\
\hline
Suposição de fonte & Emissão de um fóton por pulso & Distribuição de Poisson; intensidades múltiplas \\
Ataque crítico & PNS inoperante (ausente) & PNS detectável/limitado por estimativas de $Q_1,e_1$ \\
Estimadores & $Q,\,QBER$ globais & Ganhos por intensidade, $Q_1$ e $e_1$ por \emph{bounds} \\
Alcance/taxa & Limitados por perdas/detectores & Maior alcance prático em fibra; taxas estáveis \\
Complexidade & Menor (hipótese ideal) & Maior (modulação de intensidade e análise) \\
\hline
\end{tabular}
\end{table}

\vspace{0.5cm}

Em síntese, as diretrizes de projeto resumidas na Tabela~\ref{tab:bb84_param} orientam a operação do BB84 em enlaces reais, enquanto a comparação da Tabela~\ref{tab:bb84_ideal_decoy} justifica a adoção de \textit{Decoy-State} como prática dominante quando se empregam fontes coerentes atenuadas. Em conjunto com a Figura~\ref{fig:bb84_schematic}, esses elementos oferecem um quadro coeso que conecta a prova de segurança ao desempenho observável em campo.

O BB84 constitui-se não apenas no primeiro, mas também no mais consolidado protocolo de QKD: estabeleceu as bases conceituais e experimentais para variantes mais robustas, permanece referência em demonstrações de laboratório e redes metropolitanas, e continua a ser implementado em sistemas comerciais e pilotos de campo, servindo de padrão de comparação para novas gerações como MDI-QKD e Twin-Field.

\subsection{E91 (Entanglement-based).}

O protocolo E91, proposto por Artur Ekert em 1991 \cite{E91}, representa um avanço conceitual significativo em relação ao BB84. Enquanto este utiliza o princípio da complementaridade e a impossibilidade de clonagem de estados arbitrários, o E91 fundamenta-se de maneira explícita no emaranhamento quântico e na violação de desigualdades de Bell, elementos que fornecem garantias de segurança enraizadas na própria não-localidade da mecânica quântica. No esquema canônico, uma fonte central \(S\) gera pares de qubits em estado de Bell, tipicamente:
\[
|\Phi^+\rangle = \frac{1}{\sqrt{2}} \big( |00\rangle + |11\rangle \big),
\]
enviando um qubit para Alice (A) e o outro para Bob (B), mesmo que estejam espacialmente separados. A figura \ref{fig:e91_anexo} ilustra o cenário: \(S\) emite partículas correlacionadas em direções opostas; Alice e Bob escolhem, de forma aleatória e independente, diferentes configurações de medição (eixos \(x,y,z\) ou ângulos equivalentes em polarização). Em virtude do emaranhamento, os resultados exibem correlações que não admitem explicação por teorias clássicas de variáveis ocultas locais.

A certificação dessas correlações é formalizada pelo teste de CHSH. Definindo o parâmetro:
\[
S = E(a,b) + E(a,b') + E(a',b) - E(a',b'),
\]
onde \(E(\cdot,\cdot)\) denota a correlação entre os resultados de Alice e Bob para escolhas de observáveis \(a,a'\) e \(b,b'\), o limite clássico é \(|S| \le 2\), ao passo que a mecânica quântica permite \(|S| \le 2\sqrt{2}\). Uma violação estatisticamente significativa acima de 2 certifica a presença de não-localidade. No E91, essa certificação não é apenas um “selo” de quanticidade: ela vincula a quantidade de informação potencialmente acessível a Eve. Em termos operacionais, o protocolo segue os passos:
\begin{enumerate}
    \item \(S\) gera pares emaranhados e distribui um qubit para cada usuário.
    \item Alice e Bob escolhem aleatoriamente seus ajustes de medição dentre três opções; algumas combinações são reservadas para o cálculo de \(S\) (verificação) e outras para a geração de chave (medidas “compatíveis”).
    \item Um subconjunto de resultados é revelado publicamente para estimar \(|S|\) e taxas de erro; apenas estatísticas (não os bits de chave) são divulgadas.
    \item Se a violação de CHSH persiste acima do limiar e o erro é baixo, procede-se ao pós-processamento (reconciliação de erros e amplificação de privacidade) para extrair a chave.
\end{enumerate}

Comparado ao BB84, em que a segurança decorre primariamente de incompatibilidade de bases e da não-clonagem, o E91 “eleva” a garantia ao ancorá-la em correlações não-locais observáveis. Isso aproxima o protocolo do paradigma \textit{device-independent}: a validade não exige um modelo detalhado do dispositivo, mas sim a observação de uma violação de Bell suficientemente forte, desde que condições de fechamento de \emph{loopholes} (detecção e localidade) sejam atendidas. Na prática, a implementação impõe desafios: é preciso gerar pares emaranhados com alta taxa e fidelidade, preservar coerência por longas distâncias, sincronizar bases de medição e alcançar eficiências de detecção elevadas para superar perdas e fechar o \emph{loophole} de detecção. Tais requisitos motivaram avanços em fontes paramétricas (SPDC), acoplamento eficiente a fibras e detectores SNSPD de baixa contagem escura.

Uma forma útil de ver o E91 é pela partição entre “rodadas de verificação” e “rodadas de chave”. Nas primeiras, escolhem-se combinações de observáveis que maximizam sensibilidade à violação (por exemplo, ângulos a 0°, 45°, 22,5° e 67,5° em uma realização prática), estimando \(|S|\) e QBER; nas segundas, utilizam-se configurações que fornecem correlações quase perfeitas para extrair bits. Essa divisão explicita o custo estatístico de certificação: quanto maior a fração dedicada a verificação, mais robusta é a detecção de adulteração, ao preço de menor taxa líquida, um compromisso típico em cenários de chave finita. Em \textit{finite-key}, a taxa segura efetiva é reduzida por termos que refletem incertezas de estimação e vazamento na reconciliação; intuitivamente, confiança maior na violação observada requer amostras maiores, especialmente em canais ruidosos.

Do ponto de vista de modelagem, o E91 admite versões discretizadas que se alinham a implementações com bases finitas, inclusive o \textbf{BBM92}, frequentemente visto como “BB84 via emaranhamento”. Nessa leitura, a equivalência entre prepare-and-measure e esquemas baseados em emaranhamento é explicitada: ao traçar parcialmente a fonte, recuperar-se-ia um cenário efetivo de estados preparados do ponto de vista de cada usuário. Ao mesmo tempo, o E91 reforça o vínculo com protocolos \textbf{DI-QKD}, nos quais a chave é extraída apenas quando a violação de Bell (e.g., CHSH) é suficientemente alta para limitar a informação de Eve independentemente da caracterização detalhada dos dispositivos.

A implementação prática em fibras longas e enlaces espaço-livre exige atenção a estabilidade de fase/polarização, filtragem espectral e temporal, e equalização de perdas nos braços. Em fibras, codificações por graus de liberdade de tempo e fase podem ser preferíveis à polarização devido à birrefringência; em espaço-livre, a polarização permanece conveniente. Em ambos os casos, o controle de coincidências e a janela de temporização determinam a rejeição de ruído de fundo e a taxa de eventos válidos, impactando diretamente a estatística de \(S\) e a taxa de chave.

\begin{figure}[h!]
\centering
\includegraphics[width=0.6\textwidth]{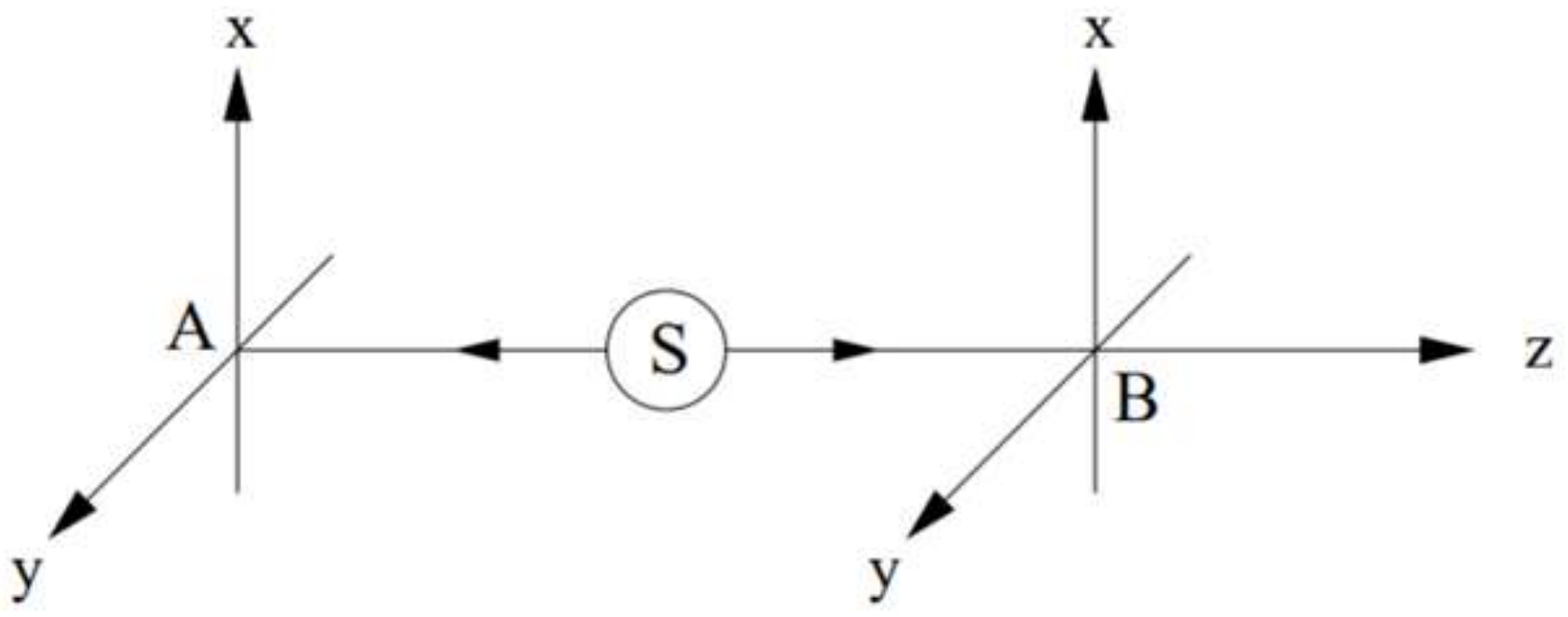}
\caption{Esquema do protocolo E91.}
\label{fig:e91_anexo}
\end{figure}

Como é possível ver na Tabela \ref{tab:e91_contexto} a fonte $S$ produz pares de partículas emaranhadas enviadas a observadores distantes (Alice e Bob), que realizam medições em ajustes distintos ($x,y,z$). A violação de CHSH ($|S|>2$) certifica não-localidade e limita a informação de um adversário, permitindo extração de chave após reconciliação e amplificação de privacidade. Para contextualizar o papel do E91 frente a variantes próximas usadas em prática e em pesquisa, a Tabela~\ref{tab:e91_contexto} resume as diferenças essenciais quanto ao fundamento de segurança e às implicações experimentais. 

\vspace{0.5cm}

Em particular, observa-se que o E91 exige a verificação explícita da violação de Bell (via CHSH), enquanto o BBM92 opera de forma operacionalmente equivalente ao BB84 emaranhado sem medir $|S|$ a cada rodada, reduzindo o custo estatístico de certificação. Já o DI-QKD representa o limite superior de garantias, mas demanda fechamento simultâneo de \emph{loopholes} e eficiências elevadas de detecção, o que explica sua atual maturidade mais restrita.

\begin{table}[h!]
\centering
\scriptsize
\caption{Contexto resumido: E91 versus variantes relacionadas.}
\label{tab:e91_contexto}
\begin{tabular}{p{3.6cm} p{4.9cm} p{4.9cm}}
\hline
\textbf{Esquema} & \textbf{Fundamento de segurança} & \textbf{Observações práticas} \\
\hline
E91 (CHSH) & Violação de Bell limita informação de Eve; verificação explícita de $|S|$ & Requer emaranhamento de alta fidelidade e eficiências elevadas de detecção \\
BBM92 & Equivalente a BB84 via emaranhamento (sem teste explícito de Bell) & Menor custo de certificação; aproximação prática do E91 \\
DI-QKD (geral) & Segurança sem confiança em dispositivos, apenas com $|S|$ alto e loopholes fechados & Desafiante hoje; forte motivador tecnológico (SNSPD, fontes brilhantes) \\
\hline
\end{tabular}
\end{table}

Como se depreende da Tabela~\ref{tab:e91_contexto}, escolher entre E91, BBM92 e DI-QKD envolve um compromisso entre custo de certificação, requisitos físicos e nível de confiança assumido nos dispositivos. Projetos que priorizam taxa e simplicidade tendem ao BBM92; cenários que demandam certificação explícita de não-localidade adotam E91; e arquiteturas com garantias máximas almejam DI-QKD, à medida que avanços em fontes emaranhadas e SNSPDs tornem viáveis as eficiências necessárias.

\subsection{SARG04}

O protocolo SARG04 (Scarani--Acín--Ribordy--Gisin, 2004) surgiu como uma modificação do BB84 voltada para enfrentar vulnerabilidades práticas decorrentes do uso de fontes não ideais, em particular os ataques de \textit{photon-number splitting} (PNS). No BB84 padrão, quando se utilizam lasers coerentes atenuados em vez de verdadeiras fontes de fóton único, existe probabilidade não nula de emissão multiphotônica. Nesse cenário, uma adversária (Eve) pode extrair um fóton do pulso e encaminhar os demais a Bob, preservando compatibilidade com o detector e aprendendo o bit sem elevar significativamente o QBER, o que compromete a segurança em longas distâncias se nenhuma contramedida for empregada.

A inovação central do \textbf{SARG04} é substituir a revelação de \emph{bases} (Z ou X) por um anúncio público de \emph{conjuntos de estados candidatos}. Com isso, mesmo na presença de pulsos multiphotônicos, a informação pública não permite a Eve distinguir deterministamente o estado real. O protocolo usa os mesmos quatro estados do BB84, $\{\ket{0},\ket{1},\ket{+},\ket{-}\}$, porém reorganiza a etapa clássica de sifting. Em linhas gerais: (i) Alice prepara e envia um dos quatro estados; (ii) após a medição de Bob (em base aleatória), Alice não revela a base, mas sim um \textbf{par de estados} que contém o estado enviado (por exemplo, $\{\ket{0},\ket{+}\}$, $\{\ket{0},\ket{-}\}$, $\{\ket{1},\ket{+}\}$, $\{\ket{1},\ket{-}\}$); (iii) Bob, à luz do seu resultado, decide se a informação é \emph{conclusiva}, isto é, se pode excluir um dos candidatos e inferir o bit, ou \emph{inconclusiva}, caso em que descarta o evento; (iv) Eve, mesmo conhecendo o par anunciado, permanece limitada pela não ortogonalidade dos estados, que impede uma discriminação sem erro em regime determinístico.

Essa mudança sutil na \emph{informação pública} altera a superfície de ataque: no BB84, a revelação de base combinada a pulsos multiphotônicos torna o PNS especialmente efetivo, pois o conhecimento da base reduz o problema de detecção de Eve à distinção entre dois estados ortogonais; no SARG04, a revelação por pares não ortogonais preserva ambiguidade residual mesmo quando há múltiplos fótons, reduzindo a vantagem de um adversário que armazena um fóton e mede depois. Em termos de desempenho, isso se traduz em maior tolerância a perdas e alcance ampliado com fontes coerentes fracas, ao custo de uma fração maior de eventos inconclusivos e, portanto, de menor taxa líquida em cenários ideais.

Do ponto de vista teórico, o SARG04 pode ser analisado como um protocolo de \emph{discriminação mínima de erro} ou \emph{discriminação sem ambiguidade} sob restrições impostas pelo anúncio público. As regiões de decisão de Bob, condicionadas tanto pela base medida quanto pelo par anunciado, garantem que os eventos aceitos tenham correlações fortes suficientes para sustentar a extração de chave após correção de erros e amplificação de privacidade. Já em \textit{finite-key}, a presença de eventos inconclusivos demanda planejamento de orçamento estatístico: é preciso reservar amostras para estimar taxas de erro condicionais e calibrar a fração de rodadas dedicadas à verificação, mantendo confiança nas estimativas sem sacrificar em demasia a vazão de chave.

Na prática, o SARG04 tem duas vantagens adicionais: (i) a \emph{compatibilidade de hardware} com plataformas BB84, moduladores de fase/polarização, atenuadores e detectores permanecem os mesmos,, concentrando as mudanças na lógica de sifting e no processamento clássico; (ii) a \emph{robustez incremental} sem introduzir, de início, as camadas de análise estatística de intensidades \emph{decoy}. Ainda que, hoje, o \textbf{Decoy-State QKD} ofereça taxas e alcances superiores com fontes coerentes, o SARG04 permanece relevante quando se busca mitigar PNS com mínima reengenharia e como referência conceitual de como a semântica do anúncio público molda a segurança. Em enlaces de fibra longos, combinar SARG04 com estabilização ativa de fase/polarização, filtragem temporal estreita e detectores de baixa contagem escura ajuda a manter o QBER sob controle e a maximizar a fração de eventos conclusivos, aproximando o desempenho do limite analítico.

\begin{figure}[h!]
\centering
\begin{tikzpicture}[
  >=Stealth, font=\small,
  box/.style={draw, rounded corners, align=center, inner sep=3pt, minimum width=36mm, minimum height=9mm},
  wide/.style={draw, rounded corners, align=center, inner sep=3pt, minimum width=80mm, minimum height=9mm},
  lbl/.style={font=\footnotesize, inner sep=1pt}
]

\node[box] (alice) at (0,0) {Alice\\Escolhe um entre\\$\{\ket{0},\ket{1},\ket{+},\ket{-}\}$};
\node[box] (bob)   [right=32mm of alice] {Bob\\Mede em base aleat\'oria\\$Z$ ou $X$};

\draw[->, thick] (alice.east) -- node[above, lbl]{canal qu\^antico} (bob.west);

\coordinate (midq) at ($(alice.east)!0.5!(bob.west)$);
\node[box] (eve) [below=9mm of midq] {Eve\\PNS/medi\c c\~ao imperfeita\\(n\~ao-ortogonalidade limita USD)};
\draw[->, dashed] (midq) -- (eve.north);

\node[box, fill=white, minimum width=56mm] (hint) [below=4mm of eve] {pares n\~ao ortogonais $\Rightarrow$ USD limitada};
\draw[densely dotted] (eve.south) -- (hint.north);

\node[wide] (ann) [below=9mm of hint] {An\'uncio de par (p\'os-medi\c c\~ao)\\
\footnotesize $\{\ket{0},\ket{+}\},\ \{\ket{0},\ket{-}\},\ \{\ket{1},\ket{+}\},\ \{\ket{1},\ket{-}\}$};

\coordinate (tapL) at ($(ann.west)+(-5mm,0)$);
\coordinate (tapR) at ($(ann.east)+( 5mm,0)$);
\draw[->, thick] (alice.south) |- ++(0,-4mm) -| (tapL) -- (ann.west);
\draw[->, thick] (bob.south)   |- ++(0,-4mm) -| (tapR) -- (ann.east);
\node[lbl, anchor=east] at ($(tapL)+(-2mm,0)$) {cl\'assico};

\node[lbl] (rule) [below=4mm of ann] {regra de decis\~ao};
\node[box] (concl)   [below left =7mm and 14mm of rule] {Caso conclusivo\\(Bob exclui 1 candidato)};
\node[box] (inconcl) [below right=7mm and 14mm of rule] {Caso inconclusivo\\(descarta evento)};

\draw[->, thick] (ann.south) |- ++(-6mm,-3mm) -| (concl.north);
\draw[->, thick] (ann.south) |- ++( 6mm,-3mm) -| (inconcl.north);

\node[box, minimum width=54mm] (raw) [below=9mm of concl] {Chave bruta (\textit{sifting})};
\draw[->, thick] (concl.south) -- (raw.north);

\node[box, minimum width=74mm] (pp) [below=9mm of raw] {Corre\c c\~ao de erros \,+\, Amplifica\c c\~ao de privacidade};
\draw[->, thick] (raw.south) -- (pp.north);

\node[box, minimum width=46mm] (sk) [below=9mm of pp] {Chave secreta final};
\draw[->, thick] (pp.south) -- (sk.north);

\end{tikzpicture}
\caption{SARG04: diagrama vertical, compacto e contido, sem texto sob linhas e sem extens\~oes para fora das margens.}
\label{fig:sarg04_vertical}
\end{figure}
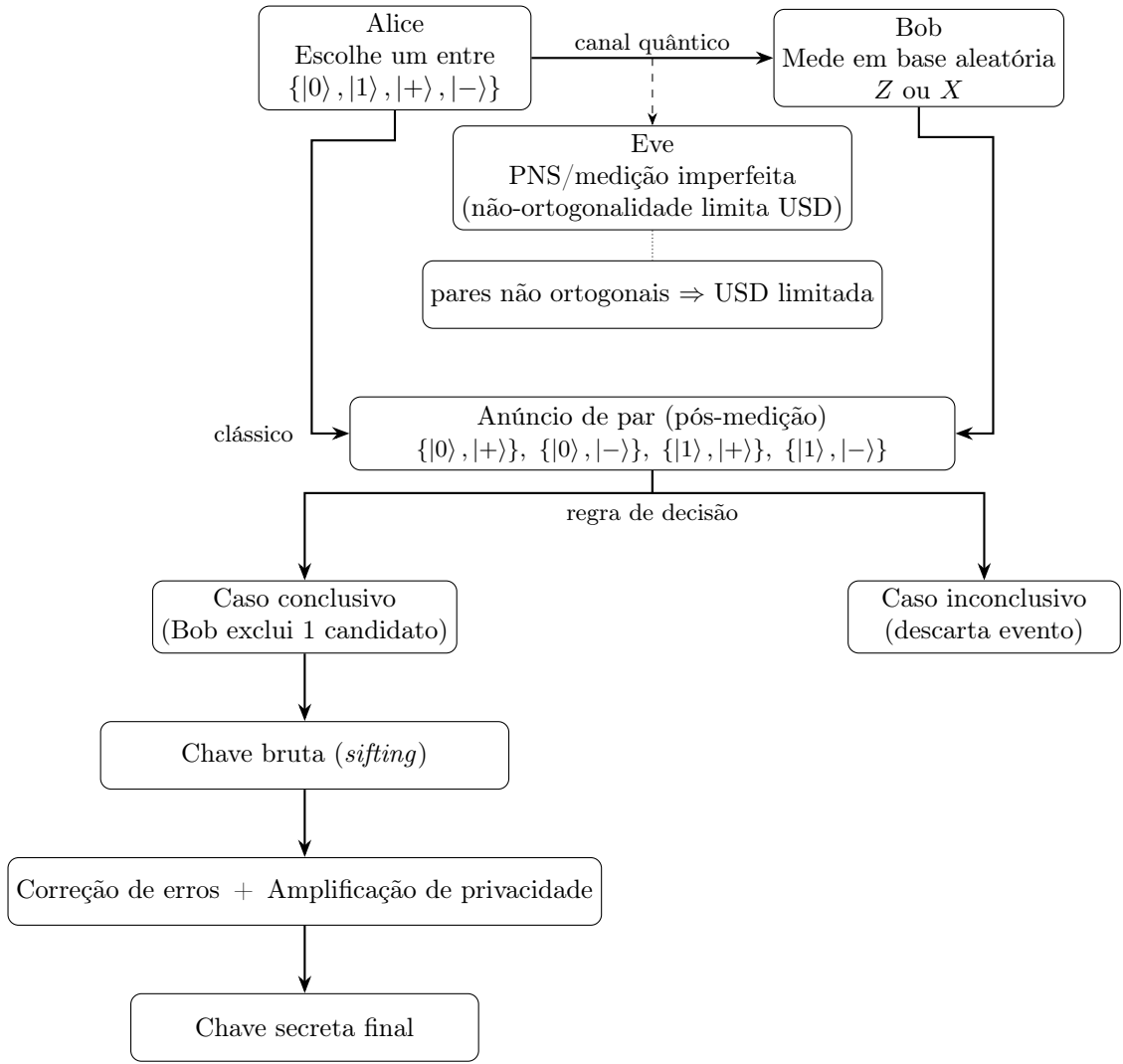

\subsection{Decoy-State QKD}
\label{subsec:decoy}

O \textit{Decoy-State QKD}, proposto por Lo, Ma e Chen em 2005~\cite{Decoy2005}, introduz um \emph{controle estatístico ativo} sobre a fonte coerente atenuada: ao randomizar a intensidade óptica entre um conjunto finito de níveis, torna-se possível separar as contribuições dos diferentes números de fótons e, com isso, impor limites contra estratégias adversariais como o ataque de \textit{photon-number splitting} (PNS). Em dispositivos reais, a fonte não emite fótons únicos ideais: usa-se um laser fraco com intensidade média $\mu<1$, cujo número $n$ de fótons por pulso segue a distribuição de Poisson:
\[
P_n(\mu)=\frac{\mu^n}{n!}\,e^{-\mu},
\]
o que abre espaço para PNS quando $n\ge 2$. A ideia central do Decoy-State é intercalar, aleatoriamente, pulsos de \textbf{intensidades diferentes} (os \emph{decoys}) entre os pulsos de sinal. Esses pulsos auxiliares não carregam bits úteis, mas permitem estimar com precisão as estatísticas de detecção por número de fótons, expondo manipulações de Eve.

Do ponto de vista de modelagem, cada intensidade $\lambda\in\{\mu,\nu,\omega\!\approx\!0\}$ induz uma mistura de estados de número de fótons com probabilidades $P_n(\lambda)$. Denotando por $Y_n$ a \emph{yield} (probabilidade de clique condicional a $n$ fótons) e por $e_n$ a taxa de erro condicionada, obtemos, para cada intensidade $\lambda$:

\[
Q_\lambda=\sum_{n\ge 0}P_n(\lambda)\,Y_n,
\qquad
E_\lambda Q_\lambda=\sum_{n\ge 0}P_n(\lambda)\,e_n Y_n.
\]

\vspace{1cm}
O truque do decoy é “fechar” um sistema em $Y_1$ e $e_1$ com poucas intensidades, usando também um nível de vácuo para ancorar $Y_0$. Com dois decoys ($\nu$ e $\omega\!\approx\!0$) e um sinal $\mu$, têm-se limites:
\[
Y_0 \approx Q_\omega,
\qquad
Y_1 \ge \frac{\mu e^{\mu}Q_\nu - \nu e^{\nu}Q_\mu - (\mu-\nu)Y_0}{\mu\nu(\mu-\nu)},
\qquad
Q_1=\mu e^{-\mu}Y_1,
\]
e um limite superior típico para o erro de fóton único:
\[
e_1 \le \frac{E_\nu Q_\nu e^{\nu} - e_0 Y_0}{\nu e^{\nu} Y_1},
\quad \text{com } e_0\simeq \tfrac{1}{2}.
\]
A taxa secreta resultante (na formulação padrão assimétrica) é:
\[
R \;\ge\; q\Big[-\,Q_\mu\,f(E_\mu)\,h_2(E_\mu)\;+\;Q_1\big(1-h_2(e_1)\big)\Big],
\]
onde $f(E_\mu)$ é a ineficiência da correção de erros, $E_\mu$ é a QBER observada para $\mu$, e $h_2$ é a entropia binária. A Figura~\ref{fig:qkd_decoy_alinhado} mostra um diagrama com ramificações (preparo, canal, coleta e estimação com loops de validação). A Tabela~\ref{tab:decoy_levels} resume uma configuração prática com três níveis e o papel estatístico de cada um.

Em implementações reais, usa-se tipicamente: (i) três níveis de intensidade, $\mu$ (sinal, p.ex. 0.45–0.6), $\nu$ (decoy, 0.05–0.2) e $\omega\!\approx\!0$ (vácuo); (ii) escolha assimétrica de bases (mais eventos em $Z$ para geração do que em $X$ para teste), reduzindo o custo de amostragem de QBER; (iii) correção de erros adaptativa com $f(E_\mu)\!\approx\!1.12$--$1.18$ e amplificação de privacidade por hashing universal. Para distâncias longas, otimiza-se conjuntamente intensidades e frações de uso levando em conta ruído escuro, afterpulsing e \textit{jitter}. Em MDI-QKD e Twin-Field QKD, o mecanismo de decoy migra para o nó de interferência, mas a lógica de “linearização estatística por Poisson” permanece idêntica.

A Figura~\ref{fig:qkd_decoy_alinhado} organiza visualmente esse fluxo em quatro blocos: preparo (programação de intensidades e bases), transmissão pelo canal e medição de Bob, estimação decoy de $Y_0$, $Y_1$ e $e_1$, e pós-processamento clássico (sifting, correção de erros e amplificação de privacidade). As duas entradas laterais para a estimação correspondem justamente às estatísticas por intensidade, $Q_\lambda$ e $E_\lambda$, cuja coleta é realizada de forma independente para cada nível $\lambda\in\{\mu,\nu,\omega\}$. Já a Tabela~\ref{tab:decoy_levels} resume uma configuração prática de três níveis, destacando o papel estatístico de cada um: o vácuo ancora $Y_0$, o decoy permite limitar $Y_1$ e $e_1$, e o sinal maximiza a contribuição positiva $Q_1\!\big(1-h_2(e_1)\big)$ na taxa $R$.

Sem decoys, ataques dependentes do número de fótons, em particular PNS, podem vazar informação sem alterar significativamente a QBER. Ao medir $(Q_\lambda,E_\lambda)$ para diferentes intensidades conforme ilustrado na Figura~\ref{fig:qkd_decoy_alinhado}, o protocolo efetua uma “tomografia” do canal por classes de fótons usando as identidades:
\[
Q_\lambda=\sum_{n\ge 0}P_n(\lambda)\,Y_n
\quad\text{e}\quad
E_\lambda Q_\lambda=\sum_{n\ge 0}P_n(\lambda)\,e_n Y_n,
\]
o que habilita os limites para $Y_1$ e $e_1$ apresentados no texto. Esses parâmetros entram de forma conservadora em $R$, garantindo segurança composicional quando combinados com autenticação e correção de erros.

Na prática, os valores de $\mu$ e $\nu$ (e suas frações de uso) são otimizados em função da distância, perdas do enlace e taxa de \emph{dark counts}. A Tabela~\ref{tab:decoy_levels} fornece faixas típicas adequadas a enlaces metropolitanos; para distâncias maiores, reduz-se tipicamente $\mu$ e aumenta-se a fração de eventos em $\nu$ para manter precisão nas estimativas. Além disso, é comum empregar bases assimétricas (maior probabilidade para $Z$ do que para $X$) para reduzir o custo do termo $f(E_\mu)h_2(E_\mu)$, como evidenciado no bloco de amplificação de privacidade da Figura~\ref{fig:qkd_decoy_alinhado}.

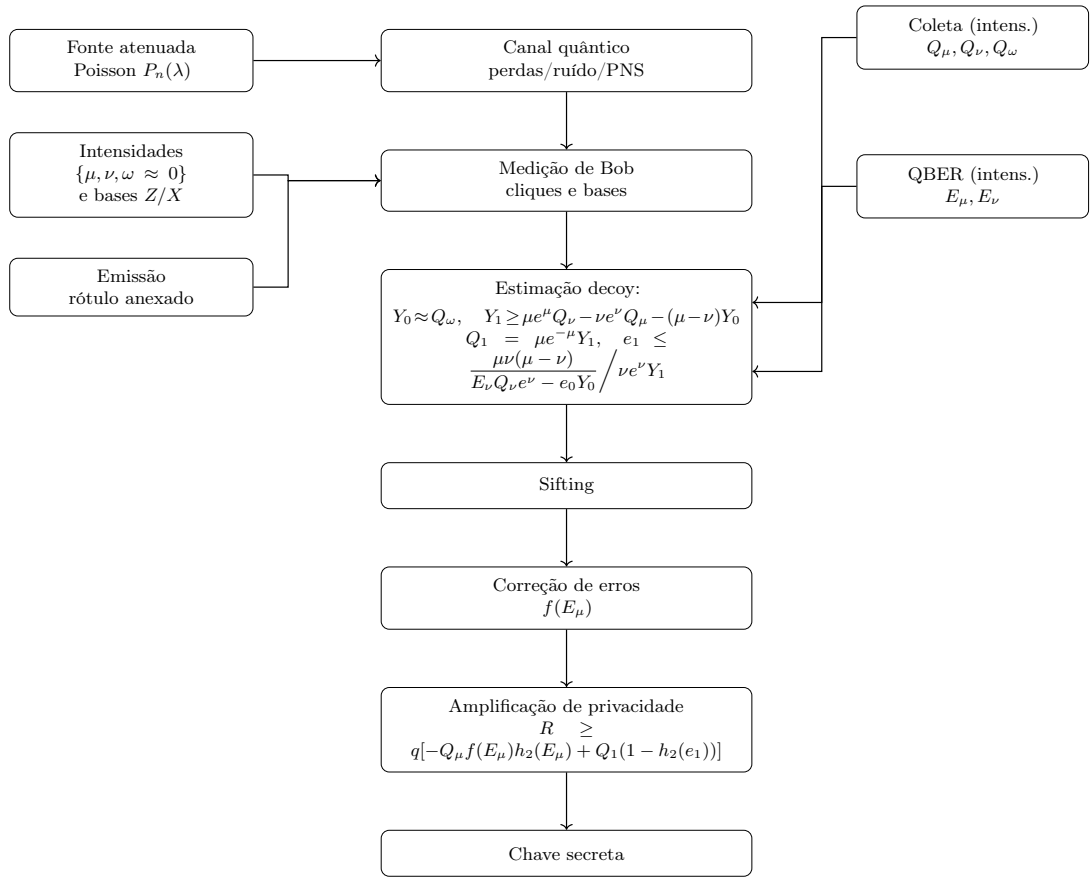
\begin{figure}[t]
\centering
\resizebox{0.9\linewidth}{!}{
\begin{tikzpicture}[
  node distance=7mm and 10mm,
  box/.style={draw, rounded corners, align=center, inner sep=2mm, minimum height=8mm, text width=60mm},
  side/.style={draw, rounded corners, align=center, inner sep=2mm, minimum height=8mm, text width=36mm},
  lbl/.style={font=\small},
  arr/.style={->, line width=0.6pt},
  every node/.style={font=\footnotesize}
]

\node[box, text width=38mm] (fonte) {Fonte atenuada\\Poisson $P_n(\lambda)$};
\node[box, below=7mm of fonte, text width=38mm] (intens) {Intensidades\\$\{\mu,\nu,\omega\!\approx\!0\}$\\e bases $Z/X$};
\node[box, below=7mm of intens, text width=38mm] (emissao) {Emissão\\rótulo anexado};


\node[box, right=22mm of fonte] (canal) {Canal quântico\\ perdas/ruído/PNS};
\node[box, below=10mm of canal] (medicao) {Medição de Bob\\ cliques e bases};
\node[box, below=10mm of medicao] (estim) {%
  Estimação decoy:\\[1mm]
  $Y_0\!\approx\!Q_\omega,\quad Y_1\!\ge\!\mu e^\mu Q_\nu- \nu e^\nu Q_\mu-(\mu-\nu)Y_0$\\
  $Q_1=\mu e^{-\mu}Y_1,\quad e_1\!\le\!\dfrac{\mu\nu(\mu-\nu)}{E_\nu Q_\nu e^\nu - e_0 Y_0}\bigg/\nu e^\nu Y_1$
};
\node[box, below=10mm of estim] (sifting) {Sifting};
\node[box, below=10mm of sifting] (corr) {Correção de erros\\ $f(E_\mu)$};
\node[box, below=10mm of corr] (priv) {Amplificação de privacidade\\
$R \ge q\!\left[-Q_\mu f(E_\mu)h_2(E_\mu)+Q_1(1-h_2(e_1))\right]$};
\node[box, below=10mm of priv] (chave) {Chave secreta};

\node[side, right=18mm of canal, yshift=4mm, anchor=west] (coleta) {Coleta (intens.)\\ $Q_\mu,Q_\nu,Q_\omega$};
\node[side, right=18mm of medicao, yshift=-1mm, anchor=west] (qber) {QBER (intens.)\\ $E_\mu,E_\nu$};

\draw[arr] (fonte.east) -- ++(6mm,0) |- (canal.west);
\draw[arr] (intens.east) -- ++(6mm,0) |- (medicao.west);
\draw[arr] (emissao.east) -- ++(6mm,0) |- (medicao.west);

\draw[arr] (coleta.west) -- ++(-6mm,0) |- ($(estim.east)+(0,6mm)$);
\draw[arr] (qber.west)   -- ++(-6mm,0) |- ($(estim.east)+(0,-6mm)$);

\draw[arr] (canal.south) -- (medicao.north);
\draw[arr] (medicao.south) -- (estim.north);
\draw[arr] (estim.south) -- (sifting.north);
\draw[arr] (sifting.south) -- (corr.north);
\draw[arr] (corr.south) -- (priv.north);
\draw[arr] (priv.south) -- (chave.north);

\end{tikzpicture}
}
\caption{Decoy-State QKD compacto e alinhado: rótulos laterais elevados e conexões ortogonais estáveis.}
\label{fig:qkd_decoy_alinhado}
\end{figure}

Pequenas imprecisões nas intensidades alvo ($\lambda\to\lambda\pm\delta$) deslocam as probabilidades Poissonianas $P_n(\lambda)$ e podem enviesar os limites para $Y_1$ e $e_1$. Implementações modernas tratam esse efeito por “decoys imprecisos”, propagando intervalos de tolerância na estimação estatística. Esse cuidado, aliado à calibração contínua e ao controle de \emph{afterpulsing}/\emph{jitter}, mantém a consistência entre as medições por intensidade (lado direito da Figura~\ref{fig:qkd_decoy_alinhado}) e os limites inseridos em $R$.

O mesmo mecanismo de amostragem por intensidades se estende a MDI-QKD e Twin-Field QKD: a coleta $(Q_\lambda,E_\lambda)$ passa ao nó de interferência, mas a lógica de linearização estatística por Poisson permanece a mesma. Assim, a Figura~\ref{fig:qkd_decoy_alinhado} pode ser lida como um diagrama canônico de estimação por decoys, enquanto a Tabela~\ref{tab:decoy_levels} serve como referência rápida de parametrização para sistemas de três níveis.

\begin{table}[h!]
\centering
\scriptsize
\caption{Configuração típica de Decoy-State QKD (três intensidades) e papel estatístico.}
\label{tab:decoy_levels}
\begin{tabular}{lccc p{6.6cm}}
\hline
Categoria & Intensidade & Fração de uso & Observáveis & Papel na estimação \\
\hline
Sinal  & $\mu \in [0.4,0.6]$   & 70--85\% & $Q_\mu,\,E_\mu$   & Contribui diretamente para a taxa $R$; combinado com decoys limita $Y_1$ e permite estimar $Q_1$ \\
Decoy  & $\nu \in [0.05,0.2]$  & 10--25\% & $Q_\nu,\,E_\nu$   & Quebra estratégias PNS; define limites para $Y_1$ e $e_1$ em função de $\mu,\nu$ \\
Vácuo  & $\omega \approx 0$     & 5--10\%  & $Q_\omega$        & Estima ruído de fundo ($Y_0$), refinando limites para $Y_1$ e $e_1$ \\
\hline
\end{tabular}
\end{table}

\subsection{DPS, COW e RRDPS}
\label{subsec:dps_cow_rrdps}

Além dos protocolos BB84-like, esquemas diferenciais em pulsos coerentes oferecem compatibilidade nativa com infraestrutura de telecom.  Os três mais maduros são o \emph{Differential Phase Shift} (DPS), o \emph{Coherent One-Way} (COW) e o \emph{Round-Robin DPS} (RRDPS).  A \autoref{fig:dps_cow_rrdps} resume seus fluxos ópticos essenciais—de Alice até Bob—enquanto o texto aprofunda modelagem estatística, provas de segurança, vulnerabilidades práticas e resultados de campo.

No \textbf{DPS}, bits são codificados pela fase relativa entre pulsos consecutivos, $\Delta\phi\in\{0,\pi\}$.  A probabilidade de clique em cada detector vale $P_{\mathrm{click}}^{\mathrm{DPS}}=1-e^{-\eta\mu}-Y_0$, onde $\eta$ inclui todas as perdas e $Y_0$ é o \emph{dark count}.  A visibilidade interferométrica relaciona-se à QBER via $Q=(1-\mathcal V)/2$ e, sob ataques coletivos, a informação de Eve obedece:
\[
I_{AE}^{\mathrm{DPS}}\le h_2(Q),\qquad
R_{\mathrm{DPS}}\ge P_{\mathrm{click}}^{\mathrm{DPS}}\bigl[1-2h_2(Q)\bigr].
\]
Ensaios de campo em fibras de 90 km já atingiram taxas superiores a $500\,$kbit/s usando lasers de banda C e detectores \emph{self-differencing} InGaAs.

No \textbf{COW}, bits são definidos pela presença ou ausência de intensidade em janelas temporais consecutivas, enquanto a coerência é monitorada por um interferômetro que mede a visibilidade $V=\tfrac{I_{\max}-I_{\min}}{I_{\max}+I_{\min}}$.  Admitindo ruído predominantemente de decoerência, a bound de chave assume:
\[
R_{\mathrm{COW}}\ge P_{\mathrm{sig}}\!\left[1-h_2(Q)\right]-p_m h_2\!\Bigl(\tfrac{1-\!V}{2}\Bigr),
\]
com $P_{\mathrm{sig}}=(1-p_m)(1-e^{-\eta\mu}-Y_0)$ e $Q\approx (1-\sqrt V)/2$.  Demonstrações comerciais alcançam $\sim1\,$Mbit/s em 25 km e mantêm chaves positivas em linhas de 150 km quando $V\gtrsim0.98$.

No \textbf{RRDPS}, a segurança é independente da QBER: Bob escolhe aleatoriamente um atraso $r\in\{1,\dots,L-1\}$ somente após receber todo o trem, limitando a informação adversária a:
\[
I_{AE}^{\mathrm{RRDPS}}\le\frac{1}{L-1},\qquad
R_{\mathrm{RRDPS}}\ge P_{\mathrm{click}}^{\mathrm{RRDPS}}\!\Bigl[1-h_2(Q)-\tfrac{1}{L-1}\Bigr],
\]
onde $P_{\mathrm{click}}^{\mathrm{RRDPS}}=1-e^{-\eta\mu L}-Y_0$.  Blocos $L\!=\!128$ já geraram $\sim10\,$kbit/s após 160 km sem decoy; $L\!>\!10^3$ viabiliza chaves mesmo com QBER acima de 8\%.

Todos os esquemas podem empregar intensidades decoy $\{\mu,\nu,\omega\}$ para restringir ataques \emph{photon-number splitting}.  No caso de DPS, define-se o ganho interferométrico por intensidade, $Q_\lambda=\sum_n P_n(\lambda)Y_n^{\mathrm{DPS}}$, e obtêm-se limites para $Y_1^{\mathrm{DPS}}$ análogos ao BB84-decoy.  Em COW, decoys calibram o ruído de fundo e apertam limites sobre $V$, enquanto no RRDPS são opcionais, mas úteis para estimar $Y_0$ com precisão.

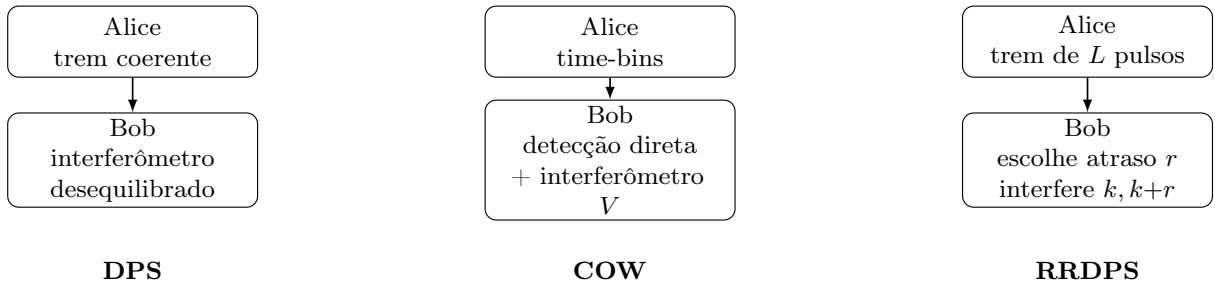
\begin{figure}[t]
\centering
\resizebox{\linewidth}{!}{
\begin{tikzpicture}[>=latex, font=\footnotesize,
  box/.style={draw, rounded corners, align=center, inner sep=2pt, minimum height=9mm, text width=3.0cm},
  arr/.style={->, line width=0.6pt}]
\coordinate (x0) at (0,0);
\coordinate (x1) at (6,0);
\coordinate (x2) at (12,0);

\node[box] (dA) at ($(x0)+(0,1.2)$) {Alice\\trem coerente};
\node[box] (dB) at ($(x0)+(0,-0.3)$) {Bob\\interferômetro\\desequilibrado};
\draw[arr] (dA) -- (dB);
\node at ($(x0)+(0,-1.7)$) {\textbf{DPS}};

\node[box] (cA) at ($(x1)+(0,1.2)$) {Alice\\time-bins};
\node[box] (cB) at ($(x1)+(0,-0.3)$) {Bob\\detecção direta\\+\ interferômetro\\$V$};
\draw[arr] (cA) -- (cB);
\node at ($(x1)+(0,-1.7)$) {\textbf{COW}};

\node[box] (rA) at ($(x2)+(0,1.2)$) {Alice\\trem de $L$ pulsos};
\node[box] (rB) at ($(x2)+(0,-0.3)$) {Bob\\escolhe atraso $r$\\interfere $k,k{+}r$};
\draw[arr] (rA) -- (rB);
\node at ($(x2)+(0,-1.7)$) {\textbf{RRDPS}};
\end{tikzpicture}}
\caption{Visão comparativa de DPS, COW e RRDPS sem sobreposições nem setas “perdidas”.}
\label{fig:dps_cow_rrdps}
\end{figure}

Sob o ponto de vista de ataques práticos, \emph{phase-remapping} e \emph{detector blinding} exigem contramedidas: (i) monitoramento de potência incidente nos APDs, (ii) varredura de fase piloto para revelar remapeamento e (iii) autenticação ativa de calhas óticas para coibir injeção maliciosa.  Estudos de \emph{finite-key} mostram que, para $10^{11}$ bits enviados, a penalização estatística $\Delta_{\mathrm{FK}}\sim10^{-3}$ é dominada por flutuações de $Q$ em DPS/COW e por estimação de $Y_0$ em RRDPS.

Integrações em fotônica de silício reduzem volume óptico e estabilizam interferômetros: PICs com moduladores termo-ópticos ($<2$ mm) já reproduzem DPS a 10 Gbit$/$s modulando $\mu\approx0.25$; redes de anéis ressonantes oferecem chaves de atraso reconfiguráveis para RRDPS.  Ensaios \emph{chip-to-chip} em fibra curta superaram 100 Mbit$/$s (DPS) e validaram COW em enlaces intra-datacenter. Para operação multiplexada WDM, filtros $\Delta\lambda<0.4\,\text{nm}$ e orçamentos de potência < -30 dBm por canal clássico mitigam espalhamento Raman que elevaria $Y_0$.

Comparando desempenho: DPS maximiza taxa bruta via interferometria simples; COW adiciona diagnóstico de integridade em tempo real com modesto sobre-custo óptico; RRDPS provê a barreira conceitual mais forte contra ruído, mas requer receptores multi-caminho de baixa perda e memória óptica curta para manter estabilidade de fase nos $L$ atrasos.  Com estabilização ativa (PID), correção LDPC eficiente $(f\lesssim1.15)$ e autenticação clássica robusta, esses esquemas diferenciais já figuram em projetos-piloto de redes metropolitanas e troncais, estendendo a camada quântica de segurança a infra-estruturas ópticas de alta velocidade.

\subsection{MDI-QKD (Measurement-Device Independent)}
\label{subsec:mdi}

O Measurement-Device-Independent QKD (MDI-QKD), introduzido por Lo, Curty e Qi em 2012 \cite{MDIQKD}, foi concebido para extinguir o ponto mais vulnerável dos sistemas de chave quântica: o estágio de detecção.  Nos protocolos anteriores, todo clique de fôton único era interpretado como veredicto absoluto da natureza; porém, em hardware real, os APDs podem ser cegados por luz forte, sofrer \emph{time-shift} ou ser induzidos a comportar-se de forma determinística, permitindo que Eve reconstrua a chave sem alterar a QBER visível.  A reengenharia do MDI-QKD desloca a fotodetecção para um nó intermediário, doravante chamado Charlie, que pode estar em posse do adversário.  Alice e Bob comportam-se apenas como transmissores, enviando pulsos coerentes fracos codificados em duas bases complementares.  A segurança deixa de depender de qualquer hipótese sobre o detector: mesmo que Charlie divulgue resultados fabricados, as estatísticas globais não podem violar as desigualdades derivadas das leis quânticas para pares de fôton único.

O ciclo fundamental é o seguinte.  Cada usuário sorteia um bit $b\in\{0,1\}$, uma base $\beta\in\{Z,X\}$ e uma intensidade $\lambda\in\{\mu,\nu,\omega\approx0\}$.  O pulso emitido é um estado coerente com amplitude $\sqrt\lambda$ e fase $\phi$ de acordo com a codificação binária da base escolhida (polarização ou fase).  Os pulsos propagam-se por fibras independentes até Charlie, que realiza uma Bell-State Measurement (BSM) por meio de interferência Hong--Ou--Mandel num divisor 50:50 seguido de dois detectores.  Uma coincidência em canais opostos identifica, por exemplo, o estado $\ket{\Psi^{+}}$, enquanto uma coincidência no mesmo canal identifica $\ket{\Psi^{-}}$.  Charlie publica somente o rótulo da projeção; não revela qualquer informação de tempo ou intensidade além do necessário para o sifting.

A interpretação do protocolo baseia-se em equivalência de quadros.  Se considerarmos uma fonte virtual de estados entre Alice e Bob que emite pares emaranhados $\ket{\Phi^{+}}$, as preparações de cada usuário podem ser vistas como medições dessa fonte em suas estações.  A BSM de Charlie, por sua vez, implementa \emph{entanglement swapping}, colapsando o par partilhado em um estado correlacionado condicionado ao resultado publicado.  A prova de segurança segue então o tratamento de entanglement-based QKD: basta estimar a fração de eventos genuinamente formados por um fóton de Alice e um fóton de Bob, denotada $S_{11}$, e o erro de fase correspondente $e_{11}$.  Sob a hipótese de fontes cujos números de fóton seguem Poisson, o método de três intensidades fornece os limites:

\[
S_{11}\ge
\frac{\mu^{2}e^{\mu}\bigl(Q_{\nu\nu}-Q_{\mu\nu}\bigr)-
      (\mu^{2}-\nu^{2})e^{\mu+\nu}Q_{\omega\omega}}
     {\mu\nu(\mu-\nu)},
\qquad
e_{11}\le
\frac{E_{\nu\nu}Q_{\nu\nu}-E_{\omega\omega}Q_{\omega\omega}}
     {S_{11}}.
\]

As quantidades $Q_{\lambda\kappa}$ e $E_{\lambda\kappa}$ são, respectivamente, ganho e QBER observados quando Alice (Bob) usam intensidades $\lambda$ ($\kappa$).  No limite assintótico, a taxa segura é:

\[
R\ge S_{11}\bigl[1-h_2(e_{11})\bigr]-Q_{\mu\mu}f(E_{\mu\mu})h_2(E_{\mu\mu}),
\]

onde $h_2$ é a entropia binária e $f(E_{\mu\mu})$ mede a ineficiência do código de correção de erros.  Para blocos finitos, adiciona-se uma penalização $\Delta_{\mathrm{FK}}\approx\sqrt{\frac{\ln(2/\varepsilon_{\mathrm{sec}})}{N}}$ que faz $R$ decrescer suavemente com o logaritmo inverso da tolerância de segurança.

A viabilidade experimental exige que os pulsos de Alice e Bob sejam indistinguíveis em largura temporal, frequências ópticas e polarização.  Controladores de temperatura mantêm a deriva de frequência abaixo de 100 MHz; compensadores de atraso acionados por piezo estabilizam diferenças de chegada $|\Delta\tau|$ a menos de 20 ps; e moduladores de polarização automáticos recuperam alinhamento sempre que a birefringência da fibra varia.  Em campo aberto, usa-se canal piloto em 1570 nm com modulação de chaveamento Manchester para estimar fase e atraso a cada centena de microssegundos, realimentando um laço PID no transmissor.

Detectores de próxima geração -- SNSPDs de NbN ou WSi -- apresentam eficiência quântica superior a 80\%, jitter sub-40 ps e ruído escuro abaixo de $10^{-8}$ por janela, elevando a visibilidade Hong--Ou--Mandel para além de 0.50.  Nesses regimes, demonstrações de laboratório já ultrapassaram 500 km de fibra standard, embora a taxa de chave caia para poucas dezenas de bits por segundo quando a atenuação total excede 70 dB.  Em enlaces metropolitanos de 10 a 40 km, FPGAs implementam decodificadores LDPC que operam a 100 Mbaud e realizam amplificação de privacidade via hashing Toeplitz em $O(n\log n)$ tempo, sustentando chaves na faixa de 10 Mbit/s.

O desenho em estrela confere escalabilidade: um único Charlie pode servir dezenas de pares Alice-Bob num esquema de múltiplos comprimentos de onda compatível com DWDM, transmitindo chaves paralelas em canais com espaçamento de 100 GHz.  Como o nó central não requer confiança, basta protegê-lo contra sabotagem física, não contra inspeção lógica.  A topologia encaixa-se em datacenters, onde transceptores quânticos podem ficar nas bordas dos racks enquanto o módulo BSM habita um armário óptico comum.  Estudos de roteamento sugerem que, em redes densas, a sobrecarga de dois enlaces em vez de um aumenta a atenuação total apenas em 3 dB na média -- sacrífico compensado pela eliminação completa de ataques a detectores.

Variações do protocolo incluem MDI-QKD assíncrono, que remove a necessidade de clock compartilhado usando pacotes de largura muito maior, e Twin-Field-MDI, que insere um laser de referência comum e obtém dependência de chave $\propto\sqrt\eta$ em vez de $\propto\eta$, estendendo a distância útil além de 600 km.  Outra linha de pesquisa usa modos de tempo-freqüência de ordem elevada para codificar qudits, aumentando a taxa por pulso sem alterar o cerne da prova de segurança.

\begin{figure}[t]
\centering
\resizebox{0.96\linewidth}{!}{%
\begin{tikzpicture}[>=latex, font=\footnotesize,
  box/.style ={draw, rounded corners, align=center,
               text width=38mm, inner sep=4pt},
  arr/.style ={->, line width=.6pt, shorten >=2pt},
  darr/.style={->, densely dashed, line width=.5pt, shorten >=2pt}]

\node[box] (alice) at (-6,  3.4)
  {Alice\\laser DFB\\PM/IM moduladores\\bases $Z/X$, decoy $\lambda$};
\node[box] (bob)   at ( 6,  3.4)
  {Bob\\laser DFB\\PM/IM moduladores\\bases $Z/X$, decoy $\lambda$};

\node[box] (synA) at (-6,  1.6) {síncrono 1570 nm};
\node[box] (synB) at ( 6,  1.6) {síncrono 1570 nm};

\node[box] (charlie) at (0,  0.9)
  {Charlie (não confiável)\\divisor 50:50\\interferência HOM\\SNSPD};

\node[box] (claA) at (-6, -0.9) {canal clássico\\TLS + MAC};
\node[box] (claB) at ( 6, -0.9) {canal clássico\\TLS + MAC};

\node[box] (sift)  at ( 0, -2.8)
  {sifting\\LDPC EC\\hash Toeplitz\\chave final};


\draw[arr] (alice.south) -- (charlie.160);
\draw[arr] (bob.south)   -- (charlie.20);

\draw[darr] (synA.south) -- (charlie.200);
\draw[darr] (synB.south) -- (charlie.-20);

\draw[arr] (charlie.south west) to[out=-120,in=90] (claA.north);
\draw[arr] (charlie.south east) to[out=-60 ,in=90] (claB.north);

\draw[arr] (claA.south) |- (sift.west);
\draw[arr] (claB.south) |- (sift.east);

\end{tikzpicture}}
\caption{Arquitetura de referência para MDI-QKD sem sobreposições.}
\label{fig:mdi_pipeline}
\end{figure}
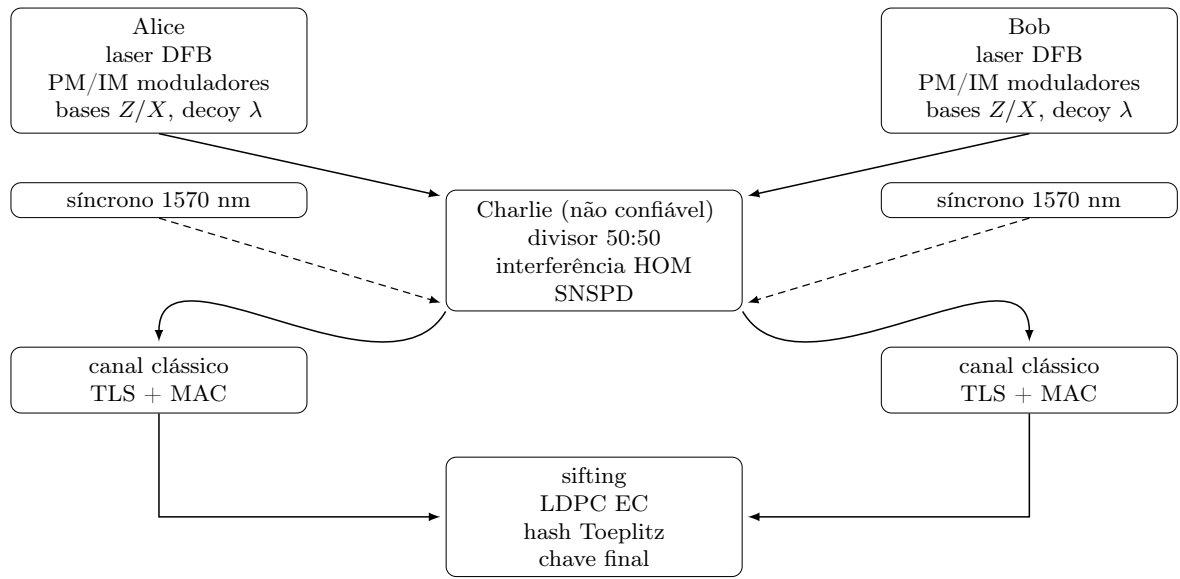

Graças à robustez contra ataques de detecção, à compatibilidade com componentes de telecom e à facilidade de integração em topologias em estrela, o MDI-QKD desponta como coluna vertebral das redes quânticas em construção, servindo tanto a enlaces backbone de centenas de quilômetros quanto a interconexões \emph{chip-to-chip} em centros de processamento de dados.

\subsection{Twin-Field QKD (TF-QKD)}
\label{subsec:tfqkd}

Twin-Field QKD (TF-QKD), proposto em 2018 por Lucamarini \emph{et al.} \cite{TFQKD2018}, mostrou que é possível ultrapassar o \gls{plob} em canais ópticos sem repetidores quânticos ideais.  Nesse protocolo, Alice e Bob nunca trocam fótons diretamente; cada um gera um pulso coerente fraco, com fase aleatória \(\phi_A\) e \(\phi_B\), que percorre apenas metade da distância total até um nó intermediário (Charlie).  Os dois pulsos interferem num divisor 50:50, e a simples divulgação pública do detector que clicou permite que os usuários, munidos de informações sobre intensidades e fatias de fase escolhidas, extraíam uma chave segura mesmo que todo o equipamento de Charlie, inclusive detectores, esteja sob controle de um adversário.  

Como cada pulso só atravessa metade do percurso, a fração útil de eventos de fóton-único passa a escalar como \(\sqrt{\eta}\), deslocando a distância de chave nula de aproximadamente \(250\;\text{km}\) (BB84-Decoy) para além de \(600\;\text{km}\) em fibra padrão; em 2024 já se demonstraram chaves positivas depois de \(1\,000\;\text{km}\) usando três enlaces TF concatenados.

\medskip
O ciclo operacional inclui três sorteios independentes em cada repetição: (i) a intensidade \(\{\mu,\nu,\omega\}\) para o método decoy, (ii) o bit lógico \(\theta\in\{0,\pi\}\) e (iii) a fase contínua \(\phi\in[0,2\pi)\), quantizada em \(M\) fatias angulares.  Só se mantêm eventos em que \(\phi_A-\phi_B\) cai na mesma fatia; a estatística resultante gera os ganhos \(Q_k\) e erros \(E_k\) que, por técnicas decoy, isolam \(Y_{11}^{(k)}\) (rendimento real de pares fóton-único) e o erro de fase \(e_{11}^{\varphi}\).  No regime de bloco finito obtém-se:
\[
R \ge \sum_{k=1}^{M}\frac{N_k}{N}
      \Bigl[\,Y_{11}^{(k)}\bigl(1-h_2(e_{11}^{\varphi})\bigr)
      - Q^{(k)}_{\mu\mu}\,f\!\bigl(E^{(k)}_{\mu\mu}\bigr)
        h_2\!\bigl(E^{(k)}_{\mu\mu}\bigr)\Bigr]
      - \sqrt{\frac{\ln(2/\varepsilon_{\mathrm{sec}})}{N}},
\]
onde \(f(E)\) é o overhead do código LDPC e a raiz final cobre estatísticas finitas.

\medskip
A interferência é extremamente sensível a flutuações de caminho: em \(400\;\text{km}\) a diferença de fase deve permanecer abaixo de \(50\;\text{mrad}\) em janelas de microssegundos.  Na prática empregam-se
\begin{enumerate}
    \item pulsos piloto TDM a \(1310\;\text{nm}\) que percorrem o mesmo enlace e retornam por refletores Faraday, fornecendo a fase absoluta local;
    \item moduladores de fase \(\text{LiNbO}_3\) controlados por servos PID de \(20\;\text{kHz}\);
    \item lasers DFB travados a cavidades de silício resfriadas, reduzindo o \emph{linewidth} a \(\lesssim 500\;\text{Hz}\).
\end{enumerate}

As demonstrações mais relevantes até hoje incluem 6 kbit/s em 300 km (2019), 1,4 kbit/s em 509 km subterrâneos (2020), 14 kbit/s após 615 km (2022) e 3 kbit/s depois de 1 000 km com três nós TF (2024).  Testes WDM paralelizaram 16 canais em 100 km, entregando 30 kbit/s, confirmando compatibilidade com infra-estrutura DWDM existente. Estão em investigação: (i) a variação \emph{sending-not-sending}, que reduz o número de intensidades; (ii) versões \emph{dual-band}, deslocando o piloto para \(2\,\upmu\text{m}\) e atenuando espalhamento Raman; (iii) integração fotônica em \(\mathrm{Si}_3\mathrm{N}_4\), permitindo emissores “plug-and-play”; e (iv) esquemas híbridos que intercalam nós TF e repetidores de memória limitada para alcançar enlaces transcontinentais antes da maturidade dos repetidores quânticos completos.

\begin{figure}[t]
\centering
\resizebox{0.9\linewidth}{!}{%
\begin{tikzpicture}[>=latex, font=\footnotesize,
  box/.style ={draw, rounded corners, align=center,
               text width=43mm, inner sep=4pt},
  arr/.style ={->, line width=.6pt, shorten >=2pt},
  darr/.style={->, densely dashed, line width=.5pt, shorten >=2pt}]

\node[box] (alice) at (-6, 2.8)
  {Alice\\laser ultraestável\\PM/IM moduladores\\fase $\phi_A$};
\node[box] (bob)   at ( 6, 2.8)
  {Bob\\laser ultraestável\\PM/IM moduladores\\fase $\phi_B$};

\node[box] (charlie) at (0, 0)
  {Charlie (não confiável)\\beam splitter 50\!:\!50\\D0 \quad D1\\\scriptsize(estima $\phi_A-\phi_B$)};

\node[box] (key) at (0, -2.5)
  {sifting + EC/PA\\chave final};

\draw[arr] (alice.south) -- (charlie.165);
\draw[arr] (bob.south)   -- (charlie.15);

\draw[darr] (alice.east)++(0.4,0.1)
            to[out=-10,in=180]
            node[midway, above, sloped] {\scriptsize pulso de referência}
            (charlie.120);

\draw[arr] (charlie.south) -- (key.north);
\end{tikzpicture}}
\caption{Esquema mínimo do TF-QKD.}
\label{fig:tfqkd_setup}
\end{figure}
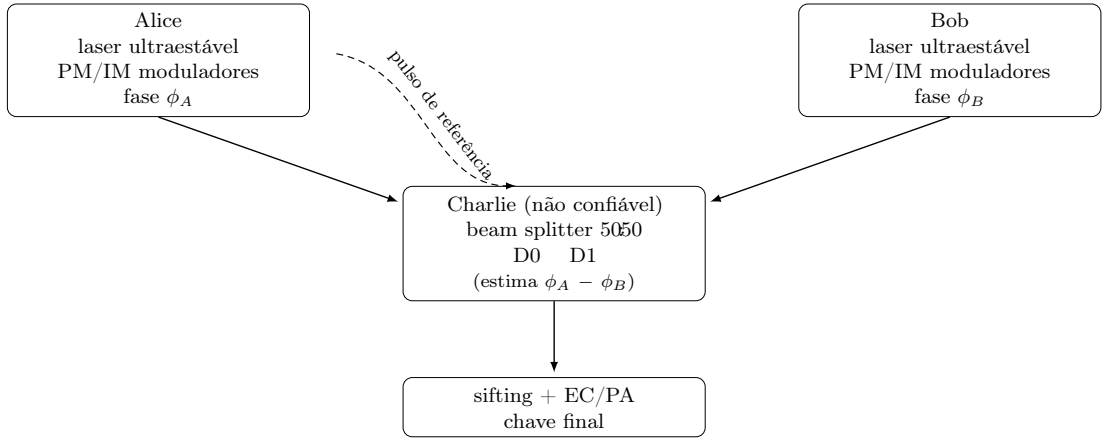

Na Figura \ref{fig:tfqkd_setup} é possível visualizar que os sinais de Alice e Bob percorrem metade da distância total e interferem em Charlie; um pulso de referência (seta tracejada) percorre o mesmo caminho e permite estimar a diferença de fase sem confiança no nó intermediário. Graças à dependência em \(\sqrt{\eta}\), à imunidade contra detectores comprometidos (herdada do modelo MDI) e à compatibilidade com fibras de telecom, o TF-QKD desponta como solução prática para \emph{backbones} quânticos de longo alcance, servindo de ponte até que repetidores plenamente quânticos sejam viáveis comercialmente.

\subsection{CV-QKD (Continuous-Variable)}
\label{subsec:cvqkd}

A criptografia quântica de variáveis contínuas (CV-QKD) foi concebida no começo dos anos 2000, notadamente com o protocolo de Grosshans–Grangier (GG02), como alternativa natural aos esquemas de variáveis discretas (BB84, E91).  A motivação principal reside no reaproveitamento de componentes já comuns em redes ópticas de telecomunicações: lasers de onda contínua, moduladores eletro-ópticos (EOM), amplificadores EDFA e receptores coerentes homódinos ou heteródinos.  Essa compatibilidade reduz custos e facilita a adoção em redes metropolitanas e \emph{backhaul} existentes.

\medskip
Nos protocolos CV a informação binária é codificada nas quadraturas \(X\) (amplitude) e \(P\) (fase) do campo eletromagnético, grandezas que obedecem à relação de comutação \([X,P]=i\hbar\).  Alice aplica uma modulação gaussiana de variância \(V_A\) a estados coerentes \(\ket{\alpha}\), com \(\alpha=x+ip\) e \(x,p\sim\mathcal{N}(0,V_A)\).  Os pulsos seguem por uma fibra óptica até Bob, que efetua:

\begin{itemize}
  \item \emph{detecção homódina}, escolhendo no momento da medição qual quadratura registrar, ou
  \item \emph{detecção heteródina}, medindo ambas simultaneamente com acréscimo de meio fóton de ruído quântico.
\end{itemize}

O resultado é um par de variáveis contínuas correlacionadas \((x_A,x_B)\) ou \((p_A,p_B)\).  Para extrair uma chave binária secreta empregam-se códigos de reconciliação de baixa densidade (LDPC) sob o paradigma de \emph{reconciliação reversa}, em que Bob publica informação de paridade e Alice o corrige; isso torna a taxa de chave crescente com a eficiência de correção \(\beta\) mesmo em canais muito atenuados.

De um ponto de vista de segurança, assume-se que Eve controla totalmente o canal gaussiano (perdas \(\eta\) e ruído excessivo \(\xi\)) mas não pode acessar a detecção de Bob.  A taxa secreta composta é limitada por:

\[
K \;\ge\; \beta I_{AB} - \chi_{BE},
\qquad
I_{AB} = \tfrac{1}{2}\log_2\!\bigl[\tfrac{V+\chi_\text{tot}}{\chi_\text{tot}}\bigr],
\]

onde \(V = V_A + 1\) é a variância total (o termo “1’’ corresponde ao ruído de vácuo) e 
\(\chi_{\text{tot}} = 1 + \xi + \tfrac{1-\eta}{\eta}\) o ruído efetivo percebido por Bob.  
O termo de Holevo \(\chi_{BE}\) obtém-se a partir dos autovalores da matriz de covariância global;  
para canais puramente atenuadores, com ruído de fase dentro das recomendações ITU-T, uma eficiência de reconciliação típica \(\beta \simeq 0{.}97\) possibilita chaves positivas até \(25\,\mathrm{dB}\) de perda (\(\approx 125\,\mathrm{km}\) em fibra SMF-28) em laboratório e \(\approx 80\,\mathrm{km}\) em enlaces urbanos sujeitos a ruído térmico. Os esquemas atualmente mais investigados são:
\begin{enumerate}
  \item o GG02 com detecção heteródina e modulação gaussiana, já operando a \(100\,\mathrm{MHz}\) e gerando dezenas de \(\mathrm{kbit/s}\) em \(40\,\mathrm{km}\);
  \item variantes baseadas em \emph{estados comprimidos}, que reduzem o ruído quântico efetivo e ampliam o alcance em cerca de \(3\text{--}4\,\mathrm{dB}\), à custa de fontes ópticas não-clássicas estáveis;
  \item o CV-MDI-QKD, que transfere a detecção para um nó não confiável, suprimindo ataques em fotodetectores, mas acrescentando \(4\text{--}5\,\mathrm{dB}\) ao orçamento de perdas.
\end{enumerate}

\medskip
Persistem, entretanto, obstáculos relevantes. O desempenho de códigos LDPC despenca quando a SNR cai abaixo de \(0{.}1\), exigindo blocos de \(\sim 10^{9}\) amostras e algoritmos de reconciliação multiescala.  Além disso, a necessidade de manter o oscilador local em fase com a portadora recebida requer distribuição remota do LO ou pilotos ópticos embutidos, hoje realizados por multiplexação em polarização ou frequência.  Por fim, ataques de saturação e manipulação de ganho no receptor coerente demandam contramedidas alinhadas às recomendações do ETSI (grupo QKD-ISG) para CV-QKD.

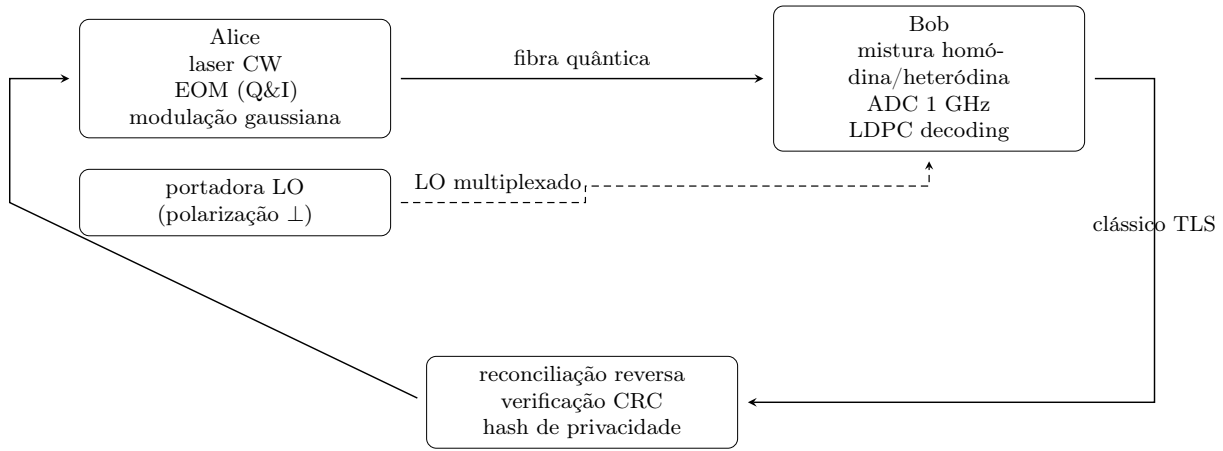
\begin{figure}[t]
\centering
\resizebox{1\linewidth}{!}{%
\begin{tikzpicture}[>=latex, font=\footnotesize,
  box/.style ={draw, rounded corners, align=center,
               text width=42mm, inner sep=4pt},
  arr/.style ={->,>=stealth, line width=.6pt,
               shorten >=4pt, shorten <=4pt},
  darr/.style={->,>=stealth, densely dashed, line width=.5pt,
               shorten >=4pt, shorten <=4pt}]

\node[box] (alice) at (-5, 3)
  {Alice\\laser CW\\EOM (Q\&I)\\modulação gaussiana};

\node[box] (lo) at (-5, 1.2)
  {portadora LO\\(polarização $\perp$)};

\node[box] (bob) at ( 5, 3)
  {Bob\\mistura homódina/heteródina\\ADC 1~GHz\\LDPC decoding};

\node[box] (class) at ( 0,-1.7)
  {reconciliação reversa\\verificação CRC\\hash de privacidade};


\draw[arr] (alice.east) -- node[midway, above]{fibra quântica} (bob.west);

\draw[darr] (lo.east) -- ++(2.8,0)                        
            node[pos=0.55, above]{LO multiplexado}
            |- ($(bob.south)+(0,-0.5)$)                   
            -- (bob.south);                               

\draw[arr] (bob.east) -- ++(1.0,0) |- 
           node[pos=0.25, above]{clássico TLS}
           (class.east);

\draw[arr] (class.west) -- ($(lo.west)+(-1.0,0)$) |- (alice.west);

\end{tikzpicture}}
\caption{Arquitetura típica de CV-QKD.}
\label{fig:cvqkd_setup}
\end{figure}

Na Figura \ref{fig:cvqkd_setup} Alice modula quadraturas \(X,P\) em pulsos coerentes; a portadora do oscilador local (LO) viaja multiplexada pelo mesmo enlace; Bob mede as quadraturas por detecção coerente e, em seguida, envia paridades autenticadas via TLS para que Alice finalize a reconciliação e a amplificação de privacidade.

Em síntese, o CV-QKD concilia teoria de informação quântica gaussiana rigorosa com um ecossistema industrial já consolidado em comunicações ópticas.  Embora a segurança composicional dependa de suposições estritas sobre ruído excessivo e linearidade dos receptores, avanços em pilotagem de fase, integração fotônica de receptores coerentes CMOS e códigos LDPC adaptativos indicam um caminho plausível para enlaces metropolitanos de centenas de quilômetro operando a taxas de megabit por segundo, preparando o terreno para redes quânticas híbridas que combinem variáveis discretas, contínuas e protocolos de campo médio.

\subsection{DI-QKD (Device-Independent)}

O Device-Independent QKD (DI-QKD) representa o limite mais forte de segurança em criptografia quântica, sendo considerado um dos objetivos finais da área. Sua proposta é eliminar a necessidade de se confiar na descrição teórica detalhada dos dispositivos usados por Alice e Bob. Em cenários convencionais (como BB84, Decoy ou mesmo MDI-QKD), assume-se que as fontes e detectores funcionam de acordo com modelos ideais ou conhecidos. No entanto, falhas ou vulnerabilidades físicas podem ser exploradas por adversários, um problema conhecido como \textit{side-channel attacks}.  

A ideia central do DI-QKD é que sua segurança não depende de pressupostos sobre como os dispositivos funcionam, mas apenas da {observação experimental de correlações quânticas não-locais}, detectadas pela violação de desigualdades de Bell. Dessa forma, mesmo que Alice e Bob utilizem equipamentos construídos por Eve, eles ainda poderão garantir segurança se observarem correlações incompatíveis com teorias de variáveis ocultas locais. Funcionamentos básicos podem ser resumidos abaixo:

\begin{enumerate}
    \item Uma fonte gera pares de partículas emaranhadas, distribuindo-as para Alice e Bob (que podem estar em locais distantes).
    \item Cada um escolhe aleatoriamente entre diferentes configurações de medição, registrando os resultados.
    \item Após muitas rodadas, eles utilizam parte dos dados coletados para calcular o parâmetro $S$ da desigualdade de CHSH:
    \[
    S = E(a,b) + E(a,b') + E(a',b) - E(a',b'),
    \]
    onde $E(a,b)$ é a correlação entre os resultados para escolhas de observáveis $a$ e $b$.
    \item Se o valor obtido satisfizer $S > 2$, há violação da desigualdade clássica e, portanto, correlações quânticas não locais foram observadas.
    \item Uma chave secreta pode ser extraída dos resultados correlacionados usando técnicas de pós-processamento semelhantes às do BB84 (correção de erros e amplificação de privacidade).
\end{enumerate}

A segurança do DI-QKD é garantida contra qualquer adversário compatível com a física quântica. O parâmetro de interesse é a {informação de Eve}, limitada pela quantidade de violação da desigualdade de Bell. Quanto maior a violação, menor a informação que um adversário poderia deter. Formalmente, a taxa de chave segura pode ser expressa por:
\[
R \geq H(A|E) - H(A|B),
\]
onde:
\begin{itemize}
    \item $H(A|E)$ é a incerteza de Eve em relação aos resultados de Alice (vinte diretamente da violação de Bell observada),
    \item $H(A|B)$ é a taxa de erro entre Alice e Bob,
    \item a diferença garante que a chave só é segura se a informação compartilhada entre Alice e Bob exceder a possível informação de Eve.
\end{itemize}

Apesar de ser o protocolo de maior segurança teórica, a implementação do DI-QKD enfrenta dificuldades consideráveis:
\begin{itemize}
    \item \textbf{Eficiência de detecção:} é necessário superar o \emph{loophole de detecção}, garantindo que as violações de Bell não sejam explicadas por falhas na coleta de dados.
    \item \textbf{Separação espacial:} Alice e Bob precisam estar suficientemente afastados para fechar o \emph{loophole da localidade}, de modo que as escolhas de medição sejam realmente independentes.
    \item \textbf{Perdas no canal:} qualquer perda significativa impede observar estatisticamente a violação necessária para prover segurança.
\end{itemize}

Nos últimos anos, experimentos de {\gls{belltest} “loophole-free”}, como os realizados em Delft (2015), NIST (2015) e Viena (2017), validaram experimentalmente a possibilidade de implementação de protocolos DI-QKD. Primeiros protótipos de DI-QKD já foram demonstrados em laboratório, com distâncias ainda limitadas (da ordem de dezenas de metros em fibras ou set-ups de laboratório).  

Com a evolução das tecnologias de fontes de fótons emaranhados de alta taxa e detectores de eficiência quase unitária, o DI-QKD caminha para aplicações práticas em redes quânticas. Uma de suas variantes mais promissoras é o {Measurement-Device-Independent QKD} (MDI-QKD), considerada uma versão intermediária que já hoje é implementável em telecomunicações, herdando princípios de segurança inspirados no DI-QKD.

\medskip
\noindent
O \textbf{DI-QKD} representa o protocolo de segurança mais forte possível, confiando apenas em correlações não-locais observadas experimentalmente. Embora ainda limitado por perdas e desafios experimentais, constitui um dos caminhos futuros mais promissores para a construção de uma rede quântica global totalmente segura e independente de dispositivos.

\subsection{QDS (Quantum Digital Signatures)}
\label{subsec:qds}

As assinaturas digitais quânticas (QDS) estendem a QKD do objetivo de confidencialidade para os requisitos de autenticidade, integridade e não–repúdio entre múltiplos receptores.  Diferentemente das assinaturas clássicas baseadas em suposições de complexidade (RSA, ECDSA), a solidez das QDS decorre de princípios físicos, teorema da não–clonagem e impossibilidade de discriminar perfeitamente estados não ortogonais, o que as torna seguras mesmo na presença de adversários com computadores quânticos universais.

\medskip
Modelos práticos de QDS usam estados coerentes atenuados e receptores com detecção direta, evitando fontes de fóton único ideais.  Em alto nível, cada mensagem curta \(m\) é associada a um vetor de sinais quânticos previamente distribuídos, com comprimento \(L\) na ordem de \(10^6\text{–}10^9\).  Para cada possível verificador \(V\in\{ \text{Bob}, \text{Charlie},\dots \}\), Alice envia uma sequência \(\mathcal{S}_V\) de estados (por exemplo, \(\{ \ket{\alpha_k},\ket{-\alpha_k}\}\) com fases binárias aleatórias), e mantém um registro clássico privado das escolhas.  Na fase de distribuição, cada verificador mede não–destrutivamente e armazena apenas resultados “compatíveis” (por exemplo, clique/no–clique ou decisão de fase com limiar).  A robustez a transferência maliciosa é obtida com simetrização: verificadores trocam aleatoriamente subconjuntos de seus resultados via canal autenticado, de modo que nenhum deles possua sozinho informação suficiente para favorecer ou desfavorecer uma mensagem contra outro.

\medskip
Para assinar, Alice anuncia publicamente o par \((m, \sigma)\), onde \(\sigma\) é uma descrição curta (p.\,ex., a sequência de fases binárias) que permite aos verificadores confrontar suas amostras locais.  Um verificador aceita se a fração de inconsistências observadas ficar abaixo de um limiar de autenticação \(s_\mathrm{auth}\); na transferência (encaminhamento de \(m,\sigma\) de Bob para Charlie), usa-se um limiar mais frouxo \(s_\mathrm{ver}\) com \(s_\mathrm{auth} < s_\mathrm{ver}\), garantindo que: (i) a falsificação exige que um receptor acerte acima de \(1-s_\mathrm{ver}\) das fases sem possuir \(\mathcal{S}\) original (exponencialmente improvável pela não–clonagem) e (ii) o repúdio é inviável porque, se Bob aceita com margem segura, então Charlie, que possui uma amostra estatisticamente correlacionada, também aceitará com alta probabilidade. Seja \(p_\mathrm{err}\) a taxa de erro honesta por símbolo (devida a perdas, escuras e imperfeições).  Assume-se que um adversário, sem acesso aos estados originais, acerta no máximo com probabilidade \(p_\mathrm{guess}<1\) (por exemplo, limite de Helstrom para coerentes binárias \(|\pm\alpha|\)).  Para blocos de tamanho \(L\), a probabilidade de falsificação decai como:
\[
\Pr[\text{forge}] \;\lesssim\; \exp\!\left[-L\,D\!\left(s_\mathrm{ver}\,\big\|\,p_\mathrm{guess}\right)\right],
\]
enquanto o repúdio decai como:
\[
\Pr[\text{repud}] \;\lesssim\; \exp\!\left[-L\,D\!\left(p_\mathrm{err}\,\big\|\,s_\mathrm{auth}\right)\right],
\]
com \(D(\cdot\|\cdot)\) a divergência de Kullback–Leibler (ou limites de Chernoff).  Assim, escolhendo \(s_\mathrm{auth}\in(p_\mathrm{err}, s_\mathrm{ver})\) e \(s_\mathrm{ver}\in(p_\mathrm{guess},1)\), obtém-se segurança composicional \(\varepsilon\)-pequena com \(L=O(\log(1/\varepsilon))\).  Em implementações com estados coerentes \(|\alpha|\approx0.2\text{–}0.4\), detectores InGaAs de baixa taxa de escuras e simetrização entre verificadores, já se demonstraram QDS em fibras metropolitanas e interurbanas acima de \(100\text{–}200\,\mathrm{km}\), com taxas de assinatura de dezenas a centenas de mensagens por segundo para mensagens curtas.

\medskip
Em termos práticos, (i)~a orçamentação de perdas e escuras é crucial para definir $p_\mathrm{err}$ e, consequentemente, os limiares $s_\mathrm{auth}$ e $s_\mathrm{ver}$. (ii)~O pré-processamento de simetrização requer o uso de canais clássicos autenticados, que podem ser providos, por exemplo, por chaves de QKD. (iii)~Ataques como o deslocamento do ponto de operação e a saturação de detectores são mitigados através de calibração contínua e testes de sanidade, que incluem a verificação de taxas e a distribuição temporal. (iv)~A integração com QKD é vantajosa, pois as mesmas fibras e relógios podem ser compartilhados, e a QKD fornece a autenticação clássica e a renovação de chaves necessárias para os canais de simetrização e anúncio.

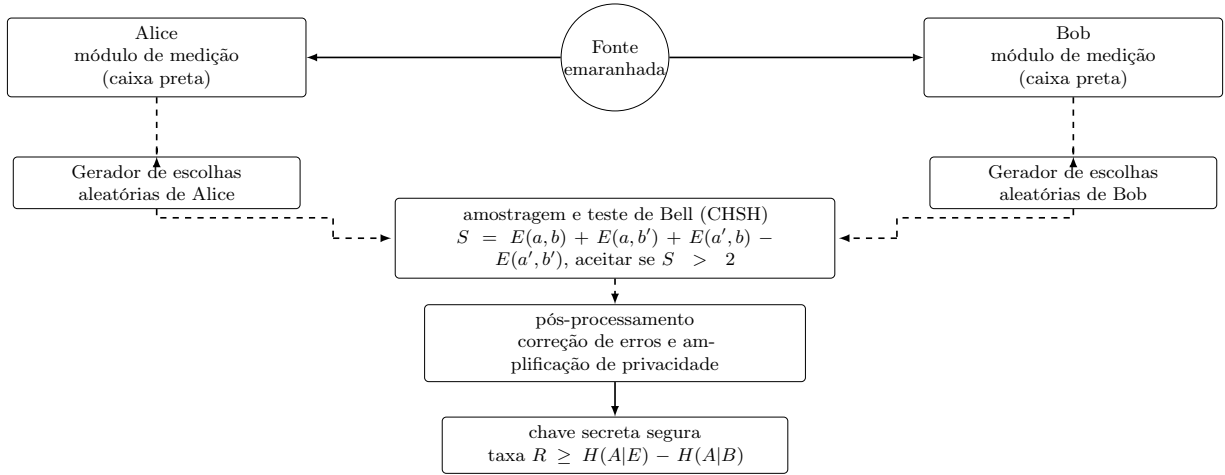
\begin{figure}[t]
\centering
\resizebox{1\linewidth}{!}{%
\begin{tikzpicture}[>=latex, font=\footnotesize,
  box/.style={draw, rounded corners=2pt, align=center, text width=48mm, inner sep=4pt},
  mid/.style={draw, rounded corners=2pt, align=center, text width=72mm, inner sep=4pt},
  circ/.style={draw, circle, minimum size=12mm, inner sep=0pt, align=center},
  qarr/.style={->, line width=.8pt},
  carr/.style={line width=.8pt, dashed}]

\node[box]  (alice) at (-7.8,  3.2) {Alice\\módulo de medição\\(caixa preta)};
\node[circ] (src)   at ( 0.0,  3.2) {Fonte\\emaranhada};
\node[box]  (bob)   at ( 7.8,  3.2)  {Bob\\módulo de medição\\(caixa preta)};

\node[box, text width=46mm] (rngA) at (-7.8, 1.05) {Gerador de escolhas\\aleatórias de Alice};
\node[box, text width=46mm] (rngB) at ( 7.8, 1.05)  {Gerador de escolhas\\aleatórias de Bob};

\node[mid] (chsh) at (0.0, 0.10)
{amostragem e teste de Bell (CHSH)\\
$S = E(a,b) + E(a,b') + E(a',b) - E(a',b')$, aceitar se $S > 2$};

\node[mid, text width=62mm] (post) at (0.0, -1.70)
{pós-processamento\\correção de erros e amplificação de privacidade};
\node[box, text width=56mm] (key) at (0.0, -3.45)
{chave secreta segura\\taxa $R \ge H(A|E) - H(A|B)$};

\draw[qarr] (src.west) -- (alice.east);
\draw[qarr] (src.east) -- (bob.west);

\draw[dashed,->,line width=.8pt] (alice.south) -- ++(0,-0.20) |- (rngA.north);
\draw[dashed,->,line width=.8pt] (bob.south)   -- ++(0,-0.20) |- (rngB.north);

\def\YbelowRNG{0.45}
\def\XgapCH{1.05}
\def\edgePad{0.06}

\coordinate (A_lane) at (-7.8,\YbelowRNG);
\coordinate (B_lane) at ( 7.8,\YbelowRNG);

\coordinate (CHL_pre) at ($(chsh.west)+(-\XgapCH,0)$);
\coordinate (CHR_pre) at ($(chsh.east)+(\XgapCH,0)$);

\coordinate (CHL_edge) at ($(chsh.west)+(0,0)$);
\coordinate (CHR_edge) at ($(chsh.east)+(0,0)$);

\draw[carr] (rngA.south) |- (A_lane) -| (CHL_pre);
\draw[carr] (rngB.south) |- (B_lane) -| (CHR_pre);

\draw[carr,->] (CHL_pre) -- ($(chsh.west)+(-\edgePad,0)$);
\draw[carr,->] (CHR_pre) -- ($(chsh.east)+(\edgePad,0)$);

\draw[carr,->] (chsh.south) -- (post.north);

\draw[qarr] (post.south) -- (key.north);

\end{tikzpicture}}
\caption{DI-QKD: módulos de medição $\to$ RNGs $\to$ CHSH $\to$ pós-processamento $\to$ chave.}
\label{fig:diqkd_flow_fixed_arrows}
\end{figure}

Em síntese, QDS fornece autenticação e não–repúdio com garantias físicas, complementando a QKD em cenários multiusuário.  Com componentes de telecom já consolidados (fontes coerentes fracas, detectores InGaAs e orquestração clássica autenticada), os esquemas modernos de QDS são compatíveis com redes metropolitanas e com integração em camadas superiores de segurança para a futura Internet Quântica.

\section{Tutorial e Metodologia}
\label{sec:tutorial-metodologia}

A fim de ilustrar de forma prática a aplicação da criptografia quântica, apresentamos tutoriais detalhados dos principais protocolos, com pseudocódigos e metodologias de implementação que podem ser aplicados tanto em simulações computacionais quanto em experimentos de laboratório. O objetivo aqui é ampliar substancialmente a parte referente ao BB84, mantendo a estrutura sequencial original, mas enriquecendo cada etapa com diretrizes de engenharia, métricas mensuráveis, estatística finita e requisitos de segurança composicional, sem introduzir subtítulos adicionais nem rótulos destacados no corpo do texto.

O protocolo BB84 pode ser estruturado nos seguintes passos fundamentais, com anotações práticas para implementação realista:

\begin{enumerate}
    \item 
    Na preparação da chave, Alice gera uma sequência de bits aleatórios $\{0,1\}$ de comprimento $N$ e, para cada posição, escolhe uma base aleatória entre $Z=\{\ket{0},\ket{1}\}$ e $X=\{\ket{+},\ket{-}\}$, com $\ket{\pm}=\tfrac{1}{\sqrt{2}}(\ket{0}\pm\ket{1})$. Para garantir imprevisibilidade, é recomendável utilizar geradores de aleatoriedade quântica (QRNG) baseados em ruído de tiro ou de fase, ou, na ausência destes, combinar fontes de entropia do sistema operacional com pós-processamento por extratores (por exemplo, matrizes Toeplitz ou SimHash) para elevar a min-entropy efetiva por bit. Na instrumentação, o mapeamento das bases para os moduladores (por exemplo, modulador de fase/polarização) deve ser calibrado periodicamente; medir e estabilizar a intensidade média por pulso, $\mu$, com fotodiodo de monitoramento (tap) e malha de realimentação ajuda a manter consistência estatística. É boa prática registrar sementes e hashes das sequências geradas exclusivamente para reprodutibilidade interna, mantendo-as isoladas e nunca divulgando conteúdo de chave.

    \item 
    Na transmissão pelo canal quântico (fibra óptica ou espaço livre), os qubits preparados sofrem atenuação e ruídos que precisam constar do orçamento. Em 1550 nm, a atenuação típica por fibra é $\alpha \approx 0.2\,\mathrm{dB/km}$; a perda total pode ser modelada como $L_{\mathrm{tot}}=\alpha L + \sum_i L_i$ (conectores, acoplamentos, componentes), com transmissividade $\eta=10^{-L_{\mathrm{tot}}/10}$. Devem-se incluir isoladores e atenuadores para mitigar retro-reflexões e ataques de injeção de luz; em co-existência WDM com tráfego clássico, filtros estreitos e controle de potência reduzem espalhamento Raman sobre o canal quântico. A estabilidade de polarização (usando controladores eletrônicos) e de fase, quando aplicável, é essencial para manter baixa taxa de erro ao longo do tempo; variações térmicas e mecânicas devem ser monitoradas.

    \item 
    Na medição, Bob escolhe aleatoriamente a base de medição ($Z$ ou $X$) para cada qubit recebido, convergindo estatisticamente para metade das coincidências de base com Alice. Na prática, é necessário sincronizar janelas de detecção com geradores de padrão de pulso e uma referência de tempo estável (por exemplo, 10 MHz/1 PPS). Em detectores APD InGaAs com operação em regime de gate, definem-se larguras de janela e deadtime para mitigar afterpulsing; em SNSPDs, obtêm-se baixíssimas contagens de escuras, mas é preciso evitar saturação e monitorar eficiência. É recomendável equalizar a resposta dos canais de detecção e registrar histogramas temporais e taxas por canal em logs assinados para posterior auditoria.

    \item 
    No sifting, Alice e Bob anunciam publicamente apenas as bases usadas (sem revelar os resultados) e descartam os eventos com bases incompatíveis, formando a chave bruta. O canal clássico usado nesse anúncio deve ser autenticado com esquemas de hash universal do tipo Wegman–Carter (consumindo uma pequena fração de chave precompartilhada que, a partir da primeira sessão bem-sucedida, pode ser reabastecida com chave QKD). É útil padronizar o formato dos pacotes (índices, bases, amostras reservadas para teste, metadados como temperatura, taxas e deriva de $\mu$) com carimbo de tempo e assinatura, de modo a permitir rastreabilidade completa.

    \item 
    Na estimativa de erros, calcula-se a taxa de erro quântico amostrando uma fração pública dos bits sintonizados em mesma base:
    \[
    Q = \frac{n_{\text{erros}}}{n_{\text{testes}}}.
    \]
    Caso $Q > 11\%$, aborta-se a sessão por falta de segurança na hipótese idealizada do BB84. Em cenários realistas com amostras finitas, é essencial associar um nível de confiança $1-\delta$ ao intervalo $[Q^{-},Q^{+}]$ por limites de concentração (Chernoff, Hoeffding ou Clopper–Pearson). É proveitoso decompor a origem dos erros: contribuições ópticas pela visibilidade/interferência (aproximando $Q_{\mathrm{opt}}\approx (1-V)/2$), escuras dos detectores ($Q_{\mathrm{dark}}$) e desalinhamentos/eletrônica ($Q_{\mathrm{misalign}}$). Em fontes atenuadas, a extensão decoy-state permite estimar parâmetros de fóton único ($Y_1$, $e_1$) com intervalos de confiança, fortalecendo o diagnóstico de segurança contra ataques PNS.

    \item 
    Na correção de erros e amplificação de privacidade, utiliza-se um protocolo de reconciliação (CASCADE para simplicidade e robustez, ou LDPC para maiores taxas) medindo a eficiência $\beta$ e contabilizando o vazamento $L_{\mathrm{EC}}$. Em seguida, aplica-se amplificação de privacidade com hash universal (por exemplo, matrizes Toeplitz) para comprimir a chave corrigida a um comprimento seguro frente ao limite de informação potencial de um adversário. Toda a sessão deve obedecer a um orçamento de segurança composicional, por exemplo:
    \[
    \varepsilon_{\mathrm{tot}} = \varepsilon_{\mathrm{sec}} + \varepsilon_{\mathrm{cor}} + \varepsilon_{\mathrm{PE}} + \varepsilon_{\mathrm{auth}},
    \]
    deixando explícitos os valores alvo por sessão, o custo de autenticação $L_{\mathrm{auth}}$ e as penalidades de finitude. Ao final, obtém-se a chave final segura, pronta para integração em um gerenciador de chaves (KMS) e uso em TLS, IPsec, SSH ou OTP.
\end{enumerate}

Para consolidar o planejamento quantitativo, expressões simples ajudam a prever desempenho e a orientar calibrações. A probabilidade de detecção por pulso pode ser aproximada por:
\[
p_{\mathrm{det}} \approx 1 - e^{-\mu\,\eta\,\eta_d} + p_d,
\]
onde $\eta_d$ é a eficiência do detector e $p_d$ a probabilidade de contagem de escuras por janela. Uma decomposição prática do erro observado é:
\[
Q \approx \frac{Q_{\mathrm{opt}}\,p_{\mathrm{sig}} + 0.5\,p_d}{p_{\mathrm{sig}} + p_d} + Q_{\mathrm{misalign}}, 
\quad \text{com} \quad p_{\mathrm{sig}} \approx 1 - e^{-\mu\,\eta\,\eta_d}.
\]
Em regime assintótico simplificado (DV sem decoy), uma forma comum para a taxa de chave líquida é:
\[
R \gtrsim q\,Q_{\mu}\,\Bigl[1 - f(E_{\mu})\,h_2(E_{\mu})\Bigr],
\]
onde $q$ é a fração de coincidências de base, $Q_{\mu}$ o ganho para intensidade $\mu$, $E_{\mu}$ o QBER e $f(\cdot)$ a ineficiência de reconciliação. Com decoy-state, substituem-se termos por limites confiáveis de fóton único, como $Y_1^{\downarrow}$ e $e_1^{\uparrow}$, resultando em expressões do tipo:
\[
R \ge q\left\{-Q_{\mu} f(E_{\mu}) h_2(E_{\mu}) + Y_1^{\downarrow}\mu e^{-\mu}\bigl[1-h_2(e_1^{\uparrow})\bigr]\right\},
\]
a serem penalizadas pelos termos de estatística finita e pelo custo de autenticação.

O funcionamento detalhado do BB84 está implementado no Algoritmo \ref{alg:BB84}, que demonstra como Alice gera sequências aleatórias de bits e bases, prepara os estados quânticos correspondentes e os transmite pelo canal quântico. A implementação de Alice deve apoiar-se em fontes de entropia verdadeiramente aleatórias e em mapeamento físico calibrado para garantir fidelidade dos estados. É imprescindível manter registros de telemetria (taxas, histogramas de chegada, temperatura, deriva de $\mu$), validar continuamente a aleatoriedade (NIST SP 800-22, Dieharder), evitar regimes de saturação e \emph{blinding} nos detectores, e autenticar todas as mensagens clássicas. Por fim, recomenda-se publicar ao término de cada sessão um sumário com $Q$, $\beta$, $L_{\mathrm{EC}}$, $L_{\mathrm{auth}}$, $\varepsilon_{\mathrm{tot}}$ e a taxa de chave líquida, com logs assinados e carimbo de tempo, assegurando reprodutibilidade e auditoria.

\begin{algorithm}[h!]
\scriptsize
\caption{Protocolo BB84 - Alice e Bob}
\label{alg:BB84}
\begin{algorithmic}[1]
\Procedure{BB84\_Alice}{$N$}
    \State $bits \gets$ \Call{GerarBitsAleatorios}{$N$}
    \State $bases \gets$ \Call{GerarBasesAleatorias}{$N$} \Comment{Z=0, X=1}
    
    \For{$i = 1$ \textbf{to} $N$}
        \State $qubit \gets$ \Call{PrepararEstado}{$bits[i], bases[i]$}
        \State \Call{EnviarQubit}{$qubit$, canal\_quantico}
    \EndFor
    
    \State \Comment{Sifting}
    \State $bases\_bob \gets$ \Call{ReceberBasesBob}{}
    \State $chave\_bruta \gets []$
    \For{$i = 1$ \textbf{to} $N$}
        \If{$bases[i] = bases\_bob[i]$}
            \State $chave\_bruta.\text{append}(bits[i])$
        \EndIf
    \EndFor
    
    \State \Comment{Teste de QBER}
    \State $amostra \gets$ \Call{SelecionarAmostra}{$chave\_bruta, 0.1$}
    \State $qber \gets$ \Call{CalcularQBER}{$amostra, bob\_amostra$}
    \If{$qber > 0.11$}
        \State \textbf{abort} "Canal inseguro"
    \EndIf
    
    \State \Comment{Pós-processamento}
    \State $chave\_corrigida \gets$ \Call{CorrecaoErros}{$chave\_bruta$}
    \State $chave\_final \gets$ \Call{AmplificacaoPrivacidade}{$chave\_corrigida$}
    \State \textbf{return} $chave\_final$
\EndProcedure

\Procedure{BB84\_Bob}{$N$}
    \State $bases \gets$ \Call{GerarBasesAleatorias}{$N$}
    \State $resultados \gets []$
    
    \For{$i = 1$ \textbf{to} $N$}
        \State $qubit \gets$ \Call{ReceberQubit}{canal\_quantico}
        \State $resultado \gets$ \Call{MedirQubit}{$qubit, bases[i]$}
        \State $resultados.\text{append}(resultado)$
    \EndFor
    
    \State \Call{EnviarBases}{$bases$} \Comment{Canal clássico}
    \State \Comment{Continua com sifting e pós-processamento...}
\EndProcedure
\end{algorithmic}
\end{algorithm}

A implementação de Bob, também mostrada no Algoritmo \ref{alg:BB84}, é complementar à de Alice, mas destaca pontos singulares que ampliam a robustez do protocolo. Ao gerar suas próprias bases de medição de modo independente, Bob introduz um grau adicional de imprevisibilidade que impede qualquer tentativa de sincronização estratégica por parte de um adversário. Esse caráter autônomo não apenas reforça a aleatoriedade intrínseca do sistema, mas também cria uma redundância estrutural que impede a manipulação unilateral do processo de geração da chave.  

Do ponto de vista prático, a atuação de Bob exige uma instrumentação sofisticada. Detectores de fótons únicos precisam operar com alta eficiência e baixa taxa de falsos positivos (\textit{dark counts}), pois esses fatores impactam diretamente na taxa de erro quântico observada (QBER). Além disso, o alinhamento temporal e a estabilidade de fase entre os pulsos recebidos constituem desafios técnicos que determinam a viabilidade da implementação em ambientes reais. Sistemas de sincronização óptica e eletrônica, aliados a técnicas de calibração automática, são fundamentais para que o protocolo atinja níveis de segurança compatíveis com aplicações críticas em redes de comunicação.  

Enquanto o BB84 destaca-se pela simplicidade conceitual e aplicabilidade experimental, o protocolo E91, apresentado no Algoritmo \ref{alg:E91}, promove uma ruptura paradigmática ao fundamentar a segurança na correlação de partículas emaranhadas. Essa abordagem introduz uma dimensão mais profunda ao problema: a verificação experimental de fenômenos quânticos não-locais. Em vez de apenas confiar em suposições sobre o funcionamento correto dos dispositivos, o E91 oferece um critério operacional objetivo, a violação das desigualdades de Bell, que permite a Alice e Bob certificarem-se da presença de correlações genuinamente quânticas antes mesmo de extrair a chave.  

Essa propriedade confere ao E91 uma relevância adicional no contexto de {segurança device-independent}, em que não é necessário assumir que os componentes de hardware funcionam de maneira ideal ou livre de manipulações externas. O protocolo, portanto, não apenas avança conceitualmente sobre o BB84, mas também antecipa a agenda contemporânea de sistemas de criptografia quântica capazes de resistir a ataques que exploram falhas de implementação. Dessa forma, o E91 inaugura uma linha de pesquisa que conecta fundamentos da física quântica às aplicações práticas em segurança da informação, estabelecendo um elo direto entre resultados experimentais de não-localidade e a garantia formal de sigilo na comunicação.

Em termos operacionais, a realização do E91 demanda uma arquitetura com fonte de pares emaranhados (tipicamente SPDC tipo-II) com alta taxa de geração, acoplamento eficiente para fibra, filtros espectrais estreitos e controle de indistinguibilidade temporal. O sincronismo entre as estações precisa ser sub-nanosegundo para que coincidências reais se distingam de acidentais. A estabilidade do alinhamento de polarização (ou equivalentes, dependendo do grau de liberdade utilizado) é crucial para sustentar correlações elevadas ao longo de horas. Além de aspectos fotônicos, a camada clássica desempenha papel central: é necessário definir claramente os subconjuntos de dados destinados ao teste de Bell e à extração de chave, com seleção aleatória e autenticação das mensagens que comunicam as escolhas de bases e resultados públicos. A instrumentação deve registrar histogramas temporais, taxas de coincidência, taxas de acidentais, visibilidade em cada base e métricas ambientais (temperatura, vibração), mantendo logs assinados para auditoria.

Sob a perspectiva de segurança, o teste CHSH implementado no E91 realiza uma forma de verificação operacional dos dispositivos, reduzindo as suposições sobre modelos internos. Ainda que o E91, em sua forma original, não seja totalmente device-independent (por exemplo, pode considerar-se fechamento parcial de brechas experimentais), ele constitui o degrau conceitual para arquiteturas DI-QKD plenas, nas quais a segurança é garantida unicamente pela violação de Bell e por suposições mínimas sobre isolamento espacial e independência de escolha de bases. Em particular, a randomicidade utilizada na escolha de bases deve ser alta e auditável; correlações sutilmente introduzidas por ruídos eletrônicos ou por fontes de entropia inadequadas degradam o parâmetro $S$ e podem induzir vieses detectáveis por um adversário.

No planejamento quantitativo, é útil acompanhar, para cada sessão, o parâmetro $S$ por janelas de tempo e por distância. Valores de referência de laboratório bem controlado costumam ficar entre $S \in [2.6,2.75]$, distantes do limite clássico $S\le 2$ e abaixo do máximo quântico $2\sqrt{2}$. A análise deve incluir incertezas estatísticas por amostragem finita, intervalos de confiança e variações sistemáticas devido a deriva térmica ou a instabilidades de alinhamento. Um critério de aceitação prático consiste em exigir $S$ acima de um limiar $S_{\min}$ com nível de confiança especificado (por exemplo, 5 desvios-padrão acima de 2), antes de permitir a extração de chave. Em sessões longas, monitorar deriva lenta de $S$ ajuda a antecipar necessidades de recalibração e a reduzir o risco de rejeição tardia de blocos.

Do ponto de vista da extração de chave, o E91 reserva subconjuntos de medições compatíveis (bases bem definidas para correlação de chave) e outro subconjunto para o teste de Bell. O tamanho relativo desses subconjuntos influencia a taxa de chave líquida: mais amostras para CHSH aumentam a confiança estatística na violação, mas reduzem a fração útil para chave; menos amostras elevam a taxa nominal, porém ampliam a incerteza e podem exigir limiares mais conservadores para segurança composicional. A engenharia do protocolo deve, portanto, otimizar a partição “teste vs. chave” para a distância e para a taxa de geração da fonte.

Em implementações reais, duas frentes técnicas sobressaem. Primeira: a qualidade e a estabilidade da fonte emaranhada. Fatores como largura de banda, correlação espectral e tempo de coerência influenciam diretamente a visibilidade e, por consequência, o valor de $S$. Ajustes finos de temperatura do cristal não-linear, seleção de filtros e compensação de \emph{walk-off} são rotinas recorrentes. Segunda: o sistema de detecção e temporização. Janelas temporais estreitas e \emph{time-taggers} de alta resolução reduzem acidentais; detectores com baixas escuras e alto \emph{count rate} efetivo maximizam coincidências verdadeiras. A calibração cruzada entre canais e a compensação de assimetrias no caminho óptico contribuem para a maximização de $S$ sem enviesar a estatística do teste.

Para a camada clássica, a segurança composicional requer contabilizar explicitamente o orçamento de $\varepsilon_{\mathrm{tot}}$ com termos de correção ($\varepsilon_{\mathrm{cor}}$), sigilo ($\varepsilon_{\mathrm{sec}}$), estimação de parâmetros ($\varepsilon_{\mathrm{PE}}$) e autenticação ($\varepsilon_{\mathrm{auth}}$). Toda a comunicação de bases e índices de amostra deve ser autenticada (por exemplo, Wegman–Carter), com consumo de chave monitorado e reabastecido pelas próprias chaves geradas quando a sessão é bem-sucedida.

Em cenários de longa distância, as perdas acumuladas e a queda de visibilidade tornam mais desafiador manter $S$ significativamente acima de 2. Estratégias de mitigação incluem aprimorar o acoplamento óptico, adotar detectores de alta eficiência (SNSPDs), aplicar filtros espectrais melhor ajustados e reduzir ruído de fundo. Em redes com multiplexação WDM, é essencial mitigar espalhamento Raman oriundo de canais clássicos de alta potência por meio de planejamento espectral e filtragem adequada. Em ambientes de campo, o \emph{runbook} operacional deve incluir rotinas de verificação automática de $S$, gatilhos de recalibração e critérios de abortamento com registro completo para auditoria.

Em síntese, o E91 articula uma ponte entre fundamentos e prática: a chave criptográfica passa a ser “condicionada” a uma evidência empírica de não-localidade. Isso redefine o ciclo de confiança, pois desloca parte das garantias de suposições internas sobre dispositivos para um verificador físico externo (o teste de Bell). A agenda atual de DI-QKD generaliza esse princípio, complementando-o com requisitos adicionais (fechamento de brechas de detecção e de localidade, independência de escolhas e isolamento de sinais). Nesse panorama, o E91 permanece como um marco metodológico e uma base pedagógica para compreender como fenômenos quânticos fundamentais podem ser utilizados de maneira operacional e auditável para prover sigilo de comunicação em cenários adversariais sofisticados.

\begin{algorithm}[h!]
\scriptsize
\caption{Protocolo E91 - Alice com teste de Bell}
\label{alg:E91}
\begin{algorithmic}[1]
\Procedure{E91\_Alice}{$N$}
    \State $bases \gets$ \Call{GerarBasesAleatorias}{$N$} \Comment{3 bases possíveis}
    \State $resultados \gets []$
    
    \For{$i = 1$ \textbf{to} $N$}
        \State $par\_emaranhado \gets$ \Call{ReceberQubitEmaranhado}{}
        \State $resultado \gets$ \Call{MedirQubit}{$par\_emaranhado, bases[i]$}
        \State $resultados.\text{append}(resultado)$
    \EndFor
    
    \State \Comment{Teste de Bell}
    \State $bases\_bob \gets$ \Call{ReceberBasesBob}{}
    \State $S \gets$ \Call{CalcularCHSH}{$resultados, bases, bases\_bob, resultados\_bob$}
    \If{$S \leq 2$}
        \State \textbf{abort} "Violação de Bell não detectada"
    \EndIf
    
    \State \Comment{Extração de chave}
    \State $chave \gets$ \Call{ExtrairChaveCompativel}{$resultados, bases, bases\_bob$}
    \State \textbf{return} $chave$
\EndProcedure

\Function{CalcularCHSH}{$res\_A, bases\_A, bases\_B, res\_B$}
    \State $E_{ab} \gets$ \Call{Correlacao}{$res\_A[bases\_A=0], res\_B[bases\_B=0]$}
    \State $E_{ab'} \gets$ \Call{Correlacao}{$res\_A[bases\_A=0], res\_B[bases\_B=1]$}
    \State $E_{a'b} \gets$ \Call{Correlacao}{$res\_A[bases\_A=1], res\_B[bases\_B=0]$}
    \State $E_{a'b'} \gets$ \Call{Correlacao}{$res\_A[bases\_A=1], res\_B[bases\_B=1]$}
    \State $S \gets E_{ab} + E_{ab'} + E_{a'b} - E_{a'b'}$
    \State \textbf{return} $S$
\EndFunction
\end{algorithmic}
\end{algorithm}

A função \texttt{CalcularCHSH} presente no Algoritmo \ref{alg:E91} implementa o teste da desigualdade de Clauser-Horne-Shimony-Holt. O valor $S$ calculado deve exceder 2 para demonstrar correlações quânticas não-locais, com o máximo teórico sendo $2\sqrt{2} \approx 2.828$ para estados maximamente emaranhados. Na prática, entretanto, o valor medido de $S$ sofre com perdas no canal, imperfeições de alinhamento e ineficiência nos detectores. Esses fatores reduzem o alcance experimental do protocolo e tornam seu desempenho altamente dependente de técnicas avançadas de correção de erros e de calibração. Além disso, a necessidade de gerar e distribuir pares de fótons emaranhados estáveis ao longo de longas distâncias constitui um desafio tecnológico central, levando à pesquisa intensa em fontes de fótons baseadas em cristais não-lineares e em interfaces eficientes entre fótons e memórias quânticas. O papel dessa função no protocolo E91 não é apenas computacional, mas conceitual: ela fornece a evidência experimental de que a segurança da chave está fundamentada em princípios de não-localidade quântica e não apenas na suposição de funcionamento ideal dos dispositivos.

O protocolo Decoy-State surge justamente para superar uma limitação prática de protocolos teóricos como o BB84, os quais assumem o uso de fontes ideais de fóton único. No contexto experimental, tais fontes ainda são tecnologicamente restritas e de difícil integração em sistemas de larga escala. Para contornar esse limite, utiliza-se luz coerente atenuada, que inevitavelmente gera pulsos contendo múltiplos fótons com alguma probabilidade. Essa fragilidade abre espaço para ataques de interceptação e divisão de fótons (\textit{Photon Number Splitting}, PNS). A contribuição do Decoy-State, ilustrada no Algoritmo \ref{alg:Decoy}, é a introdução de pulsos adicionais com intensidades variadas (\textit{decoys}), os quais têm o papel de “sondar” estatisticamente o canal e revelar discrepâncias provocadas por tentativas de espionagem.  

A ideia central é explorar o fato de que um adversário que tenta explorar pulsos multifóton de forma seletiva (por exemplo, retendo um fóton e encaminhando o restante) não consegue, ao mesmo tempo, preservar as mesmas taxas de detecção e de erro para todas as intensidades. Ao variar as intensidades (tipicamente um nível sinal $\mu$, um ou mais níveis \textit{decoy} como $\nu$ e um \textit{vacuum} $\approx 0$), Alice e Bob estimam parâmetros fundamentais dos pulsos de fóton único: o \emph{yield} $Y_1$ (probabilidade de detecção condicional a 1 fóton) e a taxa de erro $e_1$. Esses estimadores, obtidos com garantias estatísticas de alta confiança, limitam rigorosamente a informação que um adversário pode ter adquirido, permitindo calcular uma taxa de chave segura mesmo quando parte significativa dos pulsos contém 2 ou mais fótons.

Na prática, configurações minimalistas com três intensidades ($\mu_s$ para sinal, $\mu_d$ para \textit{decoy} fraco e $\mu_0 \approx 0$ para vácuo) já oferecem um bom compromisso entre complexidade e desempenho. Proporções típicas de uso podem ser, por exemplo, $p_s \in [0.6,0.85]$, $p_d \in [0.1,0.35]$ e $p_0 \in [0.02,0.1]$, ajustadas por otimização para o enlace e a distância alvo. O feixe é estabilizado em potência média por pulso (erro < 1

Essa estratégia permite que Alice e Bob estimem com elevada precisão os eventos efetivamente gerados por fótons únicos, que são os únicos capazes de formar uma chave imune a ataques clássicos. A robustez do Decoy-State foi comprovada experimentalmente, mostrando que taxas positivas de chave podem ser alcançadas mesmo a longas distâncias, superando os limites práticos do BB84 implementado com fontes atenuadas convencionais. Com isso, o protocolo estabelece uma ponte entre a teoria e a prática da criptografia quântica, viabilizando a implementação em redes ópticas metropolitanas reais e consolidando seu papel como um dos avanços mais importantes na evolução da QKD moderna. Em operação de campo, recomenda-se: (i) recalibrar periodicamente as intensidades (drift térmico pode enviesar as frações $\mu_i$), (ii) aplicar intervalos de confiança (Chernoff/Hoeffding) por intensidade para $Q_{\mu_i}$ e $E_{\mu_i}$, (iii) definir um critério de abortamento caso as desigualdades consistentes com PNS sejam violadas (por exemplo, $Q_{\mu_s}$ e $Q_{\mu_d}$ incompatíveis com o modelo Poisson + perdas uniformes), e (iv) contabilizar rigidamente o vazamento da reconciliação $L_{\mathrm{EC}}$ e o custo de autenticação $L_{\mathrm{auth}}$ no orçamento final de segurança composicional.

\begin{algorithm}[h!]
\scriptsize
\caption{Protocolo Decoy-State QKD}
\label{alg:Decoy}
\begin{algorithmic}[1]
\Procedure{DecoyState\_Alice}{$N, \{\mu_s,\mu_d,\mu_0\}, \{p_s,p_d,p_0\}$}
    \State $bits \gets$ \Call{GerarBitsAleatorios}{$N$}
    \State $bases \gets$ \Call{GerarBasesAleatorias}{$N$}
    \State $intensidades\_escolhidas \gets []$
    \For{$i = 1$ \textbf{to} $N$}
        \State $x \gets$ \Call{RNG}{}
        \If{$x < p_s$} \State $I \gets \mu_s$
        \ElsIf{$x < p_s + p_d$} \State $I \gets \mu_d$
        \Else \State $I \gets \mu_0$
        \EndIf
        \State $intensidades\_escolhidas.\text{append}(I)$
        \State $pulso \gets$ \Call{PrepararPulsoCoerente}{$bits[i], bases[i], I$}
        \State \Call{EnviarPulso}{$pulso$}
    \EndFor

    \State \Comment{Agregação por intensidade}
    \For{\textbf{each} $I \in \{\mu_s,\mu_d,\mu_0\}$}
        \State $Q_I \gets$ \Call{CalcularGanho}{$I$} \Comment{cliques por tentativa}
        \State $E_I \gets$ \Call{CalcularErro}{$I$} \Comment{erros condicionais}
    \EndFor

    \State \Comment{Estimativa segura de fóton único (exemplo 3 intensidades)}
    \State $Y_1^{\downarrow} \gets$ \Call{EstimarYieldFotonUnico}{$Q_{\mu_s},Q_{\mu_d},Q_{\mu_0},\mu_s,\mu_d$}
    \State $e_1^{\uparrow} \gets$ \Call{EstimarErroFotonUnico}{$E_{\mu_s},E_{\mu_d},Y_1^{\downarrow},\mu_s,\mu_d$}

    \State \Comment{Taxa de chave (forma típica, DV, finito omitido)}
    \State $R \gets q\Big( -Q_{\mu_s} f(E_{\mu_s}) h_2(E_{\mu_s}) + Y_1^{\downarrow}\,\mu_s e^{-\mu_s}\,[1-h_2(e_1^{\uparrow})] \Big)$
    \If{$R \le 0$} \State \textbf{abort} "Taxa de chave insuficiente" \EndIf
    \State \textbf{return} $chave\_final$
\EndProcedure

\Function{EstimarYieldFotonUnico}{$Q_{\mu_s},Q_{\mu_d},Q_{\mu_0},\mu_s,\mu_d$}
    \State \Comment{Exemplo de limitante inferior clássico para 3 intensidades}
    \State $A \gets \mu_s e^{\mu_s} Q_{\mu_d} - \mu_d e^{\mu_d} Q_{\mu_s} - (\mu_s - \mu_d) e^{\mu_0} Q_{\mu_0}$
    \State $B \gets \mu_s \mu_d (\mu_s - \mu_d)$
    \State $Y_1 \gets \max\!\big(0, A / B\big)$
    \State \textbf{return} $Y_1$
\EndFunction

\Function{EstimarErroFotonUnico}{$E_{\mu_s},E_{\mu_d},Y_1,\mu_s,\mu_d$}
    \State \Comment{Exemplo de limitante superior usando dois níveis e Poisson}
    \State $e_1 \gets \min\!\Big( \tfrac{E_{\mu_s} Q_{\mu_s} e^{\mu_s} - E_{\mu_d} Q_{\mu_d} e^{\mu_d}}{(\mu_s - \mu_d) Y_1},\, 0.5 \Big)$
    \State \textbf{return} $\max(0, e_1)$
\EndFunction
\end{algorithmic}
\end{algorithm}

Para orientar a implementação, algumas diretrizes de engenharia e estatística finita são úteis:
\begin{itemize}
    \item Calibração e monitoramento de intensidades: manter erro relativo $< 1\%$ por nível, registrar deriva por hora e compensar em tempo real quando possível.
    \item Seleção de probabilidades: otimizar $(p_s,p_d,p_0)$ para o enlace alvo; enlaces mais longos tendem a favorecer proporções maiores de sinal para manter $Q_{\mu_s}$ mensurável, mas exigem amostras \textit{decoy} suficientes para controle de $Y_1$ e $e_1$.
    \item Tratamento de vácuo: $Q_{\mu_0}$ estima contribuições de escuras e acidentais; ser conservador na incerteza de $Q_{\mu_0}$ evita superestimar $Y_1^{\downarrow}$.
    \item Reconciliação e amplificação de privacidade: escolher códigos LDPC com eficiência $\beta$ alta em regime de baixo QBER; contabilizar $L_{\mathrm{EC}}$ e adotar matrizes Toeplitz para compressão com orçamento explícito de $\varepsilon_{\mathrm{sec}}$ e $\varepsilon_{\mathrm{PE}}$.
    \item Critérios de abortamento: definir testes de consistência entre $Q_{\mu_s}$ e $Q_{\mu_d}$ esperados pelo modelo Poisson + perda; violações apontam manipulação seletiva (PNS) e devem disparar abortamento ou reconfiguração.
\end{itemize}

O Decoy-State é a peça-chave que torna viável a QKD baseada em fontes atenuadas sob hipóteses realistas de canal e detecção. Ele eleva a segurança do BB84 prático ao introduzir controle estatístico sobre o componente de fóton único, garantindo, com confiança quantificada, que a chave final reflete correlações inacessíveis a um adversário mesmo na presença de pulsos multifóton e de perdas elevadas típicas de enlaces de longa distância.

O protocolo MDI-QKD elimina vulnerabilidades relacionadas aos detectores delegando todas as medições a um nó central potencialmente não confiável, conforme especificado no Algoritmo \ref{alg:MDI}. A grande inovação está no fato de que, mesmo que este nó central seja controlado por um adversário, a segurança da chave final permanece preservada. Essa característica resolve uma das maiores fragilidades dos sistemas práticos de QKD: os ataques de exploração de detectores, que já foram demonstrados experimentalmente contra implementações comerciais do BB84.  

\vspace{0.5cm}

Do ponto de vista técnico, o funcionamento do MDI-QKD exige que os sinais enviados por Alice e Bob interfiram de forma coerente no nó central, o que demanda sincronização temporal de nível sub-nanosegundo e estabilidade de fase óptica em longas distâncias. Esses requisitos experimentais, embora desafiadores, já foram superados em campo, com demonstrações bem-sucedidas de MDI-QKD em fibras ópticas de centenas de quilômetros. Para viabilizar a interferência, é necessário igualar espectro, polarização, fase e forma de pulso dos emissores; o balanceamento de perdas e a equalização de atrasos entre os braços até o nó central também são críticos para maximizar a taxa de projeções bem-sucedidas em estados de Bell.

Outro aspecto relevante é que o MDI-QKD aumenta a resiliência das redes quânticas a ataques internos, uma vez que transfere os dispositivos mais críticos (os detectores) para nós intermediários sem comprometer a confiança entre os usuários. Isso viabiliza a construção de redes quânticas em estrela, nas quais múltiplos usuários podem se conectar de forma segura a partir de um único nó central, mantendo garantias incondicionais de segurança. Em termos de engenharia de rede, o nó central pode agregar tráfego quântico de diversos enlaces e anunciar publicamente apenas os eventos de medição de Bell bem-sucedidos, permitindo que pares de usuários distintos estabeleçam chaves simultaneamente com autenticação clássica adequada.

O protocolo também inspirou uma linha de pesquisa voltada para o conceito de segurança independente de dispositivos (\textit{device-independent security}), que busca eliminar suposições não apenas sobre os detectores, mas sobre todos os elementos da implementação. Dessa forma, o MDI-QKD não apenas fortalece a segurança prática da QKD contra ataques conhecidos, mas também se coloca como um marco de transição entre protocolos dependentes de implementação e arquiteturas mais gerais e robustas, capazes de resistir até mesmo a adversários com controle parcial sobre os dispositivos envolvidos. Na prática, políticas de sifting, estimação de parâmetros e reconciliação devem considerar o subconjunto condicionado a detecções de Bell, com contabilização do vazamento de informação durante correção de erros e amplificação de privacidade. 
\begin{algorithm}[h!]
\scriptsize
\caption{Protocolo MDI-QKD}
\label{alg:MDI}
\begin{algorithmic}[1]
\Procedure{MDI\_Alice}{$N$}
    \State $bits \gets$ \Call{GerarBitsAleatorios}{$N$}
    \State $bases \gets$ \Call{GerarBasesAleatorias}{$N$}
    \For{$i = 1$ \textbf{to} $N$}
        \State $estado \gets$ \Call{PrepararEstadoMDI}{$bits[i], bases[i]$}
        \State \Call{EnviarParaNoCentral}{$estado$}
    \EndFor
    \State $resultados\_BSM \gets$ \Call{ReceberResultadosBSM}{}
    \State $chave\_bruta \gets []$
    \For{$i = 1$ \textbf{to} $N$}
        \If{$resultados\_BSM[i] = \text{``sucesso''}$}
            \State $chave\_bruta.\text{append}(bits[i])$
        \EndIf
    \EndFor
    \State $chave\_final \gets$ \Call{PosProcessamento}{$chave\_bruta$}
    \State \textbf{return} $chave\_final$
\EndProcedure

\Procedure{MDI\_Bob}{$N$}
    \State $bits \gets$ \Call{GerarBitsAleatorios}{$N$}
    \State $bases \gets$ \Call{GerarBasesAleatorias}{$N$}
    \For{$i = 1$ \textbf{to} $N$}
        \State $estado \gets$ \Call{PrepararEstadoMDI}{$bits[i], bases[i]$}
        \State \Call{EnviarParaNoCentral}{$estado$}
    \EndFor
    \State \textbf{return}
\EndProcedure

\Procedure{NoCentral\_MDI}{$N$}
    \For{$i = 1$ \textbf{to} $N$}
        \State $qubit\_alice \gets$ \Call{ReceberDeAlice}{}
        \State $qubit\_bob \gets$ \Call{ReceberDeBob}{}
        \State $resultado\_BSM \gets$ \Call{MedicaoBell}{$qubit\_alice, qubit\_bob$}
        \State \Call{AnunciarPublicamente}{$resultado\_BSM$}
    \EndFor
\EndProcedure

\Function{MedicaoBell}{$qubit1, qubit2$}
    \State $interferencia \gets$ \Call{InterferirQubits}{$qubit1, qubit2$}
    \State $deteccao \gets$ \Call{DetectarCoincidencias}{$interferencia$}
    \If{$deteccao = \text{``coincidencia''}$}
        \State \textbf{return} "sucesso"
    \Else
        \State \textbf{return} "falha"
    \EndIf
\EndFunction
\end{algorithmic}
\end{algorithm}

Para orientar a implementação prática, recomenda-se:
\begin{itemize}
    \item Igualar espectro, polarização, fase e temporização dos emissores de Alice e Bob; usar \textit{time-taggers} e referência comum para reduzir \emph{timing jitter}.
    \item Balancear perdas dos dois caminhos até o nó central; compensar atrasos e usar filtros estreitos para reduzir acidentais.
    \item Autenticar todos os anúncios do nó central; realizar sifting e estimação de parâmetros condicionados a eventos de Bell bem-sucedidos.
    \item Utilizar reconciliação eficiente (LDPC) e amplificação de privacidade com orçamento composicional de segurança, contabilizando $L_{\mathrm{EC}}$ e $L_{\mathrm{auth}}$.
\end{itemize}

O protocolo CV-QKD codifica informação nas quadraturas contínuas do campo eletromagnético, sendo particularmente atrativo para integração com sistemas de telecomunicações existentes, como demonstrado no Algoritmo \ref{alg:CVQKD}. Diferentemente dos protocolos de variáveis discretas (DV-QKD), que dependem de detectores de fóton único, o CV-QKD utiliza técnicas de detecção coerente amplamente empregadas pela indústria em redes de fibra óptica de alta capacidade. Essa compatibilidade tecnológica reduz custos de implementação e facilita a adoção em escala, tornando-o um candidato promissor para a consolidação de redes quânticas no futuro próximo.  

Do ponto de vista teórico, a segurança do CV-QKD é analisada no formalismo gaussiano da informação quântica, em que a informação mútua $I_{AB}$ entre Alice e Bob e o limite de Holevo $\chi_{BE}$ (cota de informação acessível a um adversário) determinam a taxa de chave segura. Essa modelagem permite avaliar resiliência em cenários ruidosos, nos quais protocolos DV podem sofrer degradação acentuada. Esquemas de modulação gaussiana conferem flexibilidade para adaptar parâmetros de transmissão às condições do canal, otimizando eficiência de reconciliação e robustez frente a perdas e ruído excessivo.

Na prática experimental, a viabilidade depende fortemente da reconciliação de informações contínuas, tarefa mais exigente que nas versões discretas. Códigos LDPC de alto desempenho têm permitido fatores de reconciliação próximos do ideal ($\beta \approx 1$) em regimes de baixa SNR, enquanto eletrônica de alta velocidade e detecção homódina/heteródina de baixa perda viabilizam taxas de repetição da ordem de centenas de megahertz. Além disso, o controle preciso do LO (local oscillator), seja transmitido ou gerado localmente (\textit{local LO}),, o balanceamento de quadraturas e a calibração de ruído técnico são cruciais para garantir estimativas confiáveis de parâmetros e estabilidade operacional prolongada.

Em aplicações, o CV-QKD se integra naturalmente a infraestruturas de fibras existentes em redes metropolitanas, permitindo distribuição de chaves paralela ao tráfego clássico. Essa coabitação é estratégica para redes híbridas, em que comunicação quântica e clássica compartilham recursos físicos com isolamento espectral e de potência apropriados. Estudos apontam compatibilidade com arquiteturas de redes quânticas de maior escala quando combinado a repetidores quânticos e técnicas de distilação de emaranhamento para estados gaussianos, abrindo caminho para topologias multiusuário e interconexões globais.

Dessa forma, o CV-QKD representa um elo entre teoria da informação quântica e engenharia de telecomunicações, aproximando a QKD das demandas de escalabilidade e interoperabilidade de sistemas modernos.

\begin{algorithm}[h!]
\scriptsize
\caption{Protocolo CV-QKD (Variáveis Contínuas)}
\label{alg:CVQKD}
\begin{algorithmic}[1]
\Procedure{CVQKD\_Alice}{$N, V_M, \beta$} \Comment{$V_M$: variância de modulação}
    \State $X \gets []$, $P \gets []$
    \For{$i = 1$ \textbf{to} $N$}
        \State $x \gets$ \Call{GaussianaAleatoria}{$0, V_M$}
        \State $p \gets$ \Call{GaussianaAleatoria}{$0, V_M$}
        \State $X.\text{append}(x)$, $P.\text{append}(p)$
        \State $|\alpha\rangle \gets$ \Call{PrepararEstadoCoerente}{$x, p$}
        \State \Call{EnviarEstado}{$|\alpha\rangle$, canal\_quantico}
    \EndFor
    \State $M_B \gets$ \Call{ReceberMedicoesBob}{} \Comment{Sequência de resultados}
    \State $(\hat{T}, \hat{\xi}) \gets$ \Call{EstimativaParametros}{$X,P, M_B$} \Comment{Transmissão, ruído excessivo}
    \State $I_{AB} \gets$ \Call{InformacaoMutua}{$X,P, M_B$}
    \State $\chi_{BE} \gets$ \Call{HolevoGaussian}{$\hat{T}, \hat{\xi}, V_M$}
    \State $K \gets \beta \cdot I_{AB} - \chi_{BE}$
    \If{$K \le 0$}
        \State \textbf{abort} "Canal muito ruidoso para CV-QKD"
    \EndIf
    \State $chave\_final \gets$ \Call{ReconciliaAmplifica}{$X,P, M_B, \beta, K$}
    \State \textbf{return} $chave\_final$
\EndProcedure

\Procedure{CVQKD\_Bob}{$N$}
    \State $medicoes \gets []$
    \For{$i = 1$ \textbf{to} $N$}
        \State $|\alpha\rangle \gets$ \Call{ReceberEstado}{canal\_quantico}
        \State $b \gets$ \Call{EscolherBaseAleatoria}{} \Comment{$X$ ou $P$}
        \If{$b = X$}
            \State $r \gets$ \Call{MedicaoHomodina}{$|\alpha\rangle, X$}
        \Else
            \State $r \gets$ \Call{MedicaoHomodina}{$|\alpha\rangle, P$}
        \EndIf
        \State $medicoes.\text{append}(r)$
    \EndFor
    \State \Call{EnviarMedicoes}{$medicoes$} \Comment{Canal clássico autenticado}
\EndProcedure
\end{algorithmic}
\end{algorithm}

Para implementação prática e avaliação segura, recomenda-se:
\begin{itemize}
    \item Calibrar $V_M$ para maximizar $K$ na SNR disponível; excesso de modulação pode aumentar $\chi_{BE}$ via ruído técnico.
    \item Utilizar LO local com fase estabilizada ou LO transmitido com filtragem e controle de potência rigorosos; suprimir \emph{leakage} e batimentos espúrios.
    \item Empregar estimação de parâmetros com intervalos de confiança (Chernoff/Hoeffding) e contabilizar a penalidade de finitude no orçamento composicional de segurança.
    \item Adotar códigos LDPC de taxa adaptativa para regimes de baixa SNR, visando $\beta \to 1$, e contabilizar o vazamento $L_{\mathrm{EC}}$.
    \item Garantir isolamento espectral em coexistência WDM e mitigar espalhamento Raman de canais clássicos; registrar telemetria (SNR, fase, drift) para auditoria.
\end{itemize}

Para implementar esses protocolos em simulação computacional, recomenda-se uma abordagem estruturada que modele adequadamente tanto os aspectos quânticos quanto os clássicos dos sistemas. A modelagem do canal quântico deve incluir perdas exponenciais $\eta = 10^{-\alpha L/10}$, onde $\alpha$ é a atenuação da fibra (tipicamente $0.2~\mathrm{dB/km}$ em $1550~\mathrm{nm}$) e $L$ a distância. Em cenários realistas, convém incorporar também ruído de detectores (contagens de escuras $p_d$), \emph{timing jitter}, visibilidade/interferência $V$, desalinhamentos de polarização/fase e flutuações de intensidade do emissor. A camada clássica deve contemplar autenticação (Wegman–Carter), reconciliação (CASCADE/LDPC) com eficiência $\beta$ mensurada, e amplificação de privacidade com orçamento composicional de segurança $\varepsilon_{\mathrm{tot}}$.

No nível de simulação, uma pipeline genérica para DV-QKD (BB84/SARG04/Decoy/MDI) e CV-QKD pode seguir as etapas: (i) geração de sequência aleatória (bits/bases ou amostras gaussianas); (ii) mapeamento físico (estados, intensidades, parâmetros de modulação); (iii) canal com perdas/ruídos e modelo de detecção; (iv) sifting/seleção de subconjuntos; (v) estimação de parâmetros com intervalos de confiança; (vi) reconciliação e verificação; (vii) amplificação de privacidade; (viii) cálculo de taxa de chave líquida e auditoria dos logs.

Para o BB84 e variantes, modelos simples porém informativos incluem:
\[
p_{\mathrm{sig}} \approx 1 - e^{-\mu\,\eta\,\eta_d}, \qquad
p_{\mathrm{det}} \approx p_{\mathrm{sig}} + p_d,\qquad
Q \approx \frac{Q_{\mathrm{opt}}\,p_{\mathrm{sig}} + 0.5\,p_d}{p_{\mathrm{sig}} + p_d} + Q_{\mathrm{misalign}},
\]
onde $\mu$ é a intensidade média por pulso, $\eta_d$ a eficiência de detecção e $Q_{\mathrm{opt}}\!\approx\!(1-V)/2$. Em \emph{decoy-state}, estimam-se $Y_1^{\downarrow}$ e $e_1^{\uparrow}$ para refinar a taxa:
\[
R \ge q\left\{-Q_{\mu} f(E_{\mu}) h_2(E_{\mu}) + Y_1^{\downarrow}\mu e^{-\mu}\,[1-h_2(e_1^{\uparrow})]\right\},
\]
com penalidades de finitude e custos de autenticação $L_{\mathrm{auth}}$ devidamente contabilizados. Em MDI-QKD, condiciona-se a análise aos eventos de projeção de Bell (“sucesso” na BSM), modelando interferência no nó central com igualação de espectro, fase e polarização entre emissores.

Para CV-QKD com modulação gaussiana, simula-se a cadeia geração–canal–detecção homódina/heteródina sob hipóteses gaussianas. A taxa-chave assintótica por uso do canal pode ser aproximada por:
\[
K \approx \beta\, I_{AB} - \chi_{BE},
\]
onde $I_{AB}$ se obtém das estatísticas das quadraturas compartilhadas e $\chi_{BE}$ de uma modelagem do canal gaussiano com transmissão $T$ e ruído excessivo $\xi$ (em unidades de ruído quântico), ajustados por estimação de parâmetros com intervalos de confiança. Em regime de SNR baixa, códigos LDPC de taxa adaptativa são essenciais para $\beta \to 1$.

Um \emph{setup} básico de laboratório para o protocolo BB84 pode ser construído utilizando um laser atenuado em torno de $1550~\mathrm{nm}$ operando com intensidade de aproximadamente $0.1$ fótons por pulso, moduladores de polarização capazes de selecionar entre as bases $Z$ e $X$, e um canal quântico constituído por fibra óptica de comprimento típico entre $10$ e $100~\mathrm{m}$ para testes. Essa configuração garante que qualquer tentativa de espionagem no canal introduza erros estatisticamente detectáveis, possibilitando a verificação rigorosa da segurança da chave final através de múltiplas implementações e protocolos complementares, como demonstrado pelos Algoritmos \ref{alg:BB84}, \ref{alg:E91}, \ref{alg:Decoy}, \ref{alg:MDI} e \ref{alg:CVQKD}. Para robustez e reprodutibilidade, recomenda-se:

\begin{itemize}
    \item Calibrar e estabilizar a intensidade por pulso $\mu$ (erro relativo $<1\%$) com fotodiodo \emph{tap} e malha de realimentação; registrar deriva térmica.
    \item Caracterizar a atenuação e perdas do enlace ($\alpha$, conectores, acoplamentos) e medir contagens de escuras $p_d$ e visibilidade/interferência $V$ por base.
    \item Sincronizar temporização com referência 10~MHz/1~PPS; ajustar \emph{gates} e janelas de detecção para minimizar \emph{jitter} e acidentais.
    \item Autenticar todo tráfego clássico (sifting, índices, paridades/LDPC, \emph{seeds} de hash) com orçamento explícito de $L_{\mathrm{auth}}$.
    \item Adotar reconciliação apropriada ao regime (CASCADE para protótipos; LDPC para altas taxas), medindo $\beta$ e contabilizando $L_{\mathrm{EC}}$.
    \item Aplicar intervalos de confiança (Chernoff/Hoeffding/Clopper–Pearson) nas estimativas de $Q$, $Q_{\mu_i}$, $E_{\mu_i}$ e, em decoy, $Y_1$, $e_1$; definir critérios de abortamento.
    \item Produzir \emph{logs} assinados com telemetria: taxas por detector, histogramas temporais, temperatura, deriva de $\mu$, parâmetros de alinhamento e resultados de teste.
\end{itemize}

Para expandir o estudo comparativo entre protocolos na simulação, é útil padronizar cenários e métricas:
\begin{itemize}
    \item Cenários de distância: $L \in \{10, 25, 50, 100, 150, 200\}\,\mathrm{km}$ (DV) e faixas metropolitanas para CV; co-existência WDM opcional com canais clássicos.
    \item Conjuntos de parâmetros fixos: $\eta_d$, $p_d$, $V$, \emph{deadtime}, largura de \emph{gate}, $V_M$ (CV), $T$, $\xi$, largura de banda de filtros e potência LO (CV).
    \item Métricas: taxa de chave líquida (bpcu e bps), eficiência de reconciliação $\beta$, vazamento $L_{\mathrm{EC}}$, custo $L_{\mathrm{auth}}$, fração de abortos, uso de CPU/tempo de execução.
    \item Protocolos-alvo: BB84 (com/sem decoy), E91 (violação de CHSH com $S$ alvo), MDI-QKD (taxa condicionada a BSM), CV-QKD (heteródino/homódino, $\beta$ versus SNR).
\end{itemize}

A seguir, uma estrutura de pseudocódigo genérica para simulação modular de DV e CV, adequada para implementação em linguagens científicas (Python/Julia/Matlab), mantendo consistência com os Algoritmos \ref{alg:BB84}, \ref{alg:E91}, \ref{alg:Decoy}, \ref{alg:MDI} e \ref{alg:CVQKD}:

\begin{algorithm}[h!]
\scriptsize
\caption{Esqueleto de Simulação Unificada (DV e CV)}
\label{alg:SimulacaoUnificada}
\begin{algorithmic}[1]
\Procedure{SimularQKD}{$\text{protocolo}, \text{parametros}$}
    \State \Comment{Inicialização}
    \State \text{ConfigurarRNG}(), \text{CarregarParametros}($\alpha, L, \eta_d, p_d, V, \mu, V_M, T, \xi, \beta, \ldots$)
    \State $\eta \gets 10^{-\alpha L/10}$
    \State \Comment{Geração de dados quânticos}
    \If{$\text{protocolo} \in \{\text{BB84}, \text{Decoy}, \text{MDI}\}$}
        \State $(\text{bits}, \text{bases}) \gets \text{GerarBitsEBases}(N)$
        \If{$\text{protocolo} = \text{Decoy}$}
            \State $(\mu_i, p_i) \gets \text{ConfigurarDecoy}()$
        \EndIf
        \State \text{detec}, \text{erros} $\gets \text{CanalEDeteccaoDV}(\eta, \eta_d, p_d, V, \mu~\text{ou}~\mu_i)$
    \ElsIf{$\text{protocolo} = \text{E91}$}
        \State $(\text{basesA}, \text{basesB}) \gets \text{GerarBasesE91}(N)$
        \State $S \gets \text{SimularCHSH}(\eta, V, p_d)$
        \If{$S \le 2$} \State \textbf{abort} \EndIf
        \State \text{detec}, \text{erros} $\gets \text{EventosCoincidencia}(S, \eta, p_d)$
    \ElsIf{$\text{protocolo} = \text{CV}$}
        \State $(X,P) \gets \text{Gauss}(\mathbf{0}, V_M)$
        \State $M_B \gets \text{CanalCV}(T,\xi) \to \text{DeteccaoHomodina/Heterodina}(X,P)$
    \EndIf
    \State \Comment{Sifting e estimação de parâmetros}
    \State \text{AplicarSifting}(), \text{SelecionarAmostrasTeste}()
    \State \text{EstimarParametrosComConfianca}()
    \State \Comment{Pós-processamento}
    \State $\beta \gets \text{Reconcilia}( \text{dados})$, $L_{\mathrm{EC}} \gets \text{VazamentoReconcilia}()$
    \State \text{CalcularTaxaChave}($R$ ou $K$) \Comment{Inclui $L_{\mathrm{auth}}$ e finitude}
    \State \text{AmplificacaoPrivacidade}() $\to \text{chave\_final}$
    \State \textbf{return} \text{Relatorio}($R/K$, $\beta$, $L_{\mathrm{EC}}$, $L_{\mathrm{auth}}$, métricas, logs)
\EndProcedure
\end{algorithmic}
\end{algorithm}

Por fim, recomenda-se um protocolo de experimentação e relatório para permitir comparações justas entre diferentes implementações e ambientes:
\begin{itemize}
    \item Fixar sementes e publicar verificação de aleatoriedade (NIST SP~800-22/Dieharder) para séries usadas em simulações replicáveis.
    \item Informar claramente o orçamento de segurança composicional $\varepsilon_{\mathrm{tot}}$ e sua distribuição entre $\varepsilon_{\mathrm{sec}}, \varepsilon_{\mathrm{cor}}, \varepsilon_{\mathrm{PE}}, \varepsilon_{\mathrm{auth}}$.
    \item Divulgar curvas $R(L)$ (DV) e $K(T,\xi)$ (CV) com barras de confiança e detalhamento de parâmetros físicos e de pós-processamento ($\beta$, $f(E)$, códigos).
    \item Disponibilizar \emph{logs} assinados, especificações de hardware (detectores, moduladores, lasers, eletrônica), calibrações e scripts de análise.
\end{itemize}

\section{Estado da Arte}

A área de criptografia quântica avançou rapidamente da teoria à prática, com implementações reais em redes metropolitanas, demonstrações de distribuição de chaves em fibras de centenas de quilômetros e até experimentos de comunicação via satélite. A Tabela \ref{tab:estado_arte_quantica} sintetiza alguns dos trabalhos mais relevantes da literatura, cobrindo quase quatro décadas de evolução.

Conforme apresentado na Tabela \ref{tab:estado_arte_quantica}, a evolução da criptografia quântica pode ser dividida em etapas que vão desde a formulação teórica de protocolos até a sua implementação em larga escala. O trabalho de Bennett e Brassard (1984) marcou o ponto de partida da área, ao propor o protocolo BB84. Embora sem aplicação imediata, lançou os fundamentos que viriam a estruturar toda a pesquisa subsequente. Poucos anos depois, Ekert (1991) avançou esse quadro ao propor o protocolo E91, utilizando emaranhamento e desigualdades de Bell para garantir segurança quântica, inaugurando a vertente dos métodos “device-independent”.  

Já em meados de 1995, Huttner et al. deram origem a técnicas precursoras contra ataques de fótons múltiplos, um passo importante para lidar com as imperfeições reais das fontes de luz. Esse caminho culminaria em protocolos robustos como o SARG04 e, posteriormente, nas variantes de \textit{decoy states}.

O início dos anos 2000 trouxe as primeiras demonstrações práticas em fibras comerciais, como a realizada pelo grupo de Genebra (Stucki et al., 2002), evidenciando a viabilidade tecnológica fora do laboratório. Essa linha foi ampliada com o trabalho de Xu et al. (2014), que demonstrou experimentalmente o uso de estados \textit{decoy}, consolidando a segurança contra ataques de divisão de fótons.  

Avanços transformadores ocorreram na década de 2010, com o surgimento do MDI-QKD (Lo et al., 2012), que resolveu vulnerabilidades críticas em detectores, e com iniciativas de larga escala, como a rede SECOQC em Viena (Tang et al., 2016).  

A trajetória rumo a comunicações globais ganhou destaque com a missão espacial Micius (Liao et al., 2017), primeiro demonstrador de QKD via satélite entre continentes, seguida por experimentos em fibras de longa distância, como o de Boaron et al. (2018), que alcançou 421 km. Mais recentemente, o desenvolvimento do Twin-Field QKD (Liu et al., 2019) mostrou potencial para dobrar o alcance seguro das redes, rompendo o limite teórico PLOB.

\begin{table}[h!]
\centering
\scriptsize
\caption{Principais artigos representativos do estado da arte em criptografia quântica}
\label{tab:estado_arte_quantica}
\begin{longtable}{p{3cm}p{2cm}p{6cm}p{3cm}}
\hline
\textbf{Autor/Ano} & \textbf{Protocolo} & \textbf{Contribuição} & \textbf{Limitações} \\
\hline
Bennett \& Brassard (1984) \cite{BB84} & BB84 & Primeiro protocolo de QKD, utilizando polarização de fótons em duas bases. & Teórico, sem implementação prática inicial. \\
Ekert (1991) \cite{E91} & E91 & Protocolo baseado em emaranhamento e desigualdades de Bell. & Complexidade experimental elevada. \\
Huttner et al. (1995) \cite{SARG04} & SARG04 & Introdução de técnicas preliminares contra ataques de fótons múltiplos. & Eficiência reduzida em longas distâncias. \\
Stucki et al. (2002) \cite{Geneva2002} & Gisin-Geneva QKD & Uma das primeiras implementações práticas em fibras comerciais (67 km). & Limitação em alcance e estabilidade. \\
Lo et al. (2012) \cite{MDIQKD} & MDI-QKD & Revolucionário: elimina vulnerabilidades associadas a detectores em QKD. & Requer infraestrutura óptica avançada. \\
Xu et al. (2014) \cite{Decoy2014} & Decoy-State QKD & Demonstração prática da técnica de decoy states, aumentando segurança contra ataques de fótons múltiplos. & Reduz a taxa de geração de chave. \\
Tang et al. (2016) \cite{SECOQC} & SECOQC (EU) & Primeira rede quântica multinó em Viena, integrando diversos protocolos. & Escala ainda experimental. \\
Liao et al. (2017) \cite{Micius} & QKD via Satélite (Micius) & Primeira demonstração de QKD intercontinental, conectando China–Europa (1200 km). & Custos elevados e limitações de alinhamento. \\
Boaron et al. (2018) \cite{Boaron2018} & Long-distance QKD & QKD implementado em fibras de 421 km com taxas de chave prática. & Perdas altas, necessidade de repetidores. \\
Yin et al. (2020) \cite{CVQKD} & CV-QKD & Implementação de QKD com variáveis contínuas utilizando estados gaussianos. & Requer detecção altamente precisa. \\
Liu et al. (2019) \cite{TwinField2019} & Twin-Field QKD & Introdução do protocolo Twin-Field QKD, dobrando o alcance teórico de redes QKD. & Complexidade experimental. \\
Chen et al. (2021) \cite{China2021} & Rede Quântica de Beijing–Shanghai & Implementação de rede quântica de 2000 km integrando nós de QKD. & Dependência de repetidores confiáveis. \\
Ankers et al. (2023) \cite{QDS2023} & QDS & Demonstração experimental de assinaturas digitais quânticas em fibras ópticas. & Escalabilidade limitada a redes pequenas. \\
Fang et al. (2023) \cite{Fang2023} & Satellite-Ground & Primeira demonstração de QKD de variáveis contínuas via satélite. & Requisitos tecnológicos ainda experimentais. \\

Zhou et al. (2024) \cite{Zhou2024TFQKD} & TF-QKD & Implementação de Twin-Field QKD de longo alcance com taxas de chave aprimoradas em fibras terrestres. & Setup óptico altamente estável e sincronização de fase complexa. \\
Li et al. (2024) \cite{Li2024CVSat} & CV-QKD Satélite & Demonstração de CV-QKD integrada a enlaces satélite–solo com protocolos de reconciliação otimizados. & Sensível a turbulência atmosférica e perdas no canal espacial. \\
\hline
\end{longtable}
\normalsize
\end{table}

Na mesma direção de inovação, implementações com variáveis contínuas (CV-QKD) ganharam espaço, como o trabalho de Yin et al. (2020), que demonstrou a eficiência de técnicas gaussianas em ambientes práticos, e de Fang et al. (2023), que levou essa abordagem ao espaço. Em paralelo, Chen et al. (2021) implementaram a ambiciosa rede Beijing–Shanghai com 2000 km de alcance, exibindo a maturidade do campo. Finalmente, novas aplicações emergem, como as assinaturas digitais quânticas (QDS) apresentadas por Ankers et al. (2023), expandindo o leque de possibilidades além da distribuição de chaves.  

Esses trabalhos, tomados em conjunto, evidenciam a transição da QKD de um conceito puramente teórico no início dos anos 1980 para uma tecnologia consolidada com demonstrações reais em redes complexas, fibras de ultra longa distância e até sistemas satélite-terra. O estado da arte, portanto, sinaliza uma maturidade crescente, mas ainda acompanhada por desafios de eficiência, escalabilidade e integração em infraestruturas globais.

\section{Questões para Reflexão e Pesquisa Futura}

Ao concluir o capítulo sobre criptografia quântica, o leitor deve ser estimulado a examinar não apenas os conceitos teóricos e as demonstrações de laboratório, mas também as implicações práticas, as fragilidades de implementação e as raízes estratégicas e regulatórias dessa tecnologia. Pergunte-se inicialmente quais garantias de segurança são de fato oferecidas por cada variante de QKD (DV-QKD, CV-QKD, MDI, DI, TF-QKD) no contexto operacional que você conhece: até que ponto a prova de segurança idealizada permanece válida quando confrontada com limitações de detector, fontes não ideais e ruído ambiente? Quais são as hipóteses explícitas (p. ex. modelos de detector, estatísticas de decoy states, bloqueio de perda) nas quais se baseiam as provas de segurança, e como essas hipóteses se traduzem em requisitos de instrumentação e monitoramento para uma implantação real?

Interrogue a lacuna entre segurança teórica e segurança prática. Quais vetores de ataque exploráveis em campo, por exemplo, blinding de detectores, manipulação de temporização, back-flash óptico, interações não lineares em detectores ou side-channels eletromagnéticos, são mais críticos e como podem ser auditados sistematicamente? Em que medida mitigações propostas (detecções ativas, monitoramento de parâmetros físicos, MDI-QKD) resolvem os problemas sem inviabilizar o enlace por complexidade ou custo? Pergunte-se também sobre o trade-off entre confiar em nós "trusted" versus investir em esquemas device-independent: quais cenários (restrições operacionais, custos, requisitos de confiança entre domínios administrativos) tornam aceitável o uso de nós confiáveis, e quando a necessidade de garantias físicas independentes justifica projetos mais caros e complexos?

Considere as implicações de integração com infraestruturas clássicas. Como autenticar o canal clássico necessário para QKD sem introduzir pontos de fragilidade que anulem os benefícios físicos da distribuição de chave? Quais esquemas de autenticação (pré-compartilhada, HMAC com rotação, PQC híbrido) oferecem melhores trade-offs entre segurança e praticidade, especialmente quando se considera a necessidade de proteger o canal clássico contra ataques que possam comprometer a sessão quântica? Pense também em como a QKD se encaixa em arquiteturas heterogêneas: em redes metropolitanas com múltiplos domínios administrativos, quais modelos de key management, orquestração e provisionamento resolvem a necessidade de escalabilidade sem criar gargalos de confiança?

Analise os desafios de engenharia e métricas de avaliação. Que experimentos de campo são necessários para quantificar a viabilidade operacional de QKD em diferentes topologias (fibras metropolitanas, enlaces longos com repetidores clássicos, enlaces por satélite)? Quais métricas (taxa líquida de chave segura por tempo, latência de estabelecimento, sensibilidade a variações de perda, tempo médio entre falhas, custo por bit de chave entregue) devem ser padronizadas para permitir comparações justas entre implementações e fornecedores? Ao projetar PoCs, como incluir análises de finite-key para entender o impacto real de janelas de coleta limitadas e eventuais ataques estatísticos?

Questione ainda a relação entre QKD e outras proteções: em que cenários a combinação PQC+QKD é realmente vantajosa em relação ao uso isolado de PQC robusta? Para ativos com horizonte de proteção extremamente longo, a QKD fornece uma vantagem prática suficiente para justificar custos de instrumentação, manutenção e operação? E se a autenticação final continuar sendo feita por algoritmos cuja segurança futura é incerta, a adição de QKD resolve o problema de confiden cialidade de longo prazo ou apenas desloca o ponto fraco para outro componente do sistema?

Explore implicações tecnológicas emergentes. Como os desenvolvimentos em repetidores quânticos, memórias quânticas e satélites influenciarão a economia e topologia de futuras redes QKD? Quais são os requisitos técnicos e padrões de interoperabilidade necessários para que backbones quânticos federados possam ser operados entre organizações e países sem dependência excessiva de fornecedores únicos? Considere também o papel dos QRNGs e de módulos quânticos embarcados: até que ponto a incorporação massiva de fontes de aleatoriedade quântica em dispositivos finais altera o modelo de risco e como certificá-los de forma confiável?

Não deixe de abordar aspectos de cadeia de suprimentos, governança e regulação. Quais políticas de certificação, auditoria e conformidade são necessárias para atestar dispositivos QKD, detectores e módulos de sincronização? Como evitar que a dependência de hardware especializado crie pontos de estrangulamento geopolíticos ou comerciais? E em que medida regimes de exportação de tecnologia quântica e requisitos de soberania digital moldarão a adoção internacional de QKD?

Investigue questões econômicas e operacionais: qual é o custo total de posse (TCO) de um enlace QKD comparado com alternativas baseadas apenas em PQC, considerando aquisição, instalação, manutenção, substituição de hardware e custos de operação? Em quais modelos de negócio (serviços gerenciados, infraestrutura compartilhada, consórcios setoriais) a QKD pode ser viável economicamente além de nichos governamentais ou financeiros?

Projete experimentos e pesquisas práticas que o leitor possa executar para transformar reflexão em evidência. Exemplos concretos incluem: implantar um enlace QKD metro-local sobre fibra já em uso e medir taxa de chave útil versus perda e ruído ao longo de semanas; integrar a chave QKD com um sistema TLS existente via mecanismo de derivação híbrida e medir tempo de handshake e comportamentos de fallback sob degradação de enlace; realizar testes controlados de ataques de detector (blinding) em bancada para avaliar eficácia de contramedidas; executar análises de finite-key em cenários reais de coleta de dados para quantificar a segurança estatística efetiva; e comparar custo e latência de distribuição de chave usando satélite versus enlace terrestre em cenários de emergência/disponibilidade geográfica.

Finalmente, plante questões éticas, estratégicas e de política pública: qual o papel do Estado na promoção ou regulação de infraestruturas quânticas críticas? Quais contingências legais e de privacidade devem ser previstas caso capacidades de criptanálise retroativa sejam acionadas no futuro? Como balancear transparência técnica com proteção de conhecimento sensível em programas nacionais de infraestrutura quântica? E, de forma prática, quais linhas de pesquisa e quais métricas de sucesso (tecnológicas, econômicas e de confiança pública) o leitor considera prioritárias para que a QKD ultrapasse o patamar de tecnologia experimental para ferramenta operacionalmente útil e escalável?

Estas perguntas destinam-se a incentivar investigações experimentais e reflexões estratégicas. Recomenda-se que o leitor documente hipóteses, protocolos de teste, métricas coletadas e procedimentos de mitigação testados, de modo a transformar indagações teóricas em evidências operacionais que possam orientar decisões institucionais e contribuir para a maturação responsável da criptografia quântica.

\chapter{Criptografia Pós-Quântica}

A rápida evolução da computação quântica torna imperativa uma transição planeada dos sistemas criptográficos clássicos para alternativas resistentes a ataques quânticos. Este capítulo introduz os princípios e a motivação da criptografia pós-quântica (PQC) e apresenta, em texto corrido, a estrutura lógica que guiará a discussão nas seções seguintes. Inicialmente serão abordados os conceitos fundamentais e a motivação prática para a PQC, incluindo a análise da ameaça representada por adversários capazes de realizar ataques \emph{harvest-now-decrypt-later} e a necessidade de horizontes de proteção longos para ativos críticos. Em seguida, será desenvolvido um panorama das famílias de algoritmos candidatas, com ênfase em problemas baseados em reticulados (LWE, Ring-LWE, \gls{ntru}), códigos corretivos (como variantes de McEliece), \gls{hashbased} (por exemplo SPHINCS+), esquemas multivariáveis e propostas baseadas em isogenias, discutindo vantagens, limitações, trade-offs em termos de segurança, eficiência, tamanhos de chave/assinatura e exposição a ataques práticos (incluindo canais laterais).

O capítulo prossegue com uma análise do estado da arte e dos esforços de padronização, destacando decisões recentes, lições aprendidas e recomendações de organismos como o NIST, além de referências a implementações de referência e trabalhos relevantes. Uma seção dedicada examina os desafios de engenharia e segurança operacional associados à adoção da PQC: overheads de largura de banda e armazenamento, impacto em latência e CPU, requisitos para implementações em tempo-constante, mitigação de canais laterais, e considerações sobre licenciamento e propriedade intelectual. A partir daí, serão propostas estratégias práticas de migração e governança, incluindo princípios de crypto-agility, uso de abordagens híbridas (dual KEMs / assinaturas duplas), priorização de ativos críticos, roteiros de testes e pilotos, e métricas para avaliação de progresso, com recomendações concretas para operações, auditoria e conformidade normativa.

Por fim, o capítulo situará a PQC no contexto mais amplo das estratégias quânticas, discutindo sua complementaridade com QKD e arquiteturas híbridas (por exemplo, integração de QKD para reabastecimento de chaves em KMS de alta sensibilidade), e encerrará com conclusões executivas e técnicas que resumem recomendações para uma transição ordenada, auditável e resiliente rumo a um ecossistema criptográfico resistente aos desafios emergentes.

\section{Conceitos e Motivação}

A ameaça representada pelos computadores quânticos universais ao paradigma clássico da criptografia de chave pública é profunda e multifacetada. Protocolos amplamente utilizados atualmente, como RSA, Diffie--Hellman e ECC (Elliptic Curve Cryptography), baseiam sua segurança em problemas matemáticos que, no modelo clássico, são considerados intratáveis para dimensões práticas (fatoração inteira, cálculo do logaritmo discreto). A descoberta do algoritmo de Shor (1994) demonstrou que essas mesmas estruturas algébricas podem ser resolvidas em tempo polinomial em um computador quântico ideal, destruindo a base de segurança desses esquemas caso existam máquinas quânticas escaláveis com número suficiente de qubits e fidelidade operacional. Consequentemente, a confidencialidade, integridade e autenticidade de grande parte das comunicações e infraestruturas digitais modernas ficam potencialmente comprometidas.

O impacto dessa vulnerabilidade é sistêmico. Infraestruturas críticas, serviços bancários, redes governamentais, saúde, cadeias de suprimento, internet das coisas (IoT), e infraestruturas de certificação eletrônica, dependem fortemente de primitivas de chave pública para estabelecimento de canais seguros, assinatura de transações e ancoragem de identidades digitais. Ademais, protocolos de ampla difusão, como TLS/SSL, VPNs, PKI, SSH e mecanismos de assinatura em blockchains, repousam na confiança de que a quebra dos problemas matemáticos subjacentes é impraticável. A possibilidade de um adversário armazenar hoje comunicações cifradas e, no futuro, quebrá-las com um computador quântico (o risco conhecido como \emph{harvest-now-decrypt-later} ou “store now, decrypt later”) aumenta dramaticamente a urgência por estratégias de proteção que sejam resistentes a ataques quânticos, sobretudo para dados com horizonte de confidencialidade longo.

A criptografia pós-quântica (PQC, do inglês Post-Quantum Cryptography) surge exatamente nesse contexto como uma abordagem pragmática: desenvolver primitivas e protocolos que possam ser executados em hardware clássico existente, mas cuja segurança se baseie em problemas matemáticos para os quais, até o conhecimento atual, não há algoritmos quânticos eficientes conhecidos. Entre os problemas explorados destacam-se reticulados (LWE, Ring-LWE, NTRU), códigos corretivos, construções baseadas em funções hash, esquemas multivariáveis e isogenias. Cada família traz um perfil distinto de propriedades, resistência conjectural a ataques quânticos, custos computacionais, tamanhos de chave/assinatura e suscetibilidade a ataques de implementação, tornando necessário um desenho de políticas que privilegie diversidade e resiliência.

A motivação prática para a adoção antecipada de PQC baseia-se em vários argumentos objetivos. Primeiro, a janela temporal entre a investigação, padronização e ampla implantação de novos esquemas é longa; portanto, iniciar o processo de migração antes da disponibilidade de computadores quânticos plenamente funcionais reduz o risco de exposição de ativos longamente sensíveis. Segundo, muitos sistemas críticos utilizam assinaturas e chaves que devem permanecer válidas por décadas (certificados raiz, registros legais, dados arquivados); proteger esses ativos apenas quando a tecnologia quântica se tornar madura pode ser tarde demais. Terceiro, a migração para PQC, por suas características (tamanhos maiores de chave, diferentes requisitos computacionais e de protocolo), exige planejamento arquitetural e operacional significativo, desde atualizações de HSMs/TPMs até adaptações em protocolos (por exemplo, TLS, SSH e formatos de mensagens), o que justifica uma abordagem faseada e testada.

É necessário, entretanto, distinguir corretamente os papéis de PQC e QKD (Quantum Key Distribution). Enquanto a QKD oferece garantias fundamentadas em princípios físicos para a distribuição de chaves (sob hipóteses físicas específicas), ela depende de infraestrutura quântica especializada e tem limitações práticas e de escalabilidade que ainda restringem sua adoção generalizada. A PQC, por sua vez, opera inteiramente no domínio clássico e é imediatamente aplicável em larga escala. Assim, as duas abordagens são complementares: QKD pode ser vantajosa em enlaces de altíssima sensibilidade ou em arquiteturas híbridas, enquanto PQC fornece proteção ampla e de fácil integração para a maior parte das aplicações existentes.

Do ponto de vista estratégico, a resposta à ameaça quântica requer princípios claros de governança e engenharia. A \emph{crypto-agility}, ou seja, a capacidade de trocar algoritmos e parâmetros criptográficos sem rearquitetar sistemas inteiros, deve ser um objetivo de projeto prioritário. Abordagens híbridas (por exemplo, encapsular chaves com um KEM pós-quântico e um clássico simultaneamente) permitem reduzir riscos durante a transição, combinando compatibilidade retroativa com proteção adicional. Políticas de priorização de ativos são essenciais: migrar primeiro chaves e serviços com maior horizonte de proteção e maior impacto em caso de comprometimento (CA roots, certificados de assinatura de código, bases de dados de saúde, registros legais).

A incerteza científica quanto à resistência absoluta de qualquer candidato post-quantum impõe a adoção de diversidade algorítmica e vigilância contínua: fomentar múltiplos esquemas com diferentes bases matemáticas diminui o risco sistêmico caso vulnerabilidades sejam descobertas. Implementações seguras, auditadas e resistentes a ataques de canais laterais (tempo, consumo de energia, emissões eletromagnéticas) são igualmente críticas, especialmente em ambientes embarcados e IoT. Finalmente, fatores práticos, como overheads de largura de banda e armazenamento, latência adicional e implicações para HSMs, TPMs e infraestruturas legadas, devem ser avaliados e mitigados por meio de testes, pilotos e atualizações incrementais.

\section{Principais Classes de Algoritmos}

O desenvolvimento da criptografia pós-quântica levou à consolidação de diferentes classes de algoritmos, cada uma baseada em problemas matemáticos considerados resistentes mesmo diante de computadores quânticos universais. Essas classes refletem décadas de pesquisa em teoria da complexidade e criptografia, e representam distintas abordagens para garantir a segurança de longo prazo em sistemas digitais. A Tabela \ref{tab:classes_pqc} apresenta uma síntese comparativa entre as principais famílias de algoritmos, destacando exemplos, pontos fortes e limitações.

A primeira e mais consolidada é a dos algoritmos \emph{baseados em lattices}. Estes algoritmos exploram a dificuldade de problemas como o \textit{Learning With Errors} (LWE), o Ring-LWE e o Module-LWE, todos considerados intratáveis para algoritmos quânticos conhecidos. O atrativo dessa família está no equilíbrio entre eficiência, segurança formal e aplicabilidade prática, sendo atualmente a principal aposta do NIST. Destacam-se o CRYSTALS--Kyber e o CRYSTALS--Dilithium, já padronizados pelo NIST em 2022; o \gls{falcon}, esquema de assinaturas digitais extremamente compacto, mas de difícil implementação; e o NTRU, proposto em 1996 e ainda seguro. Outros algoritmos dessa família incluem NTRUEncrypt, NTRUSign e o FrodoKEM, que opta por uma abordagem mais conservadora ao evitar estruturas adicionais. Essa diversidade faz dos lattices a família mais promissora para aplicações práticas em sistemas embarcados, nuvem e internet.

Outra classe de grande relevância é a dos algoritmos \emph{baseados em códigos corretores de erros}. Desde o esquema de McEliece (1978), essa linha demonstra resiliência impressionante. O \gls{mceliece} permanece como um dos candidatos mais fortes no NIST, sem ataques práticos em mais de 40 anos. Apesar das chaves públicas gigantescas, variantes como \gls{bike} (Bit Flipping Key Encapsulation), \gls{hqc} (Hamming Quasi-Cyclic) e LEDAsig (algoritmo de assinaturas baseado em códigos esparsos) surgem como alternativas mais modernas. Esses esquemas são especialmente investigados para aplicações de longo prazo em que robustez é mais crítica que desempenho.

A família baseada em funções hash representa a abordagem mais conservadora, pois sua segurança repousa em primitivas já amplamente analisadas. O exemplo mais relevante é o SPHINCS+, aprovado no processo de padronização do NIST. Outros trabalhos dessa linha incluem variantes como XMSS (eXtended Merkle Signature Scheme, padronizado pelo IETF) e \gls{lms} (Leighton–Micali Signatures). Embora possuam assinaturas maiores e processamento mais lento em comparação com outras alternativas, esses protocolos oferecem segurança baseada em suposições muito sólidas.

Os algoritmos baseados em isogenias exploram a dificuldade de encontrar isogenias entre curvas elípticas supersingulares. Essa linha foi liderada pelo SIKE (Supersingular Isogeny Key Encapsulation), que oferecia chaves públicas muito pequenas e baixo custo de transmissão. Entretanto, ataques recentes demonstraram que a dificuldade matemática subjacente pode ser explorada, resultando na quebra prática de SIKE em 2022. Outras variantes em estudo, como CSIDH (Commutative Supersingular Isogeny Diffie–Hellman), tentam contornar essas limitações, mas ainda permanecem em estágio inicial de maturidade. Entre os algoritmos multivariados, baseados na dificuldade de resolver sistemas de equações quadráticas multivariadas (problema MQ), destacam-se esquemas como Rainbow, GeMSS, UOV (Unbalanced Oil and Vinegar) e HFE (Hidden Field Equations). O Rainbow chegou à fase final do processo de padronização do NIST, mas foi quebrado pouco antes da decisão final, evidenciando o risco inerente a essa linha.

Além das famílias assimétricas anteriores, também é essencial mencionar a adaptação de algoritmos simétricos e funções de hash à era pós-quântica. Embora não ameaçados pelo algoritmo de Shor, sistemas como AES e SHA sofrem impacto parcial do algoritmo de Grover, que reduz a segurança efetiva em fator quadrático. A contramedida é simples: utilizar chaves e saídas de hash maiores. Assim, recomenda-se AES-256 em vez de AES-128 e SHA3-512 em vez de SHA-256 para garantir longevidade da segurança.

Em conjunto, essas famílias compõem o panorama atual da criptografia pós-quântica, cada uma com suas vantagens e desafios. A estratégia dominante é a padronização de algoritmos baseados em lattices, complementados por opções baseadas em códigos e hash, enquanto multivariados e isogenias permanecem como frentes de pesquisa. A diversidade apontada na Tabela \ref{tab:classes_pqc} é fundamental para assegurar resiliência frente a novas descobertas em criptanálise.

\begin{table}[h!]
\centering
\scriptsize
\caption{Resumo das principais classes de algoritmos de criptografia pós-quântica, destacando exemplos, vantagens e limitações.}
\label{tab:classes_pqc}
\begin{tabular}{p{1.5cm}p{2.5cm}p{6cm}p{4cm}}
\hline
\textbf{Classe} & \textbf{Exemplos} & \textbf{Vantagens} & \textbf{Limitações} \\
\hline

Base em Lattices & Kyber, Dilithium, Falcon, NTRU, FrodoKEM, SABER, NewHope & Alta segurança, bom desempenho, já padronizados pelo NIST; flexíveis em hardware/software; aplicações práticas em nuvem e IoT & Algumas variantes complexas de implementar; desafios em dispositivos de baixo consumo; necessidade de mitigação contra ataques de canal lateral \\

Base em Códigos & Classic McEliece, BIKE, HQC, LEDAsig & Robustez histórica; mais de 40 anos sem ataques práticos bem-sucedidos; maturidade criptanalítica elevada & Chaves públicas extremamente grandes (100 KB–1 MB+), dificultando uso em dispositivos móveis e protocolos de rede \\

Base em Hash & SPHINCS+, XMSS, LMS, GMSS & Segurança conservadora; independência de problemas matemáticos específicos; padronizados pelo IETF/NIST; alta confiabilidade em longo prazo & Tamanho excessivo de assinaturas; baixa velocidade de verificação; uso restrito a cenários que demandam máxima segurança \\

Base em Isogenias & SIKE, CSIDH, SQISign & Chaves públicas curtas (centenas de bytes); baixo custo de transmissão; elegância matemática que permite aplicações compactas & Ataques recentes quebraram SIKE; eficiência computacional baixa; segurança sob revisão ativa \\

Base em Multivariados & Rainbow, GeMSS, UOV, Oil-Vinegar, HFE, TTS & Boas assinaturas digitais; operações rápidas em alguns cenários; variedade de variantes & Diversos esquemas já quebrados (ex. Rainbow); segurança altamente dependente da escolha de parâmetros \\

Simétricos adaptados & AES-256, SHA-512, SHA3-512, Blake3 & Já consolidados; pouca ou nenhuma perda de desempenho na transição; proteção contra Grover com aumento do tamanho da chave & Apenas mitigam ameaças em primitivas simétricas; não resolvem a vulnerabilidade em esquemas assimétricos como RSA ou ECC \\

\hline
\end{tabular}
\end{table}

\subsection{Lattices (Reticulados)}

Algoritmos com base em lattices constituem hoje a família mais estudada e adotada na prática para construção de KEMs e esquemas de assinatura pós-quânticos. A intuição central repousa na dificuldade de problemas geométricos sobre reticulados inteiros: encontrar vetores curtos (\gls{svp}: Shortest Vector Problem) ou resolver sistemas ruidosos lineares (LWE: Learning With Errors) são tarefas que, nas melhores técnicas conhecidas, exigem recursos exponenciais em parâmetros de segurança crescentes. Muitos esquemas modernos (Ring-LWE, Module-LWE, NTRU e variantes) exploram versões estruturadas desses problemas para obter eficiência adicional sem, idealmente, sacrificar a base de dificuldade. Importante notar que algumas construções admitem reduções teóricas que ligam a segurança em média a instâncias de pior caso de problemas como SVP, o que fornece um argumento formal complementar às evidências empíricas de resistência.

Do ponto de vista algorítmico, a construção típica de um KEM base em LWE envolve uma matriz pública \(A\), um segredo \(\mathbf{s}\) e um erro pequeno \(\mathbf{e}\), com o encapsulamento produzindo algo da forma \(\mathbf{t}=A\mathbf{s}+\mathbf{e}\) (modulo alguma base). A presença de \(\mathbf{e}\) é a fonte do “ruído” que torna a recuperação de \(\mathbf{s}\) difícil; o decapsulamento utiliza o conhecimento de um trapdoor ou de uma estrutura algébrica para recuperar a informação útil apesar do ruído. Essa mesma ideia se traduz em esquemas de assinatura via técnicas de trapdoor sampling ou por transformações do problema de aprendizado com erro em operações de assinatura verificáveis. A Figura \ref{fig:lattice_overview} apresenta uma ilustração didática desse princípio: uma rede de pontos de reticulado, um vetor alvo deslocado por um pequeno ruído e a dificuldade de associar o alvo ao vetor gerador correto sem o conhecimento secreto.

A escolha de parâmetros é crítica e multidimensional. Devem ser considerados: a dimensão do reticulado (ou grau do polinômio em construções anelares), a distribuição e variância do ruído (discretized Gaussian, centered binomial ou outra), o módulo aritmético, e margens de segurança contra ataques clássicos e quânticos. As estimativas de segurança atuais baseiam-se em modelos de custo para algoritmos de redução de base (BKZ com blocos \(\beta\)), enumeração otimizada e técnicas de sieving, incluindo variantes quânticas de sieving com velocidades assintoticamente melhores, fatores que impõem que parâmetros sejam selecionados com margens conservadoras. Por essa razão, a comunidade recomenda seguir parâmetros padronizados (por exemplo, os conjuntos de parâmetros do NIST para Kyber/Dilithium) e manter monitoramento ativo das melhorias em criptoanálise de lattices.

No aspecto de implementação, lattices oferecem um bom equilíbrio entre velocidade e tamanho de chave/assinatura quando comparados a outras famílias: polinômios sobre anéis permitem multiplicações via transformadas rápidas (NTT/FFT-like) que aceleram operações essenciais, enquanto variantes sem estrutura (ex.: FrodoKEM) evitam potenciais superfícies de ataque algébrico à custa de maior custo computacional e de comunicação. Implementações de alto desempenho exploram vetorização (SIMD), multiplicação por NTT em tempo constante, e otimizações de memória; entretanto, a superfície de ataques por canais laterais, timing, cache, consumo energético, exige implementações cuidadosamente constant-time, com mascaramento e/ou mitigação de acesso a cache. Além disso, a geração de ruído (amostragem gaussiana discreta ou centered binomial) requer RNGs de alta qualidade e métodos de amostragem reversíveis/seguro-tempo para evitar vazamentos.

Em termos de aplicações práticas, esquemas com base em lattices já são utilizados em implementações experimentais de TLS, VPNs e bibliotecas criptográficas integradas a sistemas operacionais; sua viabilidade em dispositivos embarcados está em constante avanço, com versões de NTRU/NTRU-Prime e SABER sendo otimizadas para microcontroladores. No entanto, em contextos de recursos muito restritos (certos dispositivos IoT), o custo de armazenamento e processamento pode ainda ser um limitador, exigindo trade-offs e, eventualmente, uso de arquiteturas híbridas ou aceleração por hardware (por exemplo, suporte em HSMs/TPMs que incluam instruções para NTT).

A Figura \ref{fig:lattice_overview} mostra que o vetor público (ou ciphertext) é um ponto alvo deslocado por um ruído pequeno \(\mathbf{e}\); sem o segredo/trapdoor, associar o alvo ao gerador correto do reticulado é computacionalmente difícil.Por fim, do ponto de vista de segurança prática, é imprescindível adotar estratégias de defesa em profundidade: combinar diversidade algorítmica (não depender exclusivamente de uma única família), aplicar parâmetros padronizados com margens conservadoras, executar auditorias e testes de criptoanálise paramétrica, e exigir implementações certificadas com contramedidas contra canais laterais. A adoção de lattices na indústria (Kyber para encapsulamento e Dilithium para assinatura são exemplos proeminentes) reflete esse equilíbrio entre segurança teórica, eficiência prática e maturidade de implementação.

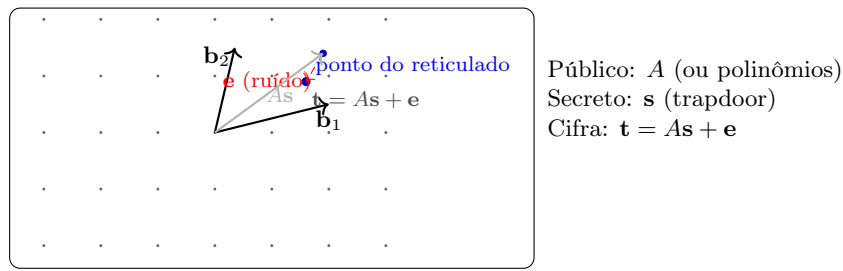
\begin{figure}[h!]
\centering
\begin{tikzpicture}[scale=0.75,
    every node/.style={font=\footnotesize, inner sep=0.6pt},
    baseline=(current bounding box.north)]
  \foreach \x in {-3,-2,-1,0,1,2,3}
    \foreach \y in {-2,-1,0,1,2}
      \filldraw[black!60] (\x,\y) circle (0.45pt);

  \draw[->,thick] (0,0) -- (2.0,0.5) node[pos=0.86,below right] {$\mathbf{b}_1$};
  \draw[->,thick] (0,0) -- (0.35,1.5) node[pos=0.9,left] {$\mathbf{b}_2$};

  \coordinate (target) at (1.6,0.9);
  \filldraw[blue!80!black] (target) circle (2.0pt);
  \node[blue, anchor=south west, xshift=2.8pt, yshift=1.8pt] at (target) {\scriptsize ponto do reticulado};

  \coordinate (nearest) at (1.9,1.4);
  \filldraw[black!20!blue] (nearest) circle (1.6pt);

  \draw[->,dashed,red] (nearest) -- (target) node[pos=0.55, below left] {\scriptsize $\mathbf{e}$ (ruído)};

  \draw[->,thick,gray!60] (0,0) -- (1.9,1.4) node[pos=0.6, below] {\scriptsize $A\mathbf{s}$};

  \node[black!70, below right=1mm and 0.5mm of target] (tlabel) {\scriptsize $\mathbf{t}=A\mathbf{s}+\mathbf{e}$};

  \node[anchor=west] at (5.8,0.6) {\begin{minipage}{4.0cm}\footnotesize
    Público: \(A\) (ou polinômios)\\
    Secreto: \(\mathbf{s}\) (trapdoor)\\
    Cifra: \(\mathbf{t}=A\mathbf{s}+\mathbf{e}\)
  \end{minipage}};

  \draw[rounded corners=4pt] (-3.6,-2.4) rectangle (5.6,2.2);
\end{tikzpicture}
\caption{Visão esquemática de uma construção baseada em lattices.}
\label{fig:lattice_overview}
\end{figure}

\subsection{Códigos Corretivos (Code-based)}

Esquemas baseados em códigos fundamentam sua segurança na dificuldade da decodificação de códigos lineares gerais sem conhecer uma estrutura secreta que permita a correção eficiente de erros. No paradigma clássico de McEliece, o titular da chave privada conhece um código estruturado (por exemplo, um código de Goppa) e transformações secretas, tipicamente uma permutação e transformações lineares, que geram uma matriz geradora disfarçada \(G_{\text{pub}}\). A chave pública é precisamente essa matriz disfarçada; para encapsular/criptografar, um remetente codifica uma mensagem \(m\) usando \(G_{\text{pub}}\) e adiciona um vetor de erro \(e\) de peso apropriado, produzindo o vetor \(c = mG_{\text{pub}} + e\). Sem a estrutura secreta, recuperar \(m\) a partir de \(c\) e \(G_{\text{pub}}\) equivale, essencialmente, a resolver um problema de decodificação geral, tarefa que permanece intratável nos parâmetros recomendados. O receptor, por sua vez, aplica o procedimento inverso usando o conhecimento do código secreto e da permutação para decodificar \(c\) e recuperar \(m\).

Exemplos modernos incluem Classic McEliece (o candidato NIST mais representativo desta família), BIKE (baseado em QC-MDPC), HQC (Hamming Quasi-Cyclic) e LEDAsig (assinaturas baseadas em códigos esparsos). Essas propostas exploram diferentes trade-offs: Classic McEliece privilegia robustez conservadora com chaves muito grandes; BIKE e HQC reduzem o tamanho de chave utilizando estruturas quasi-cíclicas (o que facilita compressão e implementação), enquanto LEDAsig busca eficiência prática em esquemas de assinatura baseados em códigos esparsos.

Os parâmetros típicos nesta família apresentam características marcantes. As chaves públicas podem alcançar centenas de kilobytes até megabytes, dependendo do esquema e do nível de segurança; em contrapartida, as operações de encapsulamento/decapsulamento costumam ser muito rápidas, o que torna esses esquemas atraentes quando latência de operação é crítica e largura de banda/armazenamento não são restrições severas. Otimizações de implementação concentram-se em compressão de chaves (representações compres­sas de matrizes/quasi-cíclicas), decodificação eficiente (algoritmos de decodificação específica do código) e estruturas QC-MDPC que permitem trade-offs entre tamanho e segurança. A resistência a ataques estruturais (por exemplo, ataques que exploram a estrutura oculta do código) é determinante para a escolha de parâmetros e para as variantes adotadas.

Na prática, as limitações mais evidentes estão relacionadas ao overhead de armazenamento e transmissão das chaves públicas, que pode dificultar a utilização direta em protocolos com restrições de MTU, em dispositivos móveis com memória limitada ou em cenários IoT. Ademais, certas otimizações que reduzem tamanhos de chave, como introduzir estruturas quasi-cíclicas, exigem análise criptoanalítica cuidadosa, pois estruturas adicionais podem criar superfícies de ataque. Por isso, variantes conservadoras sem estrutura adicional (quando toleráveis) podem ser preferidas para ativos que exigem proteção de décadas.

Do ponto de vista de engenharia, recomenda-se considerar compressão e cache de chaves públicas, uso de HSMs para armazenamento de chaves privadas, e estratégias híbridas (por exemplo, combinar um KEM code-based com um KEM pós-quântico alternativo em TLS) para mitigar riscos e compatibilizar com restrições de protocolo. A família é especialmente indicada para cenários onde a robustez a longo prazo é prioritária e o custo de armazenamento/transmissão das chaves é aceitável, por exemplo, proteção de arquivos arquivados, infraestruturas de certificação de longo prazo ou enlaces onde as chaves são trocadas raramente.

A Figura \ref{fig:code_overview} ilustra de forma esquemática o fluxo básico de um esquema code-based: a formação do ciphertext \(c\) no lado do remetente a partir de \(m\) e \(G_{\text{pub}}\), a inserção controlada de ruído \(e\), e o decodificador detentor da chave privada que aplica a transformação inversa para recuperar \(m\). Note-se que um adversário que disponha apenas de \(G_{\text{pub}}\) e \(c\) enfrenta o problema de decodificação geral, cuja solução é computacionalmente infeasible nos parâmetros recomendados.

\begin{figure}[h!]
\centering
\resizebox{0.88\linewidth}{!}{%
\begin{tikzpicture}[node distance=9mm, auto, >=Stealth, font=\small]
  \node[draw, rounded corners, align=center] (sender) {Remetente\\(Encapsulador)};
  \node[right=8mm of sender] (m) {$m$ (mensagem / seed)};
  \node[right=of m] (Gpub) {$G_{\text{pub}}$ (chave pública)};
  \node[below=3mm of m, draw, rounded corners, align=center] (enc) {Encode: $u=mG_{\text{pub}}$\\Adicionar erro: $c=u+e$};

  \draw[->] (sender) -- (m);
  \draw[->] (m) -- (enc);
  \draw[->] (Gpub) -- (enc);

  \node[right=28mm of enc] (channel) {Canal público};
  \draw[->] (enc.east) -- ++(6mm,0) node[midway, above] {$c$} -- (channel);

  \node[right=30mm of channel, draw, rounded corners, align=center] (receiver) {Receptor\\(Decapsulador)};
  \node[below=3mm of receiver, draw, rounded corners, align=center] (dec) {Aplicar permutações e trapdoor\\Decodificar e remover $e$\\Recuperar $m$};

  \draw[->] (channel) -- (receiver);
  \draw[->] (receiver) -- (dec);

  \node[above=6mm of receiver] (priv) {$\mathrm{Priv}$: (código secreto, permutações)};
  \draw[->] (priv) -- (receiver);

  \node[below=13mm of channel, align=center] (adv) {\small Adversário conhece $G_{\text{pub}}$ e $c$\\\small Problema: decodificação geral};
  \draw[->, dashed] (channel.south) -- (adv.north);

  \draw[rounded corners] ($(sender.north west)+(-3mm,3mm)$) rectangle ($(dec.south east)+(3mm,-3mm)$);
\end{tikzpicture}}
\caption{Esquema simplificado de um KEM/cryptosystem baseado em códigos.}
\label{fig:code_overview}
\end{figure}
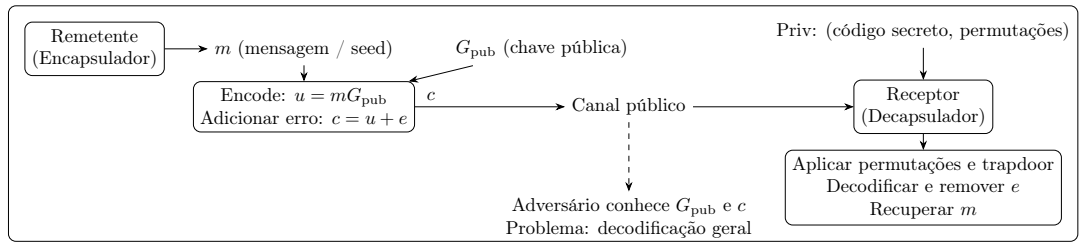

\subsection{Assinaturas Baseadas em Hash (Hash-based)}

Assinaturas com base em hash constituem a abordagem mais conservadora e fundamentada entre os candidatos pós-quânticos para assinaturas digitais, pois sua segurança repousa diretamente na resistência das funções de hash e não em suposições algebraicas complexas susceptíveis a avanços inesperados em criptoanálise. A ideia básica é simples e elegante: construir primitivas de assinatura a partir de funções de compressão e de estruturas de árvore (Merkle trees) que compõem várias assinaturas one-time (ou poucas vezes reutilizáveis) num esquema global verificável por meio de uma raiz pública compacta. Essa engenharia oferece, em princípio, uma confiança de longo prazo muito alta porque qualquer ataque efetivo exigiria comprometer a própria função de hash, um objetivo de difícil realização quando se usam funções bem analisadas e com saídas suficientemente longas.

Historicamente, as construções partem de assinaturas one-time como o esquema de Lamport e suas variantes compactas (WOTS: Winternitz One-Time Signature e WOTS+), que assinam uma única mensagem com segurança baseada em pré-imagem e colisões de hash. Para permitir assinaturas múltiplas, as chaves one-time são organizadas como folhas de uma árvore de merkle; a raiz da árvore serve como chave pública do esquema composto. Para assinar, o emissor cria uma assinatura one-time sobre a mensagem e acompanha essa assinatura com o caminho de autenticação (authentication path) necessário para provar que a chave one-time usada é de fato uma folha válida da árvore cuja raiz é a chave pública. A verificação, então, revalida a assinatura one-time e recalcula a raiz por meio do caminho de autenticação, comparando com a raiz pública conhecida. A robustez do esquema deriva da dificuldade de forjar uma assinatura válida sem conhecer a entrada de hash correspondente a uma folha genuína ou sem quebrar a função de hash utilizada para construir a árvore.

Uma distinção operacional importante nas implementações é entre esquemas stateful e stateless. Esquemas stateful, como XMSS e suas variantes (XMSS-MT), exigem que o signatário mantenha e atualize um estado interno (por exemplo, um índice de folha já utilizada) para evitar reutilização de chaves one-time; a violação dessa regra pode levar imediatamente à quebra da segurança da chave privada. Em contraste, esquemas stateless, exemplificados por SPHINCS+, buscam eliminar a necessidade de manter estado externo, geralmente por meio de camadas adicionais de hashing, estratégia de folhas pseudo-aleatórias e uso de estruturas probabilísticas que sacrificam tamanho e tempo por praticidade operacional. SPHINCS+ alcançou boa aceitação por ser stateless e por possuir argumentos de segurança baseados em funções de hash bem compreendidas, o que o tornou um candidato forte para cenários nos quais gerenciamento de estado é impraticável.

Quanto a parâmetros e desempenho, as assinaturas baseadas em hash apresentam trade-offs claros: são extremamente confiáveis do ponto de vista criptográfico, mas frequentemente apresentam assinaturas maiores e, em muitos casos, tempos de assinatura/verificação maiores quando comparados a esquemas baseados em lattices. As escolhas concretas, Winternitz parameter (W), altura da árvore, número de camadas em XMSS-MT, função de hash (SHA-2, SHA-3, Blake2/3) e técnicas de compressão de caminho de autenticação, determinam diretamente tamanho de assinatura, custo de computação e requisitos de armazenamento temporário. Em XMSS, por exemplo, o balanceamento entre profundidade da árvore e custo por assinatura pode reduzir o tamanho do caminho de autenticação mas aumenta a complexidade de gerenciamento de chaves. SPHINCS+ opta por uma topologia por camadas que combina pequenas árvores de Merkle com funções de hash e variantes de HORST/WOTS, resultando em assinaturas relativamente grandes, porém sem necessidade de estado e com parâmetros ajustáveis para vários níveis de segurança.

No domínio das implementações, os desafios práticos são múltiplos e muito diferentes daqueles de algoritmos assimétricos tradicionais. Para esquemas stateful, o principal desafio operacional é garantir a integridade do contador/estado de folha: qualquer restauração incorreta do estado (por exemplo, após restauração de imagem de disco, rollback de dispositivo ou operação concorrente mal coordenada) pode levar à reutilização de chaves one-time e à exposição total da chave privada. Assim, recomenda-se que o estado seja armazenado em memória não-volátil segura (HSM/TPM) com operações atômicas de atualização, ou que se implemente protocolos de sincronização robustos em sistemas distribuídos. Em contrapartida, esquemas stateless impõem sobrecarga de tamanho e CPU; otimizações possíveis incluem uso intensivo de instruções de hashing vetorizadas (SIMD), pré-cálculo e cache de partes recorrentes da árvore, e compressão dos caminhos de autenticação. A escolha da função de hash e sua implementação (em software puro, com vetorização, ou acelerada em hardware) tem impacto decisivo no throughput: Blake2/3 e implementações otimizadas de SHA-3 podem oferecer vantagens notáveis em dispositivos com suporte adequado.

Em termos de segurança operacional e análise de ataques, as assinaturas hash-based possuem vantagens claras: não dependem de problemas algebraicos suscetíveis a avanços futuros específicos (como fatoração ou reticulados), e seu nível de resistência é, na prática, limitado pela qualidade da função de hash e pelo comprimento de saída. Ainda assim, ataques práticos podem explorar implementações defeituosas (falhas na geração de aleatoriedade para folhas pseudo-aleatórias, vazamento de estado, falhas em atualizações atômicas) e canais laterais (tempo, consumo de energia). Por isso, contramedidas clássicas, execução constant-time onde aplicável, proteção de memória, integridade de armazenamento de estado e revisões de código, são tão essenciais aqui quanto em qualquer outra família. Além disso, a escolha de parâmetros deve considerar a evolução de ataques relacionados a pré-imagem e colisões contra as funções de hash adotadas; por isso recomenda-se usar funções com saídas mais longas e com análise criptoanalítica madura quando o horizonte de proteção for longo.

Do ponto de vista de padronização e adoção, várias iniciativas já consolidaram padrões e boas práticas: XMSS e LMS foram padronizados pelo IETF (\gls{rfc} correspondentes) com orientações para uso stateful, enquanto SPHINCS+ foi incluído no processo de padronização do NIST como um candidato stateless de grande confiança. Essas decisões implicam que bibliotecas criptográficas modernas e ferramentas de integração começam a oferecer suporte a formas interoperáveis desses esquemas, embora a integração em protocolos como TLS, S/MIME e formatos de assinatura ainda exija cuidado nas negociações de tamanho de campo e compatibilidade retroativa.

A Figura \ref{fig:hashsig_overview} ilustra de forma esquemática o princípio operativo de uma assinatura baseada em hash com árvore de Merkle: cada folha corresponde a uma chave one-time (por exemplo WOTS), a assinatura consiste na assinatura one-time sobre a mensagem mais o caminho de autenticação até a raiz, e a raiz funciona como chave pública compacta. Essa figura ajuda a compreender por que a integridade do uso de folhas (estado) é crítica em esquemas stateful e como a prova de pertença (authentication path) permite verificar a validade sem expor todas as chaves one-time.

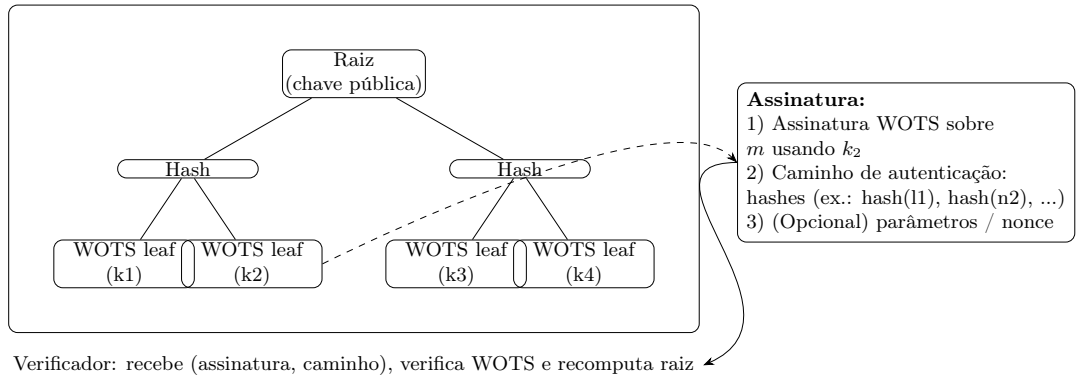
\begin{figure}[h!]
\centering
\resizebox{0.88\linewidth}{!}{%
\begin{tikzpicture}[font=\footnotesize,
  wnode/.style={draw, rounded corners, align=center, minimum width=22mm, inner sep=1pt},
  sigbox/.style={draw, rounded corners, inner sep=4pt, align=left, minimum width=36mm},
  smalltext/.style={font=\footnotesize, align=left},
  >=Stealth, baseline=(current bounding box.north)]
  \node[wnode] (root) at (0,0) {Raiz\\(chave pública)};
  \node[wnode] (n1)   at (-2.6,-1.5) {Hash};
  \node[wnode] (n2)   at ( 2.6,-1.5) {Hash};
  \node[wnode] (l1)   at (-3.6,-3.0) {WOTS leaf\\(k1)};
  \node[wnode] (l2)   at (-1.6,-3.0) {WOTS leaf\\(k2)};
  \node[wnode] (l3)   at ( 1.6,-3.0) {WOTS leaf\\(k3)};
  \node[wnode] (l4)   at ( 3.6,-3.0) {WOTS leaf\\(k4)};

  \draw (root) -- (n1);
  \draw (root) -- (n2);
  \draw (n1) -- (l1);
  \draw (n1) -- (l2);
  \draw (n2) -- (l3);
  \draw (n2) -- (l4);

  \node[draw, rounded corners=4pt, inner sep=7mm, fit=(root) (l1) (l4)] (box) {};

  \node[sigbox, anchor=west] (sig) at (6.0,-1.4) {\textbf{Assinatura:}\\
    1) Assinatura WOTS sobre \\ \(m\) usando \(k_2\)\\
    2) Caminho de autenticação: \\ hashes (ex.: hash(l1), hash(n2), ...)\\
    3) (Opcional) parâmetros / nonce
  };
  \draw[->, dashed] (l2.east) .. controls +(12mm,6mm) and +(-10mm,10mm) .. (sig.west);

  \node[smalltext] (verify) at (0,-4.6) {Verificador: recebe (assinatura, caminho), verifica WOTS e recomputa raiz};
  \draw[->] (sig.west) .. controls +(-18mm,0) and +(18mm,8mm) .. (verify.east);

\end{tikzpicture}}
\caption{Esquema simplificado de uma assinatura baseada em hash com árvore de Merkle.}
\label{fig:hashsig_overview}
\end{figure}

Por fim, em termos de estratégia de adoção, as assinaturas baseadas em hash são ideais quando a longevidade da segurança e a simplicidade criptoanalítica são requisitos primordiais, por exemplo, certificados de longo prazo, registros legais e assinaturas sobre firmware crítico. Para usos onde tamanho da assinatura é crítico (protocolos com limitações de MTU ou cadeias de blocos sensíveis a payload), pode ser necessário avaliar trade-offs com esquemas mais compactos (lattices) ou empregar estratégias híbridas (por exemplo, assinatura post-quantum combinada com esquema clássico) que permitam compatibilidade gradual sem sacrificar a segurança de longo prazo.

\subsection{Isogenias}

Os esquemas baseados em isogenias exploram propriedades profundas da teoria das curvas elípticas e das aplicações (morphisms) entre elas, conhecidas como isogenias. Uma isogenia é um morfismo não constante entre curvas elípticas que preserva a estrutura do grupo; matematicamente, uma isogenia \(\phi: E \to E'\) é um homomorfismo de grupos sobre corpos finitos, com núcleo finito. A dificuldade computacional explorada por essas construções consiste justamente em recuperar uma isogenia secreta (ou o caminho de isogenias) que conecta duas curvas conhecidas, dado apenas o par \((E, E')\), tarefa que, sob certas hipóteses e parâmetros, parecia resistente a algoritmos clássicos e quânticos conhecidos.

O paradigma inicial de maior destaque foi o SIDH/SIKE, que utilizava curvas supersingulares e trocas de pontos de torsão auxiliares para construir um KEM com chaves muito compactas em relação a outras famílias. Em síntese operacional, os participantes trocam imagens de pontos de torsão sob isogenias secretas, de modo que o compartilhamento dessas imagens permite a derivação de uma curva ou de um valor comum sem revelar as isogenias privadas. A eficiência em termos de largura de banda e a elegância matemática dessas construções chamaram a atenção da comunidade e dos processos de padronização. No entanto, ataques práticos demonstraram que certas informações auxiliares, especificamente, dados relacionados a pontos de torsão usados no protocolo SIDH, podiam ser exploradas por um adversário para recuperar a isogenia secreta, levando à quebra prática de SIKE em 2022. Esse episódio ressaltou que propriedades subtis do esquema (por exemplo, a presença de pontos auxiliares e certas simetrias) podiam introduzir superfícies de ataque não previstas inicialmente.

Além do SIDH/SIKE, outras abordagens isogenia-baseadas foram propostas com estruturas distintas, tais como CSIDH, que fundamenta a segurança em uma ação de grupo comutativa sobre curvas supersingulares definidas sobre corpos primos (class group action viewpoint). A diferença arquitetural, ausência das mesmas informações auxiliares exploradas contra SIDH, torna CSIDH e suas variantes um campo ativo de pesquisa. Entretanto, a maturidade desses esquemas é menor quando comparada a families como lattices ou códigos: tanto a criptoanálise quanto as otimizações de implementação ainda estão em intensa evolução, e novas técnicas podem afetar as conjecturas de segurança. Adicionalmente, propostas de assinaturas baseadas em isogenias (por exemplo, SQISign e variantes experimentais) exploram transformações e construções matemáticas mais complexas, com trade-offs distintos em eficiência e segurança.

Do ponto de vista de implementação, os esquemas isogenia-baseados apresentam desafios técnicos relevantes. Cálculos eficientes sobre curvas supersingulares, manipulação de pontos de torsão, verificação de pontos recebidos e validade das curvas são operações sensíveis que exigem atenção a vulnerabilidades de implementação (validação incompleta pode facilitar ataques de curva inválida, por exemplo). Além disso, operações envolvendo coordenadas projectivas, rotação de bases e outras otimizações aritméticas devem ser feitas em tempo constante e com mitigação de canais laterais (tempo, consumo de energia, emissões) para reduzir o risco de vazamento de segredos. A compacticidade das chaves, um atrativo evidente, só é útil se as implementações forem robustas e as hipóteses de segurança mantidas; quando não há garantias suficientes, o pequeno tamanho das chaves pode ser irrelevante frente à perda de segurança.

Em termos de utilidade prática, isogenias oferecem potencial para cenários em que a restrição de largura de banda ou o custo de transmissão é crítico, por exemplo, enlaces de baixa taxa, radio-comunicações restritas ou aplicações embarcadas com forte limitação de payload. A representação curta de chaves e a possibilidade de construir primitives elegantes faz com que a família continue atraente para investigação académica e para protótipos experimentais. Ainda assim, a experiência com SIKE demonstra que a adoção precoce em ambientes de produção sem ampla revisão criptoanalítica e sem variantes robustas é arriscada.

Do ponto de vista da pesquisa e da padronização, a lição trazida pelas isogenias é tripla. Primeiro, pressupostos matemáticos sofisticados podem esconder fragilidades exploráveis por técnicas novas e inesperadas, portanto, a validade de conjecturas de dificuldade deve ser testada por longos períodos e por múltiplas equipes. Segundo, a presença de componentes auxiliares no protocolo (pontos públicos, auxiliares de verificação, representações compactas) pode ser tão crítica quanto a suposta dificuldade fundamental; análise do protocolo completo, não apenas do problema matemático isolado, é essencial. Terceiro, a diversidade continua sendo uma estratégia prudente: manter alternativas de outras famílias reduz risco sistêmico caso vulnerabilidades teóricas ou práticas emerjam.

A Figura \ref{fig:isogeny_overview} ilustra, de forma esquemática e didática, o conceito de isogenia entre curvas elípticas: duas curvas \(E\) e \(E'\) são conectadas por uma isogenia \(\phi: E \to E'\) cujo núcleo é um subgrupo finito \(K\subset E\). Em muitos protocolos, a ação secreta consiste em conhecer uma descrição eficiente dessa isogenia (por exemplo, um gerador do núcleo) enquanto um adversário vê apenas as curvas ou imagens públicas de pontos e precisa reconstruir \(\phi\) a partir dessas informações. A figura destaca também o papel do kernel como trapdoor estrut, e ilustra a ideia de caminhos de isogenia (sequências compostas de isogenias) que formam a base de muitos esquemas.

\begin{figure}[h!]
\centering
\begin{tikzpicture}[scale=0.85, every node/.style={font=\scriptsize, inner sep=0.8pt},
    baseline=(current bounding box.north)]
  \draw[rounded corners] (-4.2,-1.9) rectangle (4.2,1.9);

  \draw[thick] (-2,0) ellipse (1.6 and 1.2);
  \node at (-2,1.5) {\scriptsize $E$};

  \filldraw[black] (-2.45,0.55) circle (1.5pt) node[above left=1pt] {$Q$};
  \filldraw[black] (-1.15,0.45) circle (1.5pt) node[above right=1pt] {$P$};
  \filldraw[black] (-1.9,-0.35) circle (1.5pt) node[below left=1pt] {$-P$};

  \draw[thick] (2,0) ellipse (1.2 and 1.0);
  \node at (2,1.25) {\scriptsize $E'$};

  \filldraw[black] (1.6,0.5) circle (1.5pt) node[above right=1pt, text=blue] {$\phi(P)$};
  \filldraw[black] (2.15,-0.05) circle (1.5pt) node[below right=1pt, text=blue] {$\phi(-P)$};
  \filldraw[black] (1.9,-0.35) circle (1.5pt) node[below left=1pt, text=blue] {$\phi(Q)$};

  \draw[->, very thick] (-0.35,1.35) to[bend left=10] node[midway, above, fill=white, inner sep=1pt] {\(\phi\)} (0.35,1.35);


  \node[fill=white] at (0,-1.9) {\scriptsize Núcleo \(K=\langle P,\dots\rangle\) (subgrupo finito)};

\end{tikzpicture}
\caption{Visão esquemática de uma isogenia \(\phi: E \to E'\).}
\label{fig:isogeny_overview}
\end{figure}
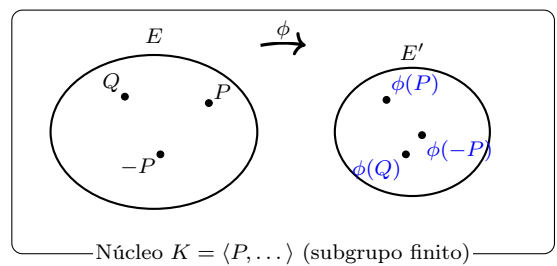

\subsection{Multivariados}

Algoritmos multivariados têm como pedra angular a dificuldade de resolver sistemas de equações polinomiais quadráticas sobre corpos finitos, o problema MQ (Multivariate Quadratic). A construção típica parte de uma função pública \(P: \mathbb{F}_q^n \to \mathbb{F}_q^m\) composta por \(m\) polinômios quadráticos em \(n\) variáveis. A dificuldade geral de encontrar uma pré-imagem \(x\) tal que \(P(x)=y\) para um dado \(y\) faz dessa família uma fonte natural de construções para assinaturas digitais e, em alguns casos, intercâmbio de chaves. Para tornar a função invertível para o titular da chave privada, os esquemas multivariados geralmente empregam um \emph{trapdoor} baseado em uma estrutura algébrica escondida: o construtor define um mapa central \(F\) (com estrutura interna que torna a inversão fácil), e compõe \(F\) com transformações lineares ou afins secretas \(S\) e \(T\) de modo que a função pública seja \(P = S \circ F \circ T\). O segredo é exatamente a decomposição \(S, F, T\), que permite ao titular da chave privada inverter \(P\) resolvendo primeiro \(S^{-1}(y)\), então invertendo \(F\) de forma eficiente e finalmente aplicando \(T^{-1}\) para recuperar a pré-imagem original. Um adversário que disponha apenas de \(P\) enfrenta, em princípio, o problema MQ sem trapdoor, tipicamente considerado intratável para parâmetros bem escolhidos.

Essa família inclui uma diversidade de propostas e variações do conceito, cada uma com escolhas específicas para o mapa central \(F\) e para as transformações de camuflagem. Exemplos históricos e recentes incluem Rainbow, GeMSS, UOV (Unbalanced Oil and Vinegar) e HFE (Hidden Field Equations); algumas dessas propostas chegaram a fases avançadas de avaliação, mas também sofreram quebras ou ataques refinados que mostram a sensibilidade ao ajuste paramétrico e ao desenho do trapdoor. Em Rainbow, por exemplo, a construção multicamada prometia desempenho atraente em verificações rápidas, mas vulnerabilidades paramétricas e melhorias em técnicas algébricas conduziram à quebra prática de variantes usadas em candidaturas. GeMSS e UOV exploram outras formas de estruturar as equações e o trapdoor, buscando um equilíbrio entre eficiência de assinatura, tamanho de chave e resistência a ataques algebraicos.

Do ponto de vista de desempenho, os esquemas multivariados são frequentemente atrativos por possuírem verificações de assinatura extremamente rápidas e, em muitos casos, latências de verificação muito baixas, o que os torna candidatos naturais para ambientes com alto volume de validações (por exemplo, servidores de autenticação em massa ou certos contextos embarcados). Por outro lado, o tamanho das chaves públicas e privadas e os custos de assinatura podem variar consideravelmente segundo a construção; alguns esquemas buscam compactar chaves, outros aceitam chaves maiores em troca de assinaturas muito rápidas.

A segurança prática depende criticamente da escolha de parâmetros e das estruturas algébricas ocultas: técnicas de análise algébrica (como ataques XL, F4/F5 e variantes de relinearização) e heurísticas de redução podem explorar regularidades implícitas na construção do mapa central ou em padrões introduzidos pelas transformações \(S\) e \(T\). Assim, enquanto a família oferece potencial para soluções muito eficientes em verificação, ela exige uma avaliação criptoanalítica rigorosa e contínua, inclusive por múltiplas equipes independentes, antes de adoção em ambientes críticos. Implementações devem também considerar contramedidas clássicas a canais laterais e garantir a correta geração de parâmetros aleatórios, visto que entropia insuficiente ou padrões em geração de coeficientes podem reduzir drasticamente a resistência do esquema.

Na prática de engenharia, recomenda-se aos projetistas que considerem a multivariada como parte de um portfólio diversificado: usar múltiplas famílias com fundamentos distintos reduz o risco sistêmico caso uma construção seja comprometida. Adicionalmente, onde se pretende empregar esquemas multivariados em produção, executar testes de interoperabilidade, pilotagens e auditorias de criptoanálise paramétrica é mandatório. Em cenários com restrições severas de latência de verificação, os multivariados podem ser uma escolha atraente, mas sempre acompanhados de estratégias de mitigação (por exemplo, assinaturas em paralelo com esquemas de outra família ou uso em funções específicas de curta validade).

A Figura \ref{fig:multivariate_overview} apresenta um diagrama esquemático do princípio estrutural típico de muitos esquemas multivariados: o mapa público \(P\) é mostrado como a composição \(P = S \circ F \circ T\); a chave privada é o triplo \((S, F, T)\). Para assinar uma mensagem, o signatário inverte \(P\) aplicando \(S^{-1}\), invertendo \(F\) (usando o trapdoor) e aplicando \(T^{-1}\); o verificador confirma a assinatura ao checar que \(P(x)=y\). A figura ajuda a visualizar por que a segurança repousa tanto na dificuldade de inverter o sistema quadrático público quanto na ocultação da estrutura interna \(F\).

\begin{figure}[h!]
\centering
\resizebox{0.95\linewidth}{!}{%
\begin{tikzpicture}[font=\footnotesize, >=Stealth]
  \node[draw, rounded corners, minimum width=32mm, minimum height=10mm, align=center] (sign) at (-6.0, -2.5) {Signatário};

  \node[draw, rounded corners, minimum width=26mm, minimum height=12mm, align=center] (T) at (-6.0, 0) {Transformação\\secreta $T$};
  \node[draw, rounded corners, minimum width=36mm, minimum height=12mm, align=center] (F) at ( 0.0, 0) {Mapa central\\$F$ (trapdoor)};
  \node[draw, rounded corners, minimum width=26mm, minimum height=12mm, align=center] (S) at ( 6.0, 0) {Transformação\\secreta $S$};
  \coordinate (Pout) at (10.5,0);

  \begin{scope}[on background layer]
    \node[draw, rounded corners, inner sep=12mm, fit=(T) (F) (S)] (privbox) {};
  \end{scope}

  \draw[->, semithick] (T.east) -- (F.west) node[midway, above, fill=white, inner sep=1pt] {$x$};
  \draw[->, semithick] (F.east) -- (S.west) node[midway, above, fill=white, inner sep=1pt] {$F(T(x))$};
  \draw[->, semithick] (S.east) -- (Pout)   node[midway, above, fill=white, inner sep=1pt] {$P(x)$};

  \node at (0,2.6) {Pública: $\displaystyle P = S\circ F\circ T$};

  \draw[dashed,->] ($(F.east)+(0,18pt)$) -- ($(S.west)+(0,18pt)$);
  \node[font=\scriptsize, fill=white] at ($(F.east)!0.5!(S.west) + (0,24pt)$) {$S^{-1}$};

  \draw[dashed,->] ($(T.east)+(0,18pt)$) -- ($(F.west)+(0,18pt)$);
  \node[font=\scriptsize, fill=white] at ($(T.east)!0.5!(F.west) + (0,24pt)$) {$F^{-1}$};

  \draw[dashed,->] ($(T.south)$) -- ($(sign.north)$);
  \node[font=\scriptsize, fill=white] at ($(T.west)!0.5!(sign.east) + (0,-24pt)$) {$T^{-1}$};

  \node[font=\scriptsize] at (0,-1.25) {inverter $F$ usando o trapdoor (etapa central)};
  \node[anchor=west, align=left, font=\scriptsize] at (10.6,0) {\textbf{Chave privada:}\\ $(S,F,T)$ \\ (permite inverter $P$)};
  \node at (0,-2.5) {Verificador: checa $P(x)=y$};
\end{tikzpicture}}
\caption{Diagrama simplificado do princípio estrutural de esquemas multivariados: $P=S\circ F\circ T$.}
\label{fig:multivariate_overview}
\end{figure}
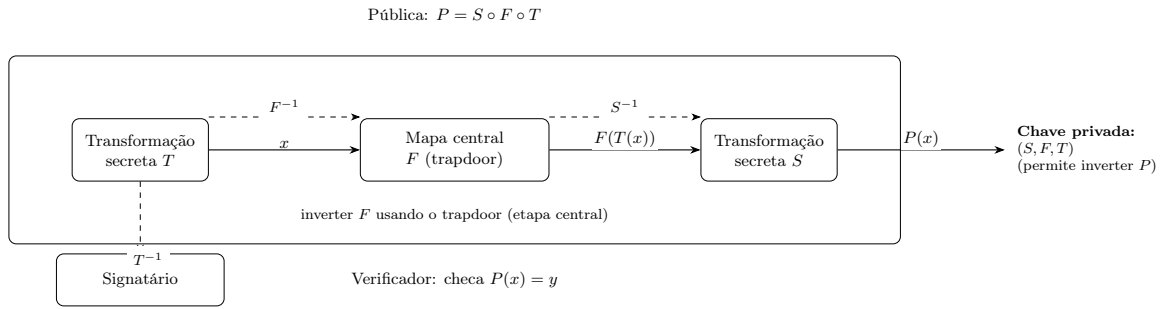

\subsection{Primitivas Simétricas e Funções de Hash}

Embora a ênfase da criptografia pós-quântica recaia sobre esquemas assimétricos, os componentes simétricos e as funções de hash permanecem críticos na arquitetura de segurança: KDFs, PRFs, MACs, funções de derivação em TLS/SSH, e primitives de cifra autenticada (AEAD) continuam a compor a espinha dorsal operacional. A principal diferença introduzida pelo modelo quântico é a existência de algoritmos genéricos (em particular, o algoritmo de Grover) que aceleram buscas em espaço não estruturado, reduzindo a complexidade exaustiva de \(2^n\) para aproximadamente \(2^{n/2}\). Em termos práticos isso significa que parâmetros clássicos que ofereciam \(n\) bits de segurança passam a ter cerca de \(n/2\) bits de segurança efetiva contra adversários com computadores quânticos que possam executar Grover em escala relevante. Portanto, a contramedida direta e robusta é aumentar a entropia e o tamanho de chaves/saídas de hash de forma que a segurança efetiva permaneça adequada ao horizonte temporal exigido (por exemplo, dobrar o comprimento da chave: AES-256 em vez de AES-128).

No caso de funções de hash e construções baseadas em pré-imagem (como HMAC e KDFs), Grover melhora a busca de pré-imagem em fator quadrático, implicando que saídas de hash de comprimento \(m\) passam a oferecer aproximadamente \(m/2\) bits de resistência a pré-imagem. Para colisões, embora existam algoritmos quânticos com vantagens (por exemplo, variantes quânticas do algoritmo de colisão), a consequência prática usual é que também se deve privilegiar funções com saídas maiores e análise criptoanalítica madura. Em consequência, recomenda-se o emprego de funções com saídas de 512 bits ou XOFs configuráveis (SHAKE, SHA3-512, BLAKE3 em modos apropriados) quando o horizonte de proteção for muito longo.

As implicações para primitivas concretas são as seguintes: para cifras simétricas, AES-256 oferece uma margem confortável frente a Grover (efeito prático: segurança ~128 bits), enquanto AES-128 teria segurança reduzida para ~64 bits, considerado insuficiente para muitos ativos críticos. Para stream ciphers e alternativas (por exemplo ChaCha20), o uso de chaves de 256 bits é preferível. Em protocolos AEAD, além do tamanho da chave, deve-se considerar o tamanho da tag de autenticação: ataques de força bruta a tags seguem a mesma diminuição quadrática; portanto, tags de 128 bits continuam relevantes como mínimo em muitos cenários, mas, para ativos com exigência de longevidade ou alto valor, avaliar tags maiores e estratégias adicionais de autenticação pode ser aconselhável.

As funções derivadoras de chaves (KDF/HKDF), PRFs e constructions usadas em protocolos (TLS, SSH, IPsec) devem produzir saídas e entropia suficientes para alimentar esquemas pós-quânticos e para resistir a ataques quânticos genéricos. Recomendações práticas incluem: usar HKDF com uma função hash de saída longa (por exemplo SHA3-512), derivar material com entropia mínima correspondente ao nível de segurança desejado (por exemplo, 256 bits de entropia para chaves de 256 bits), e evitar truncamentos agressivos de saídas quando o horizonte de proteção é longo.

A qualidade e disponibilidade de entropia tornam-se ainda mais críticas: gerar sementes de 256 bits de entropia de alta qualidade exige TRNGs validados e mecanismos robustos de coleta de entropia; RNGs determinísticos (DRBGs) devem ser inicializados com entropia suficiente e suportar reseeding seguro. HSMs/TPMs e bibliotecas criptográficas corporativas precisarão ser atualizados para suportar chaves maiores, funções hash de saída estendida e novas opções de KDF, além de fornecer contramedidas a canais laterais.

Do ponto de vista operacional e de compatibilidade, existem impactos mensuráveis: chaves maiores e saídas de hash aumentam requisitos de armazenamento, uso de banda e, em alguns casos, latência de CPU (especialmente em dispositivos com recursos limitados). Em muitos contextos, esses custos são aceitáveis frente ao benefício da segurança; em outros (dispositivos IoT com memória e energia restritas), são necessários trade-offs, por exemplo, empregar aceleração por hardware, adotar esquemas híbridos ou priorizar proteção para ativos com maior horizonte de confidencialidade. Além disso, o desenho de protocolos deve preservar propriedades como forward secrecy; isso implica combinar KEMs/assinaturas pós-quânticas com mecanismos efêmeros apropriados, ou usar KDFs que garantam derivação segura de material efêmero.

A Figura \ref{fig:grover_effect} sintetiza de modo esquemático o efeito de Grover sobre a segurança de chaves simétricas: enquanto um adversário clássico precisaria de \(\mathcal{O}(2^n)\) operações para quebrar uma chave de \(n\) bits, um adversário quântico com Grover reduz essa complexidade para \(\mathcal{O}(2^{n/2})\), implicando a necessidade prática de dobrar \(n\) para manter o mesmo nível de segurança efetiva. A complexidade de busca em espaço de chaves é reduzida de \(2^n\) para aproximadamente \(2^{n/2}\), o que justifica o uso de chaves maiores (por exemplo, migrar de AES-128 para AES-256).

\begin{figure}[h!]
\centering
\resizebox{0.88\linewidth}{!}{%
\begin{tikzpicture}[font=\small, >=Stealth]
  \node[draw, rounded corners, minimum width=40mm, minimum height=14mm, align=center] (classical) at (-3.5,1.6)
    {Adversário clássico\\Complexidade $\sim 2^{n}$};
  \node[draw, rounded corners, minimum width=40mm, minimum height=14mm, align=center] (quantum)  at ( 3.5,1.6)
    {Adversário quântico\\(Grover)\\Complexidade $\sim 2^{n/2}$};

  \begin{scope}[on background layer]
    \node[draw, rounded corners, inner sep=10mm, fit=(classical) (quantum)] (box) {};
  \end{scope}

  \draw[->, thick] (classical.east) -- node[midway, above, fill=white] {referência} (quantum.west);

  \node[align=center, minimum width=42mm] (ex1) at (-3.5,-1.2) {AES-128\\(clássico: $2^{128}$)};
  \node[align=center, minimum width=42mm] (ex2) at ( 3.5,-1.2) {AES-128\\(quântico: $\sim 2^{64}$)};
  \node[align=center, minimum width=42mm] (ex3) at (-3.5,-2.6) {AES-256\\(clássico: $2^{256}$)};
  \node[align=center, minimum width=42mm] (ex4) at ( 3.5,-2.6) {AES-256\\(quântico: $\sim 2^{128}$)};

  \draw[dashed,->,shorten >=4pt,shorten <=4pt] (ex1.east) -- (ex2.west);
  \draw[dashed,->,shorten >=4pt,shorten <=4pt] (ex3.east) -- (ex4.west);

  \node[font=\small] at (0,-3.6) {Efeito qualitativo de Grover: a complexidade de busca em espaço de chaves reduz de $2^{n}$ para aproximadamente $2^{n/2}$.};
\end{tikzpicture}}
\caption{Efeito qualitativo de Grover sobre a segurança simétrica.}
\label{fig:grover_effect}
\end{figure}

Em suma, embora as primitivas simétricas não sejam “quebradas” pelos algoritmos quânticos de maneira análoga aos esquemas baseados em fatoração ou discrete log, elas exigem ajustes de parâmetros, rigor na geração de entropia, atualizações de infraestrutura e políticas operacionais para manter níveis de segurança compatíveis com o risco quântico projetado. A combinação de boas escolhas de primitives simétricas (chaves maiores e hashes longas), práticas de engenharia (HSM/TPM atualizados, mitigação de canais laterais) e políticas organizacionais (crypto-agility, rotação de chaves, inventário de ativos) constitui a resposta prática mais robusta.

\section{Padronização (NIST PQC)}

Diante da urgência de proteger sistemas contra futuros ataques com computadores quânticos, o NIST (National Institute of Standards and Technology) iniciou em 2016 um processo internacional de padronização em criptografia pós-quântica (PQC). O objetivo foi identificar algoritmos capazes de substituir ou complementar os sistemas clássicos de chave pública, como RSA e ECC, garantindo resiliência frente a computadores quânticos universais. O processo envolveu três rodadas principais de avaliação e, atualmente, encontra-se em uma quarta fase voltada para a análise de algoritmos alternativos.

Na fase inicial, o NIST recebeu 69 propostas de algoritmos, representando diversas famílias matemáticas. Após análises teóricas e práticas, esse número foi reduzido para 26 candidatos na segunda rodada e, posteriormente, para sete finalistas e oito alternativos na terceira rodada. Esses resultados refletem não apenas avanços em segurança teórica, mas também extensos estudos sobre eficiência prática, consumo energético, tamanho de chaves e resistência a ataques de implementação. A avaliação combinou análise criptoanalítica (incluindo modelos que consideram capacidades quânticas), benchmarks de desempenho em hardware real, testes de interoperabilidade e revisões por pares, de modo a priorizar soluções com equilíbrio entre segurança e viabilidade operacional.

A Figura \ref{fig:nist_timeline} apresenta a linha do tempo oficial do processo de padronização do NIST PQC, indicando marcos importantes: o anúncio dos candidatos nas diferentes rodadas, a publicação dos padrões preliminares e a estimativa de especialistas quanto à viabilidade de ataques quânticos contra RSA-2048. Este panorama ilustra a relação entre a evolução dos padrões e a crescente preocupação com a ameaça prática dos computadores quânticos.

\begin{figure}[H]
    \centering
    \includegraphics[width=0.8\textwidth]{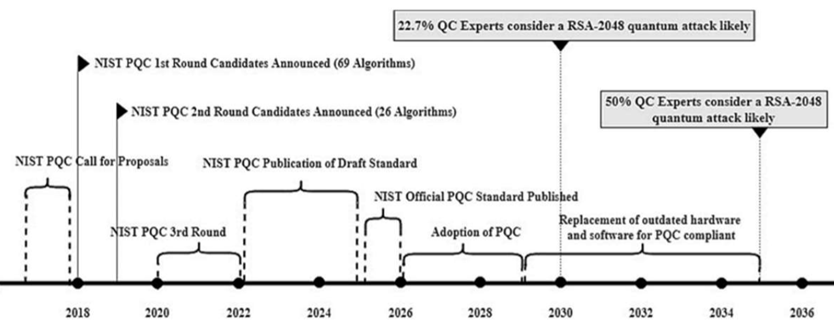}
    \caption{Linha do tempo do processo de padronização NIST PQC.}
    \label{fig:nist_timeline}
\end{figure}

Após três rodadas, em 2022 o NIST anunciou os primeiros algoritmos selecionados para padronização:
\begin{itemize}
    \item \textbf{CRYSTALS-Kyber}, KEM (Key Encapsulation Mechanism) baseado em reticulados (lattices), indicado para troca de chaves e encapsulamento;  
    \item \textbf{CRYSTALS-Dilithium} e \textbf{FALCON}, esquemas de assinatura digital baseados em lattices, escolhidos por seu compromisso entre segurança e desempenho;  
    \item \textbf{SPHINCS+}, esquema de assinatura baseado em hash, que adiciona diversidade matemática ao conjunto padronizado e possui propriedades conservadoras de segurança.
\end{itemize}

A escolha evidencia a forte confiança da comunidade científica nos algoritmos baseados em lattices, considerados os melhores no balanço entre robustez e desempenho para muitos cenários práticos. A inclusão do SPHINCS+ reforça a estratégia de diversidade algorítmica, reduzindo o risco sistêmico caso uma família matemática venha a ser comprometida no futuro. Além dessa seleção inicial, o NIST abriu uma fase complementar para continuar avaliando alternativas que podem se tornar padrões suplementares; entre os candidatos notáveis estão o \textbf{Classic McEliece} (códigos corretores de erro) e esquemas como \textbf{BIKE} e \textbf{HQC}. Apesar de limitações operacionais (por exemplo, tamanhos de chaves maiores), essas famílias mantêm interesse pelo histórico de resistência ou por adicionar diversidade matemática ao ecossistema.

A diversidade de propostas submetidas ao processo NIST PQC é ilustrada na Figura \ref{fig:pqc_families}, que organiza graficamente as principais famílias de algoritmos considerados. O diagrama destaca desde os esquemas mais tradicionais, como McEliece e variantes hash-based, até linhas mais experimentais como multivariados e isogenias, reforçando a complexidade do balanço entre segurança teórica, viabilidade prática e compatibilidade com a infraestrutura global de internet.

\begin{figure}[H]
    \centering
    \includegraphics[width=0.95\textwidth]{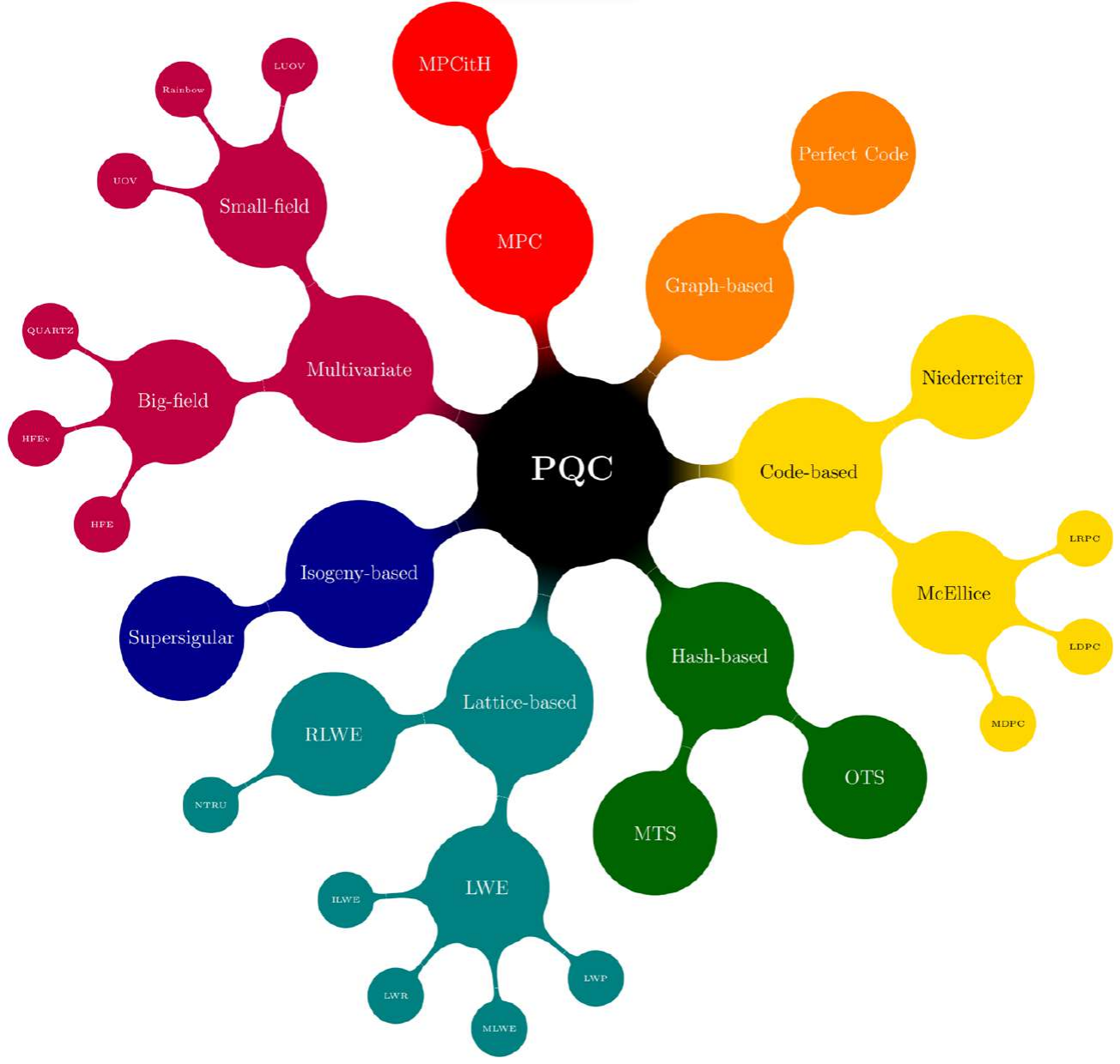}
    \caption{Principais famílias de algoritmos de criptografia pós-quântica considerados no processo do NIST PQC.}
    \label{fig:pqc_families}
\end{figure}

Do ponto de vista prático, a padronização traz implicações operacionais relevantes. Em primeiro lugar, as primitivas padronizadas cobrem funções fundamentais, encapsulamento de chave (KEM) e assinaturas, porém a adoção em protocolos (TLS, SSH, PKI, formatos de certificados) e em infraestruturas exige trabalho adicional: perfis de uso, interoperabilidade, testes de integração e guias de migração. Em segundo lugar, propriedades não criptográficas como latência, uso de memória, largura de banda (tamanhos de chaves/assinaturas) e consumo energético variam significativamente entre famílias; logo, avaliações de desempenho no hardware alvo (servidores, dispositivos móveis, HSMs, dispositivos embarcados) são imprescindíveis antes da implantação em produção.

Recomendações práticas recorrentes para equipes de segurança e engenharia incluem:
- adotar estratégias híbridas durante a transição (por exemplo, combinar um KEM pós-quântico padronizado com um KEM clássico em protocolos TLS) para reduzir riscos de quebra indevida;  
- priorizar a proteção de dados com janela de sensibilidade longa, adotando medidas de “store-now, decrypt-later” para informação que precisa permanecer confidencial por décadas;  
- realizar testes de desempenho e segurança em ambientes reais, incluindo análises de resistência a ataques por canal lateral e validações de implementações constant-time;  
- planejar cronogramas de atualização coordenados para bibliotecas, clientes, servidores, HSMs e dispositivos embarcados, porque migrações em ambientes críticos podem demandar anos;  
- participar de iniciativas de interoperabilidade (test vectors, implementações de referência e testes cross-vendor) para reduzir riscos de integração.

Além das recomendações operacionais, é importante destacar que a padronização não encerra a pesquisa: a publicação de padrões formais é um ponto de partida que tende a ser seguido por refinamentos de parâmetros, otimizações de implementação e eventuais reavaliações decorrentes de novos avanços criptoanalíticos. A comunidade de pesquisa e indústria deve, portanto, manter monitoramento contínuo das evidências técnicas e práticas, atualizando perfis de segurança e procedimentos de implantação conforme necessário.

Em resumo, o processo NIST PQC consolidou um conjunto inicial de primitivas que viabilizam o início da transição para infraestruturas resistentes a ataques quânticos, ao mesmo tempo em que preservou alternativas que garantem diversidade matemática. Para engenheiros e gestores de segurança a recomendação é: mapear ativos e janelas de sensibilidade, iniciar testes de interoperabilidade com as primitivas padronizadas, adotar modos híbridos quando possível e planejar cronogramas de atualização coordenados para reduzir riscos durante a transição tecnológica.

\section{Estado da Arte}

A literatura atual em criptografia pós-quântica não se restringe apenas ao desenvolvimento teórico de novos algoritmos, mas cresce em direção à sua validação experimental, otimização em hardware/software e integração em sistemas reais. Estudos recentes focam em desafios como consumo energético em dispositivos embarcados, ataques de canal lateral, tamanhos de chaves, além da compatibilidade com protocolos da internet (TLS, VPNs, blockchain).  

A Tabela \ref{tab:estado_arte_pqc} apresenta uma síntese de alguns trabalhos relevantes da literatura, cobrindo análises experimentais, propostas de implementação e resultados de padronização, que refletem o avanço da área nos últimos cinco anos.

\begin{table}[h!]
\centering
\scriptsize
\caption{Estado da arte em criptografia pós-quântica, com contribuições recentes e suas limitações.}
\label{tab:estado_arte_pqc}
\begin{tabular}{p{3cm}p{2cm}p{6cm}p{3cm}}
\hline
\textbf{Autor/Ano} & \textbf{Algoritmo} & \textbf{Contribuição} & \textbf{Limitações} \\
\hline

Basu et al. (2019) \cite{Basu2019} & Dilithium & Análise de assinaturas baseadas em lattices para IoT; estudo prático de desempenho. & Tamanho das chaves elevado. \\

Alkim et al. (2021) \cite{Alkim2021} & Kyber & Avaliação de Kyber em dispositivos embarcados de baixo consumo. & Vulnerabilidade a ataques de canal lateral. \\

Bernstein et al. (2021) \cite{Bernstein2021} & Classic McEliece & Revisão crítica de variantes modernas do esquema e aplicações em protocolos de rede. & Chaves públicas muito grandes, dificultando adoção em aplicações móveis. \\

NIST (2022) \cite{NIST2022} & Kyber, Dilithium, SPHINCS+ & Seleção oficial dos algoritmos vencedores no processo de padronização. & Adoção ainda em fase inicial. \\

Basso et al. (2023) \cite{Basso2023} & FALCON & Benchmarking de assinaturas em servidores de alta performance; eficiência comprovada em TLS. & Implementação complexa, sensibilidade a erros de arredondamento. \\

Yamakawa et al. (2023) \cite{Yamakawa2023} & BIKE e HQC & Avaliação de esquemas baseados em códigos contra ataques avançados. & Desempenho inferior a designs lattice-based. \\

Xagawa et al. (2024) \cite{Xagawa2024} & Diversos (Lattices e Hash) & Discussão sobre transição global para PQC e protocolos híbridos em TLS e 5G. & Desafios de padronização e migração lenta. \\

Wang et al. (2024) \cite{Wang2024} & Kyber e Dilithium & Estudo de aplicação de PQC em blockchains; análise de impacto em custos de transação. & Taxas e tempos de transação aumentados. \\

Zhang et al. (2025) \cite{Zhang2025} & Kyber e Saber & Análise comparativa de esquemas KEM lattice-based em hardware reconfigurável voltado a data centers. & Trade-off entre área, consumo de energia e latência ainda desfavorável para alguns perfis de aplicação. \\

Rodrigues et al. (2025) \cite{Rodrigues2025} & SPHINCS+ e FALCON & Estudo de implantação híbrida de assinaturas pós-quânticas em infraestruturas de PKI de larga escala. & Complexidade operacional e necessidade de atualização massiva de certificados. \\

\hline
\end{tabular}
\end{table}

Como sintetizado na Tabela \ref{tab:estado_arte_pqc}, a comunidade científica avalia não apenas a segurança teórica dos algoritmos, mas também sua aplicabilidade em cenários reais. No estudo de {Basu et al. (2019)}, o foco foi o esquema {Dilithium}, analisado em dispositivos de Internet das Coisas. O trabalho destaca que, embora as assinaturas baseadas em lattices sejam eficientes em termos de segurança, o tamanho das chaves pode comprometer o desempenho em dispositivos limitados, o que exige otimizações específicas de memória e energia.   Em {Avanzi et al. (2020)}, o algoritmo {SPHINCS+} foi estudado sob a ótica de implementação eficiente em software. O artigo confirma a robustez matemática desse esquema baseado em hash, mas evidencia que o tamanho de suas assinaturas ainda representa um gargalo, especialmente em ambientes com restrições de largura de banda, como redes móveis e sensores.  

Já {Alkim et al. (2021)} investigaram o algoritmo {Kyber} em dispositivos embarcados. O trabalho mostra que, mesmo em ambientes de baixo consumo, Kyber mantém desempenho satisfatório, desde que sejam aplicadas técnicas de mitigação contra \textit{side-channel attacks}, garantindo sua viabilidade prática em IoT e dispositivos móveis.   O estudo de {Bernstein et al. (2021)} atualizou a literatura sobre o {Classic McEliece}. Trata-se de uma linha de pesquisa histórica, extremamente resistente a avanços quânticos, mas que apresenta a desvantagem de chaves públicas gigantescas (na ordem de centenas de KB até MB), dificultando a adoção prática em protocolos de internet e sistemas embarcados.  

Em 2022, o {NIST} marcou um divisor de águas ao anunciar a seleção oficial dos primeiros algoritmos de PQC, {Kyber, Dilithium e SPHINCS+}. Esse marco institucionalizou a PQC como caminho oficial da transição para segurança quântica, embora a adoção prática ainda esteja em estágios iniciais. O trabalho de {Basso et al. (2023)} ofereceu uma visão crítica sobre o {FALCON}, um dos esquemas aprovados pelo NIST. O estudo demonstra suas vantagens em eficiência para servidores de alto desempenho, como em TLS, mas alerta para sua complexidade de implementação e vulnerabilidade a erros de arredondamento numérico, que podem afetar a segurança.  

{Chen et al. (2023)} trouxeram um resultado disruptivo ao demonstrar ataques práticos contra o algoritmo \gls{multivariado} {Rainbow}, até então considerado promissor. O artigo revelou que esse esquema, na prática, não oferece a segurança esperada, levando à sua exclusão do processo de padronização. {Yamakawa et al. (2023)} analisaram em profundidade esquemas {baseados em códigos}, como BIKE e HQC. Apesar de apresentarem altas garantias teóricas contra ataques quânticos, os testes revelaram desempenho inferior aos algoritmos lattice-based, principalmente em operações de chave pública e nas rotinas de encapsulamento.  

Já {Xagawa et al. (2024)} abordaram os desafios da {transição global para PQC}, avaliando a integração com TLS e redes móveis 5G. A contribuição central está no levantamento de barreiras práticas de migração, como compatibilidade com hardware legado e a necessidade de modelos híbridos para um período de transição seguro. Por fim, o estudo de {Wang et al. (2024)} investigou a aplicação dos algoritmos {Kyber e Dilithium} em {blockchains}, avaliando impacto direto em taxas de transação, custos computacionais e armazenamento. Embora os resultados mostrem a viabilidade da PQC nesse contexto, há aumento notável nas taxas de transação e nos tempos de verificação, o que pode restringir aplicações em sistemas de alto volume.  

Em síntese, a literatura recente demonstra que a segurança pós-quântica já está próxima da adoção prática, mas ainda enfrenta colisões entre {robustez matemática} e {viabilidade de implementação}. Assim, a pesquisa contemporânea oscila entre a validação experimental e a busca constante por equilíbrio entre eficiência, robustez e compatibilidade com sistemas existentes.

\section{Questões para Reflexão e Pesquisa Futura}

Ao finalizar o capítulo sobre criptografia pós-quântica (PQC), o leitor deve ser levado a questionar não apenas os resultados teóricos e as listagens de algoritmos, mas sobretudo as implicações práticas, os trade-offs operacionais e as fragilidades que emergem ao colocar essas soluções em produção. Comece interrogando as suposições de segurança: quais modelos formais (redução à dificuldade de um problema de lattice, código, hash, etc.) sustentam cada esquema e que lacunas existem entre essas provas teóricas e o comportamento de implementações reais? Pergunte-se, por exemplo, como mudanças de encoding, compressão de chaves ou tratamentos de erro podem invalidar premissas de prova ou introduzir vetores de oráculo; que invariantes de segurança (domain separation, KDF labels, não-reutilização de nonce) devem ser explicitamente documentadas e verificadas em todo o ciclo de desenvolvimento?

Reflita sobre seleção de parâmetros e sobre a durabilidade das escolhas: para quais horizontes temporais de proteção (5, 10, 30 anos) os parâmetros usados hoje são adequados, e como políticas de atualização/rotação devem ser definidas para ativos com diferentes exigências de confidencialidade? Indague também sobre o custo prático dessa durabilidade: tamanhos de chaves e assinaturas, impacto em MTU/fragmentação, e aumento de latência no estabelecimento de sessão (especialmente em cenários de conexões curtas massivas, como IoT). Quais métricas concretas (latência p50/p95/p99 do handshake, custo CPU por estabelecimento, uso de memória, número de bytes extra por pacote, taxa de falha sob perda de pacotes) devem orientar decisões de adoção em vez de confiar apenas em números teóricos?

Analise os desafios de implementação: até que ponto as otimizações para desempenho (vetorização, assembly hand-tuned, uso de aceleradores) ou ajustes para reduzir footprint em MCUs (parâmetros menores, compressão) afetam a resiliência a ataques por canais laterais? Que metodologias e suites de teste automatizadas (timing, cache, EM, fault injections) devem ser integradas aos pipelines CI/CD para detectar regressões de segurança quando se aplicam otimizações? Pergunte-se também como garantir que HSMs e TPMs que passaram a suportar PQC o façam de forma auditável e com cadeia de confiança minimamente exposta, quais evidências e provas técnicas (attestation, logs imutáveis) são necessárias para confiar nesses módulos?

Coloque em dúvida as estratégias de migração e interoperabilidade: como projetar algIDs, perfis X.509 e mecanismos de fallback sem introduzir vetores de downgrade inseguros? Em implementações híbridas (clássico + PQC), quais regras de composição e domain separation devem reger a derivação de chaves (por exemplo, combinação de outputs de dois KEMs em HKDF) para evitar combinações inseguras? Que procedimentos de rollout (PoC, canary por topologia, telemetria específica, rollback automático) são realmente eficazes para minimizar impacto no serviço e riscos de configuração? Além disso, que critérios objetivos usar para decidir se um determinado serviço deve migrar primeiro (por criticidade, horizonte de proteção, impacto regulatório)?

Questione o ecossistema de bibliotecas, padronização e conformance: bibliotecas de referência (liboqs, PQClean, implementações OpenSSL-OQS, RustCrypto) e padronizações (NIST, IETF) evoluem rapidamente, como manter uma cadeia de produção que acompanhe mudanças sem incorrer em risco de regressão? Que infraestrutura de testes cross-vendor (testbeds reproducíveis, vetores de teste, suites de interoperabilidade) é necessária para validar perfis e certificar conformidade antes do rollout em produção? Considere também a cadeia de suprimentos: como auditar fornecedores de firmware/ASIC/FPGA para garantir que aceleradores PQC e componentes de boot não adicionem backdoors ou fragilidades?

Explore modelos de ameaça ampliados: vá além de modelos onde o adversário apenas possui um computador quântico no futuro. Modele adversários compostos que combinam capacidades quânticas com ataques por canais laterais, corrupção de chaves por insiders, captura passiva de tráfego massivo (harvest now, decrypt later) e ataques de configuração. Como esses modelos alteram prioridades de mitigação (por exemplo, privilégios mínimos, segregação de funções, retenção de logs, rotação agressiva de chaves)? Que experimentos e simulações podem quantificar a eficácia de contramedidas frente a adversários híbridos?

Incentive experimentos reproduzíveis e mensuráveis: proponha PoCs que comparem stacks (ex.: OpenSSL-OQS vs rustls + pqm4) em cenários reais, servidores web e clientes móveis, gateways IoT, HSMs, medindo não só throughput e latência, mas também percentis de latência, impacto em sessão curta versus longa, custos energéticos em MCUs e comportamento sob perda e reconexão. Realize testes de interoperabilidade cross-vendor, benchamarks de assinaturas (latência/assinatura/verificação), KEMs (encaps/decaps) e impactos em TLS (size blowup, fragmentação, handshake time). Documente scripts, parâmetros e datasets para permitir reprodução por terceiros.

Finalmente, inclua questões de governança, regulação e ética: que requisitos legais podem forçar a migração (ex.: normas setoriais que exigem proteção contra “harvest now, decrypt later”)? Como equilibrar transparência técnica com proteção comercial em avaliações de segurança e auditorias? Que políticas de disclosure coordenado são apropriadas quando vulnerabilidades significativas em um algoritmo PQC forem descobertas? E, por fim, que linhas de pesquisa aplicadas, compressão de chaves/assinaturas, proofs-of-concept de key-management escalável para PQC, frameworks automáticos de rollout/rollback, e provas composicionais práticas incorporando encodings reais, têm maior potencial de reduzir a distância entre a pesquisa acadêmica e operações industriais seguras?

Estas questões devem ser tratadas como pontos de partida para estudos, PoCs e projetos colaborativos entre equipes de segurança, engenharia, operações e órgãos reguladores. Recomenda-se que o leitor transforme cada pergunta em um experimento prático documentado (hipótese, metodologia, métricas observadas, interpretação) para gerar evidência operável que subsidie decisões técnicas e políticas.

\chapter{Criptografia Híbrida e Emergente}

A transição entre os sistemas criptográficos clássicos, amplamente utilizados nas últimas décadas, e os novos paradigmas da era pós-quântica não deve ocorrer de forma abrupta. A necessidade de proteger dados em ambientes complexos, que incluem desde servidores em nuvem até dispositivos embarcados e redes críticas, levou ao surgimento da abordagem conhecida como {criptografia híbrida}. Este conceito busca combinar diferentes mecanismos de segurança, de modo que a quebra de um deles não comprometa a confidencialidade ou integridade da comunicação.  

O termo \textit{emergente}, por sua vez, refere-se às arquiteturas mais recentes que vão além da hibridização clássica + pós-quântica, integrando também técnicas de distribuição quântica de chaves (QKD) e soluções multicamadas para setores estratégicos como telecomunicações, finanças, defesa e blockchain.  

\section{Conceito e Aplicações}

A criptografia híbrida fundamenta-se no princípio da \textit{redundância criptográfica}, em que duas ou mais primitivas distintas são empregadas em paralelo ou em camadas, de modo a ampliar a resiliência global do sistema. O objetivo é criar um "duplo seguro": ainda que uma técnica venha a ser comprometida por avanços computacionais (como o surgimento de computadores quânticos universais) ou pela descoberta de vulnerabilidades matemáticas, uma segunda camada continua garantindo proteção. Esse conceito não é novo, versões iniciais de híbridos já estavam presentes em protocolos como o SSL/TLS (uso de criptografia assimétrica para troca de chaves e simétrica para cifragem de sessão),, mas assume caráter crítico no contexto da computação quântica e da transição para padrões pós-quânticos.

Um dos modelos mais difundidos é a adoção simultânea de algoritmos clássicos consolidados (RSA, ECC) e algoritmos pós-quânticos em estágios de negociação de chaves, assinatura ou encapsulamento. Essa forma de hibridização tem sido testada em grandes plataformas de Internet, como protocolos experimentais de \textit{TLS 1.3 híbrido}, VPNs corporativas e navegadores que implementaram versões piloto usando ECC em conjunto com Kyber, finalista da padronização do NIST. A principal vantagem é a compatibilidade com a infraestrutura atual: a rede segue reconhecendo os algoritmos clássicos, enquanto a camada PQC adiciona longevidade e resistência contra ataques quânticos. O desafio, entretanto, é lidar com o incremento de overhead computacional, principalmente em handshakes, e com mensagens ligeiramente maiores.

Uma abordagem mais recente e ainda restrita a pilotos de pesquisa é a junção da PQC com a distribuição quântica de chaves (QKD). Nesse modelo, a QKD fornece sigilo de chave com base em princípios físicos (não-clonagem, emaranhamento), enquanto a PQC assegura autenticação, assinatura digital e verificabilidade escalável. Essa arquitetura é particularmente atraente para redes 5G, enlaces diplomáticos e sistemas financeiros críticos, nos quais o risco de vazamento de dados a longo prazo é inaceitável. A limitação, por ora, está no custo elevado da infraestrutura quântica e no desafio de integração de canais ópticos dedicados em arquiteturas já existentes de rede.

Em contextos de alta criticidade, surgem modelos que utilizam múltiplas camadas de segurança: criptografia simétrica reforçada (como AES-256) para grandes volumes de dados, algoritmos clássicos para compatibilidade de legado e PQC para resiliência de longo prazo. Blockchains permissionados, sistemas bancários distribuídos e plataformas de nuvem governamental começam a discutir esses arranjos. A lógica é simples: nenhuma chave isolada deve definir a segurança do sistema; mesmo que um adversário avance num vetor de ataque, outra camada assegura a continuidade da confidencialidade e integridade. O custo, contudo, é pago em escalabilidade, interoperabilidade e eficiência.

Um campo de pesquisa que cresce rapidamente é a adaptação da criptografia híbrida a dispositivos embarcados, sensores e redes IoT. A ideia é integrar algoritmos PQC de baixa demanda computacional com mecanismos clássicos já consolidados nos microcontroladores atuais, visando mitigar riscos de ataques quânticos sem inviabilizar a operação de dispositivos com alimentação restrita. Aqui, mais do que redundância, busca-se adequação: escolher algoritmos híbridos que mantenham a segurança alinhada à realidade de hardware de baixo custo.

\begin{table}[h!]
\centering
\scriptsize
\caption{Formas de hibridização em criptografia emergente e suas principais características}
\label{tab:conceito_hibrida}
\begin{tabular}{p{2.1cm}p{3.3cm}p{5.1cm}p{3.8cm}}
\hline
\textbf{Tipo de Hibridização} & \textbf{Exemplos Práticos} & \textbf{Objetivo Técnico} & \textbf{Desafios} \\
\hline
Clássica + Pós-Quântica & TLS 1.3 híbrido (ECC + Kyber), VPNs experimentais & Garantir proteção redundante na troca de chaves e autenticação durante a transição & Overhead computacional; aumento de tempo em handshakes e largura de banda \\

PQC + QKD & Pilotos em redes 5G, enlaces diplomáticos, redes financeiras críticas & QKD assegura sigilo incondicional de chaves; PQC fornece autenticação escalável & Alto custo de infraestrutura óptica; integração em topologias complexas \\

Multicamadas (Clássica + PQC + Simétrica) & Blockchain, sistemas bancários distribuídos, nuvens soberanas & Proteção redundante em diferentes camadas; mitigação de falhas isoladas & Impacto em eficiência e escalabilidade; compatibilidade com sistemas legados \\

IoT híbrida (Clássica + PQC leve) & Redes de sensores, dispositivos médicos, sistemas embarcados autônomos & Adequar resistência quântica a hardware de baixo custo energético & Redução de desempenho; necessidade de algoritmos otimizados \\

\hline
\end{tabular}
\end{table}

A Tabela \ref{tab:conceito_hibrida} sintetiza os principais esquemas já explorados. A hibridização clássico-pós-quântica aparece como ponte prática no curto prazo; a integração PQC + QKD representa uma linha mais visionária, que alia garantias físicas e matemáticas; os modelos multicamadas são resposta a ambientes em que um único ponto de falha é inadmissível; e os arranjos emergentes para IoT sugerem que a hibridização precisa ser calibrada ao contexto de restrições computacionais. Em conjunto, essas estratégias demonstram que o futuro da criptografia não será marcado pela substituição simples de algoritmos, mas pela convivência híbrida e gradual, adaptada a diferentes cenários de risco e capacidade tecnológica.

\section{Algoritmos e protocolos existentes}

A transição prática para primitivas pós-quânticas exige decisões simultâneas sobre algoritmos (KEMs, assinaturas), níveis de segurança, formatos de chaves e assinaturas, além de ajustes em protocolos e pilhas de software já amplamente consolidados. Não se trata apenas de “trocar RSA por um KEM baseado em lattices”, mas de revisar todo o ciclo de vida criptográfico: geração, distribuição, rotação, armazenamento e revogação de chaves em ambientes heterogêneos (servidores, dispositivos móveis, IoT, nuvem e redes especializadas).

Nesta seção apresentamos uma visão consolidada dos algoritmos mais relevantes por função (troca de chaves, cifradores simétricos, funções de hash, esquemas de assinatura, KEMs pós-quânticos, primitivas homomórficas, mecanismos de QKD e construções híbridas), com ênfase em esquemas que já aparecem em recomendações de padronização ou em implementações de referência. Em seguida, discutimos as principais linhas de implementação e os protocolos que vêm sendo testados em produção ou em ambientes de pesquisa.

\subsection{Algoritmos}

A seleção e integração de algoritmos pós-quânticos em arquiteturas híbridas devem ser guiadas por critérios técnicos e operacionais além das suposições matemáticas, tais como tamanho de chave/assinatura, latência de operações de encapsulamento/assinatura, uso de memória, maturidade das implementações e sensibilidade a ataques por canal lateral, e a Tabela \ref{tab:algoritmos_pqc} sintetiza algoritmos relevantes; em seguida expõe-se, em texto corrido, motivos de escolha, trade-offs típicos e práticas de engenharia para integração segura. 

\begin{table}[h!]
\centering
\scriptsize
\caption{Algoritmos pós-quânticos importantes: família, função e status.}
\label{tab:algoritmos_pqc}
\begin{tabular}{p{32mm}p{30mm}p{28mm}p{50mm}}
\hline
\textbf{Algoritmo} & \textbf{Família / Assunção} & \textbf{Função} & \textbf{Status / Observações} \\
\hline
CRYSTALS-Kyber & Lattices (Module-LWE) & KEM & NIST: selecionado. Bom trade-off desempenho/compactação. \\
CRYSTALS-Dilithium & Lattices (Module-LWE) & Assinatura & NIST: selecionado. Forte para uso geral. \\
FALCON & Lattices (treliça / NTRU-like) & Assinatura & NIST: selecionado. Assinaturas compactas; implementação sensível a detalhes numéricos. \\
SPHINCS+ & Hash-based & Assinatura & NIST: selecionado. Segurança conservadora; assinaturas grandes. \\
Classic McEliece & Code-based & KEM / encriptação & Alternativo; chaves muito grandes, histórico robusto. \\
BIKE, HQC & Code-based & KEM & Alternativos em avaliação; trade-offs de tamanho/eficiência. \\
NTRU / NTRUPrime & Lattice variants & KEM & Uso histórico e experimental; variações em estudo. \\
SIKE (isogenias) & Isogenias & KEM & Sujeito a quebras práticas; não recomendado atualmente. \\
Rainbow & Multivariado (MQ) & Assinatura & Comprometido por ataques práticos; não recomendado. \\
Outros (multivariados, isogenias) & Diversas & KEM/assinatura & Alguns permanecem em pesquisa com reavaliações frequentes. \\
\hline
\end{tabular}
\end{table}

Famílias baseadas em reticulados (Module-LWE, NTRU-variants) oferecem hoje o melhor compromisso entre desempenho e segurança para uso generalizado, razão pela qual Kyber e Dilithium são referências; esquemas baseados em hash (SPHINCS+) são úteis para diversificação dada a ausência de suposições estruturadas, ainda que impliquem assinaturas maiores; esquemas baseados em códigos (Classic McEliece, BIKE, HQC) mantêm apelo por sua robustez histórica, sacrificando tamanho de chave; e famílias como isogenias e multivariadas exigem cautela face a reavaliações criptoanalíticas. Para mapear a integração arquitetural, a Figura \ref{fig:hybrid_architecture} apresenta uma visão conceptual de uma arquitetura híbrida onde se combinam KEM clássico e KEM PQC para derivação de chave de sessão, assinaturas híbridas para autenticação e cifragem simétrica para dados de sessão, de modo que a quebra de uma primitiva isolada não comprometa a confidencialidade ou autenticidade global.

\begin{figure}[h!]
\centering
\begin{tikzpicture}[
  font=\small,
  >=Stealth,
  node distance=12mm and 18mm,
  box/.style={draw, rounded corners, minimum width=3.6cm, minimum height=10mm, align=center, text width=3.4cm, inner sep=2mm},
  boxwide/.style={draw, rounded corners, minimum width=8.6cm, minimum height=12mm, align=center, text width=8.0cm, inner sep=3mm, fill=orange!8},
  kdf/.style={draw, rounded corners, fill=white, align=center, inner sep=2pt, font=\footnotesize},
  tinylabel/.style={font=\scriptsize, fill=white, inner sep=1pt}
]

\node[box, fill=blue!8] (kem_classic) {KEM Clássico \\ (ECDH / X25519)};
\node[box, fill=green!8, right=of kem_classic] (kem_pqc) {KEM PQC \\ (Kyber)};

\node[box, fill=blue!8, below=12mm of kem_classic] (sig_classic) {Assinatura Clássica \\ (ECDSA / RSA)};
\node[box, fill=green!8, right=of sig_classic] (sig_pqc) {Assinatura PQC \\ (Dilithium / Falcon)};

\node[boxwide, below=22mm of $(sig_classic)!0.5!(sig_pqc)$] (symmetric) {Cifragem Simétrica de Sessão \\ (AES-GCM / ChaCha20-Poly1305)};

\node[kdf, below=10mm of $(kem_classic)!0.5!(kem_pqc)$] (kdfnode) {$K_{\text{sessão}} \gets \mathrm{KDF}\big(K_{\text{clássico}}\|K_{\text{PQC}}\|\text{transcript}\big)$};

\draw[->, thick] (kem_classic.south) to[out=-90,in=120] ($(kdfnode.north west)$);
\draw[->, thick] (kem_pqc.south) to[out=-90,in=60] ($(kdfnode.north east)$);

\draw[->, thick] (kdfnode.south) -- ($(symmetric.north)$) ;

\draw[->, thick] (sig_classic.south) to[out=-90,in=160] ($(symmetric.north west)+(0.2,0.0)$);
\draw[->, thick] (sig_pqc.south) to[out=-90,in=20] ($(symmetric.north east)+(-0.2,0.0)$);

\node[tinylabel] at ($(symmetric.north)+(0,-17mm)$) {$\sigma_{\text{híbrida}}=(\sigma_{\text{clássica}},\sigma_{\text{PQC}})$};

\end{tikzpicture}
\caption{Arquitetura conceptual de solução híbrida.}
\label{fig:hybrid_architecture}
\end{figure}

A seguir apresenta-se o pseudocódigo dos fluxos essenciais de operação: o Algoritmo \ref{alg:hybrid_kem} descreve o encapsulamento híbrido de chave combinando KEM clássico e KEM PQC; o Algoritmo \ref{alg:hybrid_kem_decaps} cobre o desencapsulamento; os Algoritmos \ref{alg:hybrid_sig} e \ref{alg:hybrid_sig_verify} formalizam geração e verificação de assinaturas híbridas; e o Algoritmo \ref{alg:fail_safe} ilustra um procedimento de fail-safe para evitar transformar o receptor em um oracle de decapsulamento. 

\begin{algorithm}[h!]
\scriptsize
\caption{Encapsulamento Híbrido de Chave (Hybrid KEM)}
\label{alg:hybrid_kem}
\begin{algorithmic}[1]
\Require Chave pública clássica $pk_{\text{clássico}}$, chave pública PQC $pk_{\text{PQC}}$
\Ensure Ciphertexts $(c_{\text{clássico}}, c_{\text{PQC}})$ e chave de sessão $K_{\text{sessão}}$
\State $(c_{\text{clássico}}, K_{\text{clássico}}) \gets \text{KEM}_{\text{clássico}}.\text{Encaps}(pk_{\text{clássico}})$
\State $(c_{\text{PQC}}, K_{\text{PQC}}) \gets \text{KEM}_{\text{PQC}}.\text{Encaps}(pk_{\text{PQC}})$
\State $K_{\text{sessão}} \gets \text{KDF}(\text{label} \| K_{\text{clássico}} \| K_{\text{PQC}} \| \text{transcript})$
\State \Return $(c_{\text{clássico}}, c_{\text{PQC}}), K_{\text{sessão}}$
\end{algorithmic}
\end{algorithm}

\vspace{0.5cm}
A linha 1 declara as entradas: as chaves públicas clássica e PQC; antes do uso essas chaves devem ser validadas quanto a formato e parâmetros (por exemplo curvas, tamanhos e formatos esperados). A linha 2 indica as saídas: os dois ciphertexts e a chave de sessão; especifique o formato (ordem, length-prefix) e a codificação (raw bytes, DER, etc.). A linha 3 executa o encapsulamento clássico e produz $c_{\text{clássico}}$ e $K_{\text{clássico}}$, em implementações ECDH atenção à normalização do ponto público e à necessidade de um KDF intermédio se a API do KEM não fornecer o shared secret já processado. A linha 4 executa o encapsulamento PQC (ex.: Kyber) e produz $c_{\text{PQC}}$ e $K_{\text{PQC}}$; documente e meça latência e uso de memória para o KEM PQC. A linha 5 realiza a derivação final de $K_{\text{sessão}}$ com uma KDF; em prática recomenda-se HKDF (SHA-256/512) com domain separation (p.ex. label = \texttt{HybridKEM-v1}) e encoding canônico dos inputs (cada campo length-prefixed); a linha 6 devolve o par de ciphertexts e a chave de sessão e os segredos temporários devem ser sobreescritos após o uso.

\begin{algorithm}[h!]
\scriptsize
\caption{Desencapsulamento Híbrido de Chave (Hybrid KEM Decaps)}
\label{alg:hybrid_kem_decaps}
\begin{algorithmic}[1]
\Require Chave privada clássica $sk_{\text{clássico}}$, chave privada PQC $sk_{\text{PQC}}$, ciphertexts $(c_{\text{clássico}}, c_{\text{PQC}})$
\Ensure Chave de sessão $K_{\text{sessão}}$
\State $K_{\text{clássico}} \gets \text{KEM}_{\text{clássico}}.\text{Decaps}(sk_{\text{clássico}}, c_{\text{clássico}})$
\State $K_{\text{PQC}} \gets \text{KEM}_{\text{PQC}}.\text{Decaps}(sk_{\text{PQC}}, c_{\text{PQC}})$
\State $K_{\text{sessão}} \gets \text{KDF}(\text{label} \| K_{\text{clássico}} \| K_{\text{PQC}} \| \text{transcript})$
\State \Return $K_{\text{sessão}}$
\end{algorithmic}
\end{algorithm}

A linha 1 lista os inputs; a linha 2 é o Decaps clássico que pode falhar se o ciphertext for malformado ou os parâmetros divergirem, implemente sem early returns observáveis. A linha 3 é o Decaps PQC; dado o possível desbalanceamento de tempos entre PQC e clássico, a implementação deve normalizar o comportamento temporal e de erro para não vazar informação. A linha 4 aplica a mesma KDF usada no encapsulamento para derivar $K_{\text{sessão}}$; se uma das decaps falhar, não entregue erro visível ao atacante: em vez disso utilize o procedimento de fail-safe (Algoritmo \ref{alg:fail_safe}) que substitui a saída inválida por um fallback PRF derivado de um segredo local.

\begin{algorithm}[h!]
\scriptsize
\caption{Assinatura Híbrida (Hybrid Signature)}
\label{alg:hybrid_sig}
\begin{algorithmic}[1]
\Require Mensagem $m$, chave privada clássica $sk_{\text{clássico}}$, chave privada PQC $sk_{\text{PQC}}$
\Ensure Assinatura híbrida $\sigma_{\text{híbrida}}$
\State $\sigma_{\text{clássica}} \gets \text{Sign}_{\text{clássico}}(sk_{\text{clássico}}, m)$
\State $\sigma_{\text{PQC}} \gets \text{Sign}_{\text{PQC}}(sk_{\text{PQC}}, m)$
\State \Return $\sigma_{\text{híbrida}} \gets (\sigma_{\text{clássica}}, \sigma_{\text{PQC}})$
\end{algorithmic}
\end{algorithm}

\vspace{0.5cm}

A linha 1 apresenta os requisitos; a linha 2 produz a assinatura clássica sobre a mensagem $m$ e deve empregar RNG e práticas constant-time conforme o esquema usado; a linha 3 produz a assinatura PQC e deve observar requisitos específicos (por exemplo, cuidado com esquemas que exigem estado ou cuidados de RNG). A linha 4 combina as assinaturas no objeto híbrido; especifique o formato de empacotamento (cada assinatura com length-prefix) para interoperabilidade e documente trade-offs (AND das assinaturas garante resistência futura, mas aumenta overhead).

\begin{algorithm}[h!]
\scriptsize
\caption{Verificação de Assinatura Híbrida (Hybrid Signature Verify)}
\label{alg:hybrid_sig_verify}
\begin{algorithmic}[1]
\Require Mensagem $m$, assinatura $\sigma_{\text{híbrida}}=(\sigma_{\text{clássica}},\sigma_{\text{PQC}})$, chaves públicas $pk_{\text{clássico}}, pk_{\text{PQC}}$
\Ensure resposta \texttt{válido} / \texttt{inválido}
\State $v_{\text{clássico}} \gets \text{Verify}_{\text{clássico}}(pk_{\text{clássico}}, m, \sigma_{\text{clássica}})$
\State $v_{\text{PQC}} \gets \text{Verify}_{\text{PQC}}(pk_{\text{PQC}}, m, \sigma_{\text{PQC}})$
\If{$v_{\text{clássico}} = \texttt{válido}$ \textbf{and} $v_{\text{PQC}} = \texttt{válido}$}
  \State \Return \texttt{válido}
\Else
  \State \Return \texttt{inválido}
\EndIf
\end{algorithmic}
\end{algorithm}

Na linha 1 estão os parâmetros de entrada; as linhas 2 e 3 executam as verificações clássica e PQC, respectivamente, e ambas devem ser executadas sempre, preferencialmente de forma que o tempo total não exponha qual verificação falhou. A linha 5 impõe que ambas sejam válidas para aceitar a assinatura; registre falhas para auditoria sem incluir material sensível e normalize respostas públicas para mitigar canais laterais.

\begin{algorithm}[h!]
\scriptsize
\caption{Fail-safe Decapsulation (decapsulação resistente a oráculos)}
\label{alg:fail_safe}
\begin{algorithmic}[1]
\Require $sk_{\text{clássico}},\; sk_{\text{PQC}},\; (c_{\text{clássico}}, c_{\text{PQC}}),\; S_{\text{fb}}$
\Ensure $K_{\text{sessão}}$
\State $(v_{\text{clássico}},\ \mathsf{ok}_{\text{clássico}}) \gets \textsf{try\_decaps}(sk_{\text{clássico}}, c_{\text{clássico}})$
\State $(v_{\text{PQC}},\ \mathsf{ok}_{\text{PQC}}) \gets \textsf{try\_decaps}(sk_{\text{PQC}}, c_{\text{PQC}})$
\If{$\neg \mathsf{ok}_{\text{clássico}}$}
  \State $v_{\text{clássico}} \gets \mathrm{PRF}\big(S_{\text{fb}} \;\|\; \text{\texttt{fb\_classico}} \;\|\; c_{\text{clássico}}\big)$
\EndIf
\If{$\neg \mathsf{ok}_{\text{PQC}}$}
  \State $v_{\text{PQC}} \gets \mathrm{PRF}\big(S_{\text{fb}} \;\|\; \text{\texttt{fb\_PQC}} \;\|\; c_{\text{PQC}}\big)$
\EndIf
\State $K_{\text{sessão}} \gets \mathrm{KDF}\big(\text{label} \;\|\; v_{\text{clássico}} \;\|\; v_{\text{PQC}} \;\|\; \text{transcript}\big)$
\State \Return $K_{\text{sessão}}$
\end{algorithmic}
\end{algorithm}

As linhas 1–2 mostram que \textsf{try\_decaps} deve executar internamente todo o caminho de decapsulação e retornar um par (valor, flag ok); essa rotina não deve fazer early returns observáveis. As linhas 3–4 representam as verificações dos flags; se $\mathsf{ok}_{\text{clássico}}$ for falso (linha 5), substitui-se $v_{\text{clássico}}$ por um valor derivado do segredo local $S_{\text{fb}}$ através de um PRF (por exemplo HMAC-SHA256) que inclui um identificador literal \texttt{fb\_classico} e idealmente o HASH do ciphertext para controlar o tamanho do input; análoga lógica se $\mathsf{ok}_{\text{PQC}}$ for falso (linha 7). A linha 9 realiza a KDF final combinando os valores reais ou os fallback, com domain separation e transcript canônico; o comportamento externo (tempo, mensagens) deve ser indistinguível entre casos de sucesso e fallback para impedir oráculos de decapsulação. O segredo $S_{\text{fb}}$ deve ser protegido (HSM/enclave) e rotacionado conforme política.
\vspace{0.5cm}

A especificação da KDF e a separação de domínio são críticas: recomenda-se usar HKDF (SHA-256/512) ou uma variante SHA-3; concatenar explicitamente uma string de domain separation (por exemplo, \texttt{HybridKEM-v1}), seguir encoding canônico (cada campo length-prefixed), incluir $K_{\text{clássico}}$ e $K_{\text{PQC}}$ como IKM (ou como IKM/HKDF info conforme perfil) e o transcript completo (ou seu hash) para evitar ambiguidades e ataques de replay. Implementações reais devem garantir que ambos os decaps (clássico e PQC) sejam executados sem ramificações externas observáveis e que, em caso de falha, seja usado o fallback pseudo-aleatório derivado de $S_{\text{fb}}$ para evitar oráculos, além de aplicar práticas constant-time sempre que possível. 

\begin{figure}[h!]
\centering
\begin{tikzpicture}[font=\small, >=Stealth]
  \node (client) at (0,0) [rectangle, draw=none] {Cliente};
  \node (server) at (10,0) [rectangle, draw=none] {Servidor};

  \draw[->, thick]
    (client.north) to[bend left=50]
    node[midway, above, align=center, text width=0.58\textwidth, yshift=6mm]
    {ClientHello\\(algorithms: classical + Kyber, nonces, versions)}
    (server.north);

  \draw[->, thick]
    (server) to[bend left=20]
    node[midway, above, align=center, text width=0.58\textwidth, yshift=5mm]
    {ServerHello + ServerCert + (c\_clássico, c\_PQC) + ServerFinished}
    (client);

  \draw[->, thick]
    (client.south) to[bend right=30]
    node[midway, below, align=center, text width=0.58\textwidth, yshift=-2mm]
    {ClientFinished\\(verificação de transcript)}
    (server.south);

  \node[draw, rounded corners, fill=white, align=left, text width=0.85\textwidth]
    (kdf) at (5,-3.0) {Ambos os lados: executar Decaps clássico e PQC; derivar\\[3pt]
    $K_{\text{sessão}}=\mathrm{KDF}\big(\text{label}\,\|\,K_{\text{clássico}}\,\|\,K_{\text{PQC}}\,\|\,\text{transcript}\big)$.};

\end{tikzpicture}
\caption{Fluxo simplificado de handshake TLS com KEM híbrido.}
\label{fig:tls_hybrid}
\end{figure}
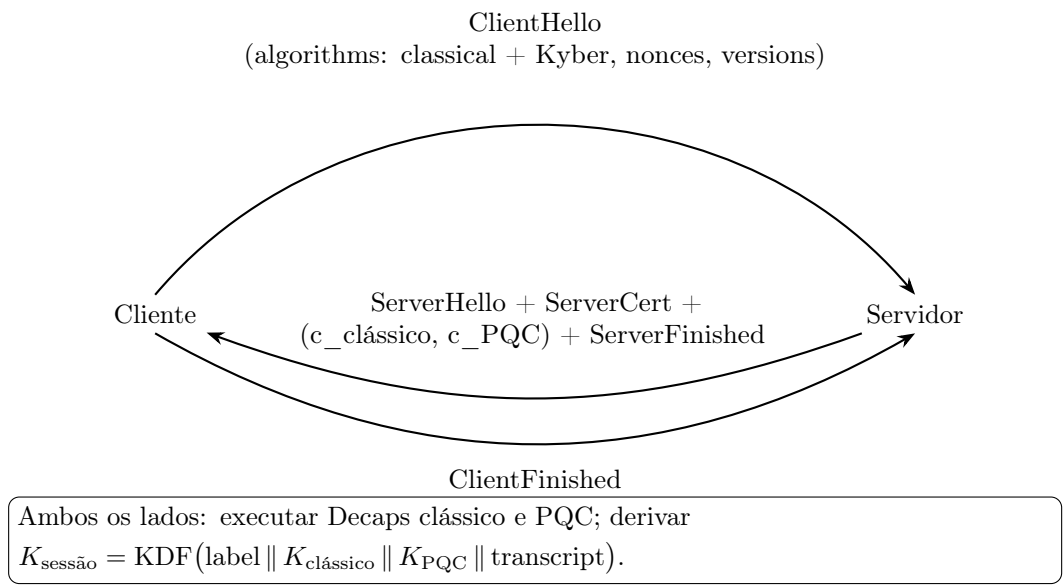

Em conclusão, os algoritmos e pseudocódigos apresentados formalizam a construção lógica da abordagem híbrida, mas a segurança prática depende de decisões detalhadas sobre KDF, tratamento atômico e resistente a oráculos de falhas, mitigação de canais laterais, perfis de parâmetros e procedimentos operacionais, todos estes aspectos devem ser documentados, testados e validados em ambientes reais antes de qualquer rollout em produção.

\subsection{Protocolos, bibliotecas e implementações}

A introdução de primitivas PQC na pilha de rede e nas bibliotecas criptográficas tem seguido duas frentes principais. A primeira é a realização de experimentos e patches em protocolos amplamente usados (por exemplo, TLS, SSH, QUIC e VPNs) para avaliar impactos práticos, esses experimentos frequentemente adotam modos híbridos, combinando KEMs clássicos (como X25519) com KEMs pós-quânticos (como CRYSTALS-Kyber). A segunda frente é o desenvolvimento de bibliotecas e implementações de referência (liboqs, PQClean, PQM4) que facilitam testes e otimizações para plataformas diversas, desde servidores x86 até microcontroladores ARM.

\begin{table}[h!]
\centering
\scriptsize
\caption{Protocolos e implementações representativas para experimentação PQC híbrida.}
\label{tab:protocolos_pqc}
\begin{tabular}{p{34mm}p{50mm}p{65mm}}
\hline
\textbf{Projeto / Protocolo} & \textbf{Descrição / Papel} & \textbf{Observações} \\
\hline
TLS 1.3 (modos híbridos) & Inserção de KEMs pós-quânticos em handshakes TLS (modo híbrido com KEM clássico + PQC) & Vários experimentos e drafts IETF; desafio de interoperabilidade e tamanho de mensagens. \\
Open Quantum Safe (liboqs) & Biblioteca open-source com implementações PQC e integrações com OpenSSL/OpenSSH & Principal referência para provas de conceito e integração (OpenSSL-OQS, OpenSSH-OQS). \\
OpenSSL / BoringSSL / NSS forks & Patches e forks para suportar KEMs/assinaturas PQC & Testes em ambiente controlado; cuidado ao migrar para produção. \\
Rustls / implementações em Rust & Integrações experimentais de PQC em stacks TLS escritas em Rust & Bom espaço para experimentação seguro e memory-safe; surgem patches comunitários. \\
OpenSSH-OQS & Integrações de PQC em SSH & Experimentos de autenticação híbrida e key-exchange. \\
WireGuard / OpenVPN (POC) & Implementações experimentais integrando KEMs PQC em troca de chaves & Demonstram impacto em latência e overhead em cenários reais. \\
Bibliotecas (PQClean, PQM4, liboqs) & Implementações otimizadas e portáveis & Essenciais para benchmarking, otimização e uso em embarcados. \\
TPM / HSM vendors & Suporte a operações de assinatura/secure storage para chaves PQC e segredos de fallback & Integração crítica para segredos de PRF/fallback e rotação segura de chaves. \\
QKD + PQC integrações & Pilotos que combinam QKD para distribuição de chave e PQC para autenticação & Estratégia de alto custo; aplicada em enlaces críticos e pilotos telco/financeiro. \\
PKI híbrida & Certificados X.509 e perfis com assinaturas/algoritmos híbridos & Abordagem de transição para compatibilidade com infraestruturas legadas; exige perfis/algIDs definidos. \\
\hline
\end{tabular}
\end{table}

Na prática, protocolos como TLS 1.3 podem operar em modo híbrido combinando um KEM clássico com um KEM PQC. Essa abordagem amplia a resistência do canal, mas impõe custos concretos: aumento do tamanho do handshake (maior uso de largura de banda e risco de ultrapassar MTU), maior consumo de CPU no estabelecimento de sessão, e requisitos adicionais de memória. Por isso, qualquer ensaio de adoção deve incluir medições sistemáticas de latência, throughput, uso de CPU, uso de memória, padrões de fragmentação e impacto em conexões de curta duração (short-lived connections). Bibliotecas como liboqs e PQClean permitem realizar estes testes de forma reprodutível e são recomendadas para provas de conceito e benchmarking; para sistemas embarcados, PQM4 e builds otimizados em assembly para ARM/ESP são pontos de partida.

Para compatibilidade com esforços de padronização e maturidade prática, use implementações e parâmetros que refletem o estado do ecossistema: CRYSTALS-Kyber (KEM), CRYSTALS-Dilithium (assinatura), Falcon (assinatura opcional) e SPHINCS+ (assinatura estática resistente a ataques futuros como fallback). Adote diferentes níveis de segurança (mapeamento aos níveis NIST) para medir trade-offs: por exemplo testar Kyber-512 vs Kyber-768/1024 para avaliar overhead vs margem de segurança.

A segurança não é apenas escolha de algoritmo: exige práticas consolidadas. Garanta RNG de qualidade certificada para geração de nonces e seeds; implemente operações constant-time quando possível; trate esquemas que exigem gerenciamento de estado (ex.: algumas variantes de hash-based) com protocolos de utilização segura; e proteja segredos de fallback (S\_fb) em HSMs ou enclaves. Realize avaliações de side-channel (timing, cache, EM) especialmente quando usar PQC em hardware com acelerações vetoriais ou instruções especiais. Para esquemas que usam amostragem (p.ex. Falcon), realize análises de correlação e testes estatísticos do RNG.

A migração exige definição de perfis: identificadores de algoritmo (algIDs), formatos de certificado (X.509 extensions/profile), codificação dos objetos híbridos (length-prefix, CBOR/DER) e vetores de teste cross-vendor. Coordene com CAs/fornecedores para publicar perfis híbridos e políticas de revogação; planeje rotação coordenada de pares (clássico, PQC) e mantenha registros claros de qual par está associado a cada certificado. Defina procedimentos de rollback seguros e testes de compatibilidade regressiva.

Recomenda-se uma progressão em três fases: (1) laboratório e PoC com liboqs/OpenSSL-OQS para validar handshakes híbridos e medir carga; (2) ambiente controlado (canary) em produção reduzida para testar comportamento sob falhas, reconexões e cargas reais; (3) rollout em larga escala com monitoramento ativo, políticas de revogação e suporte HSM/TPM. Em cada fase, execute testes de resistência (fuzzing de mensagens handshake), testes de regressão e testes de interoperabilidade com versões de outros fornecedores.

Para cada experimento, registre: configuração de teste (hardware, firmware, versão de biblioteca); combinação de algoritmos testados; latência de handshake (média, p95, p99); throughput e CPU por conexão; uso de memória máxima durante handshake; tamanho médio e máximo de mensagens de handshake; comportamento com MTU pequenos e fragmentação; tempo de reconexão e impacto em sessões long-lived; logs e telemetria de erros; resultados de testes de side-channel básicos; interoperabilidade com stacks de terceiros (OpenSSL, rustls, OpenSSH). Além disso, realize testes de “store-now, decrypt-later” para avaliar risco de dados sensíveis em retenção.

Em ambientes restritos, selecione parâmetros que equilibrem segurança e consumo energético; priorize implementações otimizadas (PQM4, builds com assembly para ARM) e, quando possível, delegue operações pesadas a um co-processador ou módulo seguro. Avalie impacto no tempo de boot, no consumo durante handshake e em limites de memória; considere esquemas híbridos onde a componente PQC seja configurável (habilitada apenas para conexões de maior sensibilidade).

Registre métricas de erro e falha sem vazar material sensível; implemente agregação segura de telemetria e alertas para padrões anômalos (aumento de falhas de verificação, tempos fora do normal). Padronize logs para incluir o perfil híbrido usado (algIDs), mas não incluir chaves/secretos. Defina políticas de rotação e revogação para pares (clássico,PQC) e procedimentos de atualização de firmware para permitir correções rápidas. Garanta que a telemetria agregada esteja anonimizável e que alertas acionem playbooks de resposta (rollback, isolamento de nós, habilitação/desabilitação de perfis) sem exposição direta de material sensível.

\begin{figure}[h!]
\centering
\begin{tikzpicture}[node distance=20mm, every node/.style={font=\small}, >=latex]
  \node[draw, rounded corners, fill=blue!8, inner sep=6pt, text width=30mm, align=center] (poc) {1. PoC\\(liboqs / OpenSSL-OQS)};
  \node[draw, rounded corners, fill=green!8, right=of poc, inner sep=6pt, text width=30mm, align=center] (canary) {2. Canary\\(ambiente controlado)};
  \node[draw, rounded corners, fill=orange!8, right=of canary, inner sep=6pt, text width=30mm, align=center] (rollout) {3. Rollout\\(produção)};
  \draw[->] (poc) -- (canary);
  \draw[->] (canary) -- (rollout);

  \node[draw, rounded corners, fill=gray!8, below=22mm of canary, inner sep=8pt, text width=0.7\textwidth, align=center] (tele) {Telemetria \& Monitoramento: logs agregados, métricas (latência, CPU, memória), alertas};
  \draw[->] (poc.south) |- (tele.north);
  \draw[->] (canary.south) -- (tele);
  \draw[->] (rollout.south) |- (tele.north);

  \node[draw, rounded corners, fill=yellow!8, below=of tele, inner sep=6pt, text width=45mm, align=center] (hsm) {HSM / TPM: proteção de S\_fb e rotação de chaves};
  \draw[->] (tele) -- (hsm);
\end{tikzpicture}
\caption{Fluxo recomendado de rollout (PoC → Canary → Rollout) com telemetria centralizada e proteção de segredos em HSM/TPM.}
\label{fig:deployment_telemetry}
\end{figure}
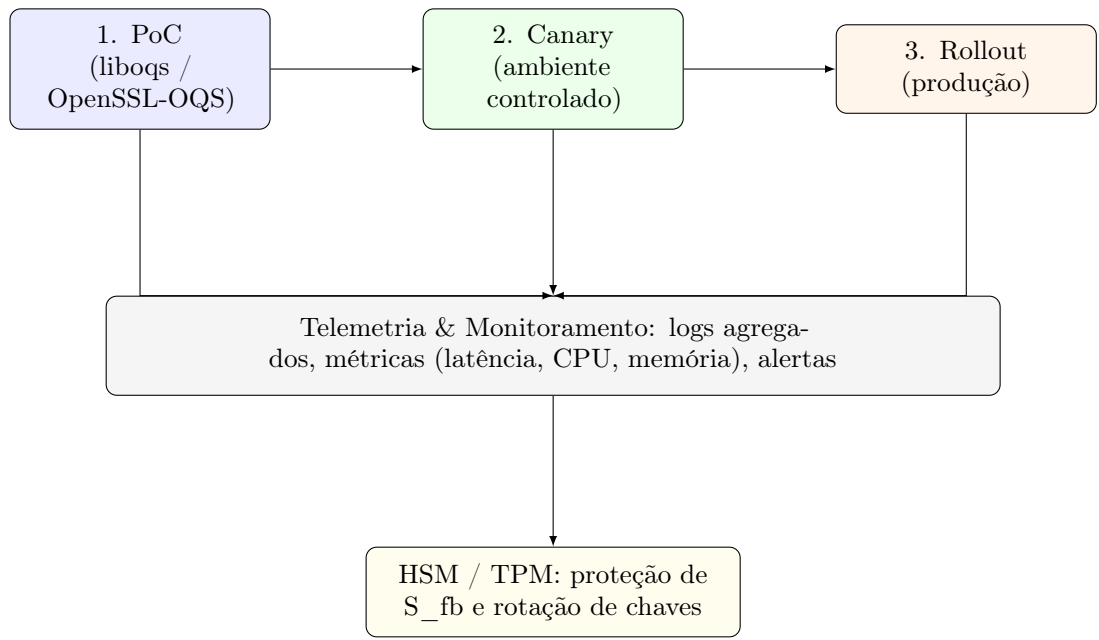

Envolva equipes de conformidade e auditoria desde o início; mantenha trilhas de auditoria para mudanças de perfil e rotação de chaves; assegure que bibliotecas usadas em produção tenham avaliações, testes e, quando necessário, certificações (ex.: processos de triagem de fornecedores de HSM).

Por fim, embora a padronização ofereça um ponto de partida estável, trata-se de um processo dinâmico: espere refinamentos de parâmetros, otimizações de implementação e eventuais reavaliações em função de novos resultados criptoanalíticos. Planeje revisões periódicas do inventário criptográfico, testes automáticos na pipeline CI/CD e capacidade de atualizar stacks criptográficos com segurança (compatibilidade regressiva, staged rollout, suporte HSM). A combinação de experimentação controlada, métricas robustas e governança clara reduz significativamente o risco durante a migração para um ambiente resistente a ataques quânticos.

\section{Estado da Arte}

A Tabela \ref{tab:estado_arte_hibrida} apresenta um resumo de trabalhos recentes que exploram abordagens híbridas em criptografia, combinando algoritmos clássicos e pós-quânticos, assim como integrações entre PQC e QKD. Os estudos cobrem desde experimentos práticos em protocolos TLS até discussões conceituais sobre integração multicamadas.

\begin{table}[h!]
\centering
\scriptsize
\caption{Estado da arte em criptografia híbrida e emergente, destacando abordagens, contribuições e limitações.}
\label{tab:estado_arte_hibrida}
\begin{tabular}{p{3cm}p{2cm}p{6cm}p{3cm}}
\hline
\textbf{Autor/Ano} & \textbf{Abordagem Híbrida} & \textbf{Contribuição} & \textbf{Limitações} \\
\hline

Croz et al. (2021) \cite{Croz2021} & PQC + TLS & Avaliação prática de implementações híbridas de TLS com algoritmos lattice-based. & Impacto em latência e overhead computacional. \\

Mosca \& Piani (2019) \cite{Mosca2019} & PQC + QKD & Discussão teórica sobre uso combinado de QKD com assinaturas pós-quânticas para autenticação. & Dependência de infraestrutura quântica dedicada. \\

Huang et al. (2023) \cite{Huang2023} & PQC + QKD em 5G & Proposta de arquitetura de rede 5G usando QKD para distribuição de chaves e PQC para autenticação. & Custos elevados e complexidade de integração. \\

\hline
\end{tabular}
\end{table}

Os trabalhos catalogados na Tabela \ref{tab:estado_arte_hibrida} atendem a diferentes objetivos: alguns enfatizam provas de conceito e medições práticas em pilhas reais (por exemplo experimentos em TLS/SSH), outros exploram arquiteturas conceituais que combinam mecanismos distintos (por exemplo QKD para estabelecimento de segredo e PQC para autenticação e interoperabilidade). O estudo de Campagna et al. (2019) \cite{Mosca2019} é representativo da primeira onda de experimentos que demonstraram um caminho pragmático para transição, ao combinar X25519 com Kyber e documentar impactos concretos, notadamente o aumento do tamanho das mensagens no handshake TLS. Trabalhos subsequentes, como Croz et al. (2021) \cite{Croz2021}, aprofundaram a avaliação de desempenho mostrando que, embora a segurança efetiva aumente, existem custos mensuráveis em latência e uso de CPU que variam conforme o algoritmo PQC escolhido e a topologia de deployment.

Concomitantemente, abordagens teóricas e arquiteturais destacam composições multicamadas nas quais QKD participa como mecanismo de distribuição de segredo em enlaces dedicados, enquanto PQC trata da autenticação e compatibilidade ampla. Mosca \& Piani (2019) \cite{Mosca2019} detalham esse tipo de composição, apontando para cenários onde QKD fornece confidencialidade física local combinada com assinaturas pós-quânticas para autenticação federada. Esses modelos apresentam elevado potencial para enlaces críticos, porém dependem de investimento em infraestrutura e apresentam desafios de integração com redes convencionais, conforme observado em trabalhos aplicados como Huang et al. (2023) \cite{Huang2023} no contexto de 5G.

Da literatura prática emergem padrões recorrentes sobre as limitações técnicas e operacionais. O aumento do tamanho do handshake e a necessidade de gerir maior uso de CPU e memória são problemas repetidos; além disso, a fragilidade frente a limites de MTU e a fragmentação de pacotes torna indispensável testes de interoperabilidade em redes heterogêneas. Em paralelo, há questões abertas sobre a normalização de encodings, domain separation e políticas de fallback: muitos trabalhos propõem combinações de KEMs seguidas de HKDF com domain - separation explícito, mas faltam provas formais que cubram as variações de encoding e os mecanismos práticos de fallback usados em implementações. A literatura sobre oráculos de decapsulação fornece contramedidas (por exemplo fail-safe com PRF baseado em segredo local), porém a padronização de comportamentos de erro e a normalização temporal entre implementações ainda são insuficientes.

Outro vetor crítico no estado da arte refere-se à engenharia de implementações: bibliotecas como liboqs, PQClean e PQM4 tornaram-se infraestrutura essencial para benchmarking, portabilidade e otimização para plataformas que vão de servidores x86 a microcontroladores ARM. Estudos de otimização evidenciam que escolhas de implementação (assembly otimizado, vetorização AVX/NEON, uso de aceleradores criptográficos) têm impacto forte tanto em desempenho quanto em vulnerabilidade a canais laterais; portanto, avaliações de segurança devem incluir testes de timing, cache e, quando aplicável, análises EM. A integração com dispositivos de proteção de chave (HSM/TPM) aparece como prática recomendada para proteção de segredos sensíveis e para suportar rotação coordenada de pares clássico/PQC em ambientes de produção.

Além de questões técnicas, o estado da arte aponta lacunas de governança e padronização que dificultam migrações em grande escala. A definição de algIDs, perfis X.509 para certificados híbridos, e procedimentos de revogação e rollback coordenados entre pares clássico/PQC exigem acordos entre CAs, fornecedores de software e operadores de infraestrutura. Trabalhos experimentais demonstram que a interoperabilidade cross-vendor (por exemplo, OpenSSL-OQS vs rustls com patches PQC) é um indicador crucial de maturidade do ecossistema, portanto iniciativas de vetores de teste e de integração entre implementações de referência são elementos de alto valor prático.

Em termos metodológicos, a literatura converge na recomendação de que estudos de PoC combinem medições de desempenho (latência média e percentis, uso de CPU e memória, tamanhos de mensagens), testes de robustez (MTU reduzida, perda de pacotes, reconexões) e avaliações de segurança práticas (fuzzing de handshake, testes básicos de side-channel). Resultados que apresentam esses três tipos de evidência oferecem maior utilidade para engenharia de produção do que benchmarks isolados. Finalmente, as pesquisas futuras mais promissoras identificadas no corpus incluem provas formais de composições híbridas que incorporem encoding e políticas de erro realistas, análises de side-channel específicas para PQC em hardware moderno, e frameworks de orquestração de migração (rollout, monitoramento e rollback) que automatizem validações e reduzam risco operacional.

Em síntese, o estado da arte confirma que a criptografia híbrida é hoje a estratégia de transição mais pragmática frente ao risco quântico: combinações de mecanismos clássicos e pós-quânticos provêm resistência adicional sem exigir ruptura imediata das infraestruturas existentes. Contudo, a adoção em produção requer cuidados rigorosos de engenharia (encodings e KDF bem especificados, proteções contra oráculos e side-channels), avaliações de desempenho por workload e governança clara (perfis, algIDs, rotação coordenada). As lacunas identificadas pela literatura oferecem um mapa de pesquisa e desenvolvimento que, se atendido, reduzirá substancialmente o custo e o risco da migração para um ecossistema resistente a ataques quânticos.

\subsection{Questões para Reflexão e Pesquisa Futura}

Ao concluir este capítulo sobre criptografia híbrida e emergente, o leitor deve ser incentivado a ir além da descrição técnica e a questionar as premissas, os trade-offs operacionais e as implicações práticas das arquiteturas propostas. Considere, primeiramente, a natureza das garantias de segurança oferecidas por combinações PQC+clássico ou PQC+QKD: em que pontos a redundância efetivamente reduz o risco e em que pontos ela introduz complexidade que pode gerar novas falhas? Se uma camada falha (por exemplo, um KEM PQC com vulnerabilidade desconhecida), quais são os cenários de degradação aceitáveis e como as políticas de fallback devem ser desenhadas para preservar propriedades essenciais como confidencialidade de longo prazo e autenticidade? Pergunte-se também como as escolhas de encodings, domain separation e KDFs afetam a segurança composicional, que invariantes devem ser formalmente verificadas para garantir que duas ou mais camadas não interajam de forma insegura?

Reflita sobre a integração entre QKD e PQC: quando é justificável empregar QKD em vez de confiar apenas em PQC bem padronizada? Quais métricas operacionais (custo total de posse, latência de estabelecimento, disponibilidade do enlace, custo de manutenção) e métricas de risco (horizonte de proteção, exposição a ataques de canal lateral, risco geopolítico) devem ser ponderadas para decidir a inclusão de QKD em uma arquitetura híbrida? Em enlaces onde QKD é usado apenas para gerar material de chaves, como garantir que a autenticação da sessão, geralmente delegada a esquemas clássicos ou PQC, não se torne o elo fraco que anula os benefícios físicos do QKD?

Questione as implicações de implementação: até que ponto otimizações para desempenho (vetorização, assembly otimizado, uso de aceleradores) impactam a segurança física de implementações híbridas? Quais processos de desenvolvimento e pipelines CI/CD são necessários para que cada alteração de build seja validada contra regressões funcionais e de segurança (incluindo testes automáticos de timing e análise de side-channel)? Como integrar verificações formais de alto nível (provas de composição) com testes práticos de implementação, de modo que ambos, teoria e prática, sirvam de evidência para decisões de implantação?

Interrogar a interoperabilidade é essencial: quais formatos de mensagens e perfis (algIDs, X.509 híbridos, protocolos de handshake) são necessários para que distintos fornecedores e stacks interoperem sem ambiguidades que possam causar downgrades inseguros? Como projetar vetores de conformance testing e testbeds cross-vendor que detectem incompatibilidades subtis, como diferenças em padding, tratamentos de erro ou KDF labels? Pense também em como garantir que políticas de rollback e compatibilidade retroativa não criem backdoors operacionais que permitam downgrades acidentais ou maliciosos.

Do ponto de vista de medição e avaliação, quais experimentos são necessários para quantificar o impacto real de abordagens híbridas? Que conjunto mínimo de métricas (latência p50/p95/p99 de handshake, custo CPU por estabelecimento, uso de memória máxima, impacto em MTU/fragmentação, taxa de falha sob perda de pacotes, consumo energético em MCUs) deve constar em um PoC representativo? Como projetar cenários de teste que reproduzam cargas do mundo real (connexões curtas massivas, sessões long-lived, dispositivos de borda) de modo a mostrar não apenas médias, mas percentis e comportamentos sob stress?

Questione os modelos de ameaça adotados: a quem estamos protegendo e por quanto tempo? Que hipóteses sobre capacidades adversariais (acesso a computadores quânticos universais, capacidade de captura passiva de tráfego, acesso físico ao equipamento) orientam a escolha entre PQC puro, QKD ou arranjos híbridos? Como modelar adversários compostos que combinam ataques quânticos com canais laterais, engenharia social e exploração de falhas de configuração? Que políticas de retenção e classificação de ativos são necessárias para priorizar proteções de longo prazo?

Considere as dimensões de governança, regulamentação e cadeia de suprimentos: quais requisitos contratuais e regulatórios deveriam guiar decisões sobre adoção de híbridos em setores regulados? Como auditar fornecedores de HSM, módulos QKD e bibliotecas PQC para reduzir risco de supply-chain? Em contextos multinacionais, que estratégias organizar para conciliar divergências regulatórias e evitar fragmentação que prejudique interoperabilidade?

Pense em experimentos de pesquisa aplicados: que PoCs, estudos comparativos e publicações de datasets e scripts reprodutíveis você poderia conduzir para contribuir com o campo? Como registrar hipóteses, metodologia, métricas e resultados de modo a fornecer evidência operacional (e não apenas teórica) que outros possam reproduzir e criticar? Que cenários de caso de uso (bancos, redes governamentais, IoT crítico) deveriam ser priorizados para avaliação?

Por fim, abra espaço para reflexões éticas e estratégicas: quais são as responsabilidades de operadores e legisladores diante do risco de “harvest now, decrypt later”? Como equilibrar a inovação (teste de QKD, integração PQC) com a necessidade de disponibilidade e continuidade de serviços essenciais? E quais linhas de investigação (por exemplo, compressão de primitives para sistemas distribuídos, frameworks automáticos de rollback, provas formais que incluam encodings reais) você considera mais urgentes para transformar as propostas híbridas em soluções industriais maduras?

Estas questões destinam-se a instigar investigação ativa e pensamento crítico. Recomenda-se que o leitor responda a cada pergunta através de pequenos projetos experimentais, estudos de caso ou revisões bibliográficas organizadas, documentando hipóteses, procedimentos e métricas coletadas para gerar evidência acionável que subsidie decisões técnicas e de governança.

\chapter{Comparação entre Abordagens}

A transição para a era quântica exige que avaliemos as alternativas criptográficas por múltiplas dimensões simultaneamente: garantias formais de segurança, resistência a adversários com capacidades quânticas, desempenho em termos de latência e throughput, viabilidade de integração com infraestruturas existentes e maturidade técnica e regulatória. Em vez de tratar cada paradigma com afirmações breves, apresentamos aqui uma análise descritiva que procura explicitar os trade-offs e os contextos nos quais cada abordagem se mostra mais apropriada, mantendo a Tabela \ref{tab:comparacao_abordagens} e a Figura \ref{fig:radar_comparacao} como pontos de referência visual e tabular.

A criptografia clássica, representada por primitivas bem estabelecidas como RSA, ECC e AES, continua sendo a espinha dorsal das comunicações e dos serviços digitais. Sua grande virtude é a maturidade: existe vasta experiência operacional, suporte hardware (TPMs, HSMs) e padrões amplamente aceitos. Em termos de desempenho, as soluções clássicas oferecem baixa latência e pequeno overhead, o que as torna ideais para aplicações de alto tráfego. Contudo, esta vantagem operacional depende do pressuposto de um adversário clássico; diante de computadores quânticos universais capazes de executar o algoritmo de Shor, primitivas baseadas em fatoração e logaritmo discreto perdem sua confiabilidade, introduzindo o risco conhecido como "harvest now, decrypt later" para dados que devem permanecer confidenciais por longos períodos. Portanto, a decisão de manter, atenuar ou migrar primitivas clássicas deve considerar o horizonte temporal de proteção exigido para cada ativo.

A criptografia quântica, na forma de QKD e suas variantes (DV-QKD, CV-QKD, MDI, DI, TF), oferece uma promessa qualitativa diferente: segurança baseada em princípios físicos. Em cenários bem instrumentados, QKD possibilita estabelecer material secreto cuja confidencialidade não depende de pressupostos sobre a dificuldade de problemas computacionais. Essa propriedade torna QKD especialmente atrativa para enlaces de altíssima sensibilidade, entre eles ligações financeiras críticas, comunicações diplomáticas e infraestrutura nacional. Na prática, no entanto, a aplicabilidade do QKD é restringida por limitações físicas e econômicas: atenuação em fibras, necessidade de detectores e fontes especializados, sincronização apurada e a ausência ainda concreta de repetidores quânticos maduros para escalabilidade intercontinental. Assim, QKD tem sido implementada de forma seletiva em enlaces dedicados e não como substituto universal das pilhas criptográficas clássicas.

A criptografia pós-quântica (PQC) surge como alternativa pragmática para adoção ampla: algoritmos baseados em lattices, códigos, funções hash e outras construções foram desenvolvidos para resistir aos ataques conhecidos de máquinas quânticas. A principal vantagem da PQC é a compatibilidade operacional com protocolos existentes, atualizações de software, sem necessidade de reforma radical de infraestrutura física, possibilitam a migração em escala. No entanto, essa vantagem operacional vem acompanhada de trade-offs concretos: aumentos no tamanho de chaves, assinaturas e ciphertexts podem impactar largura de banda e a fragmentação em redes com MTU restrito; o custo de CPU para operações de estabelecimento de sessão também tende a subir, especialmente em dispositivos com recursos limitados. A padronização em curso (NIST e outras iniciativas) reduz a incerteza, mas a seleção de parâmetros e perfis ainda exige avaliações PoC específicas por workload.

As soluções híbridas, que combinam mecanismos clássicos, pós-quânticos e, quando aplicável, elementos quânticos como QKD, propõem redundância criptográfica e diversidade de caminhos de proteção. Quando bem projetadas, arquiteturas híbridas mitigam falhas de uma única camada e diminuem a probabilidade de exposição total dos segredos. Essa característica as torna recomendáveis em ambientes de missão crítica onde a confidencialidade de longo prazo justifica o aumento de complexidade e custo. Contudo, a redundância traz desafios de engenharia: encodings e domain separation precisam ser rigorosamente especificados e implementados para evitar fragilidades; políticas de fallback e tratamento de erros requerem harmonização entre componentes; e a orquestração operacional entre elementos clássicos e quânticos aumenta a superfície de erro humano e automatização necessária. Em particular, a composição correta de KEMs e a derivação de chaves via HKDF com labels e domain separation explícitos são pontos cruciais para evitar vulnerabilidades práticas.

A comparação sistemática das abordagens exige um enquadramento claro de modelos de ameaça. Deve-se distinguir um adversário clássico de um adversário pós-quântico (capaz de executar Shor) e considerar adversários que exploram canais laterais (timing, cache, EM) ou que possuem acesso físico ao equipamento. A interpretação das métricas na Tabela \ref{tab:comparacao_abordagens} depende diretamente desses modelos: por exemplo, a mesma solução pode ser considerada adequada para um adversário clássico e inadequada para um adversário pós-quântico. Consequentemente, a seleção de uma estratégia criptográfica deve ser conduzida a partir de uma avaliação de risco que inclua probabilidade de surgimento do adversário e horizonte temporal da necessidade de confidencialidade.

Do ponto de vista experimental e metodológico, recomenda-se que comparações entre paradigmas se apoiem em medições reprodutíveis: latência de handshake (média, p95 e p99), consumo de CPU por operação criptográfica, uso de memória, variação de throughput sob carga e tamanho das mensagens (impacto em MTU/fragmentação). Adicionalmente, avaliações de robustez submetendo as implementações a perda de pacotes, reconexões frequentes e MTU reduzidas, bem como testes de interoperabilidade cross-vendor, são essenciais. Em segurança prática, fuzzing do handshake, microbenchmarks de timing e análises de side-channel devem acompanhar quaisquer PoC de PQC ou híbridas, pois otimizações de assembly ou vetorização (AVX/NEON) podem introduzir vazamentos inesperados. Resultados que combinam evidências de desempenho, robustez e segurança prática oferecem maior utilidade para a adoção industrial do que benchmarks isolados.

Em nível de engenharia, a pressão por desempenho e a exigência de segurança levam a decisões de compromisso. Implementações otimizadas em assembly melhoram latência e throughput, mas exigem controles adicionais para mitigar vazamentos por canais laterais. Em dispositivos embarcados, builds compactos (por exemplo derivados de PQM4) possibilitam adoção de PQC, mas frequentemente restringem parâmetros de segurança ou impõem delegação de operações a módulos seguros. A integração com HSM/TPM é um elemento mitigador que permite proteger chaves privadas e segredos de fallback, suportar rotação coordenada de chaves e reduzir exposição durante operações críticas, mas ela implica custos de aquisição e de adaptação de workflows operacionais.

Do ponto de vista organizacional, a migração demanda governança dedicada: inventário criptográfico detalhado, classificação de ativos por sensibilidade e horizonte de proteção, definição de políticas de rotação e revogação, treinamento operacional para novas pilhas, e pipelines CI/CD que validem compatibilidade de certificados, algIDs e perfis X.509 híbridos. A adoção responsável inclui PoC em ambiente controlado, rollouts canary acompanhados por telemetria agregada e mecanismos automatizados de rollback. Esses elementos asseguram que mudanças criptográficas não comprometam disponibilidade e desempenho dos serviços críticos.

Estudos de caso práticos tornam tangíveis os trade-offs. Experimentos com handshakes TLS híbridos combinando X25519 e Kyber tipicamente mostram aumento do tamanho de handshake e maior custo de CPU por estabelecimento de sessão; entretanto, na presença de reuse de sessões e session tickets, o overhead amortiza-se em conexões long-lived. Pilotos que integraram QKD em backbone de baixa latência demonstraram forte resiliência teórica, mas também evidenciaram requisitos de manutenção, monitoramento e sincronização que restringem sua aplicabilidade a enlaces estratégicos. Essas evidências reforçam a necessidade de PoC que reproduzam cargas e topologias reais antes de decisões de produção em larga escala.

Apesar do avanço, persistem lacunas que direcionam agenda de pesquisa e desenvolvimento. Provas formais que incorporem detalhes práticos de encoding, domain separation e políticas de erro são ainda incompletas; metodologias padronizadas para avaliação de side-channel em esquemas PQC em hardware moderno carecem de consolidação; e soluções de compressão e agregação para reduzir o overhead em sistemas distribuídos (por exemplo blockchains) necessitam de maturação. A escalabilidade do QKD depende de desenvolvimentos em repetidores quânticos e memórias quânticas, enquanto a evolução normativa—incluindo definição de algIDs, perfis X.509 híbridos e vetores de teste padronizados—será crucial para migrações coordenadas entre fornecedores.

A Tabela \ref{tab:comparacao_abordagens} resume, de forma concisa, as características centrais já discutidas e serve como ponto de referência para decisões arquiteturais. O leitor é convidado a interpretar a tabela em conjunto com a Figura \ref{fig:radar_comparacao}, que sintetiza visualmente os trade-offs entre segurança, desempenho, viabilidade e maturidade das abordagens. Essas representações são indicativas: pontuações ou qualificações numéricas dependem de parâmetros experimentais e de projeto (por exemplo, perdas de fibra, eficiência de detectores, configurações de rede) e, portanto, devem ser ajustadas ao contexto de cada organização.

A interpretação prática desses elementos deve orientar políticas que priorizem proteção pragmática hoje e flexibilidade para evolução futura. Para serviços cuja confidencialidade é sensível apenas a curto prazo, a migração coordenada para PQC acompanhada de crypto agility é uma estratégia eficaz. Para dados com necessidade de confidencialidade por décadas, recomenda-se uma estratégia em camadas, combinando PQC com medidas operacionais robustas e, quando justificável, QKD em enlaces críticos. Em todos os cenários, a adoção bem-sucedida depende de experimentação controlada, instrumentação abrangente, governança ativa e participação em iniciativas de interoperabilidade que reduzam o risco de surpresas operacionais.

\begin{table}[h!]
\centering
\scriptsize
\caption{Comparação entre diferentes abordagens criptográficas em termos de segurança, desempenho, viabilidade e maturidade tecnológica.}
\label{tab:comparacao_abordagens}
\begin{tabular}{p{1.1cm}p{3.3cm}p{3.2cm}p{3.2cm}p{3.4cm}}
\hline
\textbf{Categoria} & \textbf{Segurança} & \textbf{Desempenho} & \textbf{Viabilidade} & \textbf{Maturidade} \\
\hline
Clássica & Média–alta contra ataques clássicos; não resiliente a Shor & Excelente, baixa latência e custo/benefício elevado & Altíssima, infraestrutura consolidada & Consolidada, décadas de uso e padronização \\
\hline
Quântica & Muito alta (em teoria incondicional) quando corretamente implementada & Limitada, perdas e taxa de chave dependem fortemente de distância & Restrita, exige hardware e link dedicados & Experimental/operacional em nichos, redes piloto e soluções comerciais pontuais \\
\hline
Pós- Quântica & Alta, projetada para resistir a ataques quânticos conhecidos & Moderada, overhead em tamanho de chaves/assinaturas e processamento & Alta, atualizações de software permitem implantação ampla & Em transição: candidatos padronizados, ampla adoção em curso \\
\hline
Híbrida & Muito alta, redundância multicamada reduz superfície de risco & Variável, overhead aumenta handshakes e operação & Mediana, exige orquestração entre camadas e interoperabilidade & Emergente, pilotos e estudos de caso, mas sem ampla padronização integrada \\
\hline
\end{tabular}
\end{table}

A interpretação prática desses elementos já descritos e a adoção de um roteiro iterativo de PoC, canary e rollout, com monitoramento contínuo e capacidade de rollback automatizada, constituem o caminho mais prudente para organizações que desejam se preparar para a era quântica sem comprometer disponibilidade e desempenho. A combinação de engenharia, governança e participação ativa em iniciativas de padronização e interoperabilidade será determinante para uma transição eficaz e segura.

\begin{figure}[h!]
\centering
\begin{tikzpicture}[scale=1.1,font=\sffamily]
    \def\R{3} 
    \coordinate (seg) at (90:\R);
    \coordinate (des) at (0:\R);
    \coordinate (via) at (-90:\R);
    \coordinate (mat) at (180:\R);

    \foreach \r in {1,2,3} {
        \draw[thin,gray!30] (0,0) circle (\r);
    }

    \draw[thick,gray!70] (0,0) -- (seg) node[above, yshift=2pt]{\textbf{Segurança}};
    \draw[thick,gray!70] (0,0) -- (des) node[right, xshift=2pt]{\textbf{Desempenho}};
    \draw[thick,gray!70] (0,0) -- (via) node[below, yshift=-2pt]{\textbf{Viabilidade}};
    \draw[thick,gray!70] (0,0) -- (mat) node[left, xshift=-2pt]{\textbf{Maturidade}};

    \coordinate (C1) at (90: {2});
    \coordinate (C2) at (0: 3);
    \coordinate (C3) at (-90: 3);
    \coordinate (C4) at (180: 3);
    \draw[line width=1pt, draw=blue!80!black, fill=blue!40, fill opacity=0.35]
        (C1) -- (C2) -- (C3) -- (C4) -- cycle;
    \foreach \p in {C1,C2,C3,C4} \draw[fill=blue!80!black] (\p) circle (0.06);

    \coordinate (Q1) at (90: 3);
    \coordinate (Q2) at (0: 1);
    \coordinate (Q3) at (-90: 1);
    \coordinate (Q4) at (180: 1);
    \draw[line width=1pt, draw=orange!85!black, fill=orange!40, fill opacity=0.35]
        (Q1) -- (Q2) -- (Q3) -- (Q4) -- cycle;
    \foreach \p in {Q1,Q2,Q3,Q4} \draw[fill=orange!85!black] (\p) circle (0.06);

    \coordinate (P1) at (90: 3);
    \coordinate (P2) at (0: 2);
    \coordinate (P3) at (-90: 3);
    \coordinate (P4) at (180: 2);
    \draw[line width=1pt, draw=green!50!black, fill=green!40, fill opacity=0.35]
        (P1) -- (P2) -- (P3) -- (P4) -- cycle;
    \foreach \p in {P1,P2,P3,P4} \draw[fill=green!50!black] (\p) circle (0.06);

    \coordinate (H1) at (90: 3);
    \coordinate (H2) at (0: 2);
    \coordinate (H3) at (-90: 2);
    \coordinate (H4) at (180: 1);
    \draw[line width=1pt, draw=violet!65!black, fill=violet!35, fill opacity=0.35]
        (H1) -- (H2) -- (H3) -- (H4) -- cycle;
    \foreach \p in {H1,H2,H3,H4} \draw[fill=violet!65!black] (\p) circle (0.06);

    \node[gray] at (1,0.25) {\scriptsize 1};
    \node[gray] at (2,0.25) {\scriptsize 2};
    \node[gray] at (3,0.25) {\scriptsize 3};

\begin{scope}[shift={(0,-4.0)}] 
\node[align=center] {
  \begin{tabular}{c c}
    \tikz \draw[fill=blue!40, draw=blue!80!black] (0,0) rectangle +(0.4,0.25); & \textbf{Clássica} \\
  \end{tabular}\hspace{1.2cm}
  \begin{tabular}{c c}
    \tikz \draw[fill=orange!40, draw=orange!85!black] (0,0) rectangle +(0.4,0.25); & \textbf{Quântica} \\
  \end{tabular}\hspace{1.2cm}
  \begin{tabular}{c c}
    \tikz \draw[fill=green!40, draw=green!50!black] (0,0) rectangle +(0.4,0.25); & \textbf{Pós-Quântica} \\
  \end{tabular}\hspace{1.2cm}
  \begin{tabular}{c c}
    \tikz \draw[fill=violet!35, draw=violet!65!black] (0,0) rectangle +(0.4,0.25); & \textbf{Híbrida} \\
  \end{tabular}
};
\end{scope}
\end{tikzpicture}
\caption{Comparação visual entre abordagens criptográficas}
\label{fig:radar_comparacao}
\end{figure}
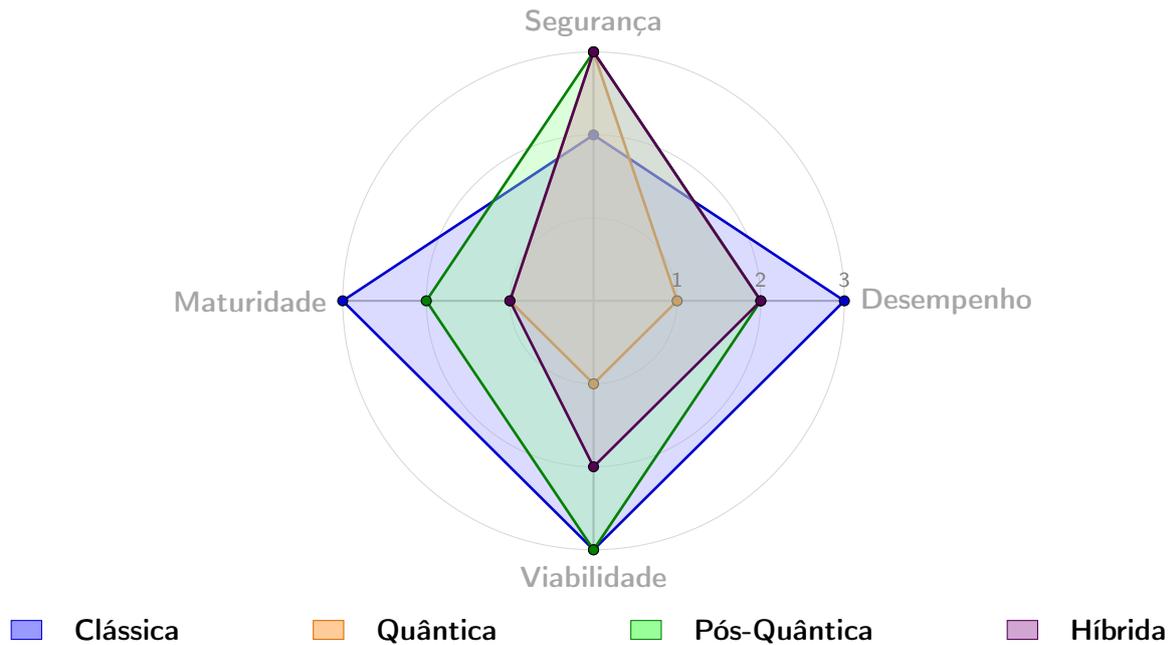

\section{Questões para Reflexão e Pesquisa Futura}

Ao concluir a comparação entre abordagens, proponho algumas perguntas curtas para estimular reflexão crítica e orientar pesquisas ou PoC rápidos:

\begin{itemize}
  \item Dado um ativo com horizonte de proteção definido, qual abordagem (clássica, PQC, QKD ou híbrida) oferece o melhor balanço entre risco residual e custo operacional nesse horizonte?
  \item Quais métricas (latência p95/p99 do handshake, custo CPU por estabelecimento, overhead de bytes, taxa de falha sob perda de pacotes) devem ter maior peso na decisão para cada classe de aplicação (IoT, data center, backbone)?
  \item Como projetar políticas de fallback e rollback que evitem downgrades inseguros em ambientes híbridos, mantendo interoperabilidade entre fornecedores?
  \item Quais barreiras de governança, padronização e cadeia de suprimentos são mais críticas para a adoção em larga escala, e que medidas práticas (conformance tests, perfis X.509 híbridos, auditoria de fornecedores) mitigam esses riscos?
  \item Considerando lacunas de pesquisa e custos, quais PoCs de curto prazo (ex.: handshake TLS híbrido, QKD integrado a PKI local, desempenho de PQC em MCUs) oferecem maior retorno de aprendizado para sua organização?
\end{itemize}

\chapter{Tendências Futuras}

As constantes ameaças advindas da computação quântica e da evolução de ataques cibernéticos indicam que a criptografia seguirá uma trajetória marcada pela inovação técnica e pela convergência de múltiplas abordagens. As tendências futuras revelam não apenas caminhos de pesquisa, mas também potenciais aplicações práticas que moldarão os próximos padrões de segurança.

A expansão das redes quânticas representa uma das transformações mais significativas no horizonte criptográfico. A distribuição quântica de chaves (QKD), que até recentemente era restrita a laboratórios e pequenos experimentos acadêmicos, caminha para cenários de larga escala. A criação de quantum backbones já foi demonstrada em alguns países, e projetos de conectividade intercontinental, combinando enlaces por fibra e por satélite, são frequentemente mencionados como metas de médio prazo. O amadurecimento de repetidores quânticos e memórias quânticas será decisivo para que o QKD deixe de ser uma solução de nicho e passe a integrar arquiteturas federadas de comunicação segura. Ainda assim, na prática, o papel esperado do QKD tende a ser especializado: enlaces de altíssima criticidade (backbones diplomáticos, financeiros, infraestruturas críticas) onde o custo e os requisitos logísticos podem ser justificados pela necessidade de sigilo de longo prazo.

Uma tendência complementar é a convergência entre soluções matemáticas e físicas por meio da integração PQC + QKD. Nesta visão, o QKD fornece material de chave de alta entropia e propriedades físicas de sigilo para enlaces selecionados, enquanto algoritmos pós-quânticos garantem autenticação, verificação de integridade e interoperabilidade ampla. Essa composição multicamada aumenta a resiliência global do sistema ao introduzir redundância de mecanismos e caminhos de verificação. Em termos práticos, arquiteturas híbridas aumentam as exigências de engenharia, exigem definições claras de encodings, domain separation, políticas de fallback e procedimentos de sincronização entre planos clássico e quântico, mas trazem benefícios relevantes em cenários onde a confidencialidade e a disponibilidade não admitem riscos significativos.

Em paralelo, a incorporação de componentes de hardware quântico acessível, como QRNGs (Quantum Random Number Generators) embutidos em chips e módulos, deve democratizar itens essenciais de segurança. QRNGs oferecem entropia de elevada qualidade e podem reduzir vulnerabilidades decorrentes de fontes pseudoaleatórias fracas. A presença de QRNGs em dispositivos móveis e IoT terá impacto direto na robustez de chaves geradas localmente, em particular quando combinada com PQC otimizada para dispositivos embarcados. Além disso, a evolução de módulos seguros que suportem operações PQC nativamente, seja em HSMs comerciais, seja em enclaves de hardware, facilitará a transição em ambientes corporativos, onde a proteção de chaves e procedimentos de rotação são requisitos operacionais.

A tendência regulatória e de padronização acompanha o desenvolvimento técnico. Espera-se aceleração em processos de homologação e certificação de esquemas PQC por organismos como NIST e ISO, bem como esforços para definir perfis X.509 que acomodem algoritmos híbridos. Políticas públicas que considerem o risco "harvest now, decrypt later" poderão induzir requisitos mínimos de proteção para setores regulados, pressionando provedores a adotar mecanismos pós-quânticos ou híbridos em prazos definidos. Nesse contexto, a adoção de princípios de crypto agility tornar-se-á mandatório: a capacidade de substituir ou combin ar algoritmos com baixo custo operacional será um requisito para responder rapidamente a descobertas criptoanalíticas ou a mudanças regulatórias.

Setores estratégicos como energia, transporte aéreo, serviços financeiros e saúde tendem a ser pioneiros na adoção de esquemas híbridos. Nestes ambientes, a redundância de mecanismos (AES com chaves geradas por QKD, autenticação por PQC, armazenamento cifrado com rotação periódica) aumenta a resistência a falhas combinadas. No entanto, a adoção em larga escala também dependerá de fatores econômicos: custo de implantação, custo de operação e o impacto sobre latência e capacidade. Por isso, é provável que vejamos um ecossistema diversificado, em que diferentes setores e organizações escolham perfis distintos conforme seu apetite de risco, restrições econômicas e requisitos regulamentares.

Do ponto de vista da pesquisa aplicada, surgem frentes bem definidas. A primeira é a otimização de PQC para recursos limitados, incluindo compressão de chaves e assinaturas, KEMs com footprint reduzido e implementações resistentes a canais laterais em arquiteturas vetoriais. A segunda envolve a integração prática de QKD e sua coexistência com redes clássicas, abordando problemas de sincronização, multiplexação e monitoramento em tempo real do estado do enlace quântico. A terceira refere-se a provas formais e modelos composicionais que incorporem encodings, domain separation e políticas de erro praticadas por implementações reais, reduzindo a lacuna entre teoria e prática. Por fim, há espaço para inovações em orquestração e automação: frameworks que habilitem rollout/rollback seguros de perfis criptográficos, validação contínua e telemetria específica para detecção precoce de regressões de segurança são áreas de impacto direto para operações.

\section{Desafios e Oportunidades para Pesquisas Futuras}

Apesar dos avanços promissores, a transição para um ecossistema criptográfico pós-quântico e híbrido não está isenta de obstáculos científicos, técnicos e regulatórios. Esses desafios não apenas definem o ritmo da adoção, como também abrem vastas oportunidades de investigação para a comunidade acadêmica e industrial. A seguir expandimos e detalhamos esses pontos, propondo direções concretas de pesquisa, experimentação e governança que podem transformar lacunas atuais em capacidades robustas.

A escalabilidade das infraestruturas quânticas permanece como um entrave técnico de primeiro plano. Além da necessidade de enlaces dedicados e de detectores sensíveis, há um conjunto de problemas práticos que dificultam a integração de QKD em topologias de backbone: multiplexação com tráfego clássico, tolerância a perdas variáveis, gerenciamento de chaves distribuídas entre múltiplos domínios administrativos e a operação em ambientes com interferências ambientais. Pesquisas aplicadas devem explorar arquiteturas híbridas de encaminhamento e agregação de chave quântica que permitam degradar graciosamente para mecanismos clássicos ou PQC quando o enlace quântico estiver degradado, sem perda de garantia de autenticação. Experimentos em campo, com monitoramento de parâmetros físicos do enlace e correlacionamento com métricas de aplicação (por exemplo, latência de estabelecimentos de sessão e taxa de erro de aplicação), são necessários para mapear limites operacionais e definir políticas automatizadas de fallback que preservem propriedades de segurança essenciais.

A eficiência da criptografia pós-quântica em dispositivos restritos constitui uma lacuna técnica ampla e multifacetada. Necessitamos tanto de avanços algorítmicos quanto de engenharia de implementação: novos KEMs e esquemas de assinatura com footprint reduzido, compilações e toolchains que produzam código constante em tempo e resistente a vazamentos, e bibliotecas de referência que ofereçam trade-offs configuráveis entre segurança, memória e latência. Estudos experimentais devem incluir medições detalhadas de consumo energético, tempos de execução em diferentes classes de processadores (MCU, ARM Cortex, SoC móveis) e impacto em ciclos de vida de bateria em dispositivos IoT. Além disso, é imperativo desenvolver métodos formais e práticos de verificação para garantir que otimizações voltadas ao desempenho não introduzam vulnerabilidades por canais laterais; isso inclui a criação de suites de testes normalizadas para timing, cache e emissões eletromagnéticas aplicadas a implementações PQC.

As vulnerabilidades por canais laterais e a interação destas com otimizações de hardware representam uma área crítica de pesquisa. Enquanto vetorizações e instruções SIMDw aceleram rotinas PQC, elas também mudam o perfil físico de execução. Pesquisa experimental deve quantificar como diferentes níveis de otimização (compilador, assembly hand-tuned, uso de AVX/NEON) afetam as superfícies de vazamento, e deve propor contramedidas viáveis para ambientes de produção. Além disso, há espaço para desenvolver metodologias automatizadas de avaliação de canais laterais que possam ser integradas em pipelines CI/CD, de modo que cada alteração de build seja submetida a verificações de regressão de segurança física antes de chegar à produção.

A interoperabilidade entre paradigmas e implementações constitui outra lacuna que tem impacto direto em adoções em escala. Questões práticas, como formatos de mensagem, encodings canônicos, domain separation em KDFs e protocolos de fallback, exigem padrões mínimos e vetores de conformance testados cross-vendor. Pesquisas nesta área devem produzir especificações detalhadas acompanhadas de testbeds reproduzíveis e conjuntos de vetores de teste (test vectors) públicos, de forma que CAs, fornecedores de HSM, stacks TLS e implementações PQC possam validar compatibilidade de ponta a ponta. Trabalhos aplicados que descrevam processos de versionamento de algIDs, migração de perfis X.509 e interoperabilidade com infraestruturas legadas são particularmente valiosos para reduzir o risco operacional durante rollouts.

A lacuna entre provas formais e implementações práticas precisa ser urgentemente reduzida. Embora existam resultados teóricos de composição de KEMs e provas de segurança no modelo abstrato, faltam provas composicionais que incorporem detalhes reais: codificações específicas, padding, tratamentos de erro, e políticas de tempo de resposta que influenciam a superfície de ataques por oráculo. Pesquisas teóricas devem focar em estender modelos formais para cobrir esses aspectos práticos e em desenvolver ferramentas que extraiam modelos automaticamente a partir de implementações reais, permitindo verificações formais que sejam aplicáveis no mundo industrial.

Outro vetor de pesquisa consiste em modelar adversários compostos e novo s cenários de ataque. Em particular, é preciso caracterizar adversários temporais que realizem estratégias de “harvest now, decrypt later”, combinadas com ataques por canais laterais, engenharia social e exploração de falhas operacionais. Além das análises criminais puramente matemáticas, estudos empíricos que simulem vetores de ataque compostos em ambientes de teste, envolvendo instrumentação física, análise de logs e correlação de telemetria, proporcionarão uma visão realista da robustez das arquiteturas híbridas e PQC em cenários de ameaça sofisticados.

No campo experimental, é indispensável a criação de infraestruturas de teste e benchmarks que promovam reprodutibilidade. Isso inclui testbeds de rede que incorporem enlaces quânticos simulados ou reais, harnesses para executar experiências de cargas realistas e conjuntos de métricas padronizadas (latência p50/p95/p99, custo CPU por handshake, uso de memória máxima, throughput por segundo e taxa de falha sob perda de pacotes). Projetos que publicarem datasets de experimentos e scripts de automação para reprodução (instruções para infraestrutura como código, containers e pipelines CI) terão impacto multiplicador na velocidade de adoção e na confiança das avaliações.

A governança, padronização e regulação representam desafios não apenas técnicos, mas também políticos e econômicos. A falta de alinhamento internacional sobre requisitos mínimos, certificações e políticas de exportação de tecnologia quântica pode gerar fragmentação. Há espaço para trabalhos interdisciplinares que proponham modelos de governança colaborativa, regimes de certificação internacional e frameworks de compliance que conciliem segurança, privacidade e interoperabilidade comercial. 

A segurança da cadeia de suprimentos tecnológica é outro ponto crítico. A integração de módulos PQC em HSMs, stacks TLS e dispositivos finais depende de componentes de terceiros; portanto, estudos sobre auditoria de fornecedores, certificação de firmware e mecanismos de verificação de integridade de cadeia (supply chain attestation) são necessários para evitar que uma migração aparente para PQC gere novas superfícies de risco. Pesquisas que proponham padrões para auditoria e transparência de fornecedores contribuirão para reduzir o risco sistêmico.

Finalmente, há um desafio cultural e operacional na adoção gradual: organizações precisam adaptar processos, formar equipes e manter continuidade de serviços durante rollouts. Investigar práticas organizacionais, modelos de treinamento, playbooks de resposta a incidentes criptográficos e métricas de prontidão operacional é tão importante quanto os avanços técnicos. Projetos pilotos que documentem, em formato reprodutível, todo o ciclo de migração, inventário, PoC, canary, rollout, rollback e revisão pós-implementação, serão recursos valiosos para a comunidade.

Como síntese prática, propõe-se que a pesquisa priorize, a curto prazo, a criação de suites de testes e testbeds reproduzíveis, medidas sistemáticas de side-channel para implementações PQC otimizadas e especificações de interoperabilidade básicas. A médio prazo, devem ser priorizados trabalhos sobre compressão e agregação de primitives para aplicações distribuídas, frameworks de orquestração de rollout e modelos regulatórios harmonizados. A longo prazo, investimentos concertados em repetidores quânticos, memórias e em infraestrutura física para QKD, bem como em provas formais composicionais que cobrem implementações reais, garantirão que as promessas teóricas se traduzam em segurança prática e escalável. Essas direções exigem colaboração estreita entre pesquisadores, indústria, operadores de telecom e órgãos reguladores, além de financiamento estável que combine recursos públicos e privados para acelerar resultados com impacto direto em produção.

\section{Casos de uso}
Nesta seção são apresentados estudos de caso práticos publicados em conferências e revistas internacionais, ilustrando como os conceitos discutidos ao longo deste livro se materializam em soluções reais de segurança em ambientes clássicos, híbridos e quânticos. Em vez de tratar apenas de algoritmos em abstrato, estes casos de uso mostram como decisões de projeto, limitações de infraestrutura e requisitos de negócio influenciam a adoção de primitivas criptográficas modernas.

Um primeiro exemplo é o modelo Prisec \cite{Prisec2019}, e Prisec II \cite{Prisec2025}, que propõe uma arquitetura abrangente de segurança para Internet das Coisas (IoT). O trabalho combina autenticação forte, monitoramento contextual e mecanismos de atualização segura, integrando primitivas criptográficas clássicas e técnicas recentes de endurecimento de dispositivos. A contribuição central está em oferecer um modelo de referência que conecta requisitos funcionais de IoT (baixa potência, conectividade intermitente, heterogeneidade de hardware) com decisões de criptografia, gestão de chaves e políticas de acesso. Por outro lado, o estudo também evidencia limitações práticas, como a dificuldade de padronizar políticas de segurança em ecossistemas com fabricantes distintos e a necessidade de balancear custo, consumo energético e latência em dispositivos extremamente restritos.

Na sequência, o trabalho PRISEC III \cite{Prisec3} aprofunda este modelo, com foco explícito em técnicas criptográficas para aumento de segurança. A proposta discute o uso combinado de algoritmos simétricos eficientes, protocolos de troca de chaves e mecanismos de autenticação adequados a redes distribuídas, aproximando o modelo conceitual de Prisec de uma implementação mais concreta. O estudo demonstra, por meio de experimentos, como diferentes combinações de primitivas impactam desempenho, consumo de energia e robustez contra ataques em cenários de IoT em larga escala. 

No contexto da criptografia quântica, o protocolo DN25 \cite{DN25} representa um caso de uso diretamente alinhado ao foco desta obra, ao propor um protocolo adaptativo de criptografia quântica para comunicação segura e eficiente. O DN25 explora protocolos de Distribuição Quântica de Chaves (QKD) em combinação com canais clássicos, adaptando parâmetros operacionais de acordo com condições de canal, taxas de erro e requisitos de qualidade de serviço. O estudo mostra, em cenários simulados e/ou experimentais, como ajustes dinâmicos de parâmetros (por exemplo, escolha de bases, taxas de repetição, janelas de reconciliação) permitem manter taxas de chave secretas úteis mesmo diante de variações de ruído e perdas. Entre as limitações, o trabalho aponta a complexidade de implantação de infraestruturas quânticas em ambientes reais, o custo dos componentes ópticos e a necessidade de integração cuidadosa com protocolos clássicos de autenticação e gestão de chaves.

Outro grupo importante de casos de uso está relacionado à migração para criptografia pós-quântica (PQC) em infraestruturas amplamente utilizadas, como TLS, VPNs e redes móveis. Estudos recentes, como o de Xagawa et al. \cite{Xagawa2024}, mostram a adoção de esquemas baseados em lattices e hash (Kyber, Dilithium, SPHINCS+) em cenários de comunicação de alta escala, propondo perfis híbridos que combinam algoritmos clássicos (por exemplo, X25519, RSA) com KEMs e assinaturas pós-quânticas. Esses trabalhos quantificam o impacto em latência, tamanho de handshake, consumo de banda e carga computacional em servidores e clientes, evidenciando que, embora a migração seja viável, há aumento de overhead e necessidade de ajuste fino de parâmetros e políticas de rotação de chaves. As limitações apontadas incluem o custo de atualização de pilhas de software em larga escala e a coexistência de múltiplos perfis criptográficos durante o período de transição.

Ainda na linha de PQC, aplicações específicas em blockchain e contratos inteligentes vêm sendo exploradas. Wang et al. \cite{Wang2024} estudam o uso de Kyber e Dilithium em plataformas de blockchain, avaliando o impacto de chaves e assinaturas maiores em métricas como throughput, tamanho de bloco e custo de transação. O caso de uso mostra que a substituição de ECDSA por esquemas pós-quânticos é tecnicamente possível, mas traz desafios de escalabilidade e armazenamento, além de exigir mudanças em carteiras, nós validadores e mecanismos de consenso. O estudo destaca que, em redes públicas, qualquer aumento de tamanho de assinatura ou chaves se traduz diretamente em maior custo econômico, o que reforça a necessidade de desenho cuidadoso de incentivos e de estratégias graduais de migração.

Casos de uso também têm emergido na integração de criptografia homomórfica (CH) em cenários de análise de dados sensíveis. Trabalhos recentes em ambientes de saúde e finanças \cite{HECaseStudy2023} mostram como esquemas como BFV e CKKS podem ser utilizados para realizar agregações estatísticas e inferência de modelos de aprendizado de máquina diretamente sobre dados cifrados. Em um cenário típico, hospitais ou instituições financeiras enviam dados criptografados para um provedor de nuvem, que executa cálculos sem acesso aos textos claros, devolvendo apenas resultados cifrados para posterior decifragem local. Esses estudos demonstram a viabilidade de cálculos como regressão linear, classificação binária e análise de risco sob sigilo, mas apontam limitações relevantes de desempenho (latências mais altas, consumo de memória) e a necessidade de especialização de modelos para reduzir profundidade de circuito e custo de bootstrapping.

Na fronteira entre criptografia quântica e redes práticas, diversos projetos de redes QKD metropolitanas implementadas em fibras ópticas comerciais podem ser vistos como casos de uso de referência \cite{China2021,Fang2023}. Essas redes interligam nós de organizações governamentais, financeiras e de pesquisa, utilizando protocolos como BB84 com decoy states, MDI-QKD e TF-QKD, e integram os enlaces quânticos a infraestruturas clássicas já existentes (por exemplo, redes IP/MPLS). Os resultados mostram chaves secretas sendo geradas continuamente para alimentar sistemas de criptografia simétrica (como AES-GCM) em camadas superiores, ilustrando o conceito de soluções híbridas: QKD fornece chaves com segurança baseada em física, enquanto algoritmos clássicos protegem o tráfego de dados. Entre as limitações, destacam-se a distância máxima em fibras sem repetidores quânticos, a necessidade de alinhamento e estabilização contínua, e os custos ainda elevados de equipamentos.

Por fim, casos de uso em nuvem e ambientes de computação distribuída têm explorado combinações de técnicas heterogêneas, como PQC, QKD, criptografia homomórfica e enclaves de execução confiável (TEE) \cite{Selvi2025Hybrid}. Em arquiteturas desse tipo, QKD pode ser utilizado para o bootstrap de chaves de longa duração entre data centers, PQC protege canais e armazenamento contra adversários quânticos futuros, e CH permite processamento de dados sensíveis em provedores terceirizados. Os estudos mostram que essa abordagem multicamadas aumenta significativamente a resiliência contra diferentes modelos de ataque (clássico, quântico e operacional), mas introduz complexidade de gestão de chaves, necessidade de padronização de interfaces e desafios de monitoramento e auditoria de segurança.

Em conjunto, esses casos de uso evidenciam que a transição para a era da computação quântica não é apenas um problema de troca de algoritmos, mas um processo de engenharia de sistemas. Modelos de segurança para IoT, como Prisec II e PRISEC III, mostram a importância de arquiteturas criptográficas pensadas de ponta a ponta; propostas como DN25 demonstram como protocolos quânticos podem ser integrados de forma adaptativa a infraestruturas existentes; e estudos em PQC, criptografia homomórfica, redes QKD e nuvem híbrida ilustram o caminho de migração gradual e multicamadas que tende a predominar na prática. Esses exemplos reforçam a mensagem central do livro: a adoção de algoritmos criptográficos para a era quântica exige não só rigor matemático, mas também uma compreensão profunda de cenários de uso, requisitos operacionais e governança criptográfica.

\section{Roteiro Prático, Recomendações e Tendências de Pesquisa}

Para organizações que planejam preparar-se para o futuro quântico, recomenda-se uma estratégia iterativa, instrumentada e alinhada com as melhores práticas de governança e inovação tecnológica. O processo deve iniciar com um inventário criptográfico detalhado, classificando ativos segundo sua sensibilidade e horizonte temporal de proteção exigido. Essa etapa é fundamental para priorizar esforços e alocar recursos de forma eficiente, evitando migrações desnecessárias ou prematuras.

Em seguida, devem ser conduzidas provas de conceito (PoC) rigorosas, que reproduzam cargas e topologias reais, permitindo avaliar impactos em desempenho, fragmentação de pacotes, interoperabilidade entre stacks e robustez sob condições adversas de rede. A instrumentação dessas PoCs deve incluir coleta de métricas detalhadas, como latência de handshake, uso de CPU e memória, além de monitoramento de falhas e regressões de segurança.

Rollouts canary controlados, com telemetria agregada e alertas automáticos, são recomendados para escalonar mudanças de forma segura, minimizando riscos operacionais. Paralelamente, o investimento em crypto agility, por meio de camadas de abstração, suporte a múltiplos perfis criptográficos e pipelines CI/CD que validem compatibilidade e segurança, reduz custos e riscos associados a futuras migrações ou atualizações emergenciais.

Para ativos que demandam confidencialidade por décadas, a adoção de arquiteturas híbridas que combinem criptografia pós-quântica (PQC), módulos de segurança hardware (HSMs) e, quando justificável, QKD em enlaces estratégicos, representa uma escolha prudente e alinhada com as melhores práticas internacionais. A participação ativa em iniciativas de interoperabilidade, bem como a contribuição para vetores de teste e bibliotecas de referência, aceleram a maturação do ecossistema e minimizam surpresas durante rollouts em produção.

Complementarmente, a agenda de pesquisa prioritária deve focar em áreas que viabilizem e fortaleçam essa transição, incluindo:

\begin{itemize}
    \item Desenvolvimento de provas formais composicionais que incorporem detalhes práticos como encodings, domain separation e políticas de erro adotadas por implementações reais, reduzindo a lacuna entre teoria e prática.
    \item Estudos aprofundados de mitigação de canais laterais específicos para PQC em hardware moderno, incluindo vetorização e uso de aceleradores, para garantir que otimizações de desempenho não comprometam a segurança.
    \item Otimizações algorítmicas e implementacionais para microcontroladores (MCUs) e dispositivos IoT, visando reduzir footprint, consumo energético e latência, ampliando o alcance da proteção pós-quântica.
    \item Arquiteturas e protocolos para integração eficiente entre QKD e redes clássicas, com mecanismos robustos de gestão de degradação, fallback seguro e orquestração automatizada.
    \item Desenvolvimento de frameworks automatizados para rollout, rollback e validação contínua de perfis híbridos, incluindo ferramentas de monitoramento e telemetria específicas para ambientes de produção.
    \item Investigação interdisciplinar sobre modelos regulatórios, governança global e políticas públicas que facilitem adoções coordenadas, promovendo interoperabilidade e mitigando riscos geopolíticos.
\end{itemize}

Esses eixos, se perseguidos de forma colaborativa entre academia, indústria e órgãos reguladores, reduzirão significativamente o custo, o risco e o tempo das transições planejadas, garantindo uma adoção segura e eficiente das tecnologias pós-quânticas e híbridas.

\begin{table}[h!]
\centering
\scriptsize
\caption{Roteiro Prático e Agenda de Pesquisa Prioritária para a Transição Criptográfica Pós-Quântica e Híbrida}
\label{tab:roteiro_pesquisa}
\begin{tabular}{p{5cm} p{9cm}}
\hline
\textbf{Roteiro Prático e Recomendações} & \textbf{Agenda de Pesquisa Prioritária} \\
\hline
Inventário detalhado de ativos e classificação por sensibilidade e horizonte de proteção & Provas formais composicionais incorporando encodings e políticas de erro reais \\
\hline
Execução de PoCs instrumentados com cargas e topologias reais para avaliação de desempenho e interoperabilidade & Estudos de mitigação de canais laterais para PQC em hardware moderno \\
\hline
Rollouts canary controlados com telemetria agregada e alertas de regressão & Otimizações algorítmicas e implementacionais para MCUs e dispositivos IoT \\
\hline
Investimento em crypto agility: camadas de abstração, suporte a múltiplos perfis e pipelines CI/CD & Arquiteturas para integração QKD + redes clássicas com fallback seguro \\
\hline
Adoção de arquiteturas híbridas para ativos com exigência de confidencialidade de longo prazo & Frameworks automatizados para rollout, rollback e validação contínua de perfis híbridos \\
\hline
Participação em iniciativas de interoperabilidade e contribuição para vetores de teste e bibliotecas de referência & Investigação interdisciplinar sobre modelos regulatórios e governança global \\
\hline
\end{tabular}
\end{table}

\chapter{Conclusão}

A trajetória da criptografia, desde os sistemas clássicos até as propostas emergentes do cenário quântico, evidencia uma evolução contínua e multidimensional que combina avanços teóricos, inovações de engenharia e pressões regulatórias. Algoritmos clássicos como RSA, ECC e AES construíram durante décadas a infraestrutura de confiança sobre a qual repousam a privacidade, a integridade e a autenticação nas comunicações digitais; entretanto, a consolidação dessa infraestrutura não garante sua imutabilidade: o advento previsível de computadores quânticos universais e o aprimoramento constante de técnicas de ataque impõem a necessidade de uma revisão estratégica abrangente. Assim, a criptografia deixa de ser apenas um conjunto de ferramentas matemáticas para tornar-se um projeto sistêmico que envolve arquitetura, operação, governança e economia de risco.

A análise comparativa entre paradigmas, clássico, pós-quântico, quântico e híbrido, mostra que cada um oferece propriedades únicas, vantagens operacionais e limitações práticas. A criptografia pós-quântica (PQC) aparece hoje como a solução com maior viabilidade de adoção em larga escala: algoritmos baseados em reticulados, códigos ou funções hash fornecem resistência conhecida contra ataques quânticos e permitem, em muitos casos, atualizações via software que não exigem revoluções físicas na infraestrutura. Ainda assim, a substituição de primitivas exige trabalho meticuloso de integração: aumento de tamanhos de chaves e assinaturas impacta overhead de rede e fragmentação; custos de processamento podem penalizar dispositivos de borda; e a compatibilidade com HSMs, certificados e perfis X.509 demanda coordenação entre fornecedores e operadores.

A criptografia quântica, em particular a QKD, traz um paradigma distinto: garantias de segurança que se assinam a princípios físicos e não à dificuldade de problemas computacionais. Experiências demonstraram a aplicabilidade em enlaces dedicados e backbones regionais, e avanços em satélites e repetidores quânticos prometem ampliar essa cobertura. No entanto, a QKD enfrenta obstáculos de escala e custo que restringem sua aplicação a nichos de alta criticidade. A decisão de empregar QKD deve considerar métricas concretas, taxa líquida de bits seguros, latência de provisão, custo de manutenção e requisitos de sincronização, e ser tomada com base em análises custo-benefício que incluam alternativas híbridas e medidas operacionais complementares.

As abordagens híbridas, que combinam elementos clássicos, PQC e, quando justificável, QKD, representam a resposta mais robusta ao caráter incerto do futuro. Redundância e diversidade criptográfica reduzem risco sistêmico e mitigam falhas de um único paradigma; por outro lado, aumentam substancialmente a complexidade de projeto, a necessidade de domain separation explícito, a urgência de políticas de fallback e a demanda por orquestração automatizada. Portanto, a adoção de arquiteturas híbridas só é efetiva quando acompanhada de especificações rigorosas de composição, suites de conformance testing cross-vendor e processos de operação que minimizem erro humano e permita rollback seguro.

Um elemento transverso e decisivo é a adoção de crypto agility como princípio de engenharia: a capacidade de trocar, combinar e atualizar primitivas com mínimo impacto operacional. Crypto agility exige camadas de abstração, interfaces claras entre componentes, perfis criptográficos configuráveis e pipelines de integração contínua que incluam testes funcionais, de performance e de segurança (incluindo verificações de side-channel). Implementar essa agilidade significa também definir políticas de governança para escolha de perfis, cadência de rotação de chaves, critérios de descontinuação de algoritmos e mecanismos de auditoria replicáveis que sustentem conformidade regulamentar e confiança institucional.

A operacionalização da transição para um ecossistema resistente ao risco quântico passa pela elaboração de um roteiro prático e pela institucionalização de capacidades técnicas e organizacionais. Esse roteiro inclui, entre outras atividades, inventário e classificação de ativos por sensibilidade e horizonte de proteção, execução de PoCs representativos, rollouts canary com telemetria coerente, integração com HSM/TPM onde aplicável, e automação de rollback. Métricas mensuráveis, latência p50/p95/p99 do handshake, custo de CPU por estabelecimento, overhead em bytes, consumo energético em endpoint e taxa de falha sob perda de pacotes, devem orientar decisões operacionais e escolhas de priorização. A documentação rigorosa de hipóteses, parametrizações e resultados empíricos é crítica para permitir avaliações repetíveis e decisões informadas.

Do ponto de vista de pesquisa e desenvolvimento, há prioridades claras que precisam ser perseguidas de modo coordenado: compressão de chaves e assinaturas para reduzir impacto em redes e dispositivos de borda; otimizações e implementações seguras para MCUs e SoCs de IoT; mitigação de canais laterais em builds vetorizados e em aceleradores dedicados; provas formais composicionais que incorporem encodings e políticas de erro praticadas em implementações reais; e frameworks automatizados de rollout/rollback e validação contínua para perfis híbridos. Esses esforços técnicos devem ser acompanhados por iniciativas práticas, testbeds reprodutíveis, conjuntos públicos de vetores de teste, benchmarks contextualizados e estudos de TCO, que permitam comparar alternativas sob cenários realistas.

A governança e a padronização são dimensões estratégicas que influenciam diretamente a velocidade e a equidade da transição. É necessário que reguladores, padrões e atores da indústria alinhem horizontes mínimos de proteção para setores sensíveis, definam critérios de certificação para componentes (HSMs, módulos QKD, bibliotecas PQC) e promovam regimes de cooperação internacional que evitem fragmentação normativa. Modelos de negócio sustentáveis, como serviços gerenciados, consórcios setoriais e infraestruturas compartilhadas, serão fundamentais para viabilizar adoção em organizações com recursos limitados, reduzindo barreiras de entrada e alinhando incentivos econômicos.

As dimensões éticas e sociais desta transição não podem ser negligenciadas. O fenômeno “harvest now, decrypt later” impõe responsabilidades sobre quem coleta, armazena e transmite dados sensíveis hoje; políticas de retenção, anonimização e exigências legais de proteção devem ser reavaliadas à luz da capacidade futura de descriptografia. Paralelamente, investimentos em capacitação e alfabetização técnica são necessários para que operadores, auditores e gestores compreendam riscos, limitem falhas operacionais e tomem decisões informadas. A transparência sobre critérios de segurança, a publicação de resultados de PoCs e a cooperação interoperável entre atores públicos e privados fomentam confiança e reduzem a assimetria informacional que pode retardar a adoção.

No plano estratégico, uma transição bem-sucedida requer coordenação entre pesquisa, indústria, operadores de rede e formadores de políticas. Laboratórios de validação, plataformas de certificação e centros de excelência podem acelerar a difusão de práticas seguras e criar recursos compartilhados de auditoria e homologação. Ao mesmo tempo, mecanismos de financiamento público-privado, incentivos regulatórios e programas de capacitação podem ser estruturados para mitigar o risco de concentração tecnológica e favorecer a interoperabilidade entre fornecedores, reduzindo a dependência de soluções proprietárias ou de cadeia de suprimentos restritas.

Em termos de cronograma pragmático, recomenda-se uma estratégia por fases: curto prazo (1–3 anos) para inventário, PoC e adoção de crypto agility; médio prazo (3–7 anos) para rollouts controlados de PQC em serviços críticos e maturação de ferramentas de orquestração e testes; longo prazo (7–15 anos) para avaliação de integração de QKD em enlaces estratégicos, evolução normativa e consolidação de práticas e padrões internacionais. Essas janelas temporais devem ser encaradas como orientativas, sujeitas a revisões conforme avanços tecnológicos, descobertas criptoanalíticas ou mudanças geopolíticas.

Como lição final, é importante reiterar que a segurança criptográfica é tanto técnica quanto institucional. As soluções mais avançadas em laboratório são inúteis se não forem implementáveis, auditáveis e sustentáveis em operações reais. Por isso, a resposta à ameaça quântica deve combinar pesquisa de ponta com engenharia prática, governança responsável e políticas que promovam interoperabilidade e compartilhamento de conhecimento. Somente através de um esforço coletivo, interdisciplinar e bem governado será possível construir um ecossistema que preserve não só o sigilo dos dados, mas também a confiança digital necessária para sustentar a economia e as instituições na próxima geração tecnológica.

\clearpage
\renewcommand{\glossaryname}{Glossário}
\addcontentsline{toc}{chapter}{Glossário}
\printglossaries

\end{document}